\documentclass[12pt,preprint]{aastex}

\usepackage{lscape}
\usepackage{wasysym}
\usepackage{natbib}
\usepackage{longtable}

\shortauthors{Fischer et al.}
\shorttitle{AGN Inclinations}

\begin{document}

\title{Determining Inclinations of Active Galactic Nuclei Via Their Narrow-Line Region Kinematics - I. Observational Results\altaffilmark{1}}

\author{T.C. Fischer\altaffilmark{2},
D.M. Crenshaw\altaffilmark{2},
S.B. Kraemer\altaffilmark{3},
H.R. Schmitt\altaffilmark{4}}

\altaffiltext{1}{Based on observations made with the NASA/ESA Hubble Space 
Telescope, obtained at the Space Telescope Science Institute, which is 
operated by the Association of Universities for Research in Astronomy,
Inc. under NASA contract NAS 5-26555. These observations are associated 
with programs 11243, 11611, and 12212}

\altaffiltext{2}{Department of Physics and Astronomy, Georgia State 
University, Astronomy Offices, 25 Park Place, Suite 600,
Atlanta, GA 30303; fischer@chara.gsu.edu}

\altaffiltext{3}{Institute for Astrophysics and Computational Sciences,
Department of Physics, The Catholic University of America, Washington, DC
20064}

\altaffiltext{4}{Naval Research Laboratory, Washington, DC 20375}

\begin{abstract}

Active Galactic Nuclei (AGN) are axisymmetric systems to first order; their 
observed properties are likely strong functions of inclination with respect to 
our line of sight. However, except for a few special cases, the specific 
inclinations of individual AGN are unknown. We have developed a promising 
technique for determining the inclinations of nearby AGN by mapping the 
kinematics of their narrow-line regions (NLRs), which are often easily resolved 
with {\it Hubble Space Telescope} ({\it HST}) [O III] imaging and long-slit 
spectra from the Space Telescope Imaging Spectrograph (STIS). 
Our studies indicate that NLR kinematics dominated by 
radial outflow can be fit with simple biconical outflow models that can be 
used to determine the inclination of the bicone axis, and hence the obscuring 
torus, with respect to our line of sight. We present NLR analysis of 53 
Seyfert galaxies and resultant inclinations from models of 17 individual 
AGN with clear signatures of biconical outflow. Our model results agree with 
the unified model in that Seyfert 1 AGN have NLRs inclined further toward our line of sight (LOS) 
than Seyfert 2 AGN. Knowing the inclinations of these AGN NLRs, and thus their 
accretion disk and/or torus axes, will allow us to determine how their observed 
properties vary as a function of polar angle. We find no 
correlation between the inclinations of the AGN NLRs and the disks of their host galaxies, 
indicating that the orientation of the gas in the torus is independent from that of the 
host disk.

\end{abstract}


~~~~~

\clearpage

\section{Introduction}
\label{sec1}
Within all massive galaxies with bulges are supermassive black holes (SMBHs) that reside 
in the center of their hosts and whose masses usually exceed 10$^6$ solar masses (M$_{\astrosun}$). 
While most galaxies in the current epoch contain SMBHs that lie dormant, a small percentage of galaxies 
with Active Galactic Nuclei (AGN) contain SMBHs that are actively 
gaining mass from the surrounding matter of their accretion disk. As the matter falls into 
a SMBH, it loses angular momentum and gravitational potential energy and as a result 
emits a massive amount of electromagnetic radiation, often outshining the rest of the galaxy.

Seyfert galaxies, a relatively moderate luminosity ($L_{bol} \approx 10^{43} - 10^{45}$ erg s$^{-1}$), 
nearby (z $\leq$ 0.1) subset of the overall collection of AGN, exhibit a dichotomy of broad and narrow hydrogen and other
permitted emission lines. This led the Seyfert class of objects to be divided into two groups \citep{Kha74}.
Seyfert 1 galaxies are defined to have spectra containing broad (full width at half-maximum [FWHM] $\geq$ 
1000 km s$^{-1}$) permitted lines, narrower (FWHM $\leq$ 1000 km s$^{-1}$) forbidden lines, and distinct, 
non-stellar optical and UV continua while Seyfert 2s contain only narrow permitted and forbidden 
lines and their optical and UV continua are dominated by the host galaxy. This difference can be attributed 
to these AGN being similar objects being viewed from different angles, where the central engine and source 
of broad line emission is visible in Seyfert 1s and obscured by a toroidal structure of gas and dust in Seyfert 2s 
\citep{Ant93}.

The NLR, a knotty, extended (1 - 1000 pc) region responsible for emitting the narrow emission lines visible 
in both Seyfert types, is the focus of our study as it is the only AGN component that can be spatially resolved 
in the optical. The narrow emission lines are generated 
by low-density (n$_H \le$ 10$^6$ cm$^{-3}$) gas clouds photo-ionized by the non-stellar continuum emission of the AGN \citep{Pet97,Ost06}.
The clouds are often driven in outflow by the central engine \citep{Hut98,Cre00a,Cre00b}, and are generally in a 
biconical structure, with the apex of the bicone residing in the central AGN \citep{Pog88,Sch94}.
Observed primarily in Seyfert 2s, these biconical formations have projected opening angles typically in the range of $\sim$ 30$^{\circ}$ 
- 100$^{\circ}$, often with well defined linear edges \citep{Sch03}, which imply the NLR is defined by collimation of ionizing 
radiation by an optically thick, torus-shaped absorbing material \citep{Ant85}, or possibly a dusty disk wind \citep{Eli06}, at 
small radial distances of a few parsecs from the SMBH. Around the same scale and position of the NLR, Seyferts 
also sometimes contain radio emitting knots of low-density plasma, which suggests that there may be 
some connection between the thermal narrow-line gas and the non-thermal plasma. At radial distances of 
$\geq$1 kpc, ionized gas often exists in an extended NLR (ENLR) likely in the plane of the host 
galaxy \citep{Ung87}. Seyfert hosts can have any morphological types, but tend to be found in early type 
galaxies \citep{Ho97}.

While it is generally accepted that Seyfert 1 AGN are viewed more face-on and Seyfert 2 AGN are viewed more 
edge-on with respect to the obscuring toroidal structure, the specific inclinations of all but a few 
AGN are generally unknown. Thus we still do not know how the properties of AGN change with inclination beyond 
comparing Seyfert 1s and 2s. By employing our NLR kinematics mapping technique, detailed below, we can for the 
first time determine our viewing angle to a sample of AGN and begin finding inclination dependencies, 
allowing us to progress beyond the simple unified model of Type 1 and 2 AGN.

We have previously shown that the NLR kinematics in Seyfert galaxies are often dominated by radial 
outflow in the approximate shape of a bicone, through kinematic modeling of the NLRs of five individual AGN: NGC 1068 
(Seyfert 2; \citealt{Cre00b,Das06}), NGC 4151 (Seyfert 1; \citealt{Cre00a,Das05}), Mrk 3 (Seyfert 2; \citealt{Rui01,Cre10b}), 
Mrk 573 (Seyfert 2; \citealt{Fis10}), and Mrk 78 (Seyfert 2; \citealt{Fis11}). 

Our current outflow model originated in the \citet{Cre00b} study while attempting to determine 
the nature of the NLR kinematics as possibly due to in-fall, rotation, outflow, or some other flow pattern. 
By observing NGC 1068, the nearest bright Seyfert 2 galaxy, with the {\it HST} STIS, they were able to 
resolve [O~III] emission-line knots that presumably accelerated out 
from the inner nucleus, reached a terminal velocity, and then decelerated due possibly to drag through interactions 
with the surrounding ambient material. Additionally, a lack of low radial 
velocities where the kinematic curves peaked suggested that the NLR was evacuated of [OIII] emission along its 
axis, which is close to the plane of the sky in NGC 1068. These data, combined with {\it HST} imaging that 
indicated a biconical geometry for the NLRs of many Seyfert 2 galaxies \citep{Sch96}, led to the postulation that 
a radial outflow shaped as a biconical shell could reproduce the NLR kinematics seen in the Seyfert 2 galaxy NGC 
1068. The resultant kinematic model assumes the knots are radially outflowing and accounted for by a simple 
velocity law close to the nucleus, $v=kr$, where $r$ is the distance from the nucleus and $k$ is a constant. In 
addition to gaining a better understanding of the NLR kinematics, their model also produced a geometry of the NLR, 
including an inclination of 5$^{\circ}$ out of the plane of the sky, which was required for the model to fit the 
observed radial velocities.

\citet{Cre00a} proceeded to apply their new kinematic model to the NLR of the brightest Seyfert 1 galaxy NGC 4151, 
in which they found, similarly to their model of NGC 1068, evidence for radial acceleration and subsequent 
deceleration of emission-line knots and a hollowed region near the axis. The largest difference between the two 
NLR models was their inclination from our line of sight. The bicone axis of NGC 4151 was inclined closer to our 
line of sight at $\sim 45^{\circ}$ out of the plane of the sky versus $\sim 5^{\circ}$ out of the plane of the sky, nearly 
perpendicular to our line of sight, for NGC 1068. By matching the observed radial velocities of a second Seyfert galaxy to 
their kinematic model, it was realized that it could be possible to determine not just the nature of the NLR kinematics, 
but the orientation and geometry of the AGN system as well.

\citet{Das05,Das06} tested the outflow model with more detailed data sets using higher-resolution spectra 
at multiple slit locations across the NLRs of NGC 4151 and NGC 1068 respectively. The original modeling code 
used by \citet{Cre00b} was also updated to include a variable $n$ in the velocity law, $v=kr^n$. They 
concluded that the models from \citet{Cre00a} and \citet{Cre00b} were consistent with their own and that the 
original velocity law ($n = 1$) remained the strongest fit to their kinematic data. Further kinematic studies 
of three additional AGN, Mrk 3, Mrk 78, and Mrk 573, confirm the success of the original velocity 
law \citep{Cre10b,Fis10,Fis11}.

In order to determine if any correlations exist 
between inclination and other AGN parameters, we require more data points than the five individual AGN 
previously modeled by our group. Thus, the motivation behind this study is to determine the inclinations 
of a larger sample of individual AGN, add them to our previous AGN inclination sample, and compare them 
with observed properties of the AGN to identify any existing correlations (to be published in 
a forthcoming paper). Section 2 describes the observations used in our work. Sections 3 and 4 describe our 
analysis technique of the spectral observations and the resultant kinematics. Section 5 describes the models 
and how we employ them, with notes on modeled AGN results in Sections 6. Finally, Sections 7 and 8 contain 
discussion and conclusions respectively.

\section{Observations}
\label{sec2}
All AGN in this work were observed using {\it HST} STIS, with the CCD detector through 
either a slit of $52'' \times 0.2''$ or $52'' \times 0.1''$, or in a slitless mode (discussed 
more in depth below). Our sample includes all Seyferts with archival G430L or G430M long-slit 
spectra of [O~III] $\lambda 5007$ in the Mikulski Archive 
for Space Telescopes (MAST; 32 AGN), as well as 11 Seyferts from our own G430M observations 
(GO-11611 and GO-12212, PI Crenshaw). Additional archival G750M observations of H$\alpha$ 
$\lambda 6563$ that exist for our sample were also harvested, as a large fraction of H$\alpha$ 
emission originates in the same gas that emits [O~III]. Both emission lines 
were chosen as they are typically the strongest optical lines originating in the NLR ([O~III]/H$\beta \approx$ 10; 
H$\alpha$/H$\beta \approx$ 3; \citealp{Pet97}). Observations using the G750M grating can contain 
emission that originates from star formation or H~II regions. However, when we compare velocities 
measured from H$\alpha$ emission to those measured from [O~III] NLR emission at the same slit position, we find 
that they agree fairly well. Specifications of each grating used are listed in Table \ref{gratspecs}. 

To expand our sample even further, we returned to results from \citet{Rui05} to analyze slitless G430M spectra 
of 6 Seyfert AGN; 4 of which were new to our sample. To quickly summarize their spectral measurement process, 
emission-line knots were identified in both the direct and grating-dispersed images. Both spatial and spectral 
images were then fit row by row with a Gaussian for each knot and the radial velocity of each knot was 
determined by subtracting the positions of the Gaussian peaks in the direct and dispersed images. This 
created a two-dimensional velocity map where each knot in the image was assigned a radial velocity, from 
which we could extract in a strip across the image to simulate a long-slit observation. 
From the slitless observations, only very bright, distinct knots of emission yielded radial velocities. Thus, 
simulated observations were taken in $2''$ wide slits. These psuedo-slit observations, which already contain 
radial velocity measurements, were then assigned a distance along the slit to the nucleus position determined via 
imaging, which produced data sets in an identical fashion as typical long-slit observations which we fit 
with our modeling program.

In total, the expanded sample, not including the five AGN previously modeled by our group, 
contains 161 spectra of 48 AGN. The STIS image scale is $0.05''$ pixel$^{-1}$ and 
the spatial resolution is $0.1''$ pixel$^{-1}$ in the cross-dispersion direction. In many cases, 
multiple observations at the same position were dithered along the slit with respect to the first 
spectrum to avoid problems due to hot pixels. Wavelength calibration lamp spectra were taken during 
Earth occultation. Observation parameters, including slit orientation, offset distance 
from the continuum flux peak, and the source of the observation, are listed in Table \ref{spectratable}. 
Spectral images of all long-slit observations are available online.

The STIS spectra were processed using Interactive Data Language (IDL) software developed at NASA's Goddard Space 
Flight Center for the STIS Instrument Definition Team. Cosmic-ray hits were removed before further processing. 
The zero points of the wavelength scales were corrected using wavelength-calibration exposures taken after each 
observation. Finally, the spectra were geometrically rectified and flux calibrated to provide spectral images 
that have a constant wavelength along each column and display fluxes in units of erg s$^{-1}$ cm$^{-2}$ \AA$^{-1}$ 
per cross-dispersion pixel. Occasionally, a number of hot pixels remained after the data were processed. In those 
cases, we performed an additional cleaning step by replacing the bad pixel with a local median value in the data.

Images with overplotted STIS slit positions for each modeled AGN are depicted in Appendix \ref{sec6}, with all other 
images available online. Table \ref{images} lists the 
{\it HST} instruments used to image each AGN, as well as their redshifts, distances, and transverse scales 
assuming H$_o$ = 73 km Mpc s$^{-1}$. Redshifts were taken from the NASA/IPAC Extragalactic Database (NED). 
Distances for nearby AGN (z $<$ .01) were retrieved from the Extragalactic Distance Database (EDD; 
\citealt{Tul09}), all other distances were derived from measured redshifts. Images taken with filters 
containing [O~III] $\lambda5007$ or H$\alpha \lambda6563$ emission were preferred, otherwise wide-band continuum 
filters, which contained some emission-line contributions, were used. 

\section{Analysis}
\label{sec3}
The described procedure to extract velocities and other information from a long-slit observation applies to every 
slit position in our sample excluding slitless observations. Each spectral image produced one 
spectrum per cross-dispersed pixel along the slit centered on the [OIII] or H$\alpha$ emission lines. Because there are 2 pixels 
per resolution element, the data are slightly oversampled. In each spectrum, 
individual [OIII] $\lambda$5007 or H$\alpha$ $\lambda$6563 emission lines were fit with Gaussians 
over an average continuum taken from line-free regions throughout the spectrum. Many spectra contained two or more 
lines, where each identifiable peak was fit with a Gaussian. If an emission line contains a peak and an asymmetric wing,
where the wing is a flux component responsible for creating a significant ($> 3\sigma$) asymmetry 
in an emission line traceable through several spectra along the slit, our fitting program fits the wing 
as a separate line.

The central peak of each Gaussian is the central wavelength from which we measured a Doppler 
shifted velocity for both the [O~III] $\lambda$5007 and H$\alpha$ emission lines, given in the rest frame of the 
galaxy and using [O~III] and H$\alpha$ vacuum rest wavelengths of 5008.2\AA~ and 6564.6\AA~ respectively. 
In solving for each velocity, we do not need the relativistic formula for velocity because our lines are always 
shifted less than 2000 km s$^{-1}$.

We employ a Gaussian fit rather than a 
direct integration across the line profile because in most cases the former is more suited to extract individual 
velocities from blended lines. Noisy spectra (S/N $< 3$ per resolution element) or broad H$\alpha$ 
lines near the nucleus of Seyfert 1s, which blended into the surrounding [NII] doublet, were not fitted. 

\begin{figure}
\centering
\begin{tabular}{cc}
\includegraphics[angle=0,scale=0.5]{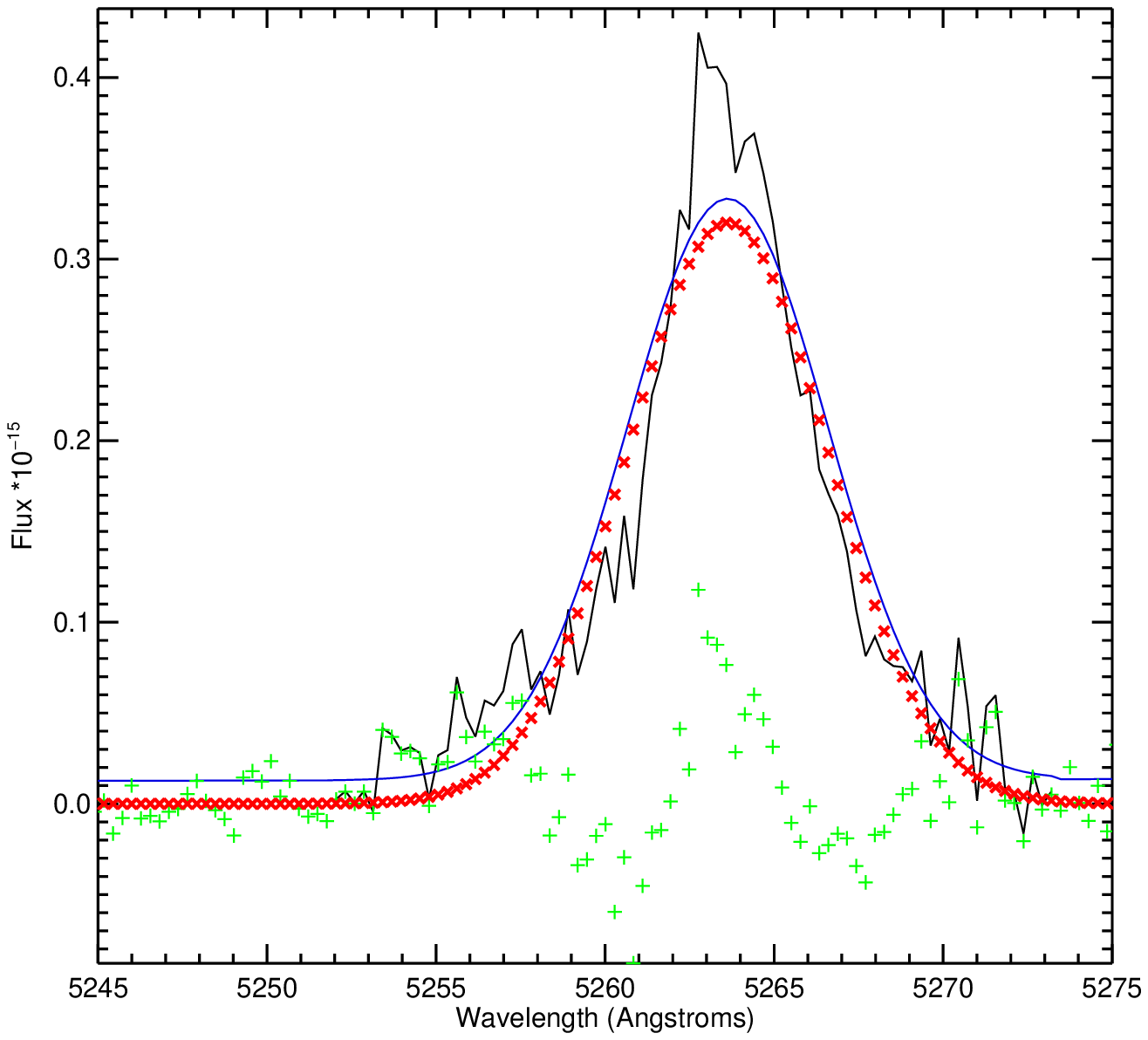}&
\includegraphics[angle=0,scale=0.5]{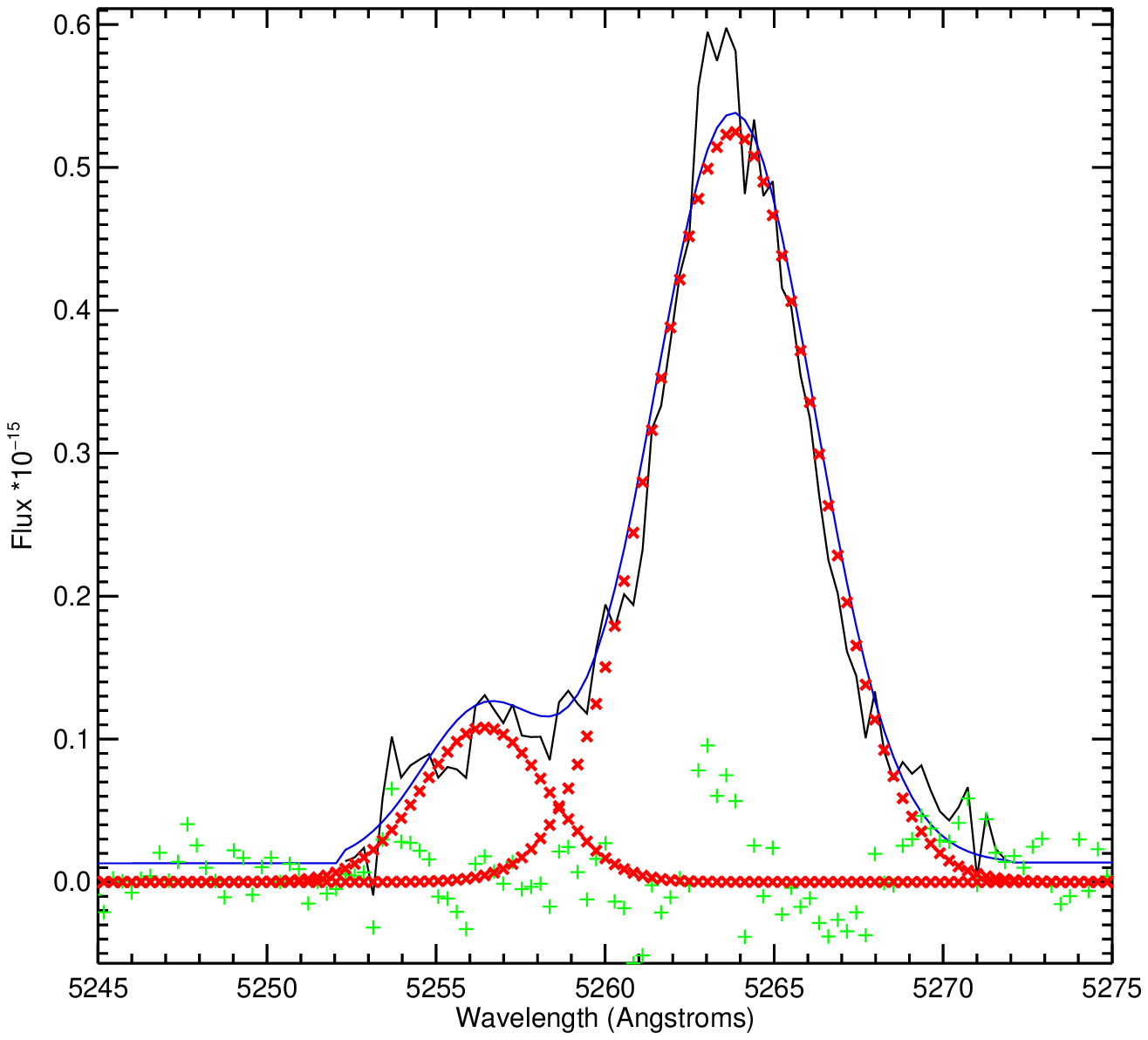}\\
\includegraphics[angle=0,scale=0.5]{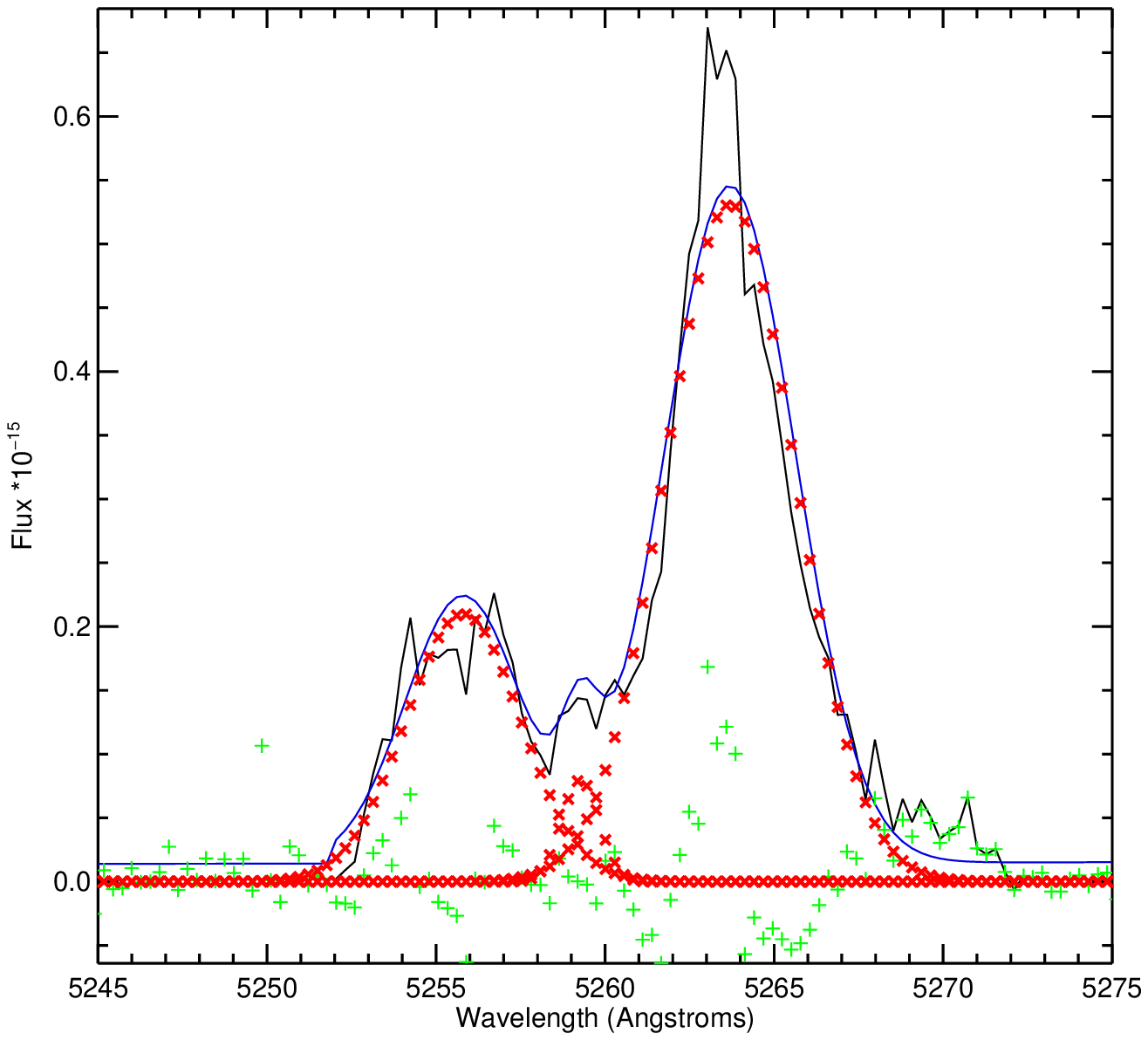}&
\includegraphics[angle=0,scale=0.5]{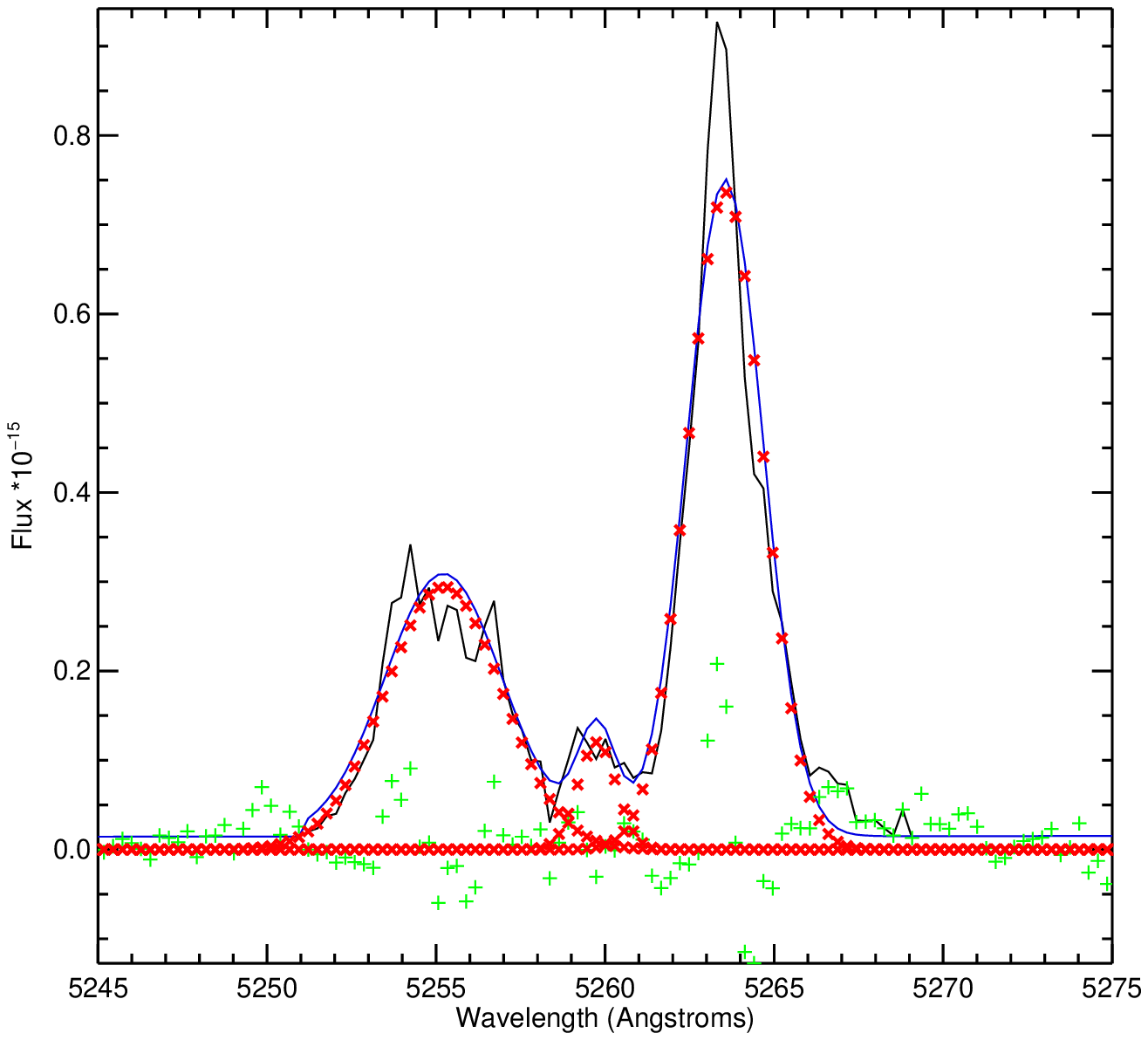}\\
\includegraphics[angle=0,scale=0.5]{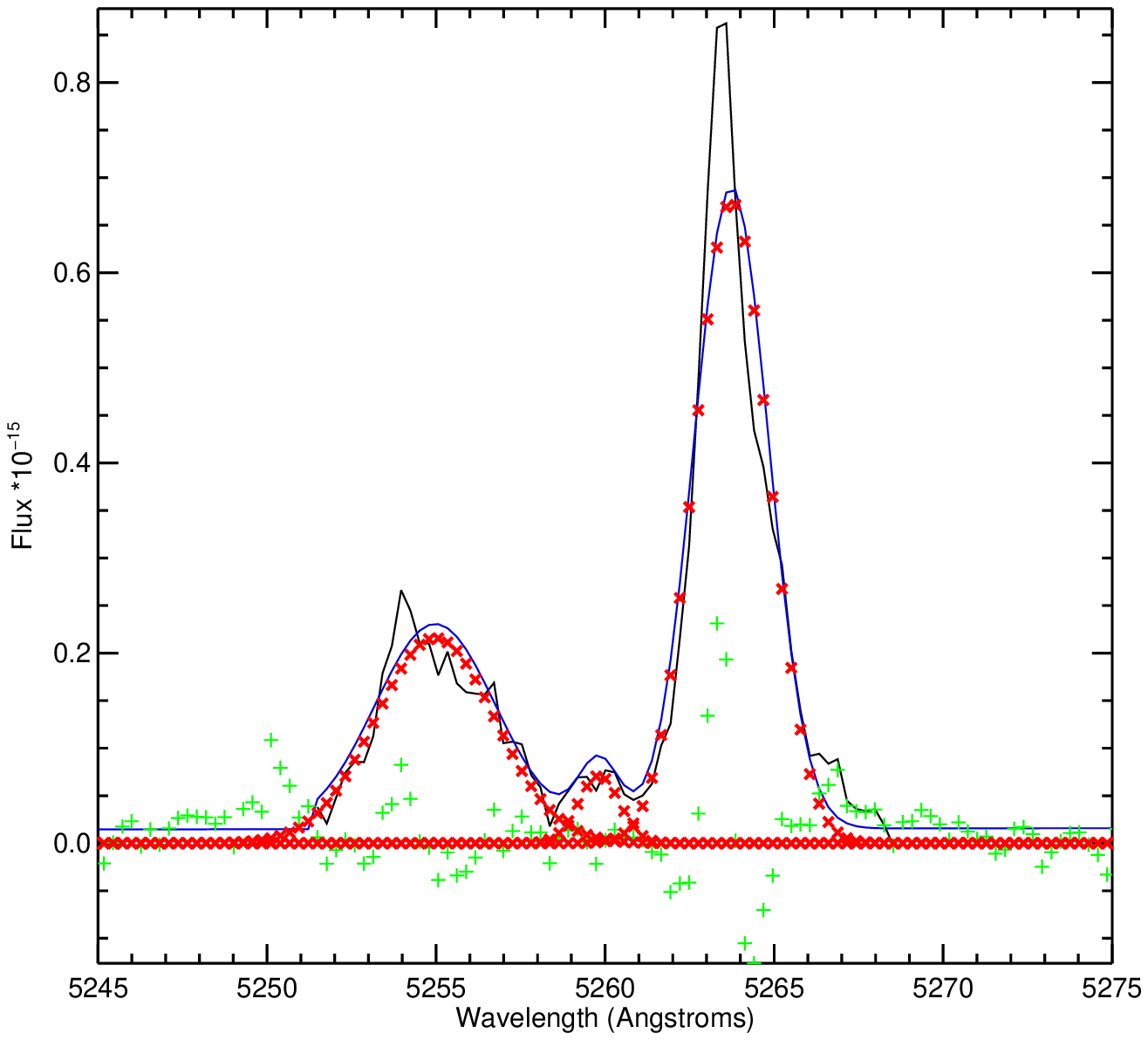}&
\includegraphics[angle=0,scale=0.5]{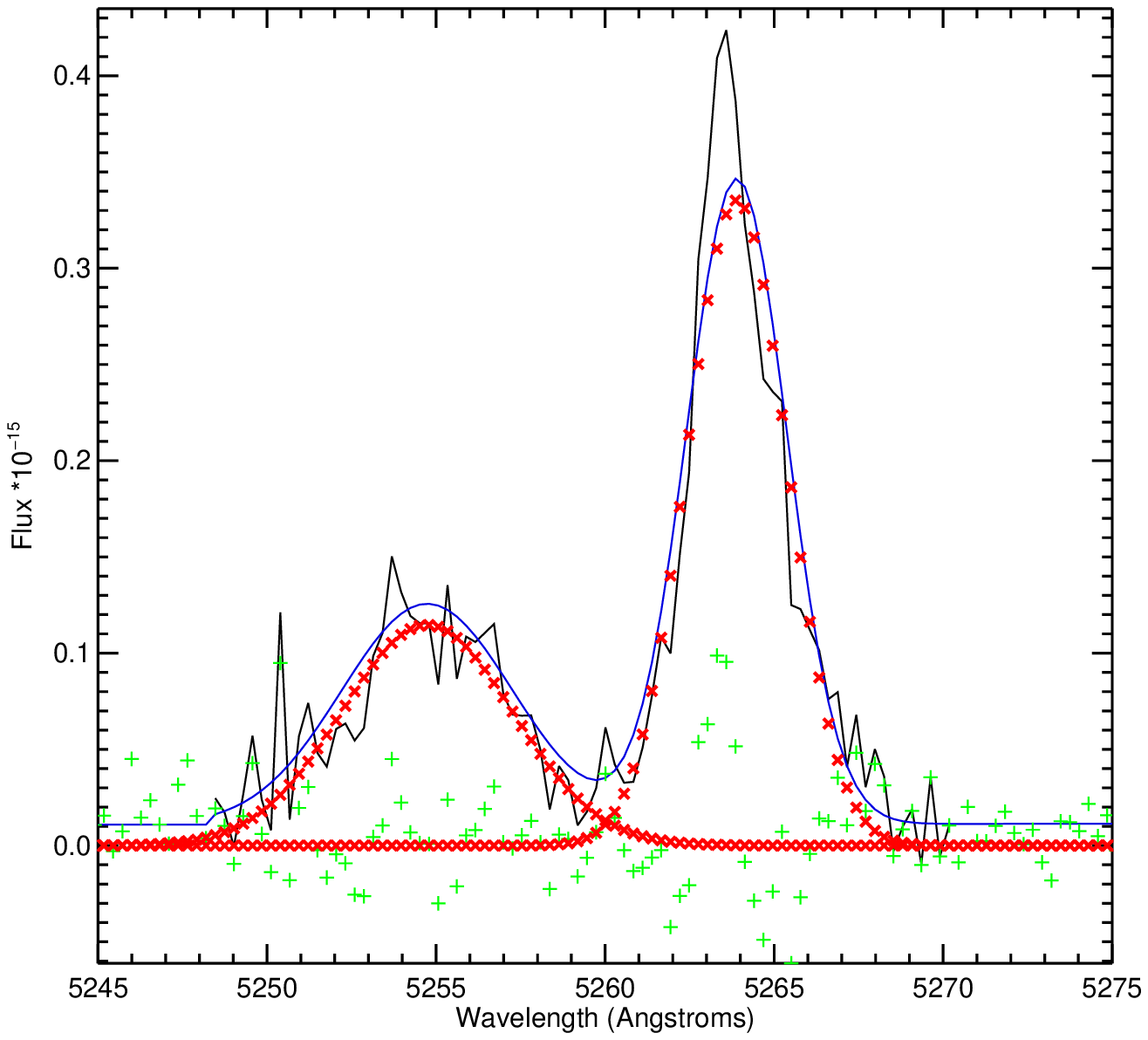}
\end{tabular}
\caption[Gaussian Fits]{Spectra in black and fits in color showing the multicomponent 
Gaussian fits. The plots, ordered from left to right and top to bottom, are from positions separated by 
$0.05''$, increasing in distance from the nucleus. Red $\times$ curves represent the Gaussian fit for each component, 
the blue line represent the model fit as the sum of the red curves and the average continuum, and green $+$ curves represent the 
difference between the black spectra and blue models. Spectra are not yet Doppler-corrected for the cosmological redshift of 
Mrk 34. Fluxes are in ergs s$^{-1}$ cm$^{-2}$ \AA$^{-1}$}
\label{gauss}
\end{figure}

Figure \ref{gauss} depicts a typical progression of spectra illustrating the fitting of multiple lines 
with change in position. The graphs represent spectra taken from the central slit position over Mrk 34 
and range from $0''$ to $0.3''$ in increments of $0.05''$, stepping away from the nucleus in the 
eastern direction. 

In the initial panel at the upper left, a central line is visible, which we fit with a Gaussian.
Besides obvious line peaks that are easy to fit, additional lines are identified by comparing 
adjacent spectra along the slit. Passing through multiple sets of emission lines, we should expect 
resolved knots of emission to be present in more than a single spectrum. 
As we step away from the nucleus, a second blueshifted line emerges to the left of the original 
line and is also fit with a Gaussian. A possible third line is visible in the fourth and 
fifth panels which is well fit with a Gaussian without altering the fits to the two adjacent,
high-flux lines. The center line ceases to exist in the bottom right panel as it can no longer 
be fit with a Gaussian and more closely resembles the increased noise seen at 5250 and 5270\AA, a 
result of this panel containing line fluxes approximately half those of the previous panels.

Figure \ref{mrk34kin} shows the entire velocity data set, as well as full-width at half-maxima (FWHM), 
and normalized fluxes across the central slit position for Mrk 34 (further kinematics for each modeled 
AGN are given in Appendix \ref{sec6}). Note that multiple velocities exist at several positions along 
the slit as shown in Figure \ref{gauss}, and that they generally agree between the two gratings. The scaled differences 
in FWHM are due to uncorrected instrumental spectral resolutions of $\sim$650, $\sim$60, and 
$\sim$130 km s$^{-1}$ for the G430L, G430M, and G750M gratings respectively. Adding the FWHM of the 
resolved G750M H$\alpha$ lines (as well as G430M [O~III] lines) to the FWHM of the line spread function 
of the G430L grating for a $0.2''$ slit in quadrature gives a value typical of the observed FWHM of the 
G430L [O~III] lines indicating the intrinsic widths from the gratings are the same. Further discrepancies 
can be attributed to blending of lines within the G430L spectra. Lumpiness of fluxes across long-slit 
observations are due to spatially resolved knots of emission.

There are two sources of uncertainty in our velocity measurements. The first is that the 
measured emission lines are not perfect Gaussians, as shown in Figure \ref{gauss}, but instead 
tend to have stronger peaks than pure Gaussians. To measure 
a typical value for this uncertainty, denoted as $\sigma_{hi}$, we found the average difference 
between Gaussians fit to a selection of 15 random, isolated high signal-to-noise emission lines 
as detailed above and the actual centroid of the same line. The second 
error comes from photon noise, $\sigma_{noise}$. By repeating the measurements performed on high S/N lines 
to 15 random, low S/N lines and subtracting in quadrature 
the subsequent averaged value from the average $\sigma_{hi}$, we can solve for the remaining variance 
due to photon noise, $\sigma_{noise}$. Both errors are given for each grating in Table 
\ref{gratspecs}, as well as the total maximum error ($\sigma_{total}$), a summation of the 
two errors in quadrature. Varied continuum placements for randomly selected spectra were also 
tested, but did not affect the central wavelengths significantly. It should be noted that $\sigma_{total}$ 
is an upper limit that accounts for the maximum value of each contributing error, as the total 
error will be lower for high signal-to-noise spectra because $\sigma_{noise}$ will be close to zero.

\section{Observational Results} 
\label{sec4}
Appendix \ref{sec6} shows a portion of the radial velocities, FWHM, and fluxes for our sample as a function of 
projected distance from the nucleus using the line analysis procedure described in the previous section, with the 
entire sample available online. Lines measured using G430L, G430M, or G750M gratings are marked as green 
diamonds, blue circles, and red squares respectively. 

\begin{figure}
\centering
\includegraphics[angle=0,scale=.65]{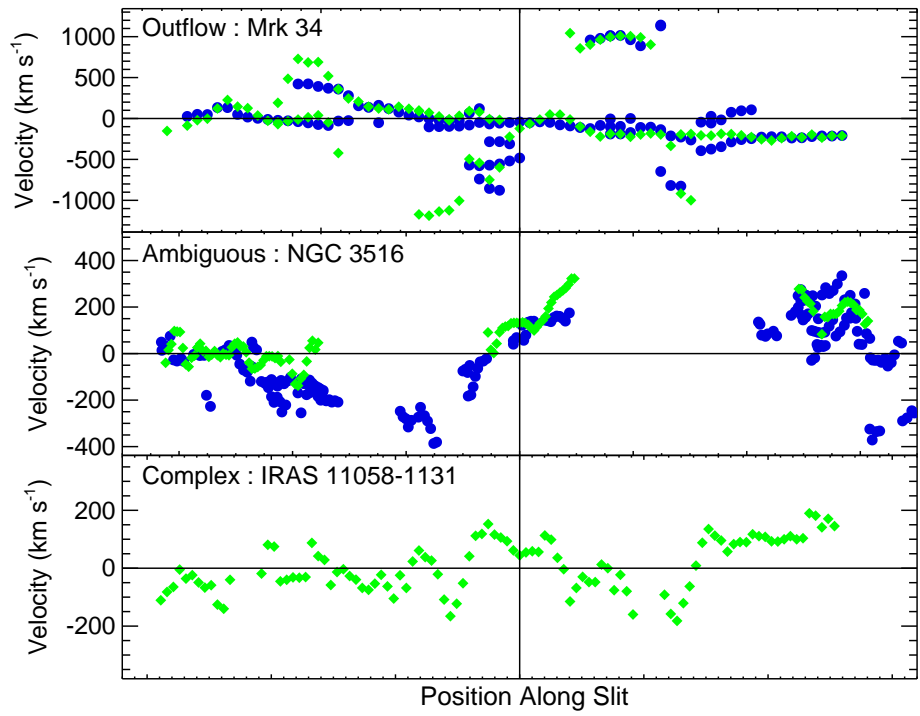}
\caption[Kinematic Types]{Radial velocities of [OIII] lines using the G430L (green diamonds) and G430M 
(blue circles) gratings for three AGN with different kinematic 'types'.}
\label{kintypes}
\end{figure}

From the observed radial velocities, we have classified the kinematics of each AGN as Outflows, Ambiguous, Complex, 
and Compact. Kinematic classifications for our entire sample, as described below and shown in Figure \ref{kintypes}, 
are listed in Table \ref{class}. 

\begin{figure}
\centering
\includegraphics[angle=0,scale=.65]{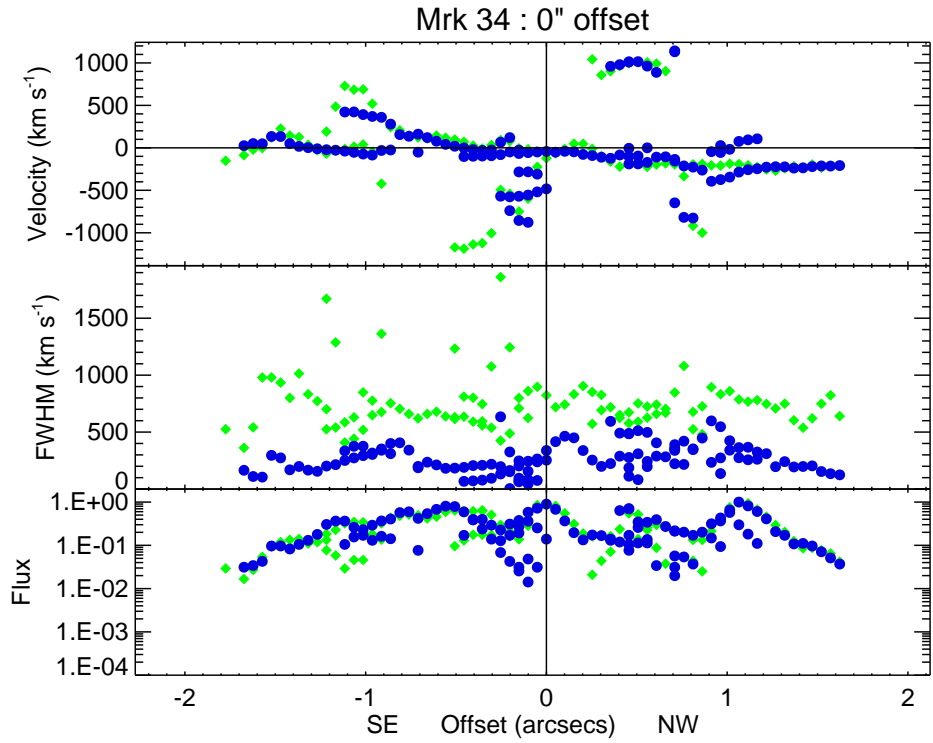}
\caption[Mrk 34 kinematics]{Radial velocities (top), FWHM (middle), and normalized fluxes (bottom) of [OIII] 
lines using the G430L (green diamonds) and G430M (blue circles) gratings for the central slit position of 
Mrk 34.}
\label{mrk34kin}
\end{figure}

Within the expanded sample, we found 12 additional AGN (including Mrk 34 in Figures \ref{mrk34kin} and \ref{kintypes}; top) 
that clearly show kinematics that are characteristic of biconical outflows seen in our previously modeled targets, and are thus 
classified as $''$Outflow$''$ kinematic targets. AGN with Outflow kinematics required multiple outflow components, sets of 
velocities that increase out to a certain radius before decelerating back to systemic velocity, to be visible to fit 
our current model. Outflow targets ideally display four components, from the near and far sides of each cone, but can 
possibly be observed with only two as discussed in Section \ref{sec5}. We do not define all outflows that originate 
from AGN by these parameters, as it is possible to have an outflow that does not exhibit decelerating velocities 
\citep{Fis10}. High radial velocities also were ideal as they discount the likelihood that the observed kinematics 
could be due to rotation, where $v$ is rarely greater than 400 km s$^{-1}$ \citep{Spa00}. 

The remaining 36 AGN within the sample do not show distinct Outflow characteristics. The sources of their kinematics remain 
unclear, and it is possible that their NLR kinematics may be due to processes other than biconical AGN outflows. 
17 AGN show kinematics that we have deemed $''$Ambiguous$''$ (Figure \ref{kintypes}; middle). Targets displaying these 
kinematics have a symmetrical component on each side of the nucleus traveling in opposite directions in 
velocity space. While some of these targets have kinematics that partially resemble those of Outflow AGN, it is unclear, 
or ambiguous, if these components possibly correspond to symmetrical sides in each 
half of a NLR bicone. As many exhibit low velocities and do not decelerate after a certain distance, they could also
be due to rotation within the host disk \citep{Spa00,Mul11} or some other factor not incorporated into our current model. 
Even with confirmation that the kinematics were half of an outflowing system, with their adjacent components either 
too faint or too compact to detect, we would be unable to create a successful kinematic model as the components are 
not adjacent to one another (as mentioned in Section \ref{sec5}).

Six additional targets show knotty, turbulent $''$Complex$''$ kinematics (Figure \ref{kintypes}; bottom) that reside near 
systemic velocity, containing large gradients in velocity possibly caused by in situ acceleration (similar to those in 
Mrk 573; \citealt{Fis10}). These targets show no signs of and cannot be explained by Outflow kinematics, though some 
exhibit signs of rotation.

There are 12 targets that do not contain enough data points in their kinematic plots to fit a model, and their kinematics are 
simply defined as $''$Compact$''$. A majority of these contain highly blueshifted velocities without corresponding 
redshifted velocities near the nucleus, suggesting outflow that cannot be resolved into individual components. 
Additionally, two targets, Mrk 348 and NGC 5347, were also deemed Compact as they were observed with the slit position 
outside the extended NLR, thus only detecting the nucleus emission. As the kinematics do not meet the criteria mentioned 
above to qualify as outflow kinematics, it is again possible that the observed velocities seen in this kinematic classification 
may be due to other processes. Finally, NGC 4303 cannot be modeled as we do not detect its NLR. Observations contain 
only continuum emission and no [O~III] lines are clearly present.

\section{Models}
\label{sec5}

The kinematic models generated to match the observations are simple, yet give good fits to radial 
velocities for both Seyfert 1s and 2s showing outflows. This simple approach stems from four basic assumptions 
concerning the characteristics of the NLR clouds:

{\noindent}{\bf 1)} The model employs a biconical geometry for the NLR, with both cones being identical. 
This geometry best explains how [O~III] images often show axisymmetric, triangular NLRs for Seyfert 2s and 
compact circular or elliptical NLRs for Seyfert 1s, consistent with the unified model \citep{Sch03}.

{\noindent}{\bf 2)} The model assumes that the biconical geometry is due to the illumination from the nucleus 
surrounded by a torus-like structure of optically thick material. Thus, the apex of the bicone originates at 
the nucleus of the AGN (i.e. the SMBH). The apex is assumed to be sharp, but this is not always the case \citep{Sch03}.

{\noindent}{\bf 3)} The model assumes a filling factor of 1 within the hollow bicone geometry and 0 outside, as 
we do not know the location of a cloud along our LOS through the bicone shell. This assumption gives a range of velocities along 
any LOS that a cloud can occupy. We assume that there is no absorption of this line within the bicone by the gas because 
the majority of our analysis is based on a forbidden line, although dust could possibly absorb some of this emission \citep{Kra11}.

{\noindent}{\bf 4)} The model assumes that the bicone edges are sharp. Thus, the inner and outer opening angles 
define edges of ionized gas. However, in reality, the observed 
bicone often has fuzzy edges \citep{Kra00b,Kra08}, but this has little effect on our results, as discussed later. 
The model also assumes a sharp edge on each end of the bicone defined by its total length, 
so that the model does not apply to clouds at greater distances. 

\begin{figure}
\centering
\includegraphics[angle=0,scale=1.0]{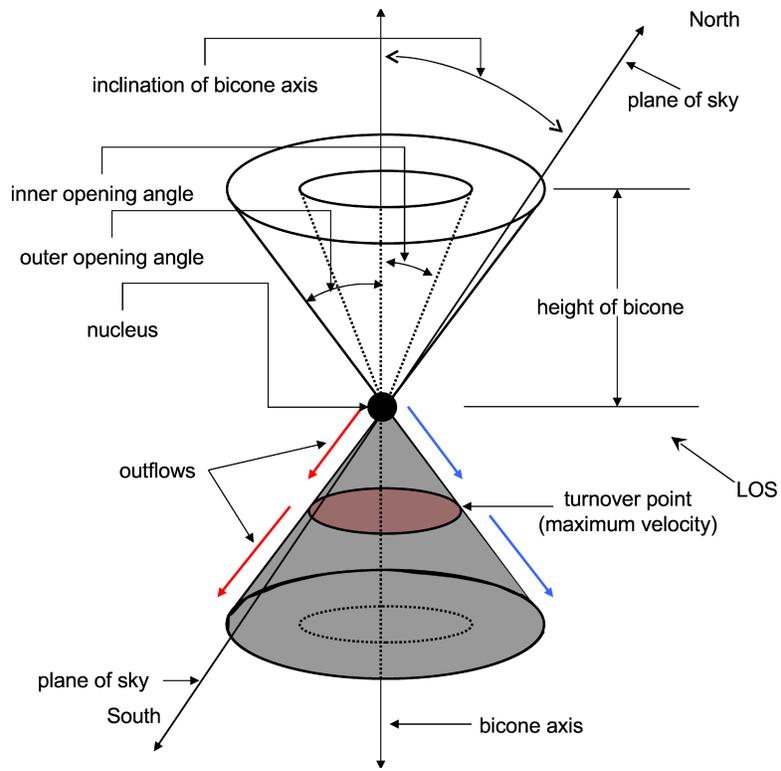}
\caption[Toy Kinematic Model]{Cartoon displaying all alterable parameters used to create a kinematic model.
}
\label{toy}
\end{figure}

The kinematic models used in this study are generated in a 3-dimensional geometry that depends on some 
basic input parameters. These parameters are listed in Table \ref{paramdef} and shown as a cartoon in Figure \ref{toy} 
as reference. The position angle ($P.A.$) is the angle between North and the bicone axis in the 
plane of the sky, measured in the eastward (counter-clockwise) direction. The inclination ($i$) of the bicone is measured out of 
the plane of the sky, with $i=0$ placing the bicone axis in the plane of the sky and $i=90$ placing the bicone axis along our line of sight. 
The inner and outer opening angles ($\theta_{min}$,$\theta_{max}$) are given as half of the total opening angles, 
measured from the bicone axis to the given opening angle. The maximum velocity ($v_{max}$)is the boundary value set 
in the velocity law $v = kr$, where the knots accelerate out (as a result of radiation pressure, for example) 
to a distance where the maximum velocity is reached before deceleration occurs. The maximum height ($z_{max}$) is the distance 
from the nucleus to one end of the bicone, measured along the bicone axis. The turnover radius ($r_{t}$) is the specific 
distance where clouds are no longer accelerated and begin to decelerate back to systemic velocity (due to, for example, 
gravitational or drag forces).

As mentioned in Section \ref{sec1}, the kinematic model used in this work originated in the 
\citet{Cre00b} study where the code generated a two-dimensional velocity map and sampled the map 
with a slit that matched the position, orientation, and width of an observed slit. The kinematic 
modeling code has since been updated, as described in \citet{Das05}, to produce a three-dimensional velocity cube 
which is sampled by extracting a two-dimensional sub-array corresponding to an {\it HST} STIS 
observation slit position, orientation, and width that contains all the radial velocity values 
within that slit. Long-slit extraction from the kinematic model results in a plot of radial velocity versus 
distance from the nucleus with up to four components. The 
model components have a width determined by $\theta_{max}-\theta_{min}$, which results in envelopes of 
shaded regions in the plots.

This model allows for different velocity laws of the form $v=kr^{n}$ at $r \le r_t$ 
(turnover radius) and $v=v_{max}-kr^{n}$ at $r > r_t$. We found that the linear form was sufficient 
as the resultant models fit the data better than the other laws in general.

\begin{figure}
\centering
\begin{tabular}{cc}
\includegraphics[angle=0,scale=0.4]{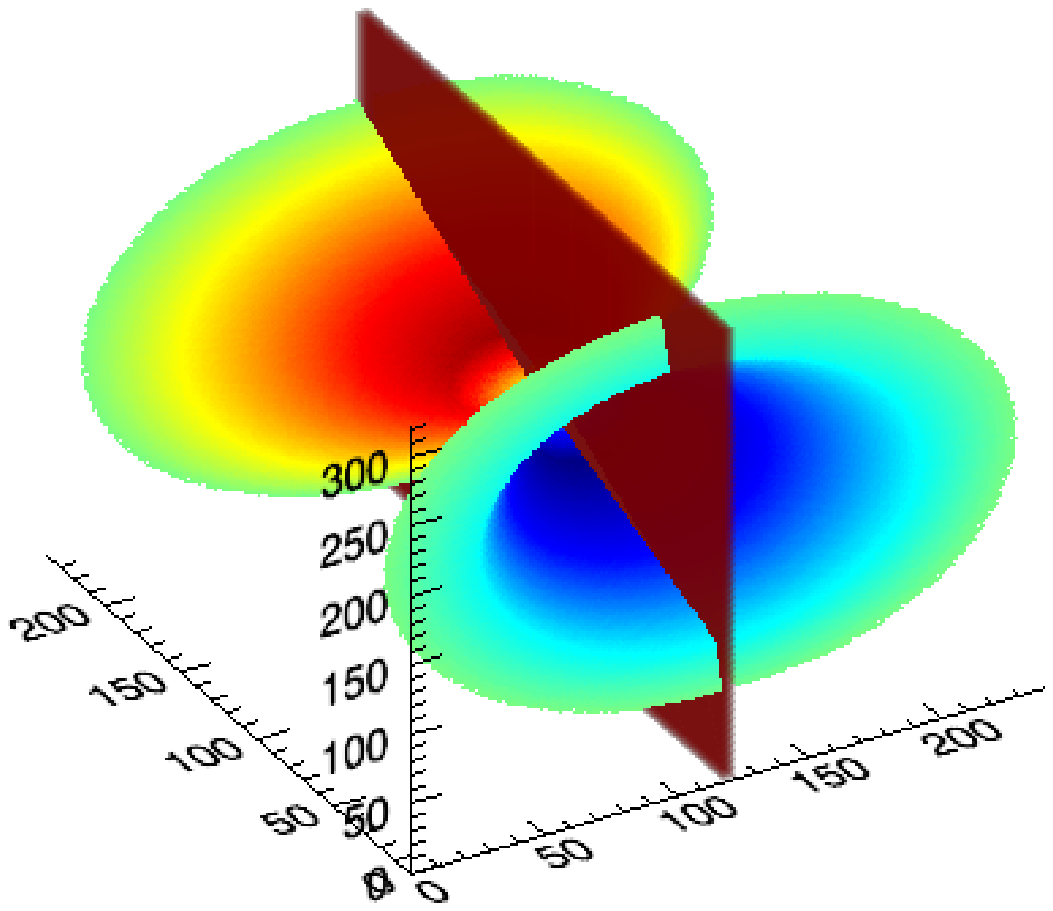}&
\includegraphics[angle=0,scale=0.4]{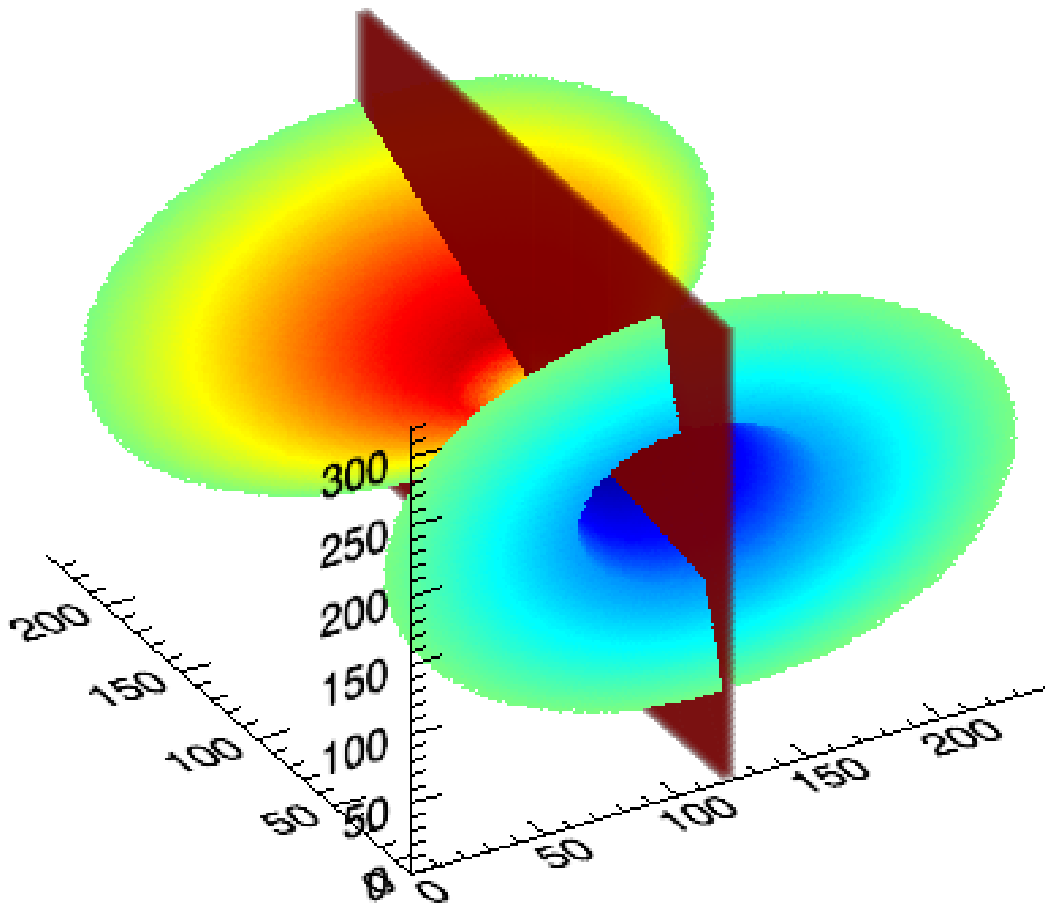}\\
\includegraphics[angle=0,scale=0.4]{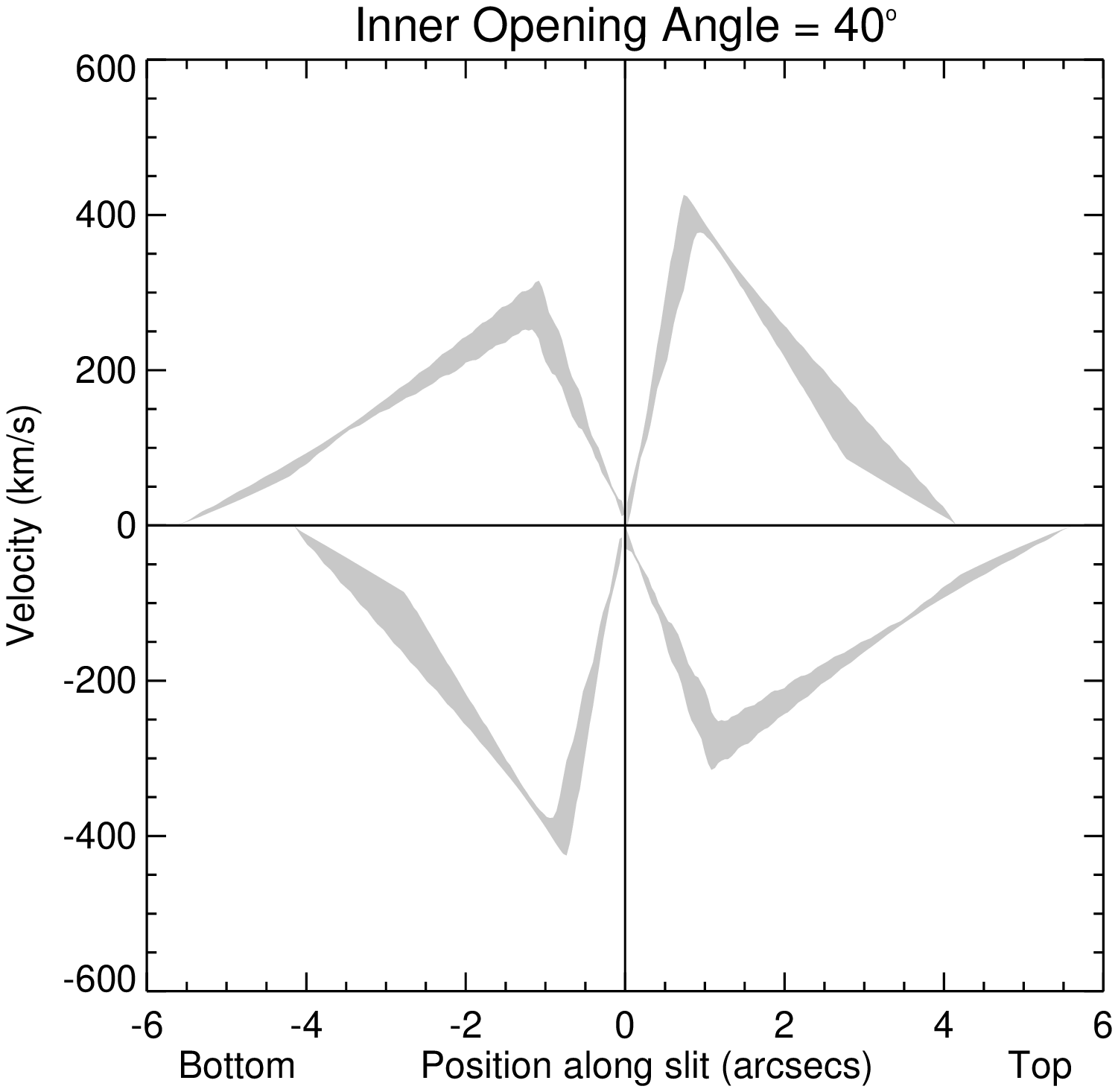}&
\includegraphics[angle=0,scale=0.4]{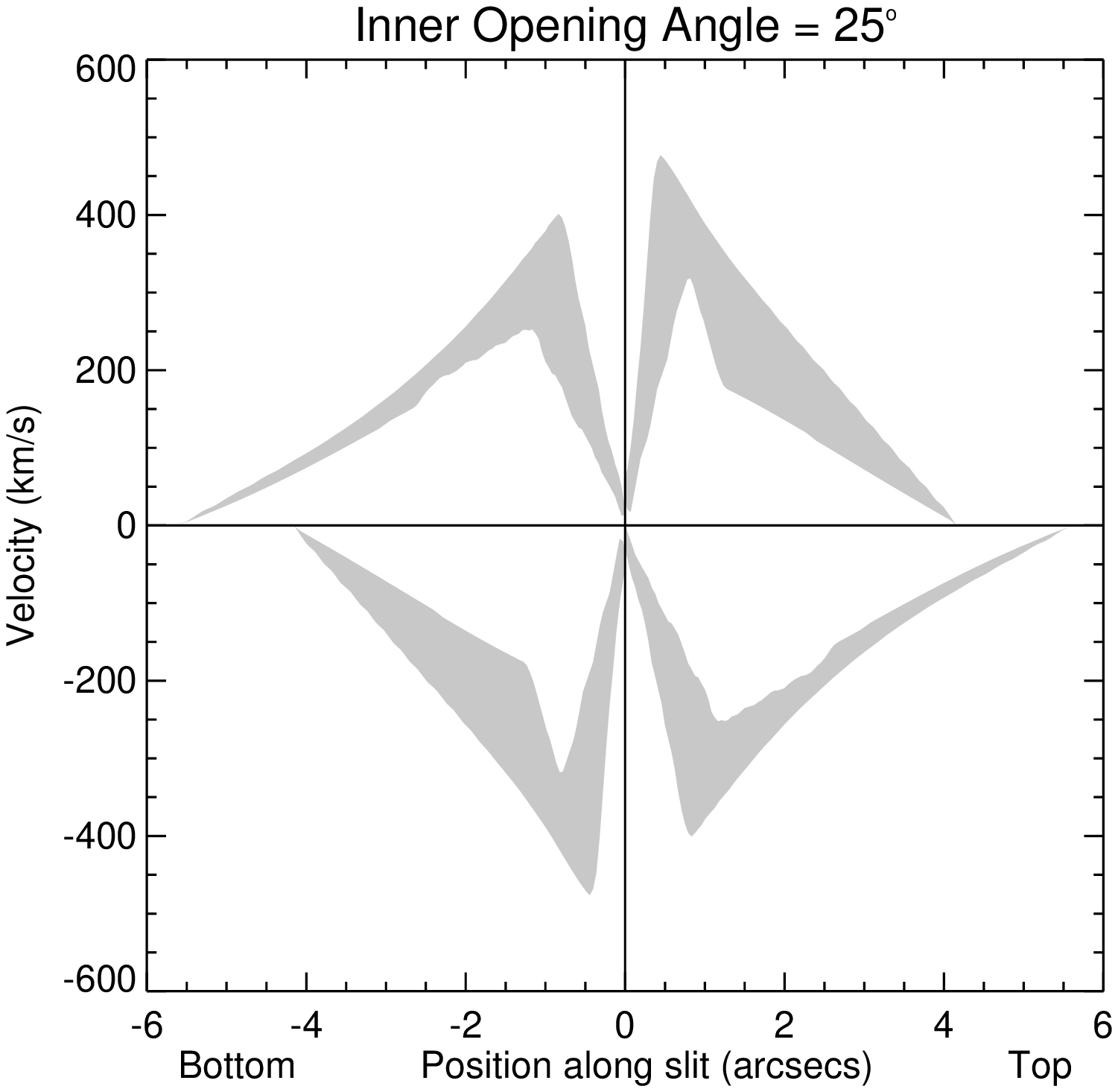}
\end{tabular}
\caption[]{Top: Kinematic models using varied inner opening angles of $40^{\circ}$ (left) and 
$25^{\circ}$ (right) and a fixed outer opening angle of $50^{\circ}$ . All other input parameters 
remain constant, inclination is fixed at $80^{\circ}$ and maximum velocity is fixed at 500 km 
s$^{-1}$. Blue colors represent blueshifted outflow velocities, red colors represent redshifted 
outflow velocities. The red plane represents the orientation of the STIS long-slit observation with 
our LOS coming from the lower right. Bottom: Resultant extracted velocity plots along the given 
long-slit position showing radial velocity (positive, redshifted velocities increasing upward) as 
a function of position along the slit. Increasing the difference between inner and outer opening angles 
results in a larger range of velocities sampled closer to the nucleus, expanding the shaded component areas.}
\label{bicone_o1}
\end{figure}

\begin{figure}
\centering
\begin{tabular}{cc}
\includegraphics[angle=0,scale=0.4]{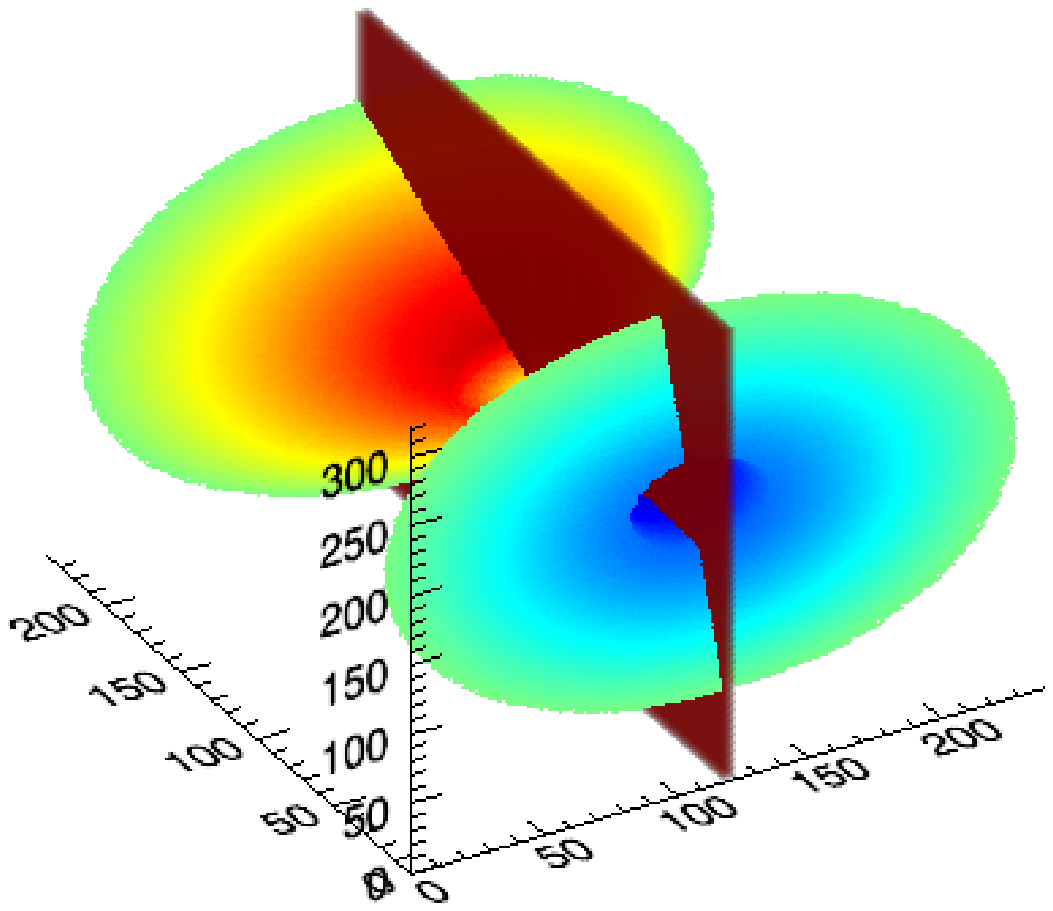}&
\includegraphics[angle=0,scale=0.4]{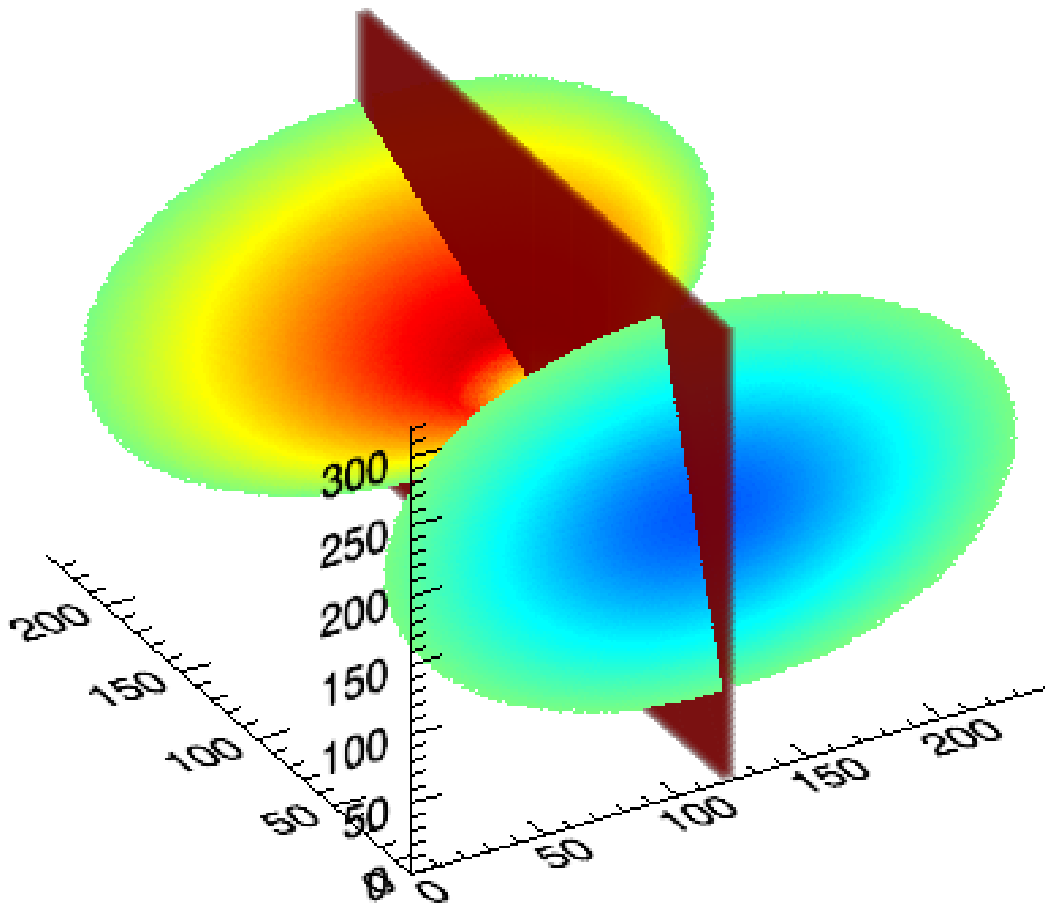}\\
\includegraphics[angle=0,scale=0.4]{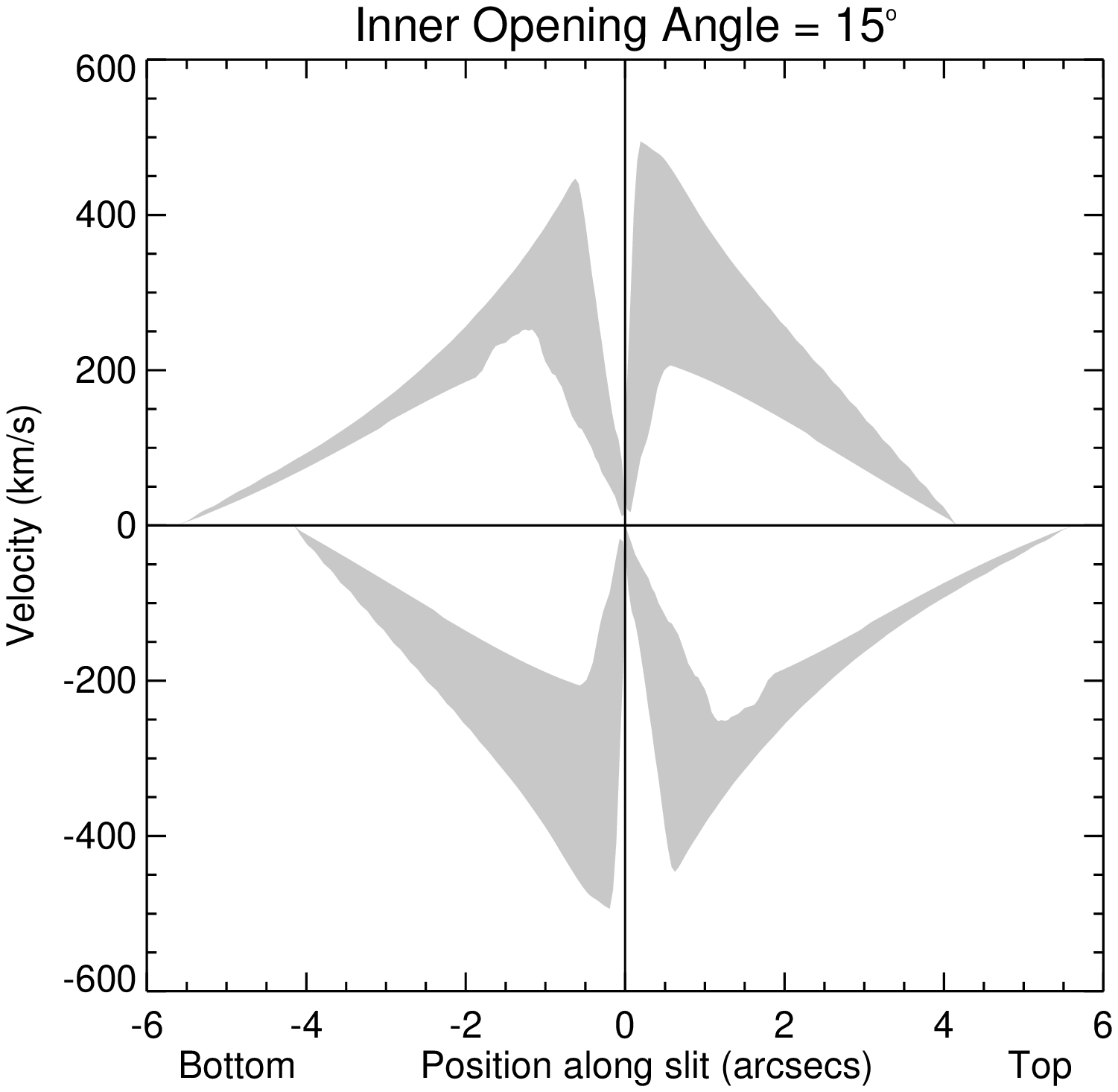}&
\includegraphics[angle=0,scale=0.4]{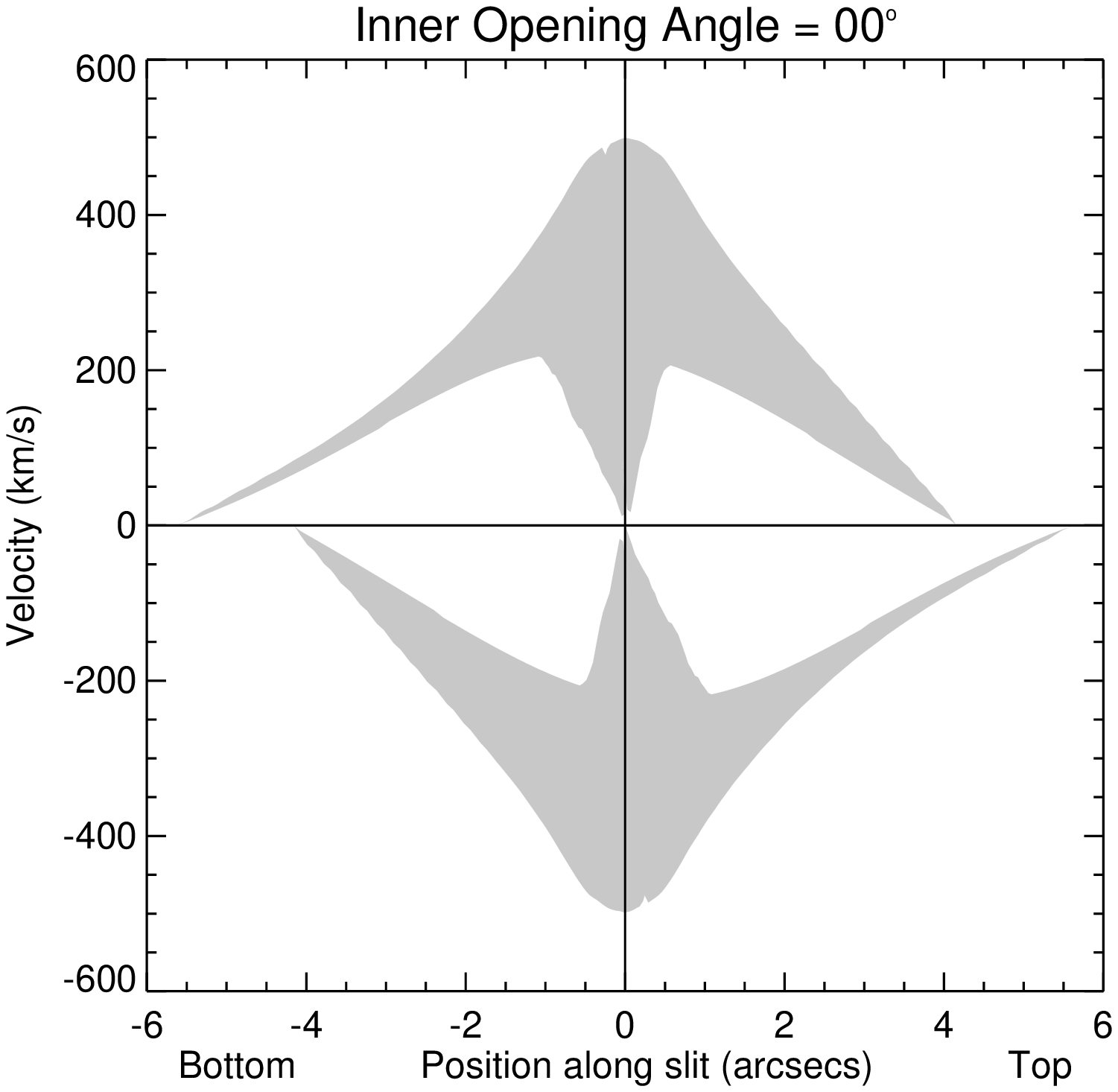}
\end{tabular}
\caption[]{Same as Figure \ref{bicone_o1}, except with inner opening angles of $15^{\circ}$ (left) and $0^{\circ}$ (right).}
\label{bicone_o2}
\end{figure}

Model parameters are initially set to observed values, taken from imaging ($z_{max}$, $\theta_{max}$, $P.A.$) 
and the kinematics data ($v_{max}$, $r_{t}$). By creating a model parameter set and extracting slit positions 
corresponding to all observational positions for a particular galaxy, we can compare the two-dimensional model 
velocity extractions to the data simultaneously. The comparison is done by eye as we have found no efficient 
statistical algorithm to compare the model and data. For example, expanding the width of the cone ($\theta_{max} - 
\theta_{min}$) will always include more observed velocities within the modeled velocity space and result in a 
lower $\chi^2$, but more empty space in the kinematic plot is also included. The process of model fitting is as follows: 

{\noindent}{\bf 1)} The best fit model parameters are obtained when a model encloses the maximum amount 
of data points within a minimum shaded region (i.e. minimum difference between $\theta_{min}$ and $\theta_{max}$, 
effects shown in Figures \ref{bicone_o1} and \ref{bicone_o2}) 
and also matches the trend of increasing and decreasing velocity in the data reasonably well.

{\noindent}{\bf 2)} Input parameters cannot change across individual slits, as models must be consistent for 
all slits for each galaxy. Additionally, each of the two cones must be identical in terms of their model parameters.

{\noindent}{\bf 3)} The shaded regions in the models should not necessarily contain all data points, because some 
points are likely not in the NLR (e.g. they may be in the host galaxy) and thusly are not incorporated in the 
bicone geometry. Furthermore, it is clear that many emission-line knots have their own peculiar velocities with 
respect to the general flow. Emphasis is placed on the higher flux emission-line knots, which generally followed a more structured curve.

\begin{figure}
\centering
\begin{tabular}{cc}
\includegraphics[angle=0,scale=0.4]{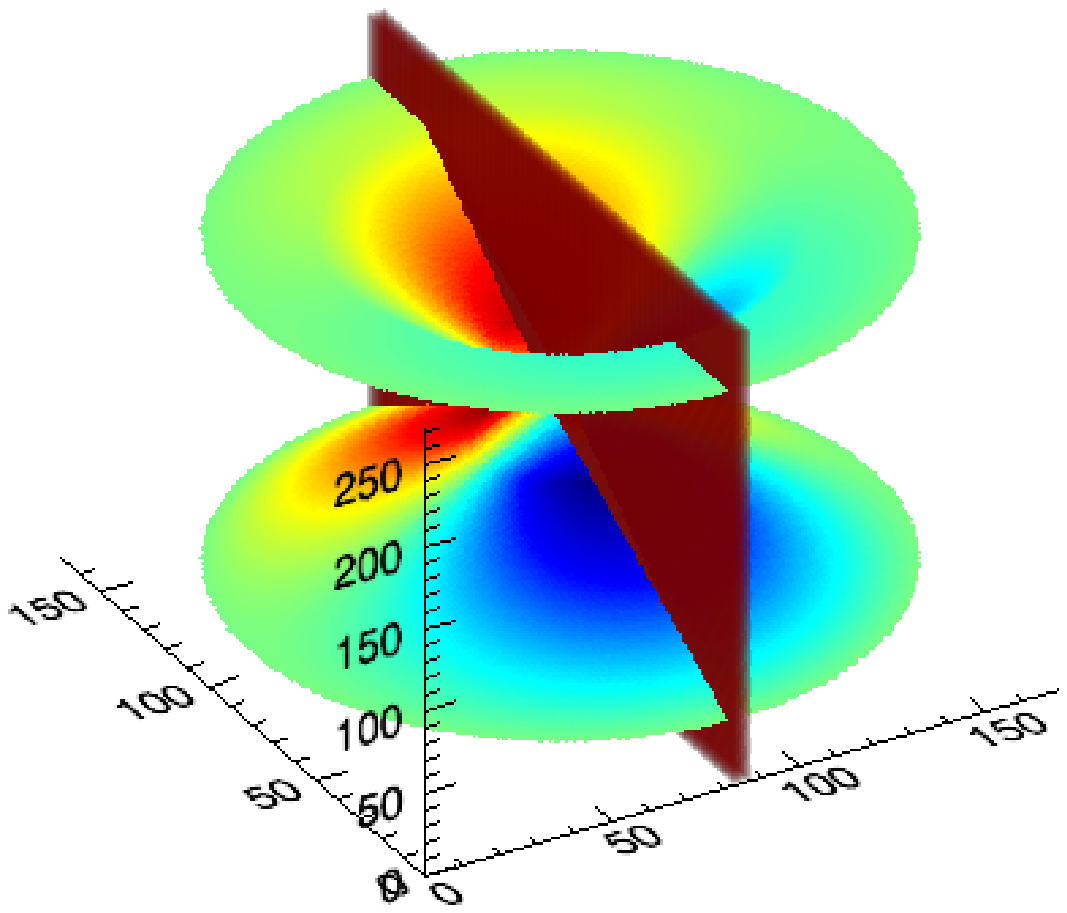}&
\includegraphics[angle=0,scale=0.4]{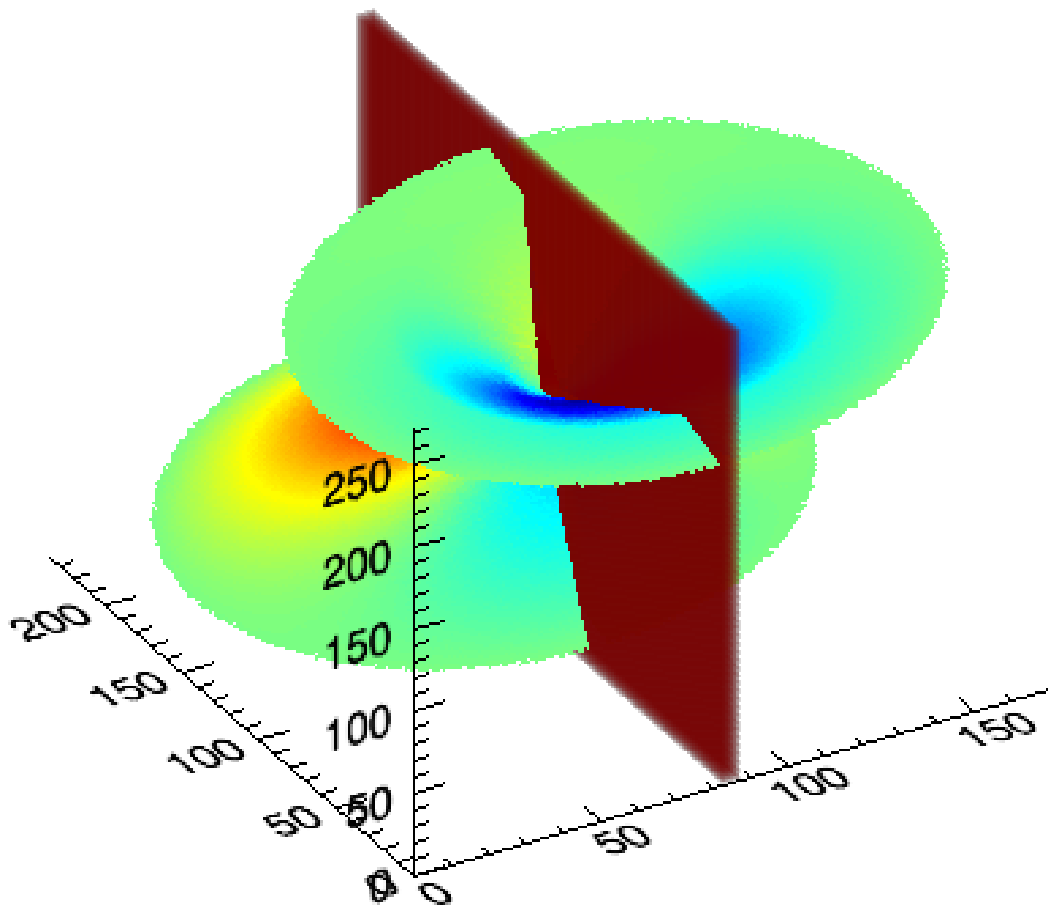}\\
\includegraphics[angle=0,scale=0.4]{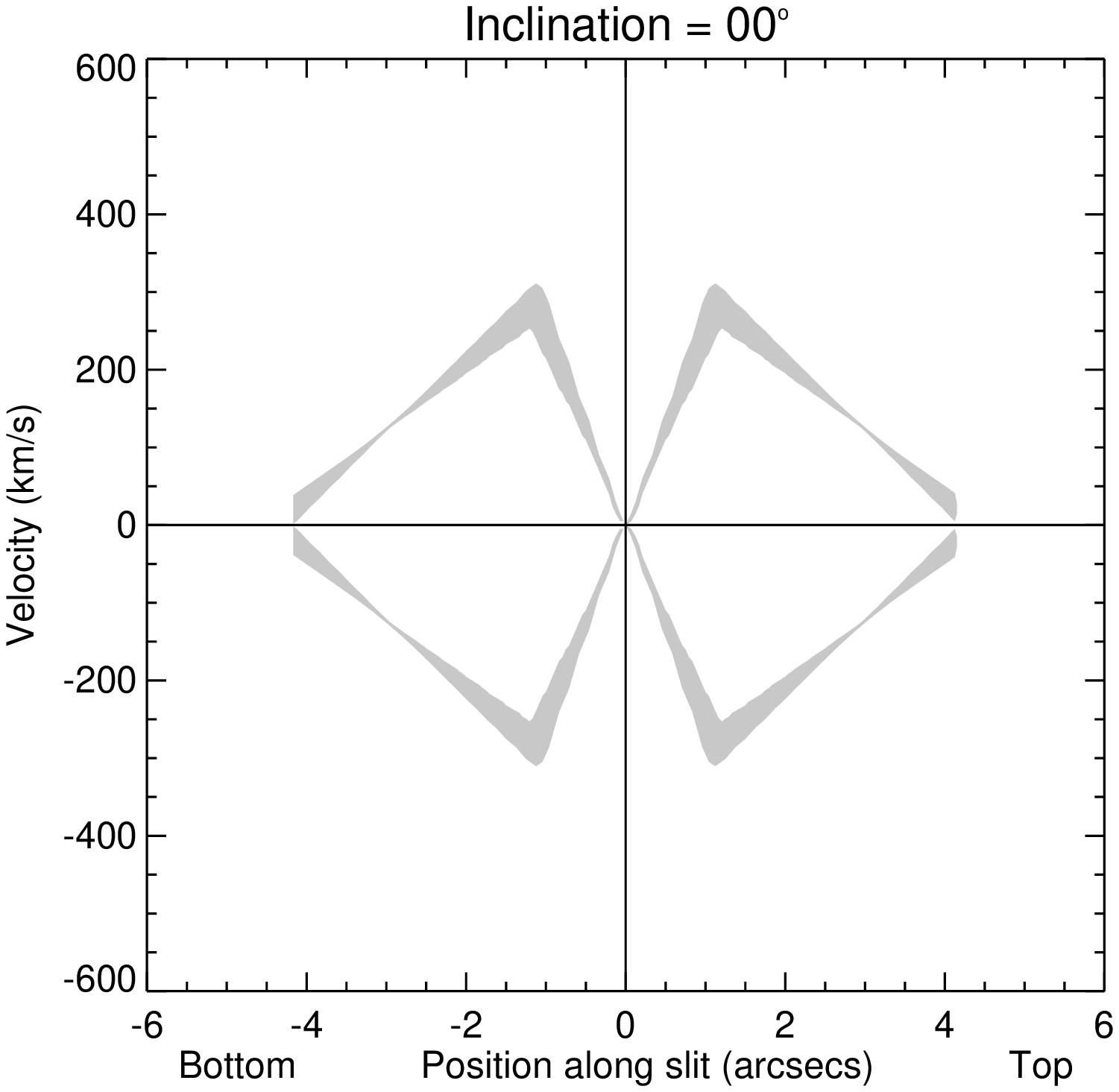}&
\includegraphics[angle=0,scale=0.4]{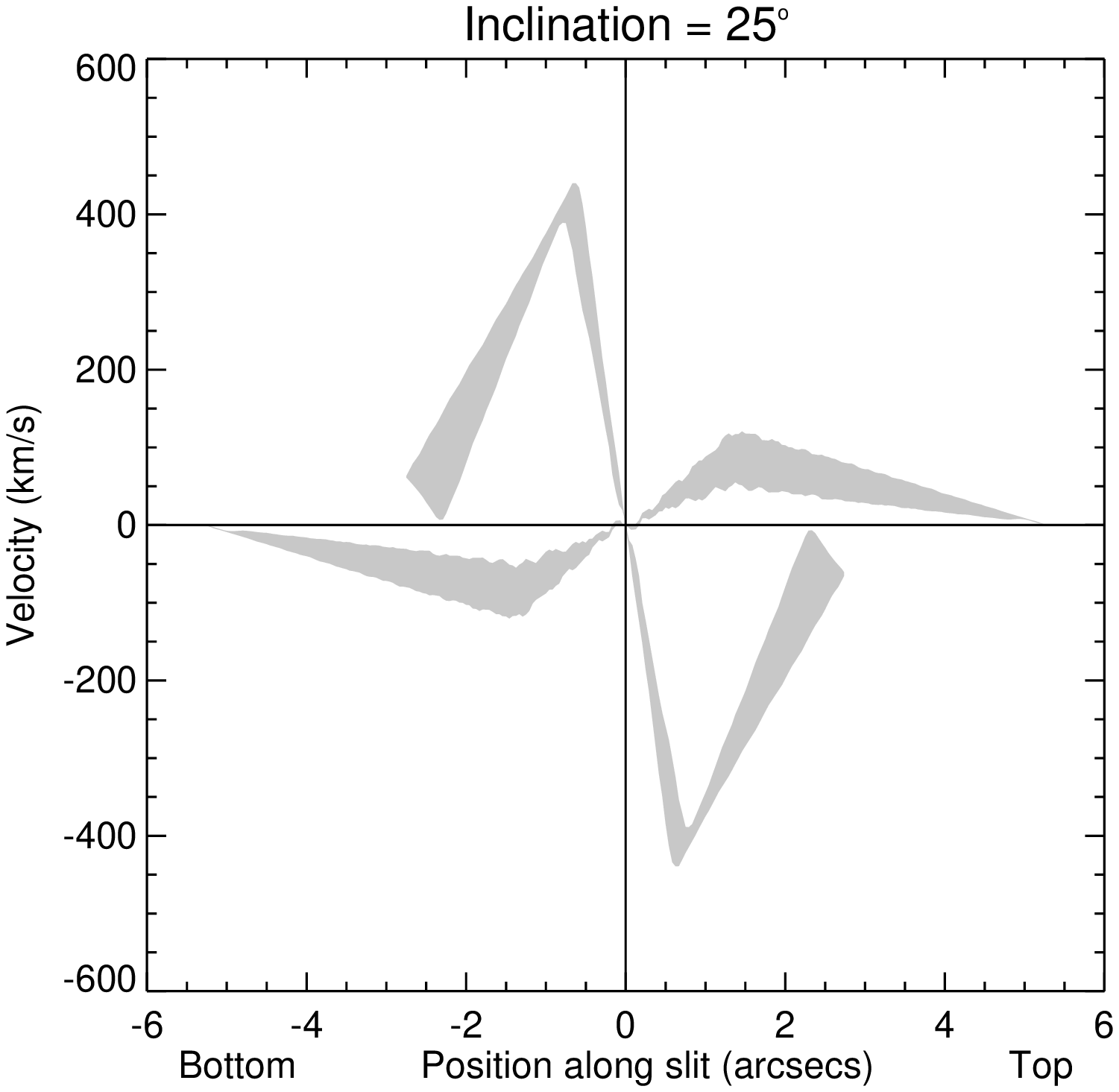}
\end{tabular}
\caption[]{Top: Kinematic models using varied inclinations of $0^{\circ}$ (left) and $25^{\circ}$ (right). 
All other input parameters remain constant. Blue colors represent blueshifted outflow velocities, red colors represent
redshifted outflow velocities. The red plane represents the orientation of the STIS long-slit observation
with our LOS coming from the lower right. Bottom: Resultant extracted velocity plots along the given long-slit position 
showing radial velocity (positive, redshifted velocities increasing upward) as a function of position along the slit.}
\label{bicone_i1}
\end{figure}

\begin{figure}
\centering
\begin{tabular}{cc}
\includegraphics[angle=0,scale=0.4]{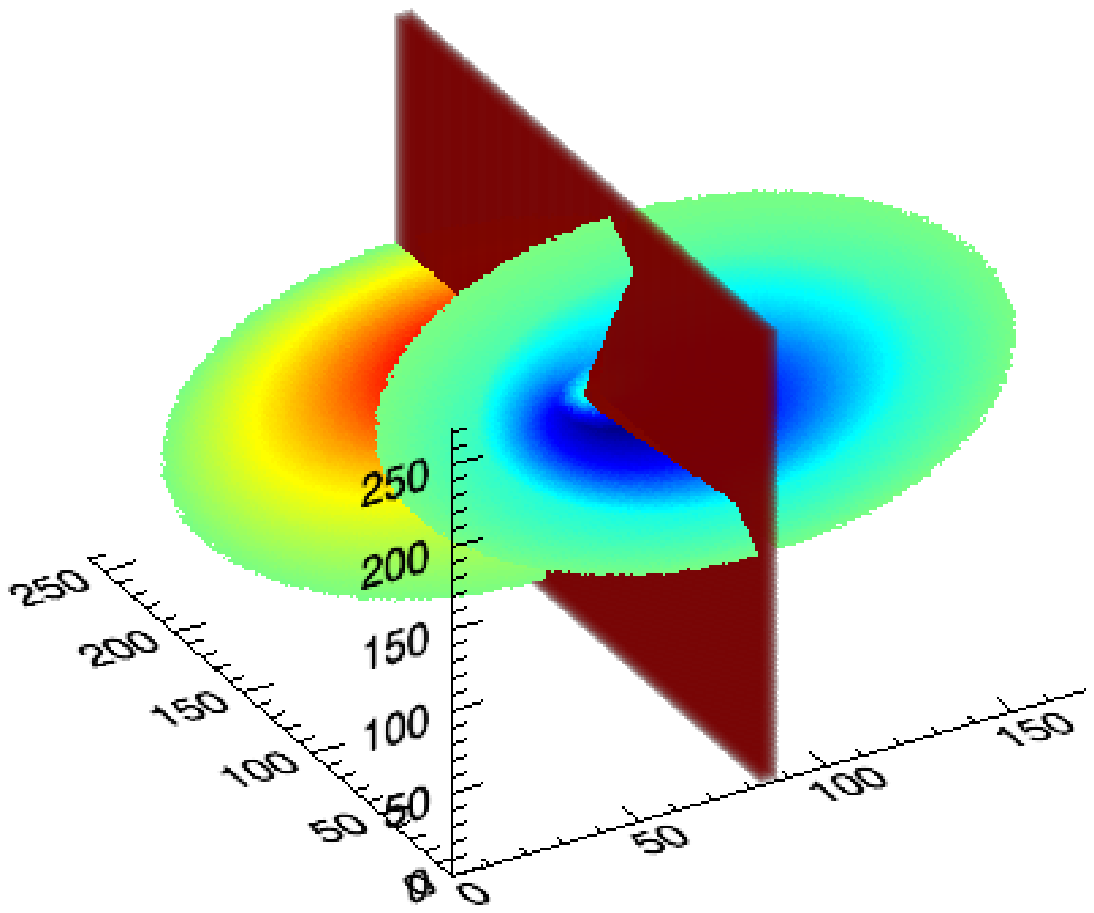}&
\includegraphics[angle=0,scale=0.4]{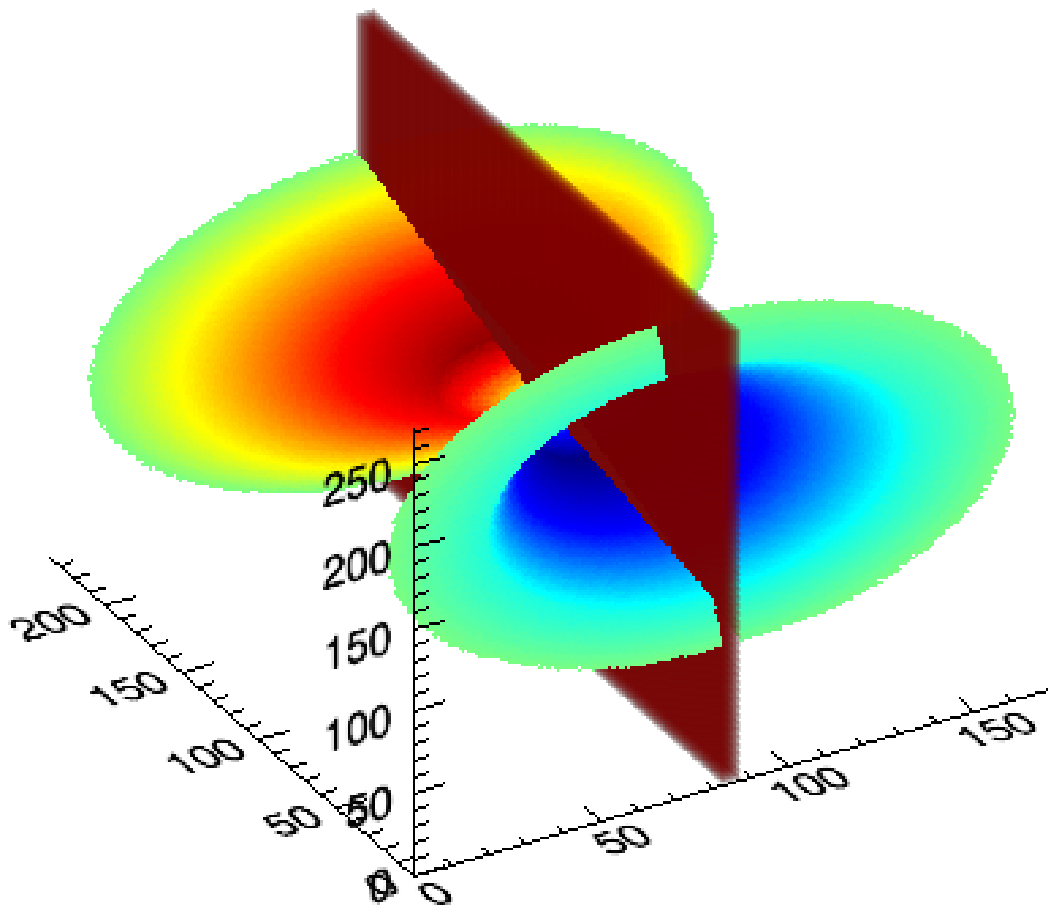}\\
\includegraphics[angle=0,scale=0.4]{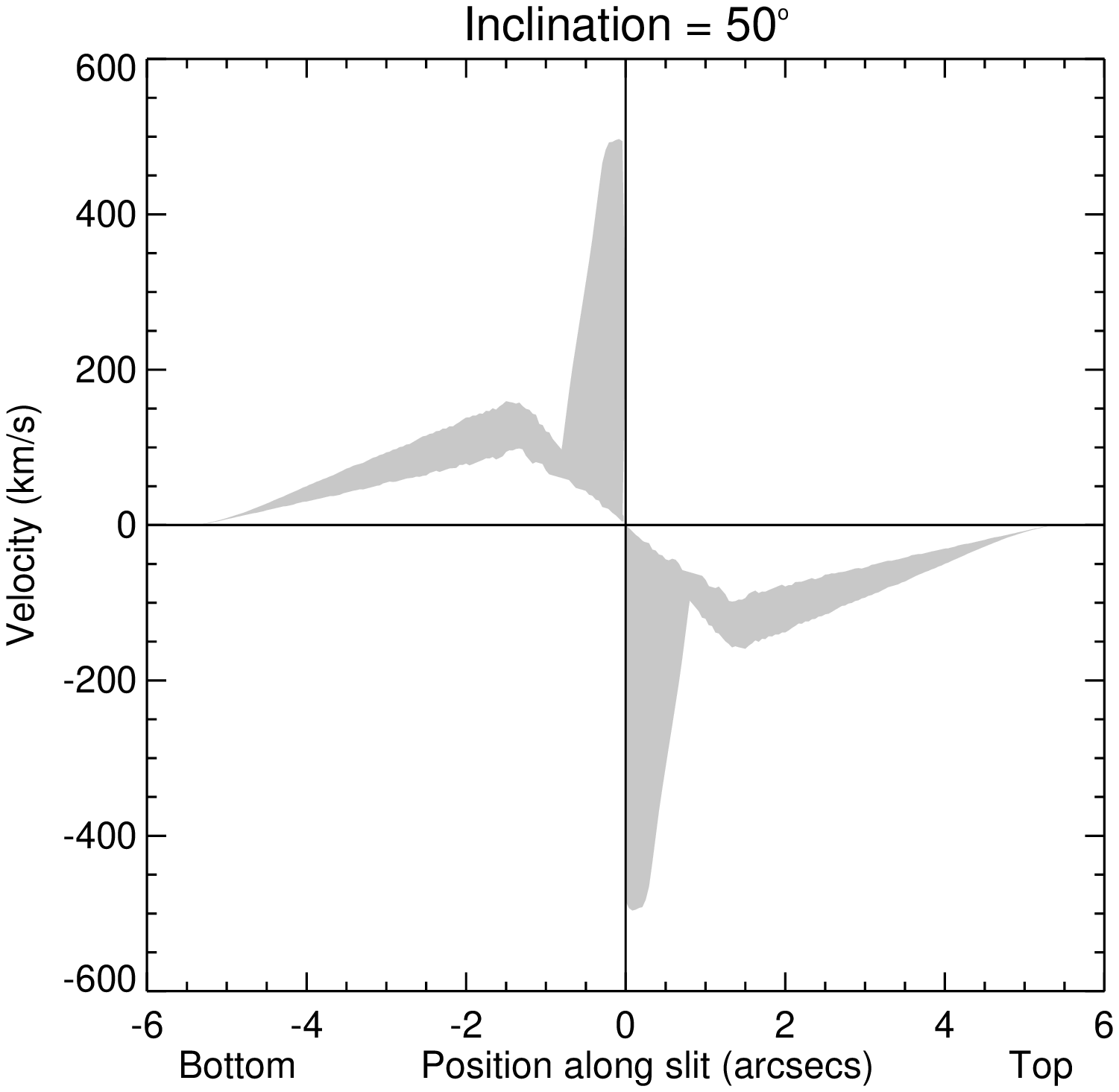}&
\includegraphics[angle=0,scale=0.4]{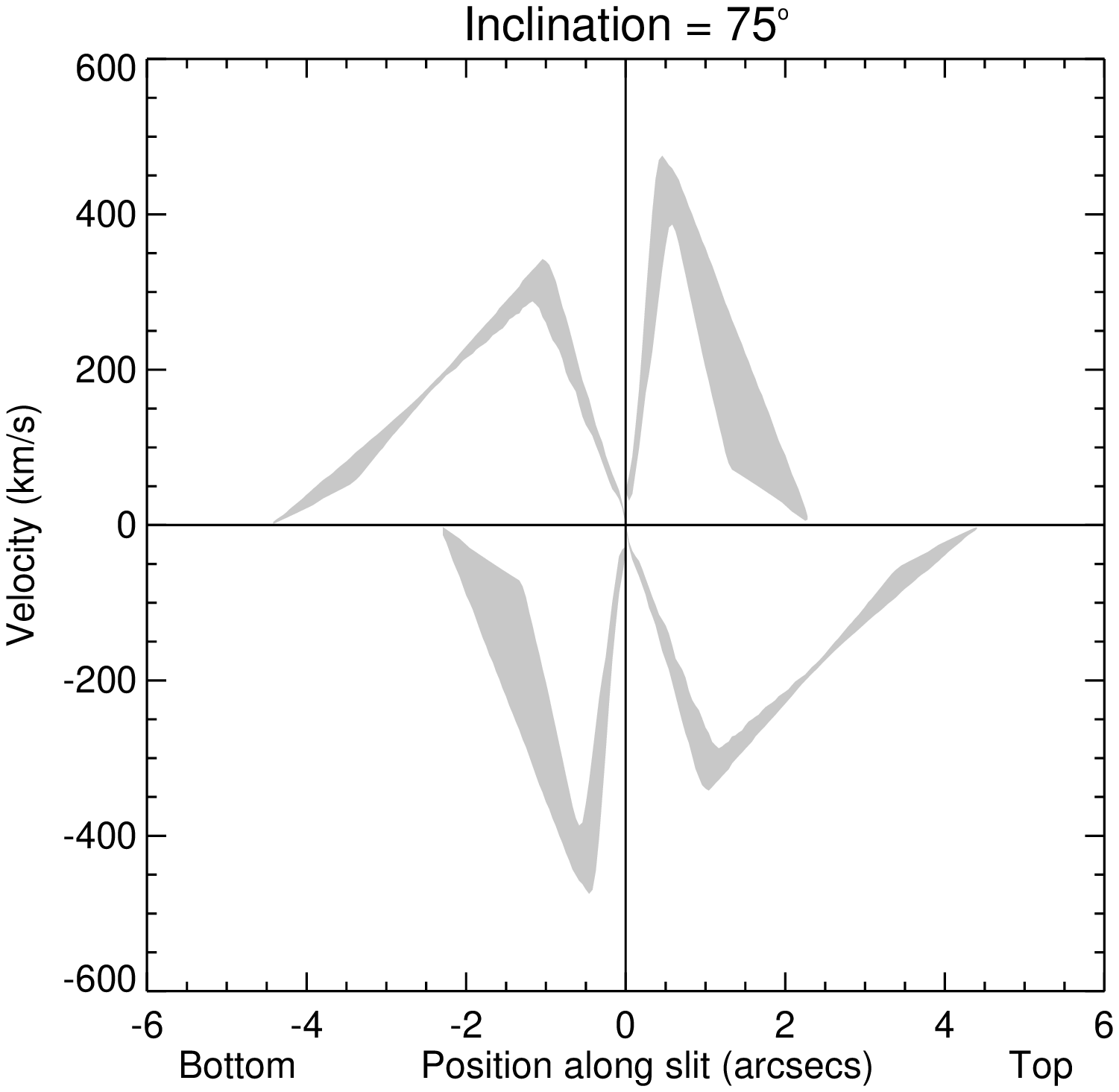}
\end{tabular}
\caption[]{Same as Figure \ref{bicone_i1}, except with inclinations of $50^{\circ}$ (left) and $75^{\circ}$ (right).}
\label{bicone_i2}
\end{figure}

{\noindent}{\bf 4)} If a fit is not suitable for the given data, the parameters are intuitively adjusted 
in an iterative process. The most important parameter is the inclination of the bicone axis as it relates to the Seyfert type 
and has the greatest influence on the position of the model components that are compared to the data. Thus, 
a frequent course of action is to vary the inclination (Figures \ref{bicone_i1} and \ref{bicone_i2}) and then make 
adjustments to the rest of the parameters to offset any resulting changes. Model generation, slit extraction, and plotting are 
repeated until an acceptable match to the data is determined. A suitable fit must account for 
most velocities and have a geometry that agrees with imaging.

{\noindent}{\bf 5)} Errors for a fit are defined as a range of values over which each individual parameter 
can vary without significantly altering the fit between model and data. For example, we can typically vary $i$ by 
$\pm5^{\circ}$ while maintaining a good fit.

We do not require our kinematic model to agree with the unified model. Should kinematics of an AGN be fit 
with two individual parameter sets that are inclined perpendicular to one another (i.e. a Type 1 vs Type 2 scenario), 
our model is constrained by available [O~III] imaging and the parameter set that best matches the NLR morphology is 
chosen. Successfully fitting a kinematic model to the velocity data does not require information from all four components. 
Only two adjacent components (i.e. two components from the same cone or two components closest to Earth) are required 
in order to see the effects of changing any of the model parameters. This means that a model can be 
obtained for targets that suffer from extinction via the host disk, which can obscure large portions of the NLR.
Without adjacent components, it is impossible to solve for a single model as the combination of one inclination 
with a multitude of opening angles (and vice versa) would form solutions matching the available kinematic data, 
which provides the main reason why $''$Ambiguous$''$ kinematics targets remain unmodeled. 

A final parameter in determining the fit of a model is the interaction between the NLR and the host disk. Adding 
the geometry of the host disk to that of the NLR allows us to check the compatibility of the kinematic model 
with available imaging. Imaging that depicts a single cone of emission, for example, must be compatible with a 
kinematic model that both fits the spectral data and places one cone of the NLR outflow behind the host disk
\footnote{In some observed cases, there is not enough dust in the disk to obscure the NLR behind it.}. 
Disk inclinations and position angles for the host galaxy of each modeled AGN are listed from the literature 
in Table \ref{modelvals} and were generally determined by fitting isophotes to the host disk \citep{Sch00}. \citet{Kin00} 
chose isophotes corresponding to a surface brightness level of 24-25 $B$ mag arcsec$^{-2}$, which is often deep 
enough to avoid bar and oval distortion problems. Assuming a circular host galaxy disk, inclinations were 
determined from the measured ellipticity. Errors for inclination and P.A. range 
between $1^{\circ}-2^{\circ}$ and $2^{\circ}-6^{\circ}$ respectively. When it was not predetermined from imaging 
which side of the inclined disk was closer to Earth, orientations that most favorably agreed with the observed 
NLR geometry were employed.

\section{Model Results}

The resultant models for each $''$Outflow$''$ AGN are detailed in Appendix \ref{sec6}, with 
the final parameter values of each modeled AGN, including the five previously observed AGN, listed in Table \ref{modelvals}. 
When comparing properties of our 17 successfully modeled versus the remaining unsuccessfully modeled AGN, we find no bias in 
the ability to model the kinematics due to Seyfert type (35$\%$ of Seyfert 1s; 37$\%$ of Seyfert 2s).

\begin{figure}
\centering
\includegraphics[angle=90,scale=0.65]{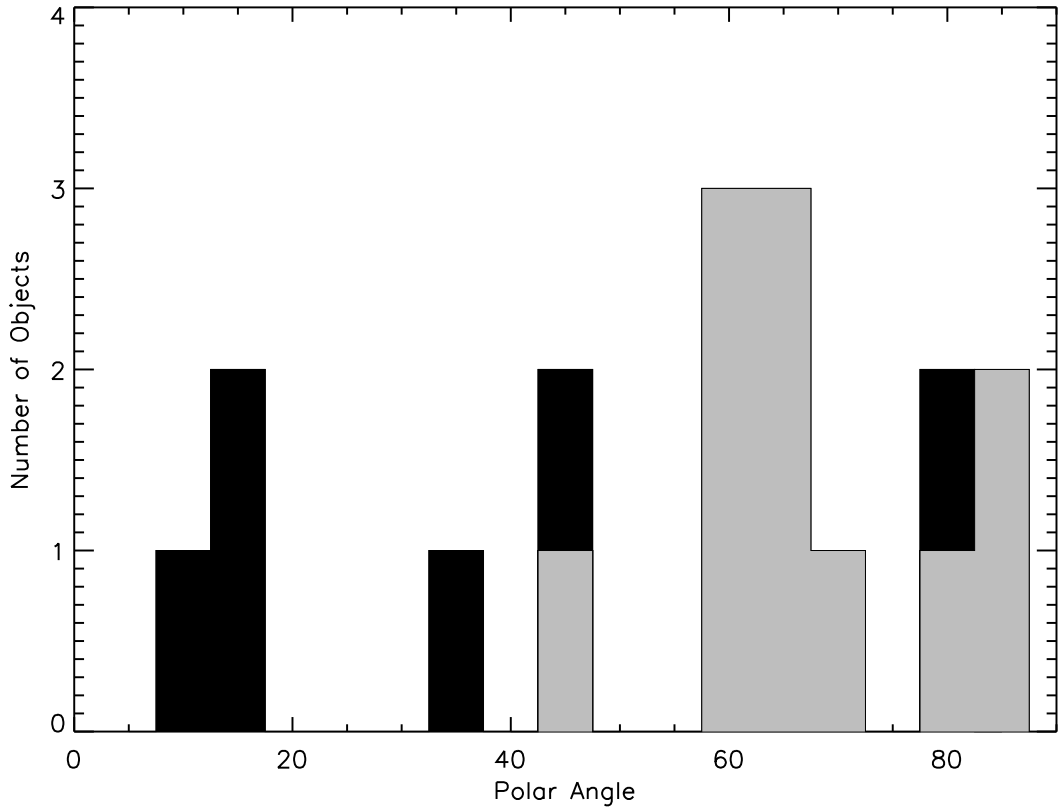}
\caption[Inclination Distribution]{Histogram displaying the distribution of inclinations for Seyfert 1s (black) and 
Seyfert 2s (grey) in our sample, where the bicone axes of Seyfert 1s tend to be less inclined than Seyfert 2s. NGC 
5506, modeled to have a high inclination, may be affected by host disk obscuration.
}
\label{histinc}
\end{figure}

\begin{figure}
\centering
\includegraphics[angle=90,scale=0.65]{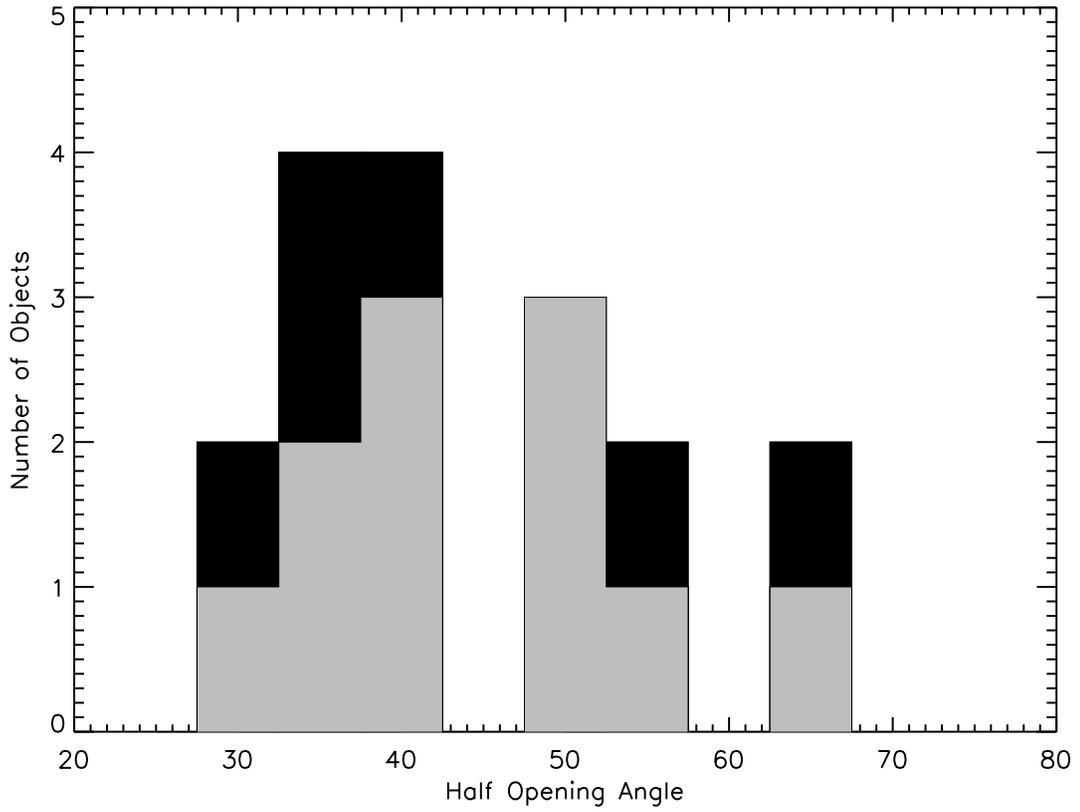}
\caption[Outer Opening Angle Distribution]{Histogram displaying the distribution of opening angles for Seyfert 1s (black) and 
Seyfert 2s (grey) in our sample. Seyfert 1s and Seyfert 2s appear to be evenly distributed.
}
\label{histopen}
\end{figure}

In Figures \ref{histinc} and \ref{histopen} we show the distribution of AGN polar angles 
($= 90^{\circ} - i$, or the angle between the bicone axis and our LOS) and opening 
angles respectively of our modeled sample. Note that while our sample size is small, 
it covers a full range in both inclination and opening angle. The distribution of polar 
angles shows a trend of Seyfert 1s having smaller inclinations from our line of sight and 
Seyfert 2s being more inclined, in nearly complete agreement with the unified model of AGN. 
The sole exception to this trend is the inclination of NGC 5506, 
a Seyfert 1.9 with an inclination near the plane of the sky, which we discuss later. The distribution of 
half-opening angles show Seyfert 1s and Seyfert 2s have no preference for specific angles,
in further agreement with the unified model. The broad range of opening angles points to 
differences in the inner torus-like geometry surrounding each AGN.

\begin{figure}
\includegraphics[angle=90,scale=0.65]{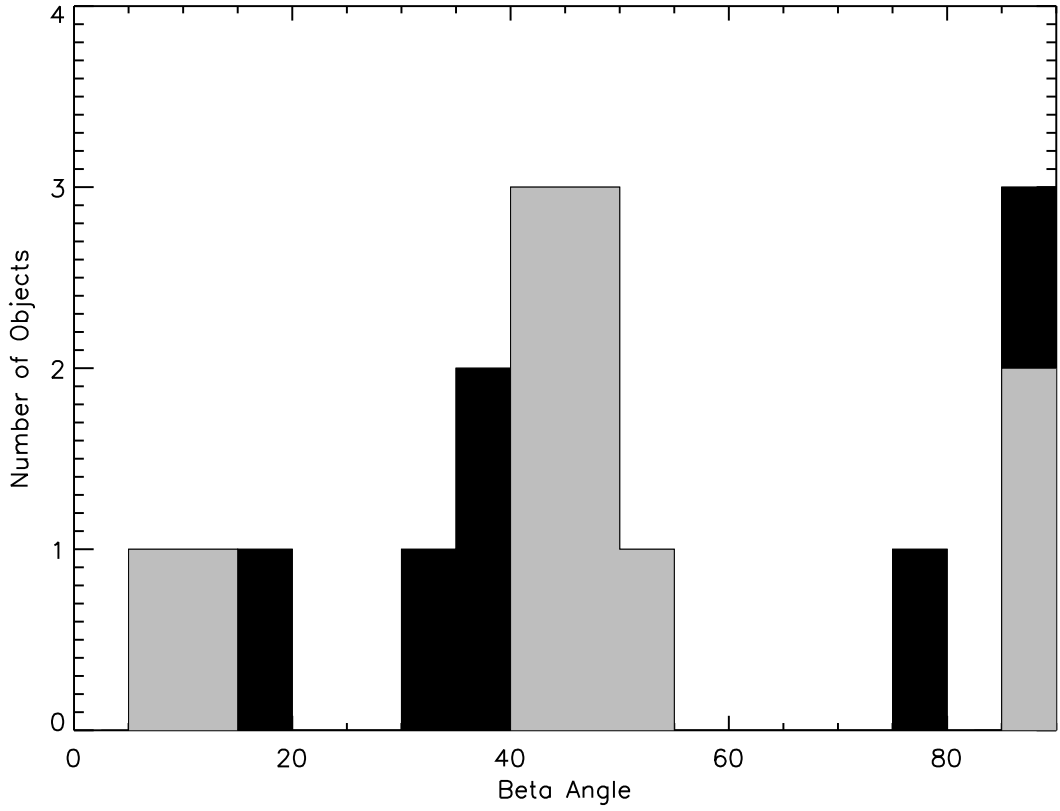}
\caption[Beta Angle Distribution]{Histogram displaying the distribution of $\beta$ angles for Seyfert 1s (black) and 
Seyfert 2s (grey) with modeled kinematics. $\beta$ angles appear homogenously distributed between $0^{\circ}-90^{\circ}$.}
\label{hist_beta}
\end{figure}

Table \ref{modelvals} also gives position angles and inclinations of each host galaxy disk 
required for our geometric models, as well as the angle between the NLR bicone axis and the 
normal to the host galaxy disk ($\beta$). The distribution of these $\beta$ values, even 
within type 1 and 2 subsets (Figure \ref{hist_beta}), shows such diversity that no alignment 
of the NLR is seen relative to the host disk. These observations concur with previous studies 
\citep{Sch97,Cla98,Kin00,Sch03} that compared the P.A. of the projected NLR to the P.A. of the host galaxy major 
axis and found that the distribution of their differences was homogeneous. Alignment between 
[O~III] NLR and radio jet emission implies that the axes of the toroidal structure surrounding 
the AGN and the accretion disk contained within are relatively well aligned, with some exceptions \citep{Sch03}. 
Thus, our $\beta$ angle distribution suggests that the orientation of the gas in the inner torus 
is controlled more by the central SMBH than the galactic disk.

\section{Discussion}
\label{sec8}

While star-formation is a common feature amongst Seyfert galaxies \citep{Cid04,Sar07,Dav07,Kau09}, it remains 
unclear how much this process contributes to the kinematics using lines associated with the NLR in our AGN sample. 
Thus, it is possible that star formation may instigate or contribute to outflows in some AGN. We have therefore 
limited our identification of AGN-driven outflows to those that show the characteristic acceleration/deceleration 
signature, identified in our previous studies of AGN, with little or no nuclear star formation \citep{Mel08}.

Excluding the 5 AGN with published results and the two galaxies
from \citet{Rui05}, the remaining ten AGN in Table 6 have $P.A._{slit} = P.A._{bicone}$ within 
an error $< 10^{\circ}$, with the exception of Mrk 279 and the long-slit observation of NGC 5643, having 
differences of $\sim 30^{\circ}$ and $\sim 50^{\circ}$ respectfully.
However, using value estimates from \citet{Sch96} and \citet{Sch03}, the position angle of the NLR axis in [OIII] imaging, 
$P.A._{[O~III]}$, is only equal to $P.A._{bicone}$ in 5 AGN (Circinus, Mrk 34, Mrk 1066, NGC 4507 and NGC 7674). 
In three galaxies (NGC 3227, NGC 3783 and NGC 4051), the difference between $P.A._{[O~III]}$ and $P.A._{bicone}$ 
is $\sim 15^{\circ}$, and in Mrk 279 the difference is greater than $50^{\circ}$ degrees. 
It is not possible to do this comparison in NGC 1667 due to the lack of [OIII] images for this AGN, 
as mentioned in Section A.5. This relationship, or lack there of, is highly dependent on the ability 
to identify the [OIII] PA in the imaging. The 5 AGN that have small differences between $P.A._{[O~III]}$ 
and $P.A._{bicone}$ are the Seyfert 2s of the sample with extended, well resolved NLRs. The remaining 4 AGN 
with larger differences between position angles are all Seyfert 1 AGN that 
contain compact elliptical or circular NLRs. These NLRs prove much more difficult to 
fit with a position angle, as no conical structure is visible and we are sometimes forced to rely 
on a single asymmetric, extended knot of emission to provide guidance. Assuming 
the morphology of these Sy1 NLRs are due to the orientation of their biconical NLR lying 
along our line of sight, these extensions may only be bright knots of emission (similar to Mrk 78; \citealt{Fis11})
within the otherwise radially symmetric NLR.

While determining inclinations for a portion of the overall AGN sample is encouraging, a question now presents itself 
as to why we cannot model the remaining 35 AGN in our sample. One answer is poor positioning of the STIS long-slit. 
It is clear that four AGN with both [O~III] spectra and imaging, Mrk 348, NGC 3081, 
NGC 5252, and NGC 5347 (Compact, Complex, Ambiguous, and Compact kinematics respectively) 
have long-slit positions that are not optimally placed inside the bicone region. Ground-based [O~III] imaging \citep{Mul96} 
of NGC 1358 and NGC 2110 (both Ambiguous kinematics) and pre-COSTAR {\it HST} WFPC \citep{Sch96} and OASIS IFU \citep{Sto09} 
[O~III] imaging of NGC 5929 also show a misalignment between NLR and STIS slit. OSIRIS IFU observations of NGC 7469 \citep{Mul11}, 
which in {\it HST} [O~III] imaging appears to have a circular NLR, show that [Si~VI] NLR kinematics have the highest 
velocities along a position angle of $\sim 90^{\circ}$ which is misaligned with the {\it HST} STIS slit position by $\sim 65^{\circ}$.
In addition, five $''$Ambiguous$''$ AGN (IC 3639, Mrk 493, NGC 5283, NGC 7682, UM 146) and three $''$Complex$''$ AGN (NGC 5427, 
NGC 5695, NGC 6300) may also have observations where the slit position is misaligned with their NLR, however this cannot be 
confirmed due to a lack of [O~III] images for these targets.

Biconical outflow may still exist in a portion of the remaining targets if the required kinematic components for our models 
are not visible because they are too compact, faint, or convolved with other components like rotation. 
Developing a rotational kinematic model to fit the Ambiguous and Complex AGN spectra \citep{Bar09,Mul11,Rif11} could 
provide insight on the true source of their kinematics as fitting and removing all AGN exhibiting rotational kinematics would allow us to search 
for possible remaining outflow components in the remaining AGN. As these data already exist, improving our 
models will be the most effective way to illuminate NLR kinematics moving forward without needing additional 
observations. AGN with truly Compact NLRs have a more clear cut dilemma: with high NLR velocities only observed in 
less than a 0.5$''$ diameter over the nucleus, we simply do not have enough spectral data to attempt fitting individual 
kinematic components. 

\section{Conclusions}

We have measured radial velocities across NLRs for 47 Seyfert AGN observed with {\it HST}/STIS 
G430L/M gratings and combined these with our published measurements of 5 Seyferts with outflows. 
From our measurements, we found an additional 12 AGN that contained kinematics characteristic of 
biconical AGN outflow. Comparing the inclination of the NLR outflow to the normal of the host disk to 
calculate a $\beta$ angle, we found a distribution that suggests that the orientation of the gas in 
the torus and that of the host disk are independent from one another.

With inclinations and geometries 
of 17 Seyfert galaxies, our model results agree with the unified model in that Seyfert 1 AGN are inclined 
further toward our LOS than Seyfert 2 AGN, save for the NLS1 NGC 5506 (Section \ref{5506}). Modeling 
the kinematics of this AGN results in a NLR cone inclined 10$^{\circ}$ from the plane of the sky. 
This contrasts with the unified model as the torus of absorbing material should obscure the BLR emission 
from our LOS at this orientation. As the host galaxy of this AGN is highly inclined (76$^{\circ}$), it is 
possible that obscuration from the host disk is extinguishing regions of extended emission and that we are 
only modeling what remains of the NLR. Additionally, we have determined a correlation between $i$ and neutral 
hydrogen column density ($N_H$) (Fischer et al. in prep) where AGN observed further from their NLR axis and 
closer to the obscuring torus are observed to have larger column densities. NGC 5506 has an $N_H$ orders of 
magnitude lower than all other AGN at high inclinations, implying that it should have an inclination closer 
to our LOS.

Similar to observations of Mrk 3 and Mrk 573, multiple observed NLRs (IRAS 11058-1131, Mrk 34, NGC 3393, NGC 3516, NGC 5252) 
have morphologies consistent with intersections between the NLR ionizing radiation and the host disk \citep{Mul96}. 
It is likely that for these targets, the emitting gas did not originate in the nucleus of the AGN, but 
rather it was accelerated off the host disk. Does NLR emitting gas primarily originate in the nucleus or is it 
mainly due to in situ acceleration? Perhaps the NLR results primarily from a stream of ionizing radiation 
unleashed upon the galaxy, ionizing and accelerating ambient material that it encounters? The NLR in Circinus, 
which is nearly perpendicular to the host galaxy, is only 35 pc in height. As our own Milky Way has a scale height 
of less than 100 pc, we could be seeing the matter-bounded ionization of an abnormally rich \citep{Fre77} 
host disk above the AGN.

Knowing the inclinations of AGN will allow us to determine how their observed properties vary 
as a function of polar angle with respect to the accretion disk and/or torus axes. As our technique for 
determining inclinations of AGN is not a simple task, correlating inclination with parameters that 
are more easily observable could provide astronomers a way to estimate the inclination of an AGN without 
modeling its NLR kinematics. Thus, the motivation for future study then is to expand beyond our current 
results by 1) determining the inclinations of the NLRs in a much larger sample of AGN, and 2) determining 
the multi-wavelength properties of these AGN to identify correlations that will probe the structure 
of the AGN components.

\acknowledgments

TCF thanks M.C. Bentz, H.R. Miller, R.J. White, and P.J. Wiita for the useful discussions, and the anonymous referee 
for their practical suggestions. Some of the data 
presented in this paper were obtained from the Mikulski Archive for Space Telescopes (MAST). STScI is 
operated by the Association of Universities for Research in Astronomy, Inc., under NASA contract NAS5-26555. 
This research has also made use of the NASA/IPAC Extragalactic Database (NED) which is operated by the Jet 
Propulsion Laboratory, California Institute of Technology, under contract with the National Aeronautics and 
Space Administration.

%TABLES%%%%%%%%%%%%%%%%%%%%%%%%%%%%%%%%
\begin{landscape}
\begin{deluxetable}{cccccccccc}
\tablecolumns{2}
\footnotesize
\tablecaption{{\it HST}/STIS Grating Specifications}
\tablewidth{0pt}
\tablehead{\colhead{Grating} & \multicolumn{2}{c}{Spectral Range (\AA)} & \colhead{Target} & \colhead{Resolving Power}             & \colhead{Resolving Power}   & \colhead{$\sigma_{hi}$\tablenotemark{c}} & \colhead{$\sigma_{noise}$\tablenotemark{b}} & \colhead{$\sigma_{total}$\tablenotemark{c}}& \colhead{$\sigma_{total}$}   \\
                             & \colhead{Complete}   & \colhead{Per Tilt}         & \colhead{Line}   & \colhead{($\lambda / \Delta\lambda$)} & \colhead{(km s$^{-1}$ FWHM)}& \colhead{(\AA)}         & \colhead{(\AA)}            & \colhead{(\AA)}           & \colhead{(km s$^{-1}$)}}
\startdata
G430L & 2900-5700   & 2800 &   [O~III] $\lambda$5007   & $\sim$900  & $\sim$330  & 0.26 & 0.79 & 0.83  & 49.7\\
G430M & 3020-5610   &  286 &   [O~III] $\lambda$5007   & $\sim$9000 & $\sim$30   & 0.11 & 0.13 & 0.17  & 10.2\\
G750M & 5450-10,140 &  570 &   H$\alpha$ $\lambda$6563 & $\sim$5900 & $\sim$50   & 0.07 & 0.34 & 0.35  & 20.9
\enddata
\label{gratspecs}
\tablenotetext{a}{Maximum error due to fitting a Gaussian to an emission line}
\tablenotetext{b}{Maximum error due to noise}
\tablenotetext{c}{Total Maximum error}
\end{deluxetable}
\end{landscape}

\clearpage

\begin{landscape}
%\pagestyle{empty}
%\hfill \thepage
\renewcommand{\thefootnote}{\alph{footnote}}
%\scriptsize
\begin{longtable}{lccccccrcr}
\caption[Expanded Sample: {\it HST}/STIS Observations]{Expanded Sample: {\it HST}/STIS Observations}
\label{spectratable}\\
\hline
\multicolumn{1}{l}{Target}          &
\multicolumn{1}{c}{STScI Data Set}   &
\multicolumn{1}{c}{Source}           &
\multicolumn{1}{c}{Date}             &
\multicolumn{1}{c}{Grating}          &
\multicolumn{1}{c}{Aperture}         &
\multicolumn{1}{c}{Central $\lambda$}&
\multicolumn{1}{c}{P.A.}             &
\multicolumn{1}{c}{Offset}           &
\multicolumn{1}{c}{Exp.}         \\
\multicolumn{1}{l}{}                &
\multicolumn{1}{c}{}                 &
\multicolumn{1}{c}{}                 &
\multicolumn{1}{c}{(UT)}             &
\multicolumn{1}{c}{}                 &
\multicolumn{1}{c}{}                 &
\multicolumn{1}{c}{(\AA)}            &
\multicolumn{1}{c}{(deg)}            &
\multicolumn{1}{c}{(arcsec)}         &
\multicolumn{1}{c}{(s)}              \\
\hline
\endfirsthead
\multicolumn{9}{l}%
{ \tablename\ \thetable{} Continued from previous page} &  \\
\hline
\multicolumn{1}{l}{Target}          &
\multicolumn{1}{c}{STScI Data Set}   &
\multicolumn{1}{c}{Source\tablenotemark{a}}           &
\multicolumn{1}{c}{Date}             &
\multicolumn{1}{c}{Grating}          &
\multicolumn{1}{c}{Aperture}         &
\multicolumn{1}{c}{Central $\lambda$}&
\multicolumn{1}{c}{P.A.}             &
\multicolumn{1}{c}{Offset\tablenotemark{b}}           &
\multicolumn{1}{c}{Exposure}         \\
\multicolumn{1}{l}{}                &
\multicolumn{1}{c}{}                 &
\multicolumn{1}{c}{}                 &
\multicolumn{1}{c}{(UT)}             &
\multicolumn{1}{c}{}                 &
\multicolumn{1}{c}{}                 &
\multicolumn{1}{c}{(\AA)}            &
\multicolumn{1}{c}{(deg)}            &
\multicolumn{1}{c}{(arcsec)}         &
\multicolumn{1}{c}{(s)}              \\ 
\hline
\endhead
\hline \multicolumn{9}{l}{Continued on next page} \\ \hline
\endfoot
\hline \hline
\endlastfoot
Akn 564        & OBGU08010 & C & 2011 Jul 31 & G430M & 52$\times$0.2 & 5093 & 159.65 &  0.0 &  695 \\
               & OBGU08010 & C & 2011 Jul 31 & G430M & 52$\times$0.2 & 5093 & 159.65 &  0.0 &  695 \\
               & OBGU08010 & C & 2011 Jul 31 & G430M & 52$\times$0.2 & 5093 & 159.65 &  0.0 &  695 \\
Circinus       & O65B01040 & A & 2000 Oct 12 & G430M & 52$\times$0.2 & 4961 & 113.65 & -0.6 & 1400 \\
               & O65B01050 & A & 2000 Oct 12 & G430M & 52$\times$0.2 & 4961 & 113.65 & -0.4 & 1443 \\
               & O65B01060 & A & 2000 Oct 12 & G430M & 52$\times$0.2 & 4961 & 113.65 & -0.2 & 1400 \\
               & O65B01070 & A & 2000 Oct 12 & G430M & 52$\times$0.2 & 4961 & 113.65 &  0.0 & 1495 \\
               & O65B01080 & A & 2000 Oct 12 & G430M & 52$\times$0.2 & 4961 & 113.65 & +0.2 & 1400 \\
               & O65B01090 & A & 2000 Oct 12 & G430M & 52$\times$0.2 & 4961 & 113.65 & +0.4 & 1492 \\
               & O65B03020 & A & 2002 Jan 04 & G750M & 52$\times$0.2 & 6581 &-145.35 &  0.0 & 2120 \\
IC 3639	       & O6BU01010 & A & 2002 Jan 23 & G750M & 52$\times$0.2 & 6581 &-104.41 &  0.0 & 1080 \\
               & O6BU01020 & A & 2002 Jan 23 & G750M & 52$\times$0.2 & 6581 &-104.41 &  0.0 & 1158 \\
               & O6BU01030 & A & 2002 Jan 23 & G750M & 52$\times$0.2 & 6581 &-104.41 &  0.0 &  900 \\ 
               & O6BU01040 & A & 2002 Jan 23 & G430L & 52$\times$0.2 & 4300 &-104.41 &  0.0 &  840 \\
               & O6BU01050 & A & 2002 Jan 23 & G430L & 52$\times$0.2 & 4300 &-104.41 &  0.0 &  823 \\
IRAS 11058     & O56C03050 & A & 1999 Apr 08 & G430L & 52$\times$0.2 & 4300 &  36.03 &  0.0 &  600 \\
MCG-6-30-15    & O5GU08010 & A & 2000 Mar 07 & G430M & 52$\times$0.2 & 4961 & -75.02 &  0.0 & 1707 \\
               & O5GU08020 & A & 2000 Mar 07 & G430M & 52$\times$0.2 & 4961 & -75.02 &  0.0 & 2511 \\
               & O5GU08030 & A & 2000 Mar 07 & G430M & 52$\times$0.2 & 4961 & -75.02 & +0.2 & 1250 \\
               & O5GU08040 & A & 2000 Mar 07 & G430M & 52$\times$0.2 & 4961 & -75.02 & +0.2 & 1192 \\
               & O5GU08050 & A & 2000 Mar 07 & G430M & 52$\times$0.2 & 4961 & -75.02 & -0.2 & 1250 \\
               & O5GU08060 & A & 2000 Mar 07 & G430M & 52$\times$0.2 & 4961 & -75.02 & -0.2 & 1189 \\
Mrk 34	       & O5G404010 & A & 2000 Feb 17 & G430M & 52$\times$0.2 & 5216 & 152.48 &  0.0 & 1500 \\
               & O5G404010 & A & 2000 Feb 17 & G430L & 52$\times$0.2 & 4300 & 152.48 &  0.0 &  627 \\
               & O5G404010 & A & 2000 Feb 17 & G430M & 52$\times$0.2 & 5216 & 152.48 & +0.28 & 1460 \\
               & O5G404010 & A & 2000 Feb 17 & G430M & 52$\times$0.2 & 5216 & 152.48 & -0.28 & 1460 \\
Mrk 279	       & OBGU05010 & C & 2011 May 10 & G430M & 52$\times$0.2 & 5093 & 124.65 &  0.0 &  712 \\
               & OBGU05020 & C & 2011 May 10 & G430M & 52$\times$0.2 & 5093 & 124.65 &  0.0 &  712 \\
               & OBGU05030 & C & 2011 May 10 & G430M & 52$\times$0.2 & 5093 & 124.65 &  0.0 &  712 \\
Mrk 348	       & O5G405010 & A & 1999 Sep 28 & G430M & 52$\times$0.2 & 5093 & 145.98 &  0.0 & 1410 \\
               & O5G405020 & A & 1999 Sep 28 & G430L & 52$\times$0.2 & 4300 & 145.98 &  0.0 &  600 \\
Mrk 463e       & O5G406010 & A & 2000 Mar 14 & G430M & 52$\times$0.2 & 5216 &-178.03 &  0.0 & 1200 \\
               & O5G406020 & A & 2000 Mar 14 & G430L & 52$\times$0.2 & 4300 &-178.03 &  0.0 &  556 \\
Mrk 493	       & O92X16010 & A & 2004 Jul 28 & G430L & 52$\times$0.2 & 4300 &  74.65 &  0.0 &  720 \\
               & O92X16020 & A & 2004 Jul 28 & G430L & 52$\times$0.2 & 4300 &  74.65 &  0.0 &  720 \\
               & O92X16030 & A & 2004 Jul 28 & G430L & 52$\times$0.2 & 4300 &  74.65 &  0.0 &  720 \\
Mrk 509        & OBGU07010 & C & 2011 Aug 07 & G430M & 52$\times$0.2 & 5093 &  74.65 &  0.0 &  695 \\
               & OBGU07020 & C & 2011 Aug 07 & G430M & 52$\times$0.2 & 5093 &  74.65 &  0.0 &  695 \\
               & OBGU07030 & C & 2011 Aug 07 & G430M & 52$\times$0.2 & 5093 &  74.65 &  0.0 &  695 \\
Mrk 705	       & OB1105010 & B & 2010 Dec 10 & G430M & 52$\times$0.1 & 5093 &-100.35 &  0.0 & 2148 \\
               & OB1106010 & B & 2011 Feb 10 & G430M & 52$\times$0.1 & 5093 &   7.92 &  0.0 & 2148 \\
Mrk 766	       & OB1101010 & B & 2010 Mar 21 & G430M & 52$\times$0.1 & 5093 & 129.65 &  0.0 & 2148 \\
               & OB1102010 & B & 2010 Dec 23 & G430M & 52$\times$0.1 & 5093 &-120.35 &  0.0 & 2148 \\
Mrk 1040       & OB1103010 & B & 2009 Jul 02 & G430M & 52$\times$0.1 & 5093 &-144.20 &  0.0 & 2148 \\
               & OB1104010 & B & 2009 Nov 14 & G430M & 52$\times$0.1 & 5093 & 119.65 &  0.0 & 2148 \\
Mrk 1044       & OBGU01010 & C & 2011 Nov 06 & G430M & 52$\times$0.2 & 5093 & -36.23 &  0.0 &  695 \\
               & OBGU02010 & C & 2011 Nov 06 & G430M & 52$\times$0.2 & 5093 & -36.23 &  0.0 &  695 \\
               & OBGU03010 & C & 2011 Nov 06 & G430M & 52$\times$0.2 & 5093 & -36.23 &  0.0 &  695 \\
Mrk 1066       & O5G407010 & A & 2000 Oct 30 & G430M & 52$\times$0.2 & 4961 & 130.65 &  0.0 & 1440 \\
               & O5G407020 & A & 2000 Oct 30 & G430L & 52$\times$0.2 & 4300 & 130.65 &  0.0 &  600 \\
NGC 1358       & O6BU03010 & A & 2002 Jan 25 & G750M & 52$\times$0.2 & 6581 &  23.89 &  0.0 & 1080 \\
               & O6BU03020 & A & 2002 Jan 25 & G750M & 52$\times$0.2 & 6581 &  23.89 &  0.0 & 1080 \\
               & O6BU03030 & A & 2002 Jan 25 & G750M & 52$\times$0.2 & 6581 &  23.89 &  0.0 &  840 \\
               & O6BU03040 & A & 2002 Jan 25 & G430L & 52$\times$0.2 & 4300 &  23.89 &  0.0 &  840 \\
               & O6BU03050 & A & 2002 Jan 25 & G430L & 52$\times$0.2 & 4300 &  23.89 &  0.0 &  805 \\
NGC 1386       & O5F402030 & D & 2000 Jun 23 & G430M & 50CCD         & 4961 & 175.82 &  0.0 & 2106 \\
NGC 1667       & O6BU04010 & A & 2001 Oct 14 & G750M & 52$\times$0.2 & 6581 &-120.21 &  0.0 & 1080 \\
               & O6BU04020 & A & 2001 Oct 14 & G750M & 52$\times$0.2 & 6581 &-120.21 &  0.0 & 1080 \\
               & O6BU04030 & A & 2001 Oct 14 & G750M & 52$\times$0.2 & 6581 &-120.21 &  0.0 &  840 \\
               & O6BU04040 & A & 2001 Oct 14 & G430L & 52$\times$0.2 & 4300 &-120.21 &  0.0 &  840 \\
               & O6BU04050 & A & 2001 Oct 14 & G430L & 52$\times$0.2 & 4300 &-120.21 &  0.0 &  805 \\
NGC 2110       & O5G401010 & A & 2000 Dec 24 & G430M & 52$\times$0.2 & 4961 & -36.20 &  0.0 & 1522 \\
               & O5G401020 & A & 2000 Dec 24 & G430M & 52$\times$0.2 & 4961 & -36.20 & +0.68 &  600 \\
               & O64F02010 & A & 2000 Dec 30 & G750M & 52$\times$0.2 & 6581 & -24.35 &  0.0 & 1440 \\
               & O64F02020 & A & 2000 Dec 30 & G750M & 52$\times$0.2 & 6581 & -24.35 &  0.0 & 1440 \\
               & O64F02030 & A & 2000 Dec 30 & G750M & 52$\times$0.2 & 6581 & -24.35 &  0.0 & 1440 \\
NGC 2273       & O6BU05010 & A & 2001 Nov 04 & G750M & 52$\times$0.2 & 6581 &-151.61 &  0.0 & 1140 \\
               & O6BU05020 & A & 2001 Nov 04 & G750M & 52$\times$0.2 & 6581 &-151.61 &  0.0 & 1226 \\
               & O6BU05030 & A & 2001 Nov 04 & G750M & 52$\times$0.2 & 6581 &-151.61 &  0.0 &  900 \\
               & O6BU05040 & A & 2001 Nov 04 & G430L & 52$\times$0.2 & 4300 &-151.61 &  0.0 &  840 \\
               & O6BU05050 & A & 2001 Nov 04 & G430L & 52$\times$0.2 & 4300 &-151.61 &  0.0 &  951 \\
NGC 3081       & O6BU06010 & A & 2001 Dec 04 & G750M & 52$\times$0.2 & 6581 &-110.52 &  0.0 & 1080 \\
               & O6BU06020 & A & 2001 Dec 04 & G750M & 52$\times$0.2 & 6581 &-110.52 &  0.0 & 1080 \\
               & O6BU06030 & A & 2001 Dec 04 & G750M & 52$\times$0.2 & 6581 &-110.52 &  0.0 &  840 \\
               & O6BU06040 & A & 2001 Dec 04 & G430L & 52$\times$0.2 & 4300 &-110.52 &  0.0 &  840 \\
               & O6BU06050 & A & 2001 Dec 04 & G430L & 52$\times$0.2 & 4300 &-110.52 &  0.0 &  812 \\
NGC 3227       & O57204010 & A & 1999 Jan 31 & G750M & 52$\times$0.2 & 6581 &-137.62 & -0.75 & 2105 \\
               & O57204020 & A & 1999 Jan 31 & G750M & 52$\times$0.2 & 6581 &-137.62 & -0.5  & 1600 \\ 
               & O57204030 & A & 1999 Jan 31 & G750M & 52$\times$0.2 & 6581 &-137.62 & -0.25 & 1884 \\
               & O57204040 & A & 1999 Jan 31 & G750M & 52$\times$0.2 & 6581 &-137.62 &  0 0 & 1890 \\
               & O57204050 & A & 1999 Jan 31 & G750M & 52$\times$0.2 & 6581 &-137.62 & +0.25 & 1600 \\ 
               & O57204060 & A & 1999 Jan 31 & G750M & 52$\times$0.2 & 6581 &-137.62 & +0.5  & 1884 \\
               & O57204070 & A & 1999 Jan 31 & G750M & 52$\times$0.2 & 6581 &-137.62 & +0.75 & 1887 \\
               & O5KP01020 & A & 2000 Feb 08 & G430L & 52$\times$0.2 & 4300 &-150.34 &  0.0 &  120 \\
NGC 3393       & O56C02010 & A & 1999 Apr 22 & G750M & 52$\times$0.2 & 6581 &  39.98 & -0.3 & 1080 \\
               & O56C02030 & A & 1999 Apr 22 & G750M & 52$\times$0.2 & 6581 &  39.98 &  0.0 &  865 \\ 
               & O56C02040 & A & 1999 Apr 22 & G750M & 52$\times$0.2 & 6581 &  39.98 &  0.0 &  600 \\
               & O56C02050 & A & 1999 Apr 22 & G430L & 52$\times$0.2 & 4300 &  39.98 &  0.0 &  600 \\ 
               & O56C02060 & A & 1999 Apr 22 & G750M & 52$\times$0.2 & 6581 &  39.98 & +0.3 & 1021 \\
NGC 3516       & O5F406030 & D & 2000 Jan 18 & G430M & 50CCD         & 5093 &-154.98 &  0.0 & 2154 \\
               & O56C01050 & A & 2000 Jun 18 & G430L & 52$\times$0.2 & 4300 &  38.98 &  0.0 &  600 \\
NGC 3783       & OBGU03010 & C & 2011 Mar 23 & G430M & 52$\times$0.2 & 4961 & -20.35 &  0.0 &  696 \\
               & OBGU03020 & C & 2011 Mar 23 & G430M & 52$\times$0.2 & 4961 & -20.35 &  0.0 &  696 \\
NGC 4051       & O5G402010 & A & 2000 Apr 15 & G430M & 52$\times$0.2 & 4961 &  89.78 & -0.05& 1796 \\
               & O5G402020 & A & 2000 Apr 15 & G430M & 52$\times$0.2 & 4961 &  89.78 & +0.2 &  600 \\
NGC 4303       & O6LC01010 & A & 2003 Mar 04 & G430M & 52$\times$0.2 & 4961 &-140.35 &  0.0 & 2156 \\
               & O6LC01020 & A & 2003 Mar 04 & G430L & 52$\times$0.2 & 4961 &-140.35 &  0.0 & 1200 \\
NGC 4395       & OBGU04010 & C & 2011 May 25 & G430M & 52$\times$0.2 & 4961 &  64.65 &  0.0 &  693 \\
               & OBGU04020 & C & 2011 May 25 & G430M & 52$\times$0.2 & 4961 &  64.65 &  0.0 &  693 \\
               & OBGU04030 & C & 2011 May 25 & G430M & 52$\times$0.2 & 4961 &  64.65 &  0.0 &  693 \\
NGC 4507       & O5DF03010 & A & 2001 Apr 04 & G430M & 52$\times$0.2 & 4961 & -34.35 &  0.0 & 1440 \\
               & O5DF03010 & A & 2001 Apr 04 & G430L & 52$\times$0.2 & 4300 & -34.35 &  0.0 &  624 \\
NGC 5135       & O6BU07010 & A & 2002 Jan 11 & G750M & 52$\times$0.2 & 6581 &-115.81 &  0.0 & 1080 \\
               & O6BU07020 & A & 2002 Jan 11 & G750M & 52$\times$0.2 & 6581 &-115.81 &  0.0 & 1104 \\
               & O6BU07030 & A & 2002 Jan 11 & G750M & 52$\times$0.2 & 6581 &-115.81 &  0.0 &  832 \\
               & O6BU07040 & A & 2002 Jan 11 & G430L & 52$\times$0.2 & 4300 &-115.81 &  0.0 &  840 \\
               & O6BU07040 & A & 2002 Jan 11 & G430L & 52$\times$0.2 & 4300 &-115.81 &  0.0 &  829 \\
NGC 5252       & O56C04010 & A & 1999 Jan 27 & G750M & 52$\times$0.2 & 6581 &-135.52 & -0.2 & 1080 \\
               & O56C04030 & A & 1999 Jan 27 & G750M & 52$\times$0.2 & 6581 &-135.52 &  0.0 &  841 \\
               & O56C04040 & A & 1999 Jan 27 & G750M & 52$\times$0.2 & 6581 &-135.52 &  0.0 &  600 \\
               & O56C04050 & A & 1999 Jan 27 & G430L & 52$\times$0.2 & 4300 &-135.52 &  0.0 &  600 \\ 
               & O56C04060 & A & 1999 Jan 27 & G750M & 52$\times$0.2 & 6581 &-135.52 & +0.2 &  997 \\
NGC 5283       & O6BU08010 & A & 2001 Oct 11 & G750M & 52$\times$0.2 & 6581 & -37.06 &  0.0 & 1200 \\
               & O6BU08020 & A & 2001 Oct 11 & G750M & 52$\times$0.2 & 6581 & -37.06 &  0.0 & 1213 \\
               & O6BU08030 & A & 2001 Oct 11 & G750M & 52$\times$0.2 & 6581 & -37.06 &  0.0 &  900 \\
               & O6BU08040 & A & 2001 Oct 11 & G430L & 52$\times$0.2 & 4300 & -37.06 &  0.0 &  900 \\
               & O6BU08050 & A & 2001 Oct 11 & G430L & 52$\times$0.2 & 4300 & -37.06 &  0.0 &  900 \\
NGC 5347       & O6BU09010 & A & 2001 Dec 24 & G750M & 52$\times$0.2 & 6581 &-102.20 &  0.0 & 1080 \\
               & O6BU09020 & A & 2001 Dec 24 & G750M & 52$\times$0.2 & 6581 &-102.20 &  0.0 & 1126 \\
               & O6BU09030 & A & 2001 Dec 25 & G750M & 52$\times$0.2 & 6581 &-102.20 &  0.0 &  840 \\
               & O6BU09040 & A & 2001 Dec 25 & G430L & 52$\times$0.2 & 4300 &-102.20 &  0.0 &  840 \\
               & O6BU09050 & A & 2001 Dec 25 & G430L & 52$\times$0.2 & 4300 &-102.20 &  0.0 &  851 \\
NGC 5427       & O6BU10010 & A & 2002 Jan 04 & G750M & 52$\times$0.2 & 6581 &-113.38 &  0.0 & 1080 \\
               & O6BU10020 & A & 2002 Jan 04 & G750M & 52$\times$0.2 & 6581 &-113.38 &  0.0 & 1080 \\
               & O6BU10030 & A & 2002 Jan 04 & G750M & 52$\times$0.2 & 6581 &-113.38 &  0.0 &  840 \\
               & O6BU10040 & A & 2002 Jan 04 & G430L & 52$\times$0.2 & 4300 &-113.38 &  0.0 &  840 \\
               & O6BU10050 & A & 2002 Jan 04 & G430L & 52$\times$0.2 & 4300 &-113.38 &  0.0 &  805 \\
NGC 5506       & O5F407030 & D & 2000 Mar 18 & G430M & 50CCD         & 4961 &-153.39 &  0.0 & 2096 \\
NGC 5548       & OBGU06010 & C & 2011 Mar 11 & G430M & 52$\times$0.2 & 5093 &-160.35 &  0.0 &  695 \\
               & OBGU06020 & C & 2011 Mar 11 & G430M & 52$\times$0.2 & 5093 &-160.35 &  0.0 &  695 \\
               & OBGU06030 & C & 2011 Mar 11 & G430M & 52$\times$0.2 & 5093 &-160.35 &  0.0 &  695 \\
NGC 5643       & O5F408030 & D & 2000 Feb 23 & G430M & 50CCD         & 4961 & -99.98 &  0.0 & 2107 \\
               & O6BU11010 & A & 2000 Mar 12 & G750M & 52$\times$0.2 & 6581 &-128.04 &  0.0 & 1140 \\
               & O6BU11020 & A & 2000 Mar 12 & G750M & 52$\times$0.2 & 6581 &-128.04 &  0.0 & 1223 \\
               & O6BU11030 & A & 2000 Mar 12 & G750M & 52$\times$0.2 & 6581 &-128.04 &  0.0 &  900 \\
               & O6BU11040 & A & 2000 Mar 12 & G430L & 52$\times$0.2 & 4300 &-128.04 &  0.0 &  840 \\
               & O6BU11050 & A & 2000 Mar 12 & G430L & 52$\times$0.2 & 4300 &-128.04 &  0.0 &  868 \\
NGC 5695       & O6BU12010 & A & 2001 Aug 11 & G750M & 52$\times$0.2 & 6581 &  50.65 &  0.0 & 1080 \\
               & O6BU12020 & A & 2001 Aug 11 & G750M & 52$\times$0.2 & 6581 &  50.65 &  0.0 & 1158 \\
               & O6BU12030 & A & 2001 Aug 11 & G750M & 52$\times$0.2 & 6581 &  50.65 &  0.0 &  900 \\
               & O6BU12040 & A & 2001 Aug 11 & G430L & 52$\times$0.2 & 4300 &  50.65 &  0.0 &  840 \\
               & O6BU12050 & A & 2001 Aug 11 & G430L & 52$\times$0.2 & 4300 &  50.65 &  0.0 &  823 \\
NGC 5728       & O5F409030 & D & 2000 Apr 24 & G430M & 50CCD         & 5093 & -79.98 &  0.0 & 2110 \\
NGC 5929       & O5G403010 & A & 2000 Feb 07 & G430M & 52$\times$0.2 & 4961 &-134.62 &  0.0 & 1524 \\
NGC 6300       & O6BU13010 & A & 2001 Nov 08 & G750M & 52$\times$0.2 & 6581 &  90.28 &  0.0 & 1140 \\
               & O6BU13020 & A & 2001 Nov 08 & G750M & 52$\times$0.2 & 6581 &  90.28 &  0.0 & 1226 \\
               & O6BU13030 & A & 2001 Nov 08 & G750M & 52$\times$0.2 & 6581 &  90.28 &  0.0 &  840 \\
               & O6BU13040 & A & 2001 Nov 08 & G430L & 52$\times$0.2 & 4300 &  90.28 &  0.0 &  900 \\
               & O6BU13050 & A & 2001 Nov 08 & G430L & 52$\times$0.2 & 4300 &  90.28 &  0.0 &  951 \\
NGC 7212       & O5F410030 & D & 2000 Jun 04 & G430M & 50CCD         & 5093 &-174.98 &  0.0 & 2112 \\
NGC 7469       & OBGU09010 & C & 2010 Oct 05 & G430M & 52$\times$0.1 & 5093 &  34.65 &  0.0 &  697 \\
               & OBGU09020 & C & 2010 Oct 05 & G430M & 52$\times$0.1 & 5093 &  34.65 &  0.0 &  697 \\
               & OBGU09030 & C & 2010 Oct 05 & G430M & 52$\times$0.1 & 5093 &  34.65 &  0.0 &  697 \\
NGC 7674       & O5DF04010 & A & 2000 Sep 12 & G430M & 52$\times$0.2 & 5093 & 124.65 &  0.0 & 1140 \\
               & O5DF04020 & A & 2000 Sep 12 & G430L & 52$\times$0.2 & 4300 & 124.65 &  0.0 &  600 \\ 
NGC 7682       & O6BU14010 & A & 2001 Oct 23 & G750M & 52$\times$0.2 & 6581 &  17.85 &  0.0 & 1080 \\
               & O6BU14020 & A & 2001 Oct 23 & G750M & 52$\times$0.2 & 6581 &  17.85 &  0.0 & 1080 \\
               & O6BU14030 & A & 2001 Oct 23 & G750M & 52$\times$0.2 & 6581 &  17.85 &  0.0 &  840 \\
               & O6BU14040 & A & 2001 Oct 23 & G430L & 52$\times$0.2 & 4300 &  17.85 &  0.0 &  840 \\
               & O6BU14050 & A & 2001 Oct 23 & G430L & 52$\times$0.2 & 4300 &  17.85 &  0.0 &  805 \\
NGC 788	       & O6BU15010 & A & 2001 Sep 17 & G750M & 52$\times$0.2 & 6581 &-129.56 &  0.0 & 1080 \\
               & O6BU15020 & A & 2001 Sep 17 & G750M & 52$\times$0.2 & 6581 &-129.56 &  0.0 & 1080 \\
               & O6BU15030 & A & 2001 Sep 17 & G750M & 52$\times$0.2 & 6581 &-129.56 &  0.0 &  840 \\
               & O6BU15040 & A & 2001 Sep 17 & G430L & 52$\times$0.2 & 4300 &-129.56 &  0.0 &  840 \\
               & O6BU15050 & A & 2001 Sep 17 & G430L & 52$\times$0.2 & 4300 &-129.56 &  0.0 &  805 \\
UM 146	       & O6BU16010 & A & 2001 Dec 20 & G750M & 52$\times$0.2 & 6581 &  21.33 &  0.0 & 1080 \\
               & O6BU16020 & A & 2001 Dec 20 & G750M & 52$\times$0.2 & 6581 &  21.33 &  0.0 & 1080 \\
               & O6BU16030 & A & 2001 Dec 20 & G750M & 52$\times$0.2 & 6581 &  21.33 &  0.0 &  840 \\
               & O6BU16040 & A & 2001 Dec 20 & G430L & 52$\times$0.2 & 4300 &  21.33 &  0.0 &  840 \\
               & O6BU16050 & A & 2001 Dec 20 & G430L & 52$\times$0.2 & 4300 &  21.33 &  0.0 &  805 
%
%
%\LL}
\footnotetext[1]{{\bf Sources:} (A) MAST Archive (B) HST Prop ID: 11611 (C) HST Prop ID: 12212 (D) \citet{Rui05}}
%\tablenotetext{1}{{\bf Sources:} (A) MAST Archive (B) HST Prop ID: 11611 (C) HST Prop ID: 12212 (D) \citet{Rui05}}
\footnotetext[2]{Perpendicular to the slit position}
\end{longtable}
\end{landscape}
\renewcommand{\thefootnote}{\alph{footnote}}
%\pagestyle{plain}
%\normalsize

\clearpage

%\begin{deluxetable}{cccccc}
%\tablecolumns{6}
%\footnotesize
%\tablecaption{Maximum Errors In Velocity Measurements}
%\tablewidth{0pt}
%\tablehead{\colhead{Grating} & \colhead{$\sigma_{hi}$} & \colhead{$\sigma_{noise}$} & \colhead{$\sigma_{dev}$} & \colhead{$\sigma_{total}$}& \colhead{$\sigma_{total}$}    \\
%                             & \colhead{(\AA)}         & \colhead{(\AA)}            & \colhead{(\AA)}          & \colhead{(\AA)}           & \colhead{(km s$^{-1}$)}       }
%\startdata
%G430L & .26 & .79 & 1.74 &  1.93  & 115.5\\
%G430M & .11 & .13 & 1.71 &  1.72  & 102.9\\
%G750M & .07 & .34 & 0.57 &  0.67  &  40.1
%\enddata
%\label{errors}
%\end{deluxetable}
%\clearpage

\begin{deluxetable}{lccrcrccl}
\tablecolumns{8}
\footnotesize
\tablecaption{AGN Distances {\it HST} Imaging Observations and Redshifts}
\tablewidth{0pt}
\tablehead{\colhead{Target} & \colhead{z} & \colhead{Source} & \colhead{Dist.} & \colhead{Ref.} & \colhead{Scale}    & \colhead{Type} & \colhead{Instr.} & \colhead{Filter} \\
                            &             &                  & \colhead{Mpc}   &                & \colhead{(pc/$''$)}&                &                  &                  }
\startdata
Akn 564         & 0.024684      & 21 cm	 & 101.4 &NED& 491   &1 & WFC3  & FQ508N\\
Circinus 	& 0.001453	& 21 cm	 & 4.2   &EDD& 20    &2	& WFPC2	& F606W \\
IC 3639	        & 0.010924	& 21 cm	 & 44.9  &NED& 217   &2	& WFPC2 & F606W \\
IRAS 11058-1131 & 0.054828	& Stellar& 225.2 &NED& 1092  &2	& WFPC2 & F547M \\
MCG-6-30-15	& 0.007749	& Stellar& 25.5  &EDD& 123   &1	& WFPC2 & FR533N\\
Mrk 34	        & 0.051167	& Stellar& 210.1 &NED& 1019  &2	& WFPC2 & FR533N\\
Mrk 279	        & 0.030451	& Stellar& 125.1 &NED& 606   &1	& WFC3  & FQ508N\\
Mrk 348	        & 0.015034	& 21 cm	 & 61.7  &NED& 299   &2	& WFPC2 & FR533N\\
Mrk 463e	& 0.050000	& ---	 & 205.3 &NED& 996   &2	& WFPC2 & FR533N\\
Mrk 493	        & 0.031485	& 21 cm	 & 129.3 &NED& 627   &1	& WFPC2 & F606W \\
Mrk 509         & 0.034397      & Stellar& 141.3 &NED& 685   &1	& WFC3  & FQ508N\\
Mrk 705	        & 0.029150	& Stellar& 119.7 &NED& 580   &1	& WFPC2 & FR533N\\
Mrk 766	        & 0.012929	& Stellar& 53.1  &NED& 257   &1	& WFPC2 & FR533N\\
Mrk 1040	& 0.016642	& 21 cm	 & 68.3  &NED& 331   &1	& WFPC2 & FR533N\\
Mrk 1044        & 0.016451      & 21 cm  & 67.6  &NED& 328   &1 & WFC3  & FQ508N\\
Mrk 1066	& 0.011858	& Stellar& 48.7  &NED& 236   &2 & WF/PC & F492M \\ 
NGC 1358	& 0.013436	& 21 cm	 & 55.2  &NED& 268   &2	& WFPC2 & F606W \\
NGC 1386	& 0.002895	& Stellar& 16.5  &EDD& 80    &2	& WFPC2 & F502N \\
NGC 1667	& 0.015257	& 21 cm	 & 62.7  &NED& 304   &2	& WFPC2 & F606W \\
NGC 2110	& 0.007789	& Stellar& 29.0  &EDD& 140   &2	& WFPC2 & F606W \\
NGC 2273	& 0.006138	& 21 cm	 & 17.9  &EDD& 87    &2	& WFPC2 & FR533N\\
NGC 3081	& 0.007988	& 21 cm	 & 28.6  &EDD& 139   &2	& WFPC2 & FR533N\\
NGC 3227	& 0.003859	& 21 cm	 & 26.4  &EDD& 128   &1	& WFPC2 & F606W \\
NGC 3393	& 0.012509	& 21 cm	 & 51.4  &NED& 249   &2	& WFC3	& FQ508N\\    
NGC 3516	& 0.008836	& Stellar& 38.0  &EDD& 184   &1 & WFPC2 & FR533N\\
NGC 3783	& 0.009730	& 21 cm	 & 25.1  &EDD& 122   &1	& WFPC2 & FR533N\\
NGC 4051	& 0.002418	& 21 cm	 & 17.1  &EDD& 83    &1 & WFC3  & F502N \\
NGC 4303	& 0.005234	& 21 cm	 & 17.6  &EDD& 85    &2	& WFPC2 & F606W \\
NGC 4395	& 0.001064	& 21 cm	 & 4.7   &EDD& 23    &1	& WFC3  & F502N \\
NGC 4507	& 0.011829	& 21 cm	 & 48.6  &NED& 236   &2	& WFPC2 & FR533N\\
NGC 5135	& 0.013693	& 21 cm	 & 56.2  &NED& 273   &2	& WFPC2 & F606W \\
NGC 5252 	& 0.023093	& 21 cm	 & 94.8  &NED& 460   &1.9& WFPC2 & FR533N\\
NGC 5283	& 0.010404	& Stellar& 42.7  &NED& 207   &2	& WFPC2 & F606W \\
NGC 5347	& 0.007959	& 21 cm	 & 39.0  &EDD& 189   &2	& WFPC2 & FR533N\\
NGC 5427	& 0.008733	& 21 cm	 & 27.0  &EDD& 131   &2	& WFPC2 & F606W \\
NGC 5506	& 0.006084	& 21 cm	 & 21.7  &EDD& 105   &1	& FOC   & F501N \\
NGC 5548	& 0.017175	& 21 cm	 & 70.5  &NED& 342   &1	& WFC3  & FQ508N\\
NGC 5643	& 0.003999	& 21 cm	 & 11.8  &EDD& 57    &2	& WFPC2 & F502N \\
NGC 5695	& 0.014093	& 21 cm	 & 57.9  &NED& 281   &2 & WFPC2 & F606W \\
NGC 5728	& 0.009316	& 21 cm	 & 24.8  &EDD& 120   &2	& WF/PC & F492M \\
NGC 5929	& 0.008543	& 21 cm	 & 32.2  &EDD& 156   &2	& WFPC2 & F606W \\
NGC 6300	& 0.003699	& 21 cm	 & 13.1  &EDD& 64    &2	& WFPC2 & F606W \\
NGC 7212	& 0.026001	& Stellar& 106.8 &NED& 518   &2 & WFPC2	& FR533N\\	
NGC 7469	& 0.016268	& 21 cm	 & 66.8  &NED& 324   &1	& WFPC2 & FR533N\\	
NGC 7674	& 0.029030	& 21 cm	 & 119.2 &NED& 578   &2	& WFPC2 & FR533P15\\
NGC 7682	& 0.017140	& 21 cm	 & 70.4  &NED& 341   &2	& WFPC2	& F606W \\
NGC 788	        & 0.013603	& Stellar& 55.9  &NED& 271   &2	& WFPC2 & F606W \\
UM 146	        & 0.017225	& 21 cm	 & 70.7  &NED& 343   &1.9& WFPC2 & F606W 
\enddata
\label{images}
\end{deluxetable}

\clearpage

\begin{deluxetable}{l|c}
\tablecolumns{2}
\tablecaption{Kinematic Model Parameter Abbreviations}
\tablewidth{0pt}
\tablehead{\multicolumn{1}{c}{Parameter} & \multicolumn{1}{|c}{Symbol}}
\startdata
Position angle of bicone axis ($^{\circ}$) & $P.A.$        \\
Inclination of bicone axis ($^{\circ}$)    & $i$           \\
Outer opening angle ($^{\circ}$)           & $\theta_{max}$\\
Inner opening angle ($^{\circ}$)           & $\theta_{min}$\\
Maximum velocity (km s$^{-1}$)             & $v_{max}$     \\
Bicone height (pc)                         & $z_{max}$     \\
Turnover radius (pc)                       & $r_{t}$       

\enddata
\label{paramdef}
\end{deluxetable}

\begin{deluxetable}{l|ccc}
\tablecolumns{4}
\tablecaption{Total Sample AGN Kinematic Classifications}
\tablewidth{0pt}
\tablehead{\multicolumn{1}{c}{Type} & \multicolumn{3}{|c}{Target}}
\startdata
{\it Outflow}  &Circinus   & Mrk 3        & Mrk 34      \\
               &Mrk 78     & Mrk 279      & Mrk 573     \\
               &Mrk 1066   & NGC 1068     & NGC 1667    \\ 
               &NGC 3227   & NGC 3783     & NGC 4051    \\ 
               &NGC 4151   & NGC 4507     & NGC 5506    \\
               &NGC 5643   & NGC 7674     &             \\ \hline
{\it Ambiguous}&Akn 564    & IC 3639      & MCG-6-30-15 \\
               &Mrk 493    & Mrk 509      & NGC 1358    \\
               &NGC 2110   & NGC 2273     & NGC 3516    \\
               &NGC 4395   & NGC 5252     & NGC 5283    \\
               &NGC 5728   & NGC 5929     & NGC 7682    \\
               &NGC 788    & UM 146       &             \\ \hline
{\it Complex}  &IRAS 11058-1131 & NGC 1386 & NGC 3081   \\
               &NGC 3393   & NGC 5135     & NGC 7212    \\ \hline
{\it Compact}  &Mrk 348    & Mrk 463e     & Mrk 705     \\
               &Mrk 766    & Mrk 1040     & Mrk 1044    \\
               &NGC 5347   & NGC 5427     & NGC 5548    \\
               &NGC 5695   & NGC 6300     & NGC 7469    \\ \hline
{\it Poor}     &NGC 4303   &              &    
\enddata
\label{class}
\end{deluxetable}

\clearpage

\begin{deluxetable}{l|rlccrrr|rcc|r}
\tablecolumns{11}
\tablecaption{Total Sample Modeled AGN Parameters.\tablenotemark{1}}
\tablewidth{0pt}
\tablehead{
\multicolumn{1}{c|}{Target}         & 
\multicolumn{7}{c|}{NLR Bicone}     & 
\multicolumn{3}{c|}{Host Disk}      \\
\multicolumn{1}{c|}{}               & 
\multicolumn{1}{c}{$P.A.$}          &      
\multicolumn{1}{c}{$i$}             & 
\multicolumn{1}{c}{$\theta_{min}$}  & 
\multicolumn{1}{c}{$\theta_{max}$}  & 
\multicolumn{1}{c}{$v_{max}$}       &  
\multicolumn{1}{c}{$z_{max}$}       & 
\multicolumn{1}{c|}{$r_{t}$}        & 
\multicolumn{1}{c}{$P.A.$}          &   
\multicolumn{1}{c}{$i$}             & 
\multicolumn{1}{c|}{Disk}           &
\multicolumn{1}{c}{$\beta$~\tablenotemark{2}}         \\
\multicolumn{1}{c|}{}               &  
\multicolumn{1}{c}{($^{\circ}$)}    &
\multicolumn{1}{c}{($^{\circ}$)}    &
\multicolumn{1}{c}{($^{\circ}$)}    &  
\multicolumn{1}{c}{($^{\circ}$)}    &
\multicolumn{1}{c}{(km/s)}          &   
\multicolumn{1}{c}{(pc)}            &   
\multicolumn{1}{c|}{(pc)}           &  
\multicolumn{1}{c}{($^{\circ}$)}    &  
\multicolumn{1}{c}{($^{\circ}$)}    & 
\multicolumn{1}{c|}{Ref.}           &
\multicolumn{1}{c}{($^{\circ}$)}

}
\startdata
Circinus    &   -52	  & 25 (NW) &  36	   &   41  &  300 &     35  &      9  &         30	&    65      & 1   &  7\\
Mrk 3       &    89       & 05 (NE) &  ---         &   51  &  800 &    270  &     80  &        129	&    64      & 9   & 52\\
Mrk 34	    &   -32	  & 25 (SE) &  30	   &   40  & 1500 &   1750  &   1000  &         65	&    30      & 2   & 85\\
Mrk 78	    &    65	  & 30 (SW) &  10	   &   35  & 1200 &   3200  &    700  &         84 	&    55      & 2   & 87\\
Mrk 279	    &   -24	  & 55 (SE) &  59	   &   62  & 1800 &    300  &    250  &         33 	&    56      & 3,4 & 86\\
Mrk 573	    &   -36	  & 30 (NW) &  51          &   53  &  400 &   1200  &    800  &        103      &    30      & 3,2 & 44\\ 
Mrk 1066    &   -41	  & 10 (NW) &  15	   &   25  &  900 &    400  &     80  &         90 	&    54      & 3   & 45\\
NGC 1068    &    30	  & 05 (NE) &  20	   &   40  & 2000 &    400  &    140  &        115 	&    40      & 8   & 45\\
NGC 1667    &    55	  & 18 (NW) &  45	   &   58  &  300 &    100  &     60  &          5      &    39      & 4   & 46\\ 
NGC 3227    &    30	  & 75 (SW) &  40	   &   55  &  500 &    200  &    100  &        -31 	&    63      & 2,5 & 76\\
NGC 3783    &   -20	  & 75 (SE) &  45	   &   55  &  130 &    110  &     32  &        -15 	&    35      & 6   & 38\\
NGC 4051    &    80       & 78 (NE) &  10	   &   25  &  550 &    175  &     52  &         50 	&    05      & 6   & 15\\
NGC 4151    &    60	  & 45 (SW) &  15	   &   33  &  800 &    400  &     96  &         33	&    20      & 7   & 39\\
NGC 4507    &   -37	  & 43 (NW) &  30	   &   50  & 1000 &    200  &     90  &         65 	&    28      & 2   & 12\\
NGC 5506    &    22	  & 10 (SW) &  10	   &   40  &  550 &    220  &     65  &        -89      &    76      & 3   & 32\\
NGC 5643    &    80	  & 25 (SE) &  50	   &   55  &  500 &    285  &     70  &        136 	&    30      & 3   & 42\\
NGC 7674    &   -63       & 30 (NW) &  35	   &   40  & 1000 &    700  &    200  &         76 	&    40      & 2   & 42
\enddata
\tablenotetext{1}{Inclination direction specifies which end of the NLR bicone is inclined out of the plane of the sky toward Earth}
\tablenotetext{2}{Angle between the NLR bicone axis and the normal to the host galaxy disk\\
{\bf References}: (1) \citet{Fre77}, (2) \citet{Sch00}, (3) \citet{Kin00}, (4) NED, (5) \citet{Xil02}, (6) \citet{Hic09}, (7) \citet{Das05}, (8) \citet{Das06}, (9) \citet{Cre10b}}
\label{modelvals}
\end{deluxetable}

\clearpage

\clearpage

\bibliographystyle{apj}             % Please learn to use the
                                     %  formatting of Latex's Bibtex. It
                                     %  will make your life easier.
% apj.bst should be in this directory as well as apj-jour.bib and reference paper.bib
\bibliography{apj-jour,paper}       % "paper.bib" contains all my
                                     %  references. "apj-jour.bib"
                                     %  contains abbreviations of
                                     %  journals.
% to get references to work in paper:
% compile paper       --> latex mrk78.tex
% compile bibtex      --> bibtex mrk78
% compile paper again --> latex mrk78.tex

\clearpage

%FIGURES%%%%%%%%%%%%%%%%%%%%%%%%%%%%%%%%%%%%%%%%%%%%

% Gauss fits to spectra

\appendix

\section{Targets With Clear Outflow Signatures}
\label{sec6}
\subsection{{\it Circinus}}

\begin{figure}[h]
\centering
\begin{tabular}{cc}
\includegraphics[width=.4\linewidth]{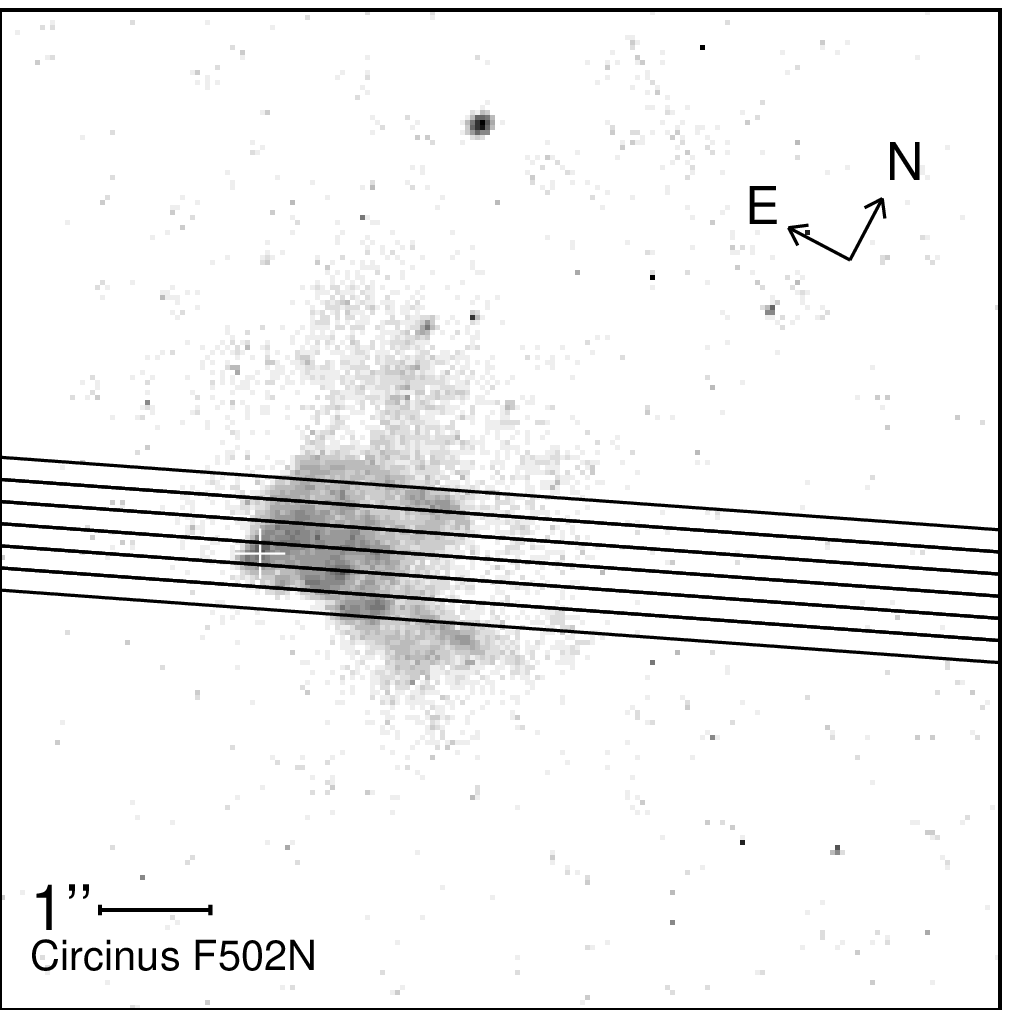} &
\includegraphics[width=.55\linewidth]{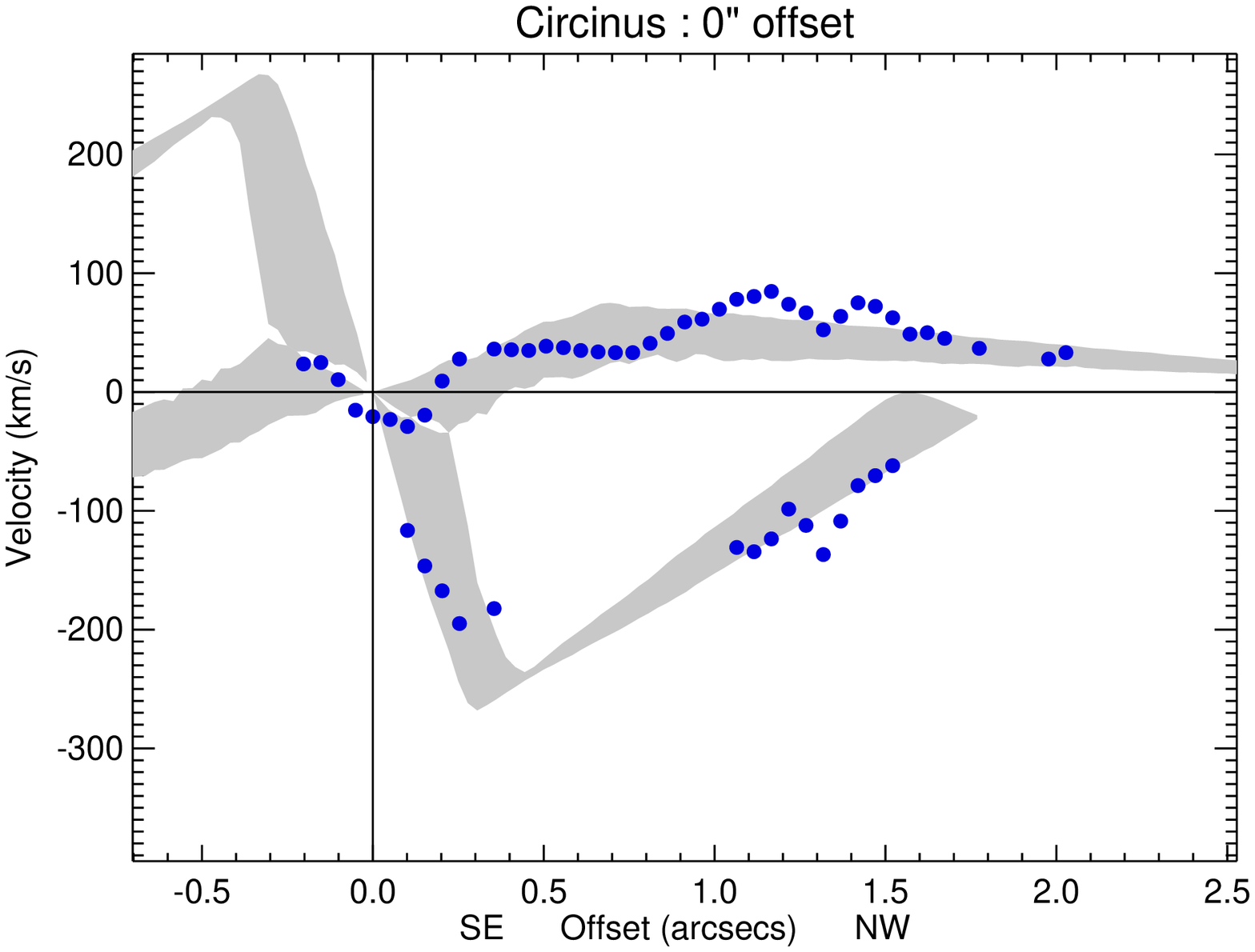} \\
\includegraphics[width=.4\linewidth]{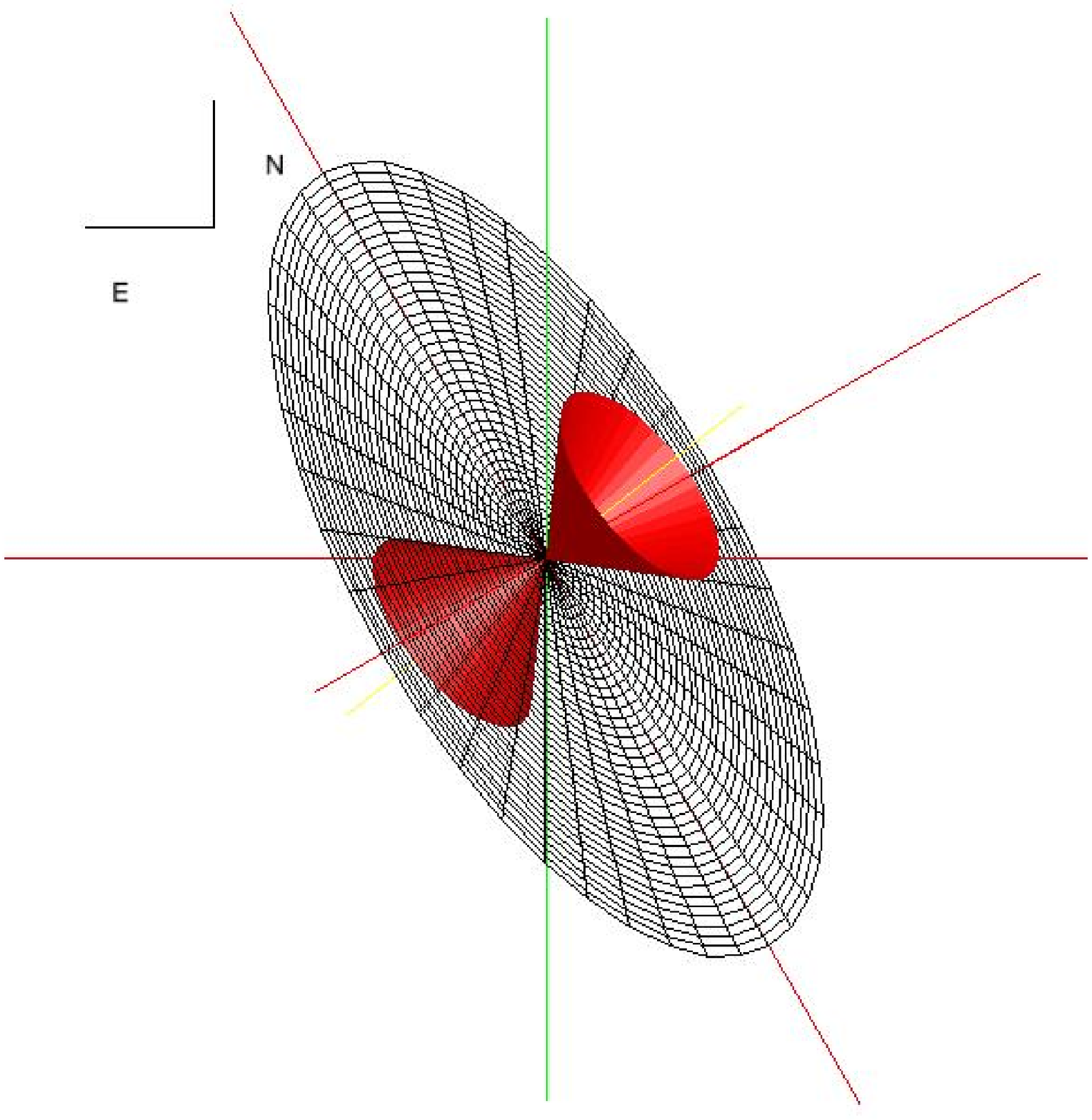} &
\includegraphics[width=.55\linewidth]{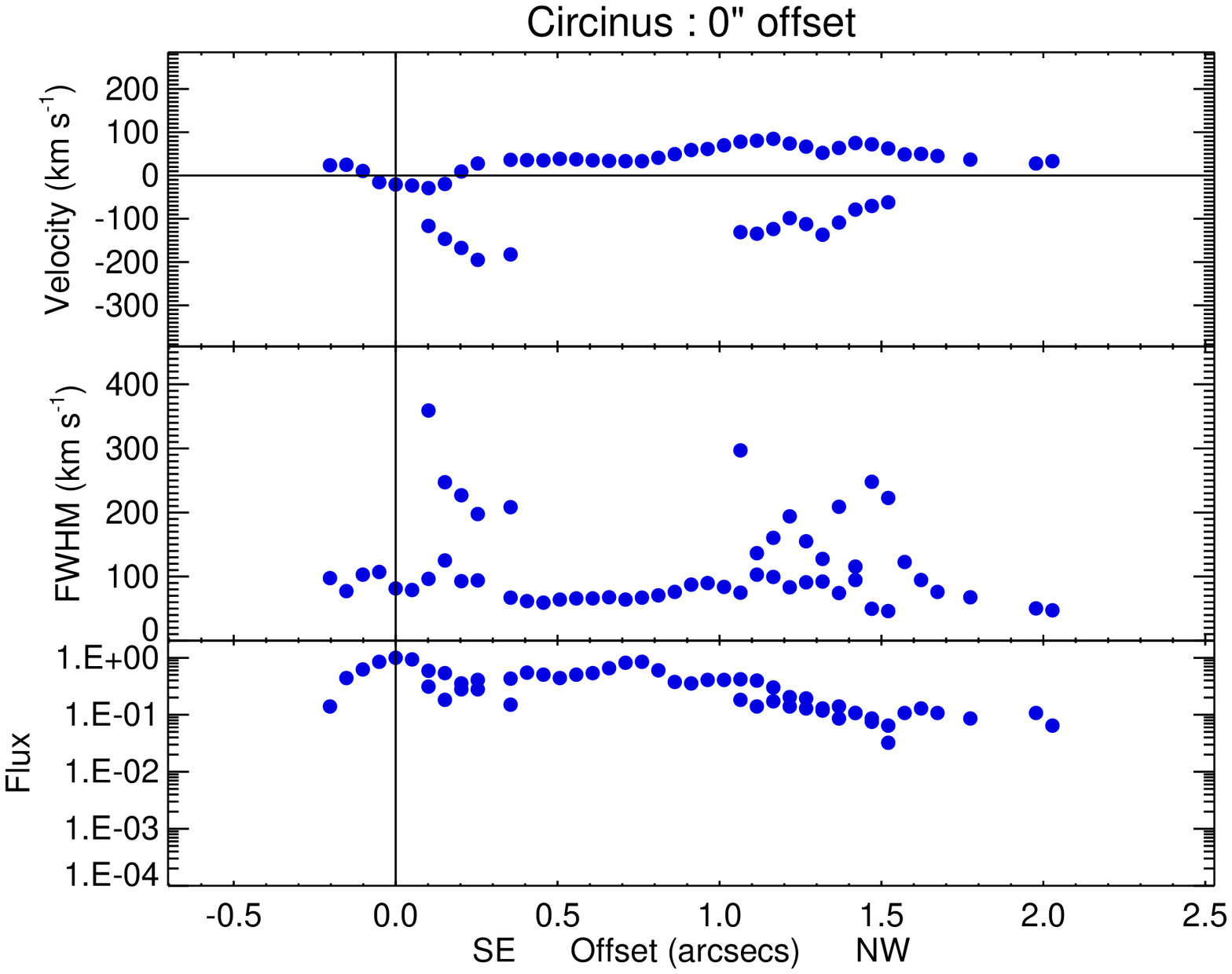}
\end{tabular}
\caption[Circinus Kinematic and Geometric Models]{Top Left: {\it HST} imaging of Circinus with STIS observation positions overplotted. 
Top Right: Kinematic model for the central slit position of Circinus fitting two kinematic 
components. Bottom Left: Corresponding geometric model of the NLR with 
disk geometry. The bicone axis is illustrated as a yellow line. Black axes illustrate the extended plane of the host galaxy. Red 
axes illustrate the plane of the sky. Bottom Right: Velocities, FWHMs, and fluxes normalized to the highest measured flux for the 
central slit position of Circinus. Green diamonds, blue circles, and red squares corresond to G430L, G430M, and G750M grating data 
respectively.
}
\label{circinus}
\end{figure}

Figure \ref{circinus} shows that this is a Seyfert 2 galaxy with a single NLR cone visible 
in [O~III] imaging that is nearly perpendicular to the major axis of the host disk. The modeled kinematics provide a 
good fit to the data, matching well to both visible outflow components. At this orientation, 
it is unlikely that an intersection exists between the extended host disk and the NLR. 
As the projected opening angle of the model NLR (91$^{\circ}$) is close to that of the 
opening angle seen in available imaging (96$^{\circ}$) and the size of the NLR in this 
very nearby AGN is only $\sim$50 pc, it is likely that the NLR is ionizing a dense medium 
in the galactic disk. The south-east cone is undetectable at optical wavelengths as it 
is hidden behind the heavy extinction of the galactic disk \citep{Rui00,Pri05}. Extended 
filimentary [O~III] emission greater than 10$''$ from the nucleus is not included in our model as 
slit positions containing this emission are displaced several arcseconds away from the 
nucleus, and the remote emission would have been difficult to accurately incorporate into 
a biconical model. An available G750M long-slit spectrum perpendicular to the NLR runs 
along the major axis and detects H$\alpha$ emission from H~II regions surrounding the 
nucleus in the plane of the disk \citep{Wil00}, peaking at an observed velocity of $\sim$ 
175 km s$^{-1}$. This allows us to determine that the rotation of the host disk is 
blueshifted to the north and redshifted to the south; combined with NLR imaging, this confirms 
that the south-east side of the host disk is closer to us, obscuring the south-east NLR 
cone. \citet{Gre03b} found that H$_2$O maser emission traces a warped, edge-on accretion 
disk at radii between $\sim$ 0.1 and 0.4 pc and suggest that the warping of the accretion 
disk collimates the NLR outflow. The resultant model from their hypothesis is similar to 
our own, though their full opening angle appears to be over 130$^{\circ}$, much larger 
than our 82$^{\circ}$. 

\clearpage

%Slit positions and corresponding spectral images of Circinus STIS 
%obserations are available in Figures \ref{fig1a}, \ref{fig2a} and \ref{fig2b}.

\subsection{{\it Mrk 34}}

\begin{figure}[h]
\centering
\begin{tabular}{cc}
\includegraphics[width=.4\linewidth]{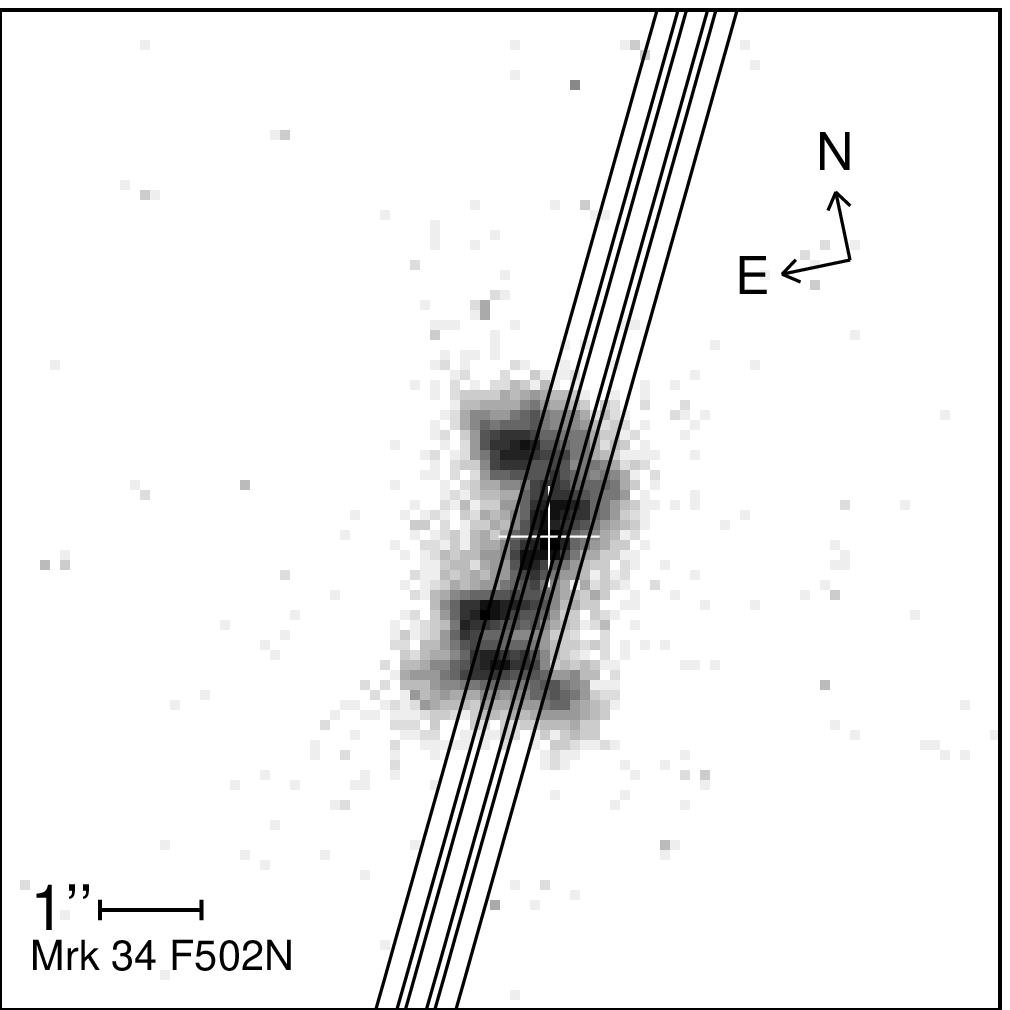} &
\includegraphics[width=.55\linewidth]{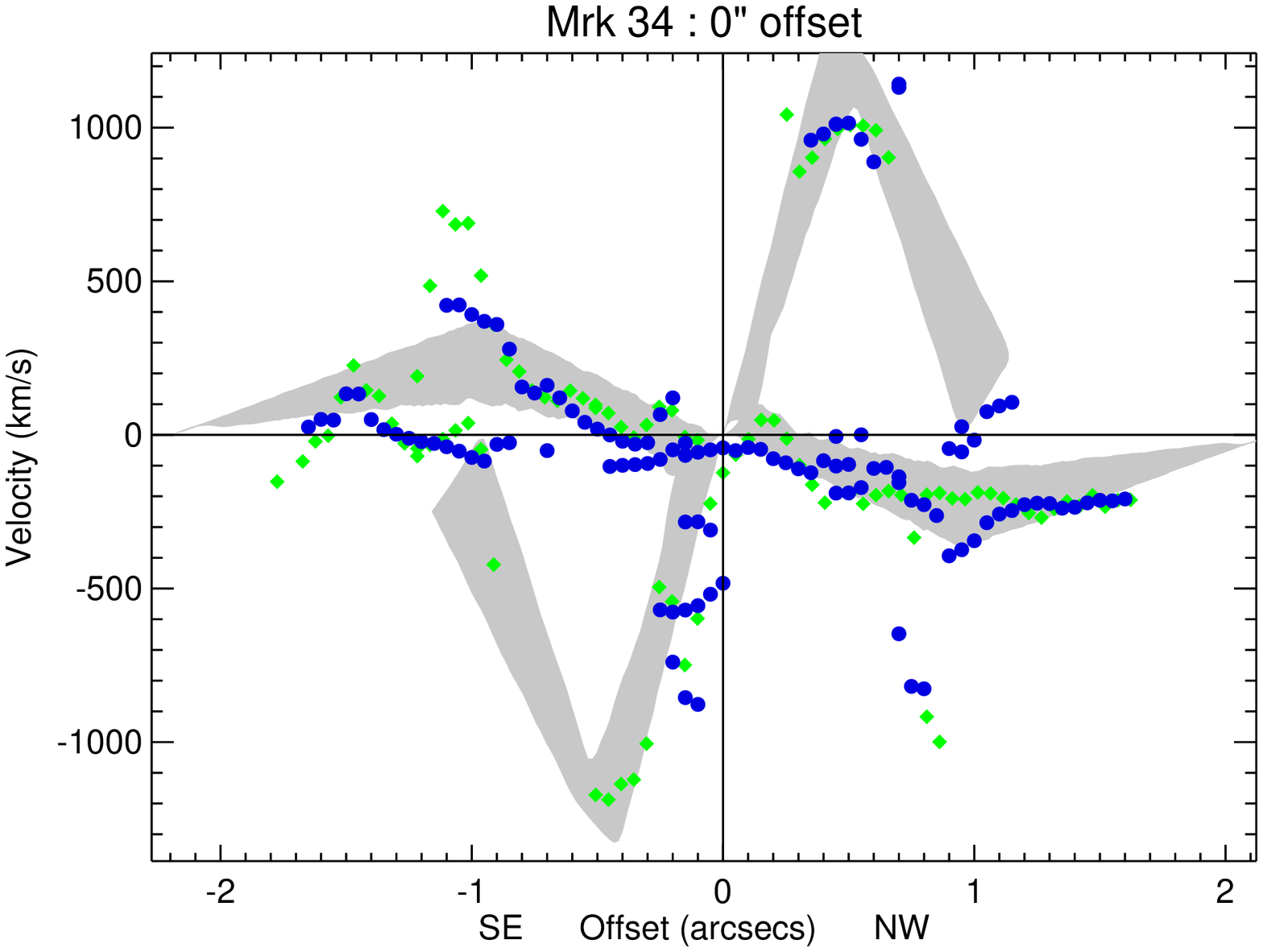} \\
\includegraphics[width=.4\linewidth]{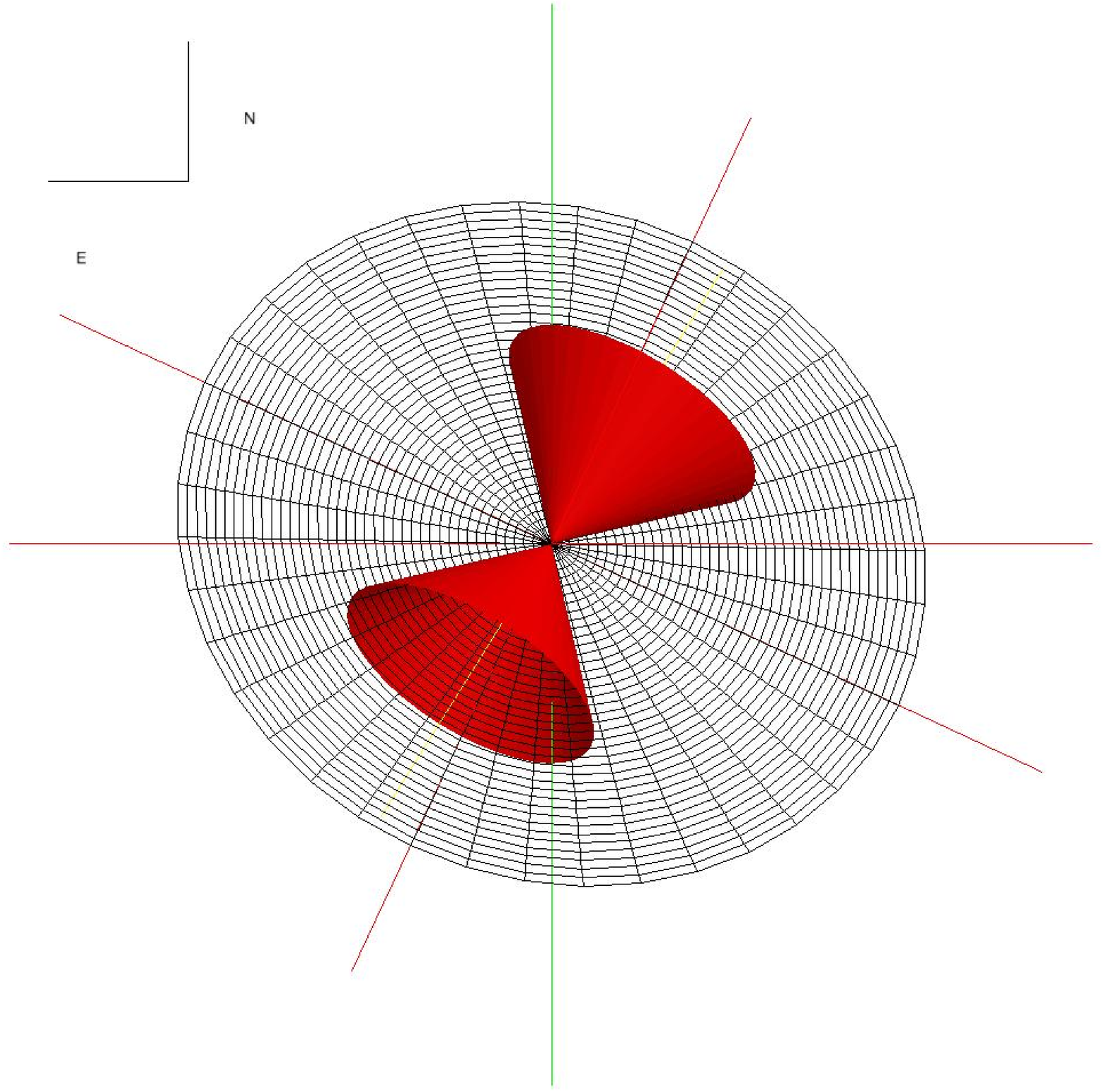} &
\includegraphics[width=.55\linewidth]{mrk34_g430l1_mul.eps}
\end{tabular}
\caption[Mrk 34 Kinematic and Geometric Models]{Same as Figure \ref{circinus}, but for Mrk 34 fitting four kinematic components..}
\label{mrk34}
\end{figure}

This is a Seyfert 2 galaxy with a backwards S-shape NLR, as shown in 
Figure \ref{mrk34}, similar to Mrk 3 \citep{Cre10b}. The modeled kinematics provide a good fit to the data, matching 
well to all four components save for a few high-velocity points. Employing our 
geometric model, we see that the S-shaped formation is likely an intersection between 
a gas spiral in the host disk and the ionization bicone as the projected area of the 
host disk enclosed by the outer opening angle matches well with emission seen in the 
[O~III] imaging. Similar to Mrk 573 \citep{Fis10}, we suggest rotation does not play a role in the 
kinematics we observe as the low velocity components would correspond to a disk 
rotating clockwise while the host disk spiral arms are winding up in the 
counterclockwise direction. \citet{Kon06} identified water maser emission in Mrk 34 
due to an edge-on accretion disk. Thus, because our model gives the inclination of 
the bicone axis to be 25$^{\circ}$ out of the plane of the sky, it is possible that 
Mrk 34 experiences the same warped disk scenario as Circinus. Alternatively, if a 
torus is collimating the ionizing radiation, it is tilted $\sim$25$^{\circ}$ with 
respect to the maser disk. This scenario would still suggest warping as the inflow 
moves from the torus to the accretion disk.

%Slit positions and 
%corresponding spectral images of Mrk 34 STIS observations are available in Figures 
%\ref{fig1b} and \ref{fig2c}.

\clearpage

\subsection{{\it Mrk 279}}

Mrk 279 is a Seyfert 1 residing in a moderately inclined host. The modeled kinematics 
provide a good fit to the data, as we are able to account for three components within 
a small area over the nucleus. Imaging of the NLR shows a compact area slightly more 
than 1$''$ in diameter. The opening angle seen in the [O~III] imaging has a similar 
width of 140$^{\circ}$ as the intersection between the host disk and NLR in the 
geometric model. While the resulting wedge geometry of the intersection matches well 
with the imaging, our model assumes that the inner portion of each NLR cone is hollow. 
This creates a discrepancy between the model and imaging 
data as we should only see a 'V' of emission across the disk without seeing the 
central portion illuminated because our model assumes the center of the cone is hollow. 
It is possible that the dust and gas within the portion of the host disk illuminated by 
the NLR has be evacuated.
%Slit positions and corresponding spectral images of Mrk 279 STIS obserations are 
%available in Figures \ref{fig1b} and \ref{fig2c}.

\begin{figure}[h]
\centering
\begin{tabular}{cc}
\includegraphics[width=.4\linewidth]{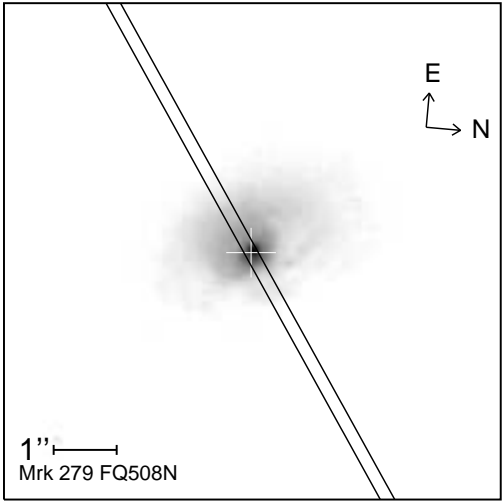} &
\includegraphics[width=.55\linewidth]{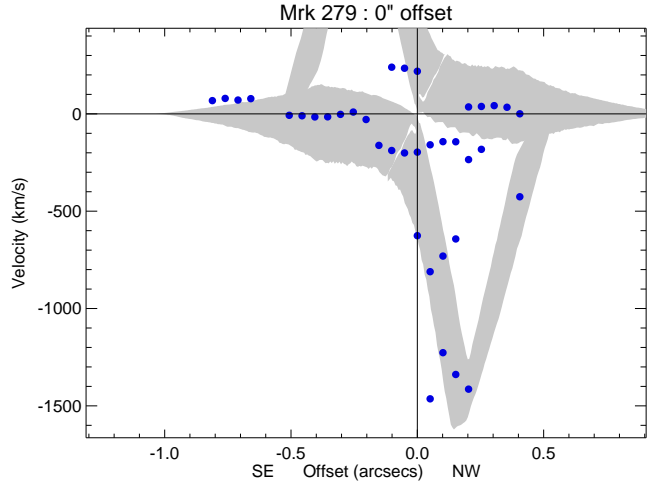} \\
\includegraphics[width=.4\linewidth]{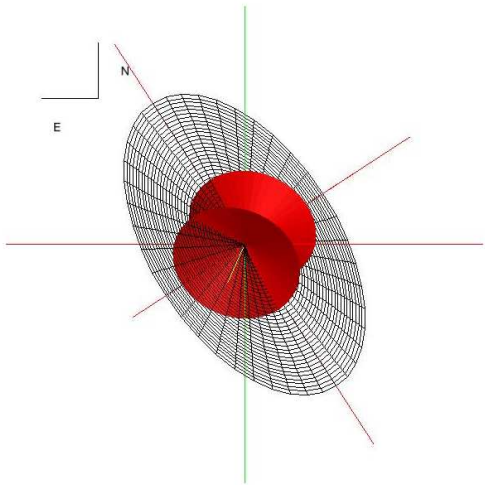} &
\includegraphics[width=.55\linewidth]{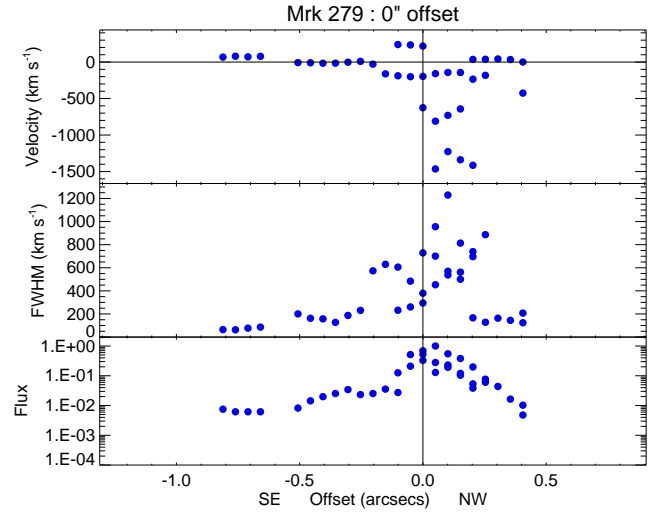}
\end{tabular}
\caption[Mrk 279 Kinematic and Geometric Models]{Same as Figure \ref{circinus}, but for Mrk 279 fitting two kinematic components.}
\label{mrk279}
\end{figure}

\clearpage

\subsection{{\it Mrk 1066}}

Mrk 1066 hosts the Seyfert 2 AGN with the smallest opening angle in our sample with 
a $\theta_{max}$ of 25$^{\circ}$. The modeled kinematics provide a good fit to the 
data. The majority of the detected emission resides in three knots northwest of the 
nucleus, which can be modeled as a single cone inclined 10$^{\circ}$ out of the plane 
of the sky. \citet{Bow95} creates an [O~III] $+$ H$\beta$ emission-line image that 
depicts a similar single cone to the northwest of the AGN nucleus with a $\theta_{max}$ 
of 23$^{\circ}$. At the modeled orientation, including the disk geometry determined 
by \citet{Kin00}, the bicone does not intersect with the host disk, which obscures the 
unobserved second cone, similar to Circinus. Likewise, the close proximity between 
the projected model opening angle (51$^{\circ}$) and the opening angle seen in 
available imaging (47$^{\circ}$) suggests that the NLR is ionizing a large medium 
above the host disk. 

\begin{figure}[h]
\centering
\begin{tabular}{cc}
\includegraphics[width=.4\linewidth]{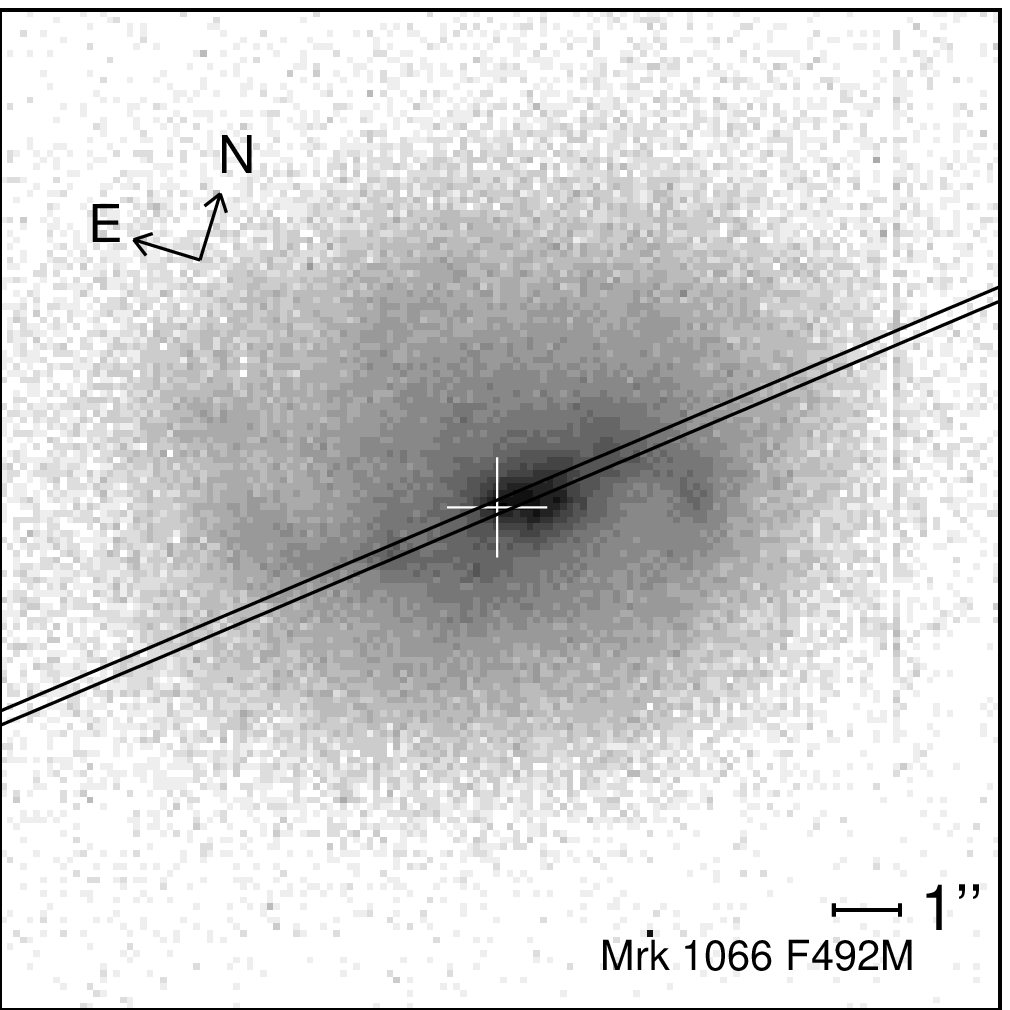} &
\includegraphics[width=.55\linewidth]{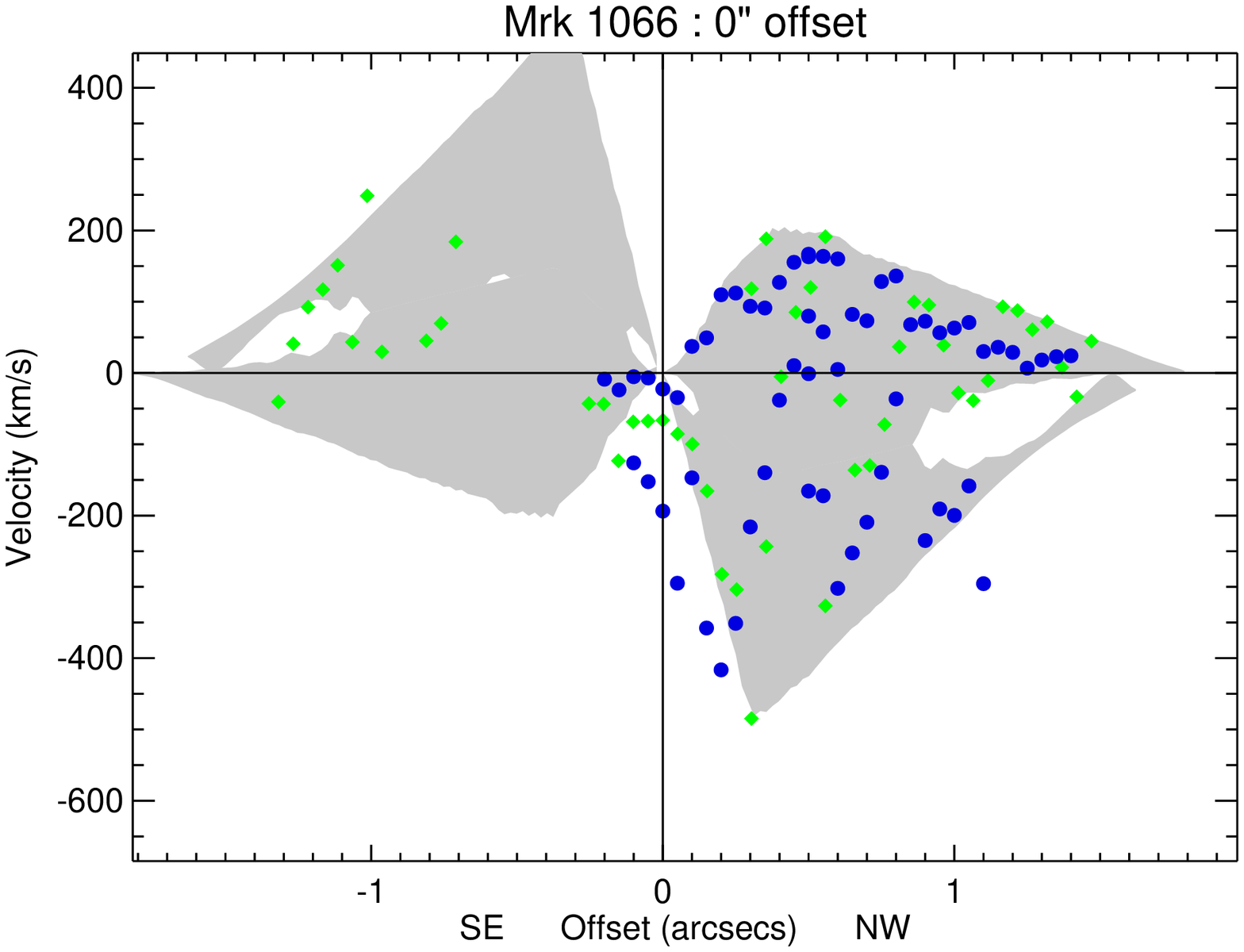} \\
\includegraphics[width=.4\linewidth]{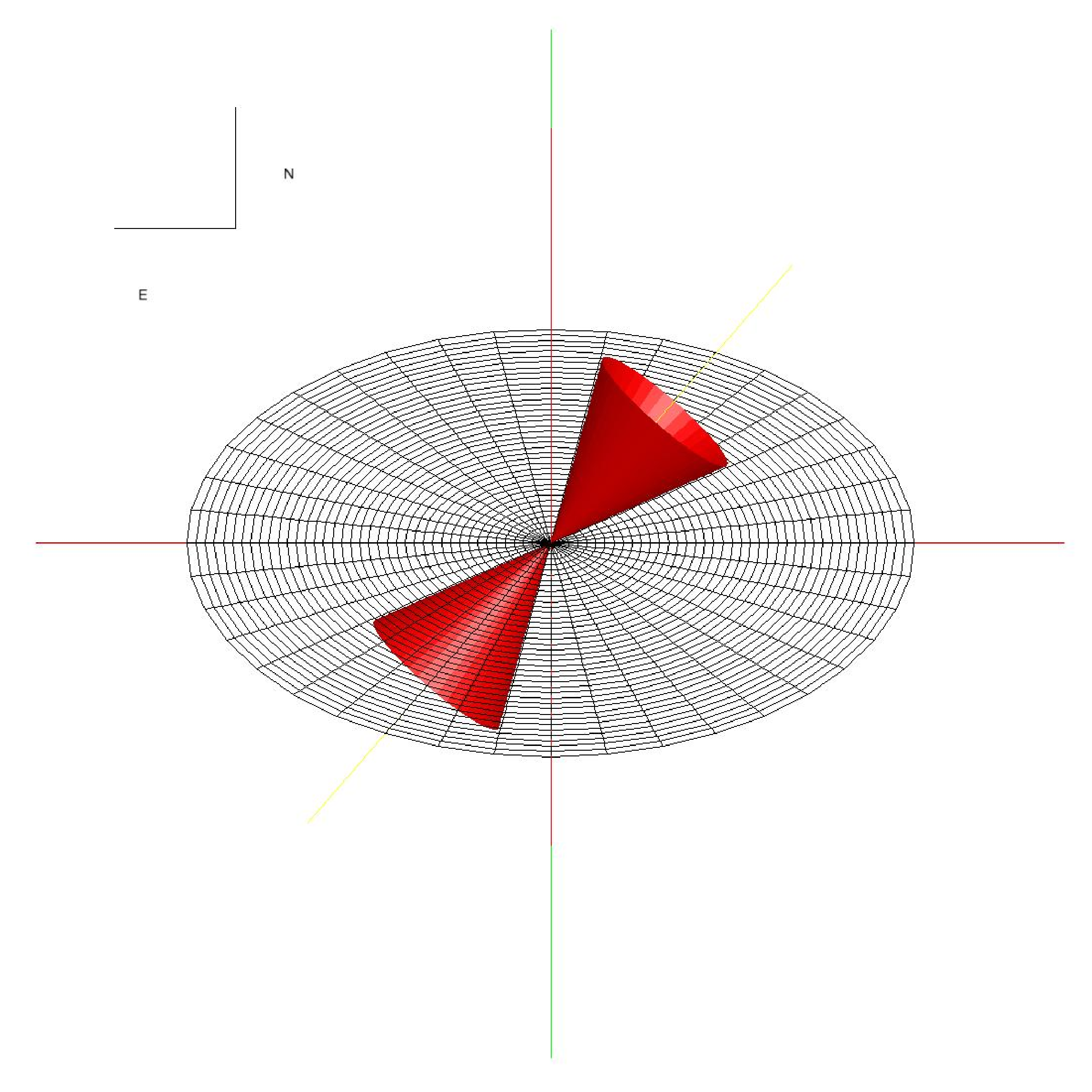} &
\includegraphics[width=.55\linewidth]{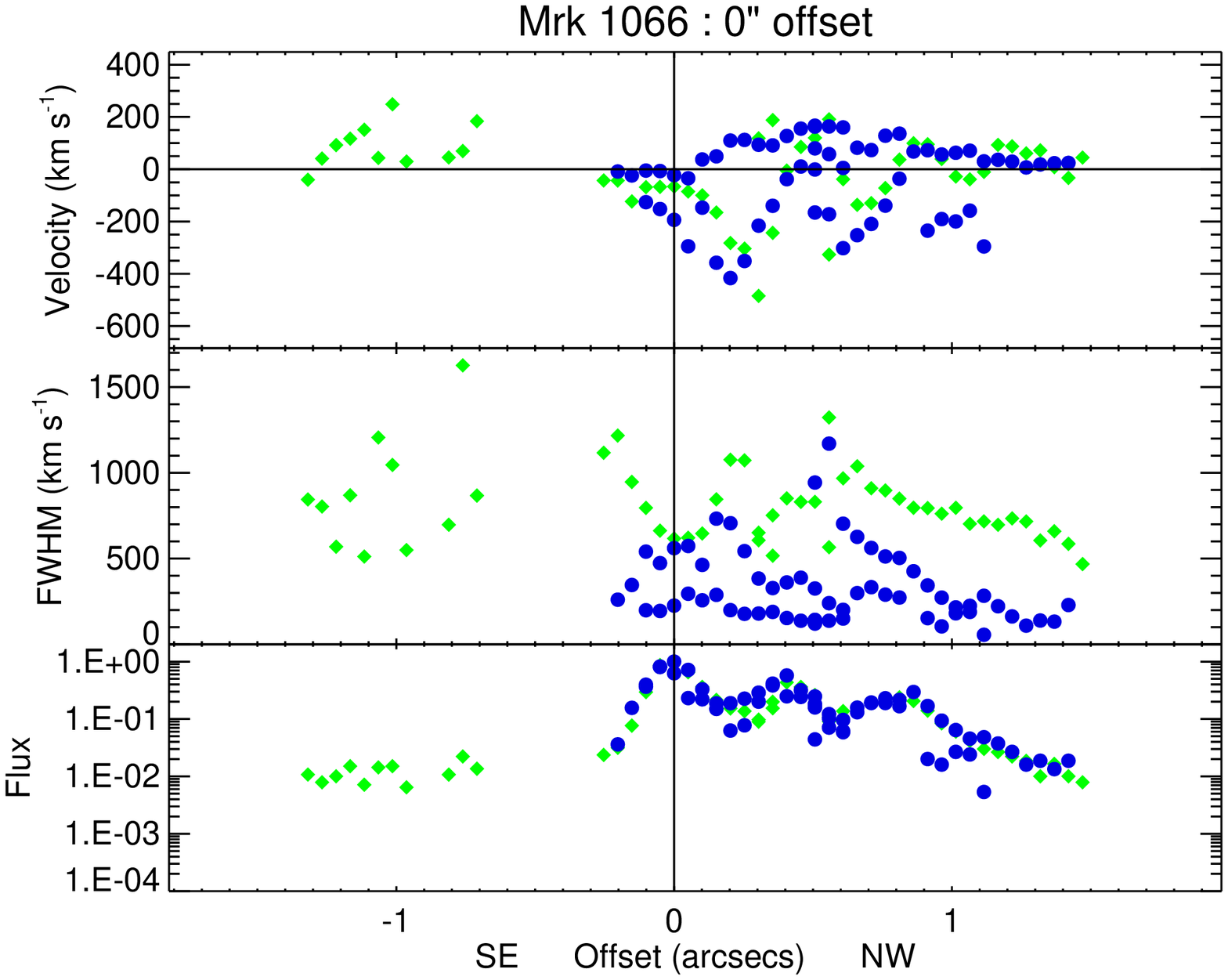}
\end{tabular}
\caption[Mrk 1066 Kinematic and Geometric Models]{Same as Figure \ref{circinus}, but for Mrk 1066 fitting three kinematic components.} 
\label{mrk1066}
\end{figure}

\clearpage

%Slit positions and corresponding spectral images of Mrk 1066 
%STIS obserations are available in Figures \ref{fig1b} and \ref{fig2c}.

\subsection{{\it NGC 1667}}

\begin{figure}[h]
\centering
\begin{tabular}{cc}
\includegraphics[width=.4\linewidth]{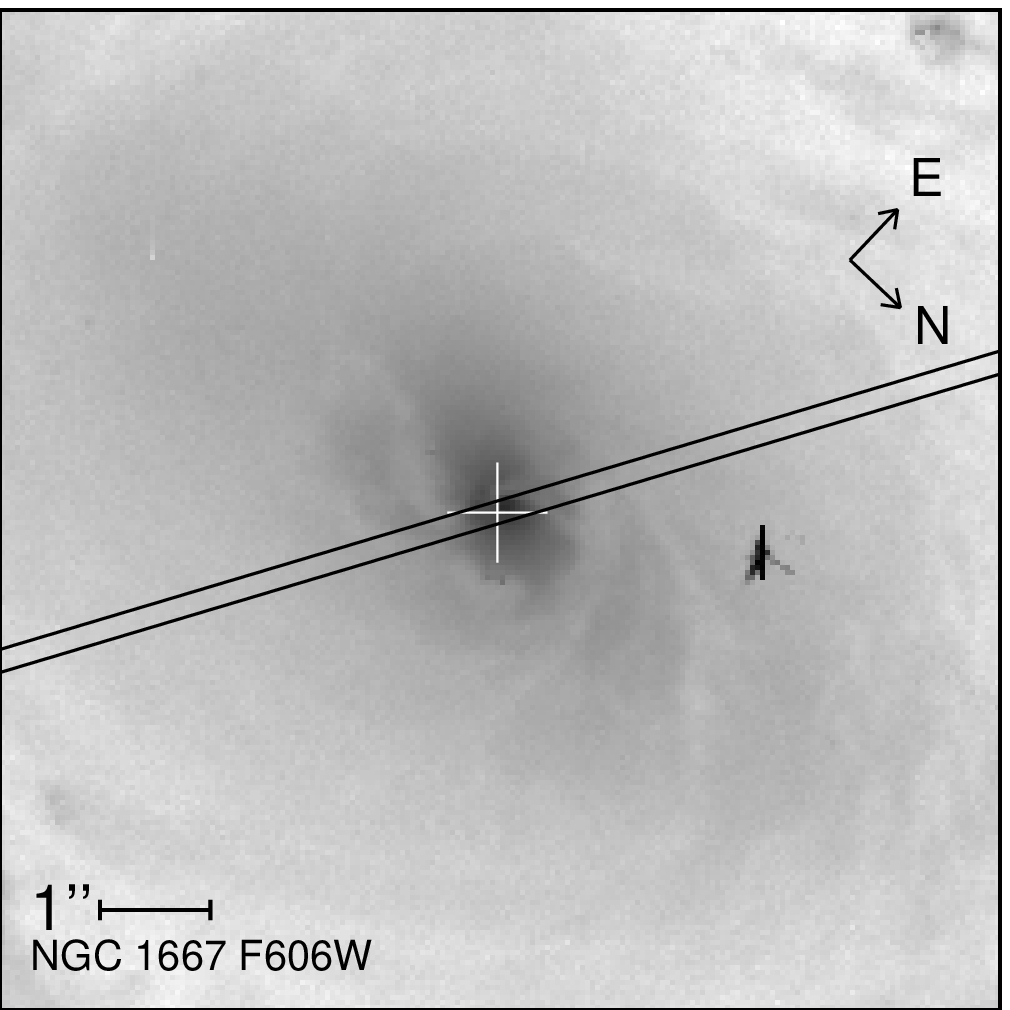} &
\includegraphics[width=.55\linewidth]{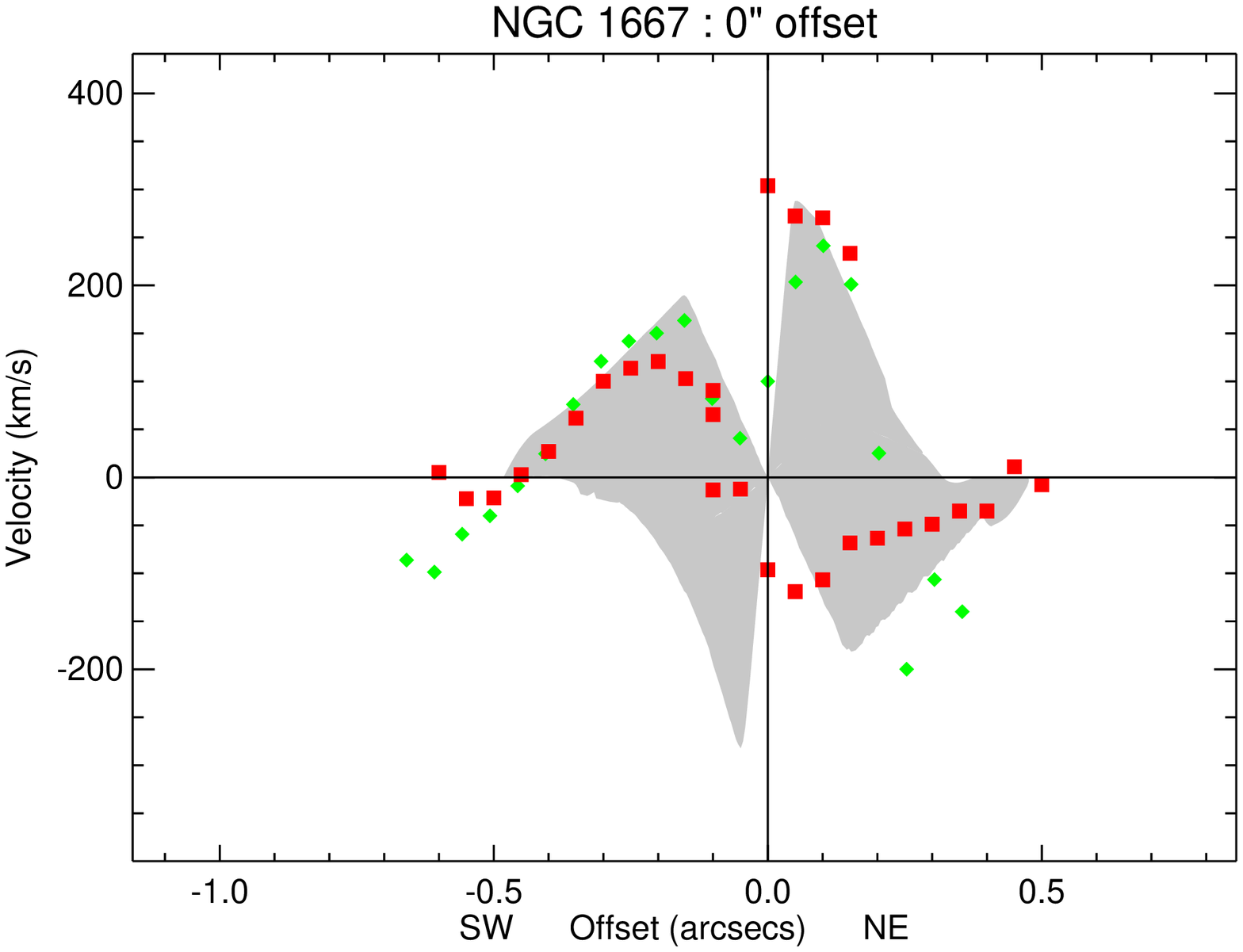} \\
\includegraphics[width=.4\linewidth]{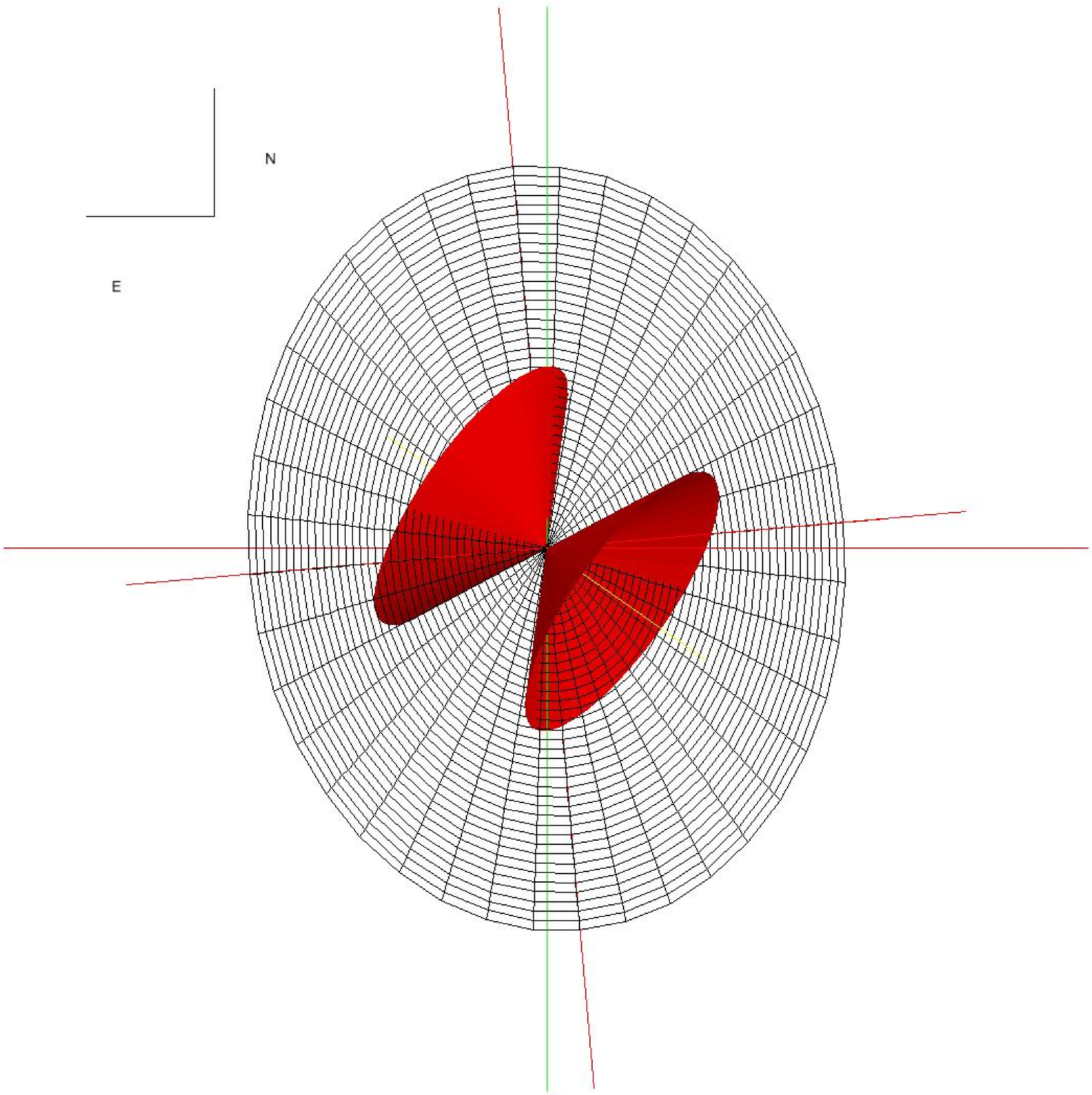} &
\includegraphics[width=.55\linewidth]{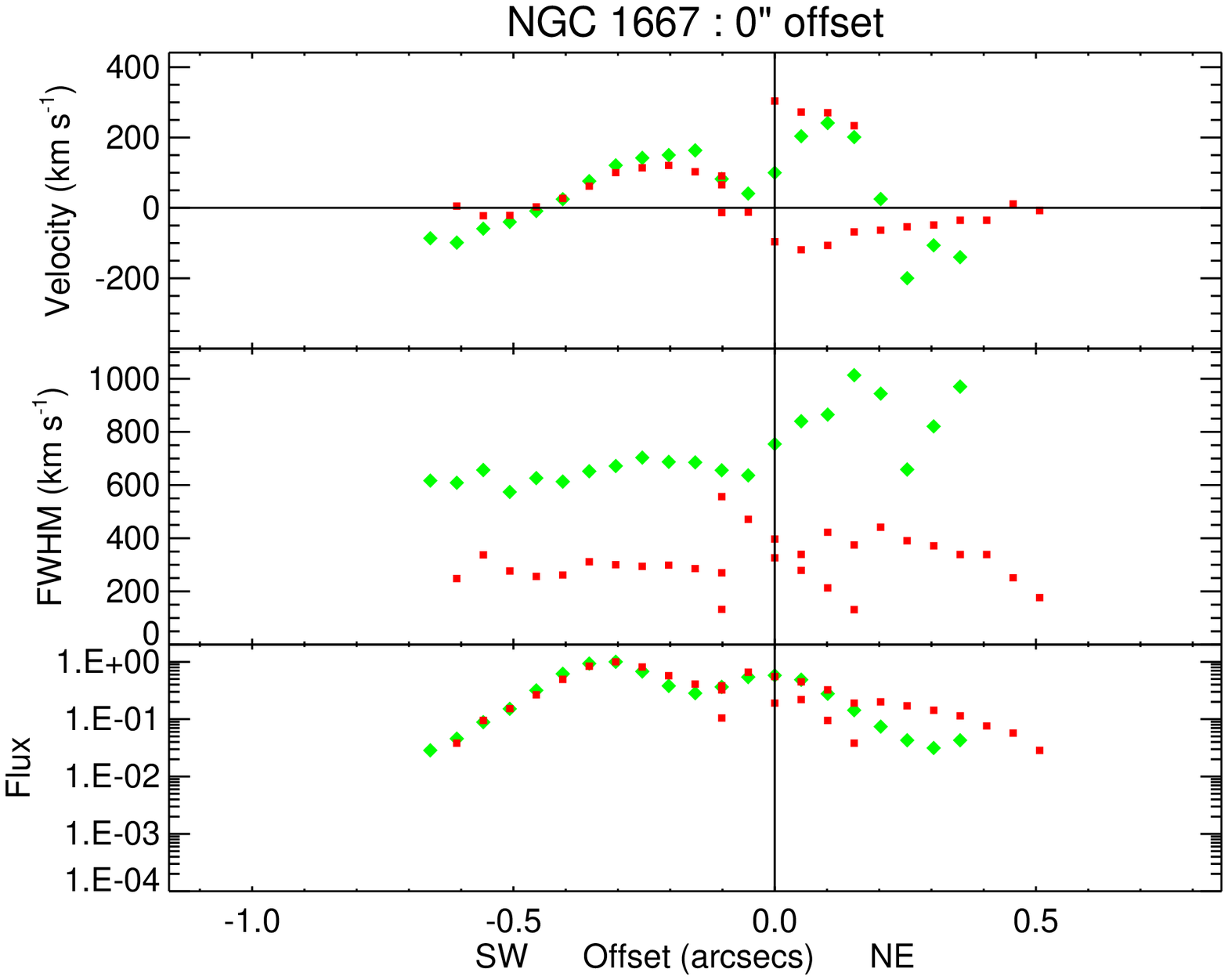}
\end{tabular}
\caption[NGC 1667 Kinematic and Geometric Models]{Same as Figure \ref{circinus}, but for NGC 1667 fitting two kinematic components.}
\label{ngc1667}
\end{figure}

NGC 1667 is a Seyfert 2 with a compact NLR spanning $\sim 1''$ in diameter in our 
STIS observation. NGC 1667 is unique in our sample in that we do not directly detect 
NLR emission in its {\it HST} images. Continuum imaging shows a fairly unspectacular nucleus, with a 
bright central region overlapped by several dustlanes. Figure \ref{ngc1667} shows 
that the modeled kinematics provide a good fit to the data, although we cannot 
compare our modeled NLR geometry with the available broad band imaging as we cannot see 
where [O~III] emission is present. As correspondance 
between imaging and our models is a parameter we use to create our most accurate fit, 
this discrepancy leaves room for improvement. 

\clearpage

\subsection{{\it NGC 3227}}

This is a highly reddened Seyfert 1 galaxy with an inclined disk that constrains 
NLR emission to the northeast of the nucleus. The modeled kinematics provide a 
fair fit to the data, with less successful fits in slit positions $\ge 0.5''$ from 
the nucleus. Modeling would not be possible using [OIII] spectra alone, as the 
single available G430L slit position does not return enough kinematic information 
to fit to a model, as shown in the online supplementary data. Fortunately, supplementary 
G750M spectra map out a large portion of the NLR, which allow us to see prominent 
doppler-shifted H$\alpha$ emission northeast of the nucleus. Though corresponding 
[O~III] emission is lacking, we can assume that the observed emission is due to 
outflows in the NLR versus ionization and rotation within the host disk as the 
kinematics are asymmetric, contain blueshifted velocities $>$ 400 km s$^{-1}$, 
and contain double-peaked profiles in the narrow H$\alpha$ emission line (see 
also \citet{Wal08}). The kinematics initially suggested that emission from both 
sides of a single cone are visible which would place the bicone axis near the 
plane of the sky. From available imaging, a conical shape that would correspond 
to such an orientation is not present, though much of the region to the east of 
the nucleus is illuminated. We find that the blue and redshifted kinematic 
components can instead be attributed to a single side of both cones, with the 
host disk extinguishing the other half of each cone, and that the bicone axis 
is near our line of sight. This alignment between bicone axis and host disk concurs 
with analysis in \citet{Cre02} which concludes that AGN reddening of NGC 3227 
is due to lukewarm absorbers coplanar with the host disk. 

\begin{figure}[h]
\centering
\begin{tabular}{cc}
\includegraphics[width=.4\linewidth]{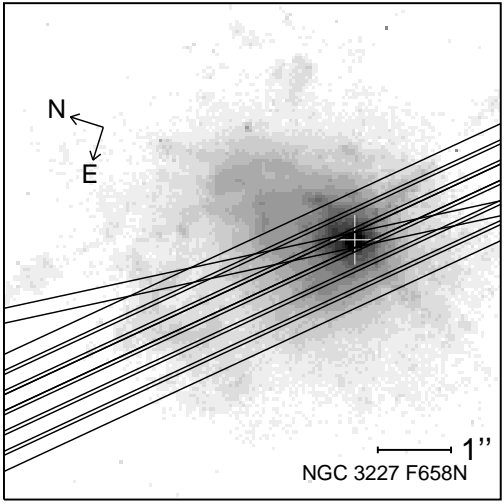} &
\includegraphics[width=.55\linewidth]{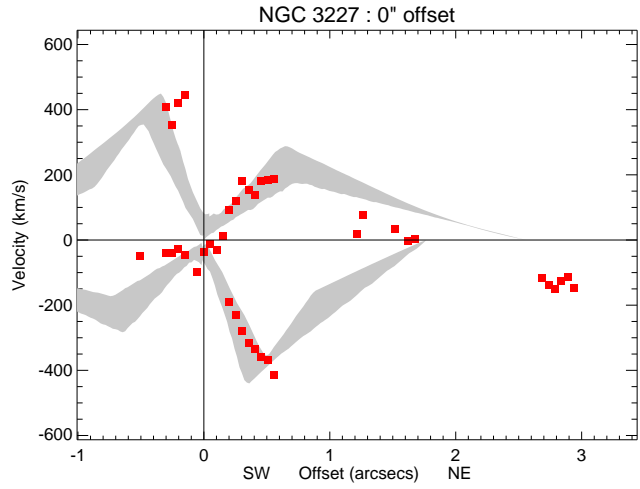} \\
\includegraphics[width=.4\linewidth]{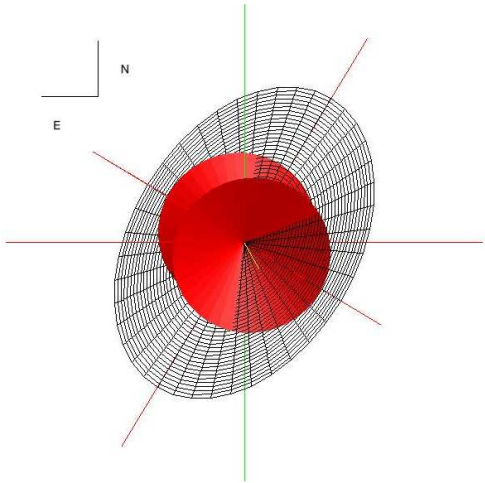} &
\includegraphics[width=.55\linewidth]{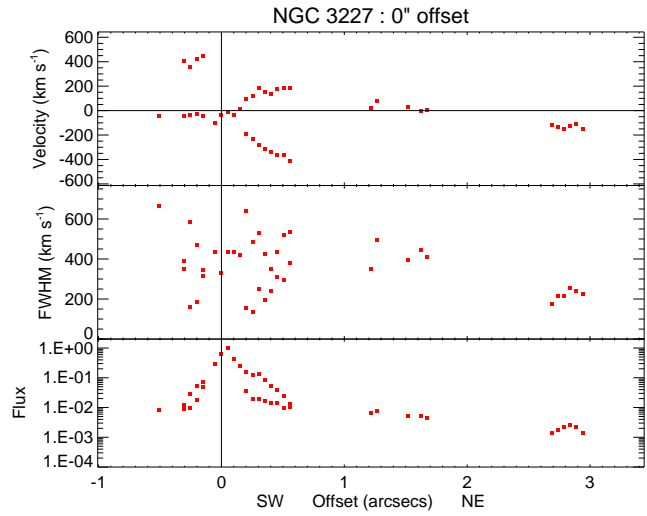}
\end{tabular}
\caption[NGC 3227 Kinematic and Geometric Models]{Same as Figure \ref{circinus}, but for NGC 3227 fitting four kinematic components.}
\label{ngc3227}
\end{figure}

\clearpage

%Slit positions and corresponding spectral 
%images of NGC 3227 STIS obserations are available in Figures \ref{fig1f}, 
%\ref{fig2i}, and \ref{fig2j}.

\subsection{{\it NGC 3783}} 

\begin{figure}[h]
\centering
\begin{tabular}{cc}
\includegraphics[width=.4\linewidth]{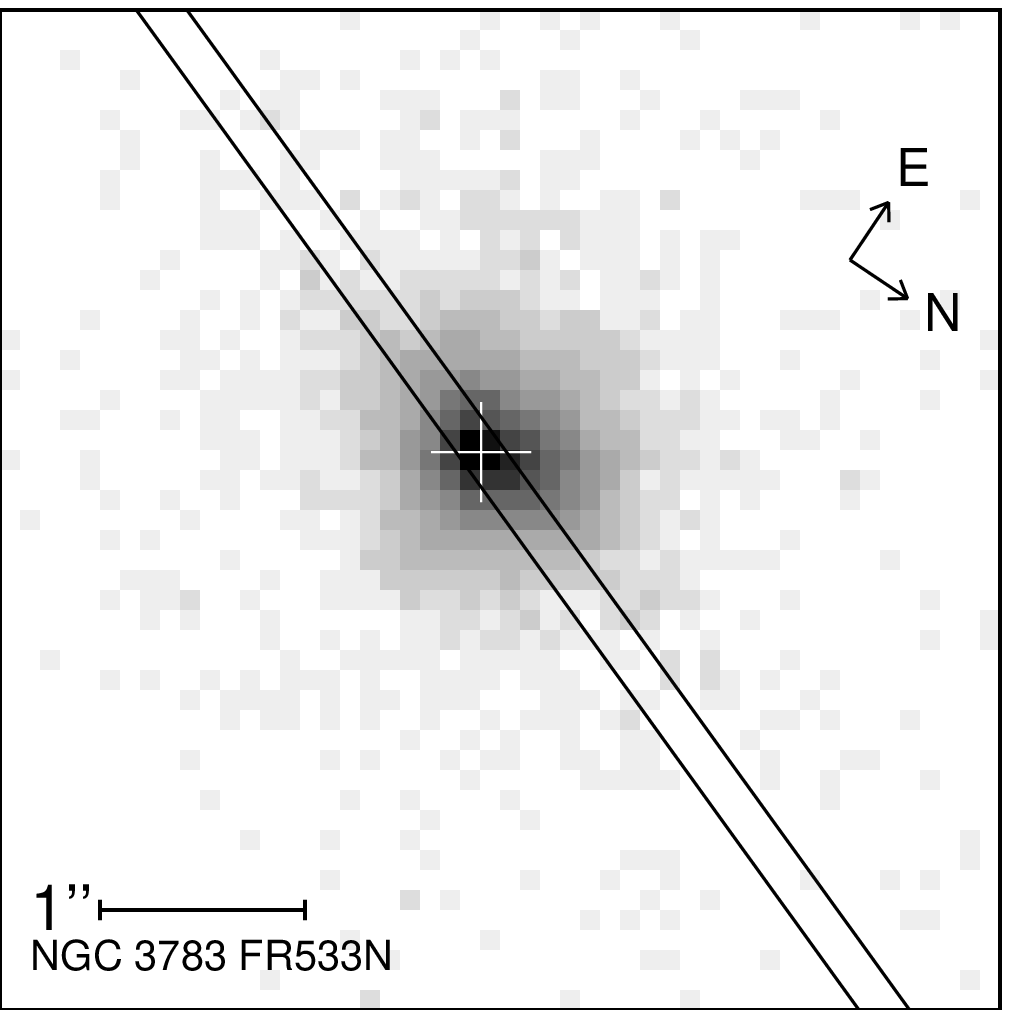} &
\includegraphics[width=.55\linewidth]{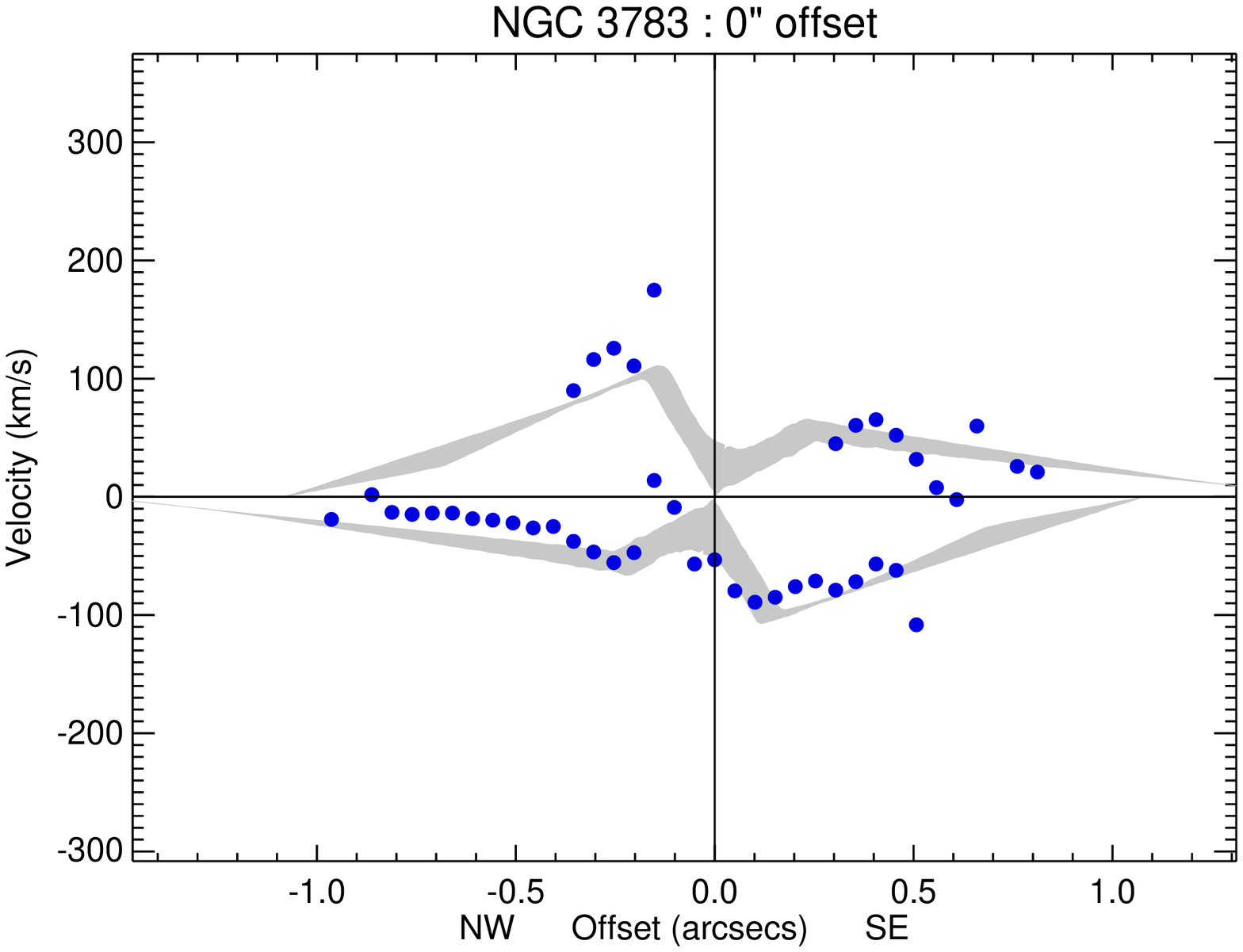} \\
\includegraphics[width=.4\linewidth]{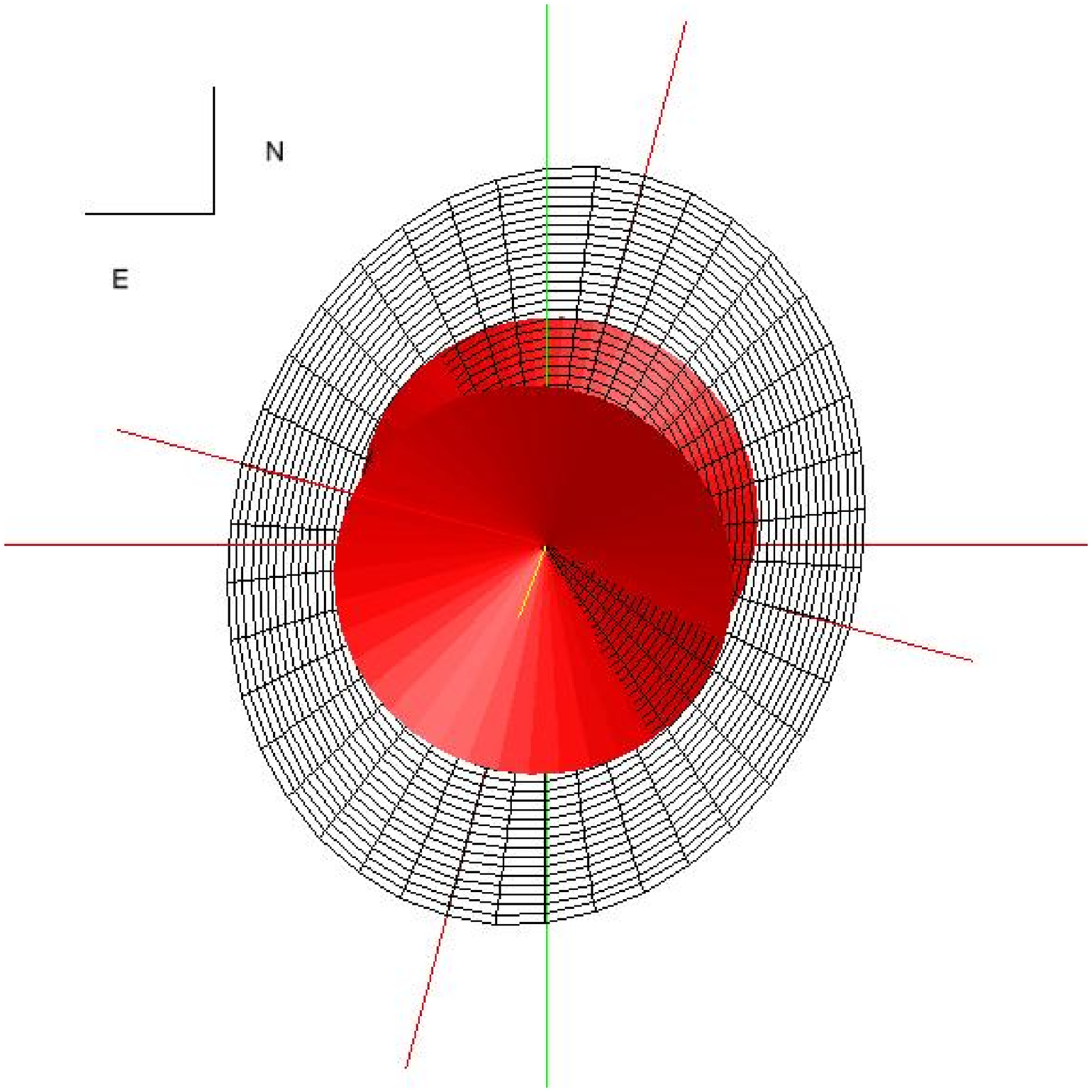} &
\includegraphics[width=.55\linewidth]{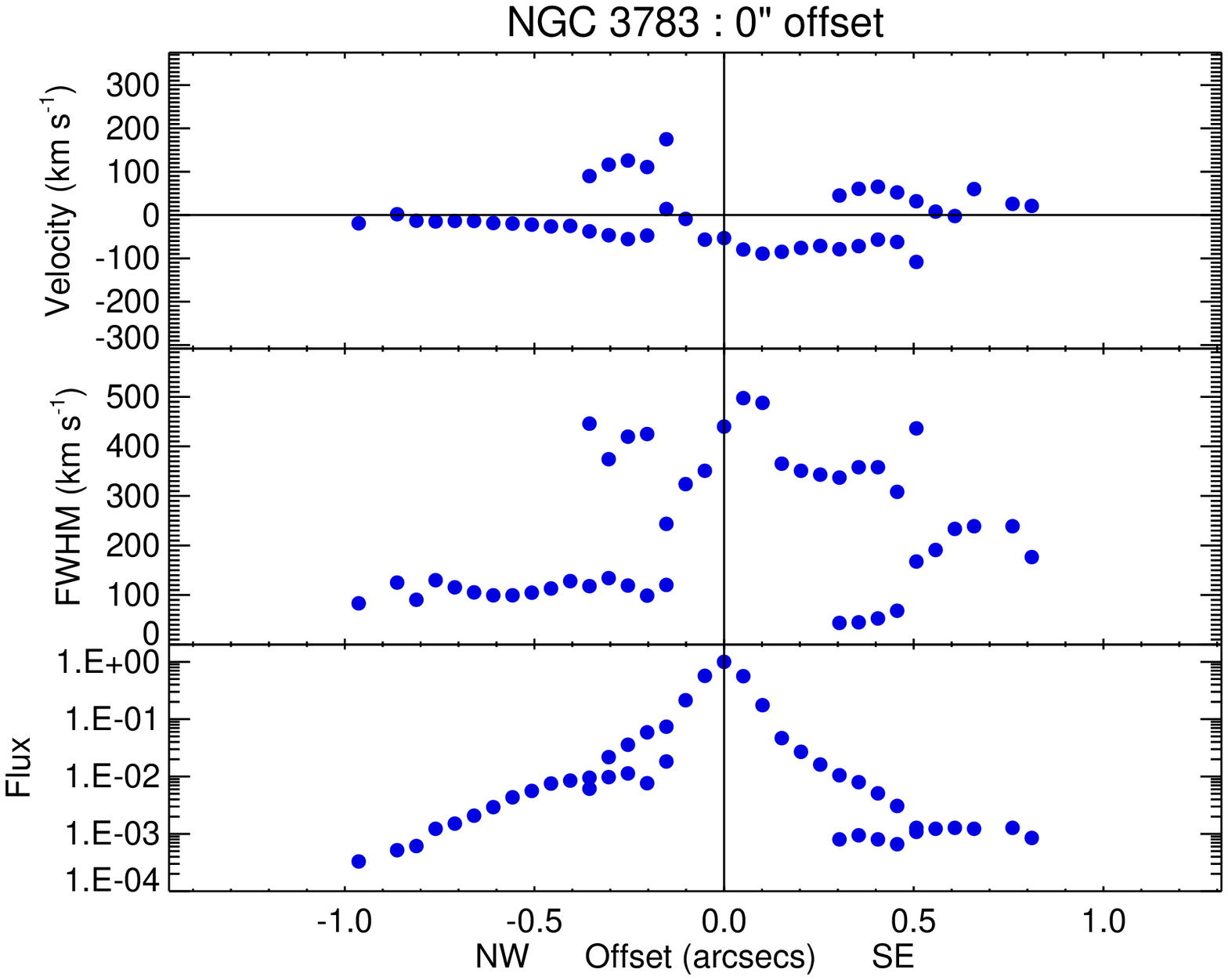}
\end{tabular}
\caption[NGC 3783 Kinematic and Geometric Models]{Same as Figure \ref{circinus}, but for NGC 3783 fitting four kinematic components.}
\label{ngc3783}
\end{figure}

{\it HST} [OIII] imaging in Figure \ref{ngc3783} depicts Seyfert 1 galaxy NGC 3783 as 
a nearly unresolved nuclear point source less than an arcsecond ($\sim$200 pc) in diameter. Emission detected 
with STIS reaches out to nearly twice that distance, with the [OIII] $\lambda$5007 
line remaining visible at fluxes less than 1\% of that at the continuum source. 
The resultant kinematics depict red and blue shifts on either side of the nucleus 
which fits well with a biconical outflow axis nearly 
perpendicular to the plane of the sky such that the kinematics for each cone are 
both redshifted or blue shifted. This pole-on geometry agrees fairly well the symmetric, 
compact point source in the [OIII] imaging and the Type 1 designation of the AGN.

This target was also studied by \citet{Mul11}, using the {\it Keck} OH-Suppressing 
Infrared Integral Field Spectrograph (OSIRIS) and kinematic models that include outflows based on the work from 
\citet{Cre00b}, where they publish a outflow parameter set with a more edge-on inclination and 
a view far outside of the bicone, invoking a clumpy torus model to explain their observations. 
Fitting their outflow $+$ rotation model parameters to our kinematics, outflows would contribute 
to the redshifted velocities to the northwest and blueshifted velocities to the southeast with 
rotational velocities accounting for the remaining kinematics. Using additional parameters of 
imaging and host disk geometry to confirm this model pose problems for this result.
[O~III] imaging shows no elongated structure along the proposed bicone axis (P.A. = -177$^{\circ}$), 
where one should expect emission on either side of the nucleus corresponding to a single cone.  
Additionally, their kinematic model states that the southern cone should be completely obscured 
by the host disk, however they provide a geometric model where the NLR outflow is instead bisected 
by the host disk, similar to our own model for NGC 3227.  

\clearpage

\subsection{{\it NGC 4051}}

Observations of NGC 4051 are of particular interest as it is a Narrow-Line Seyfert 1 (NLS1). 
NLS1s have permitted lines with widths from their BLRs that are $\leq$ 2000 km s$^{-1}$ 
(FWHM), which are narrower than those of ``normal'' broad-line Seyfert 1s (BLS1s), but still 
broader than forbidden lines from the NLR (typically $\sim$500 km s$^{-1}$ FWHM; \citealt{Ost87}). The current 
paradigm for NLS1s is that they have supermassive black holes (SMBHs) with relatively low 
masses compared to BLS1s and they are therefore radiating at close to their Eddington limits 
\citep{Pou95}, i.e. L/L$_{edd} \approx 1$. Should NLS1s be more ``pole-on'' than normal 
broad-line Seyfert 1s (BLS1s), we could determine if these properties are due instead to a 
special viewing angle for NLS1s as suggested by their compact radio morphologies \citep{Ulv95}.

\begin{figure}[h]
\centering
\begin{tabular}{cc}
\includegraphics[width=.4\linewidth]{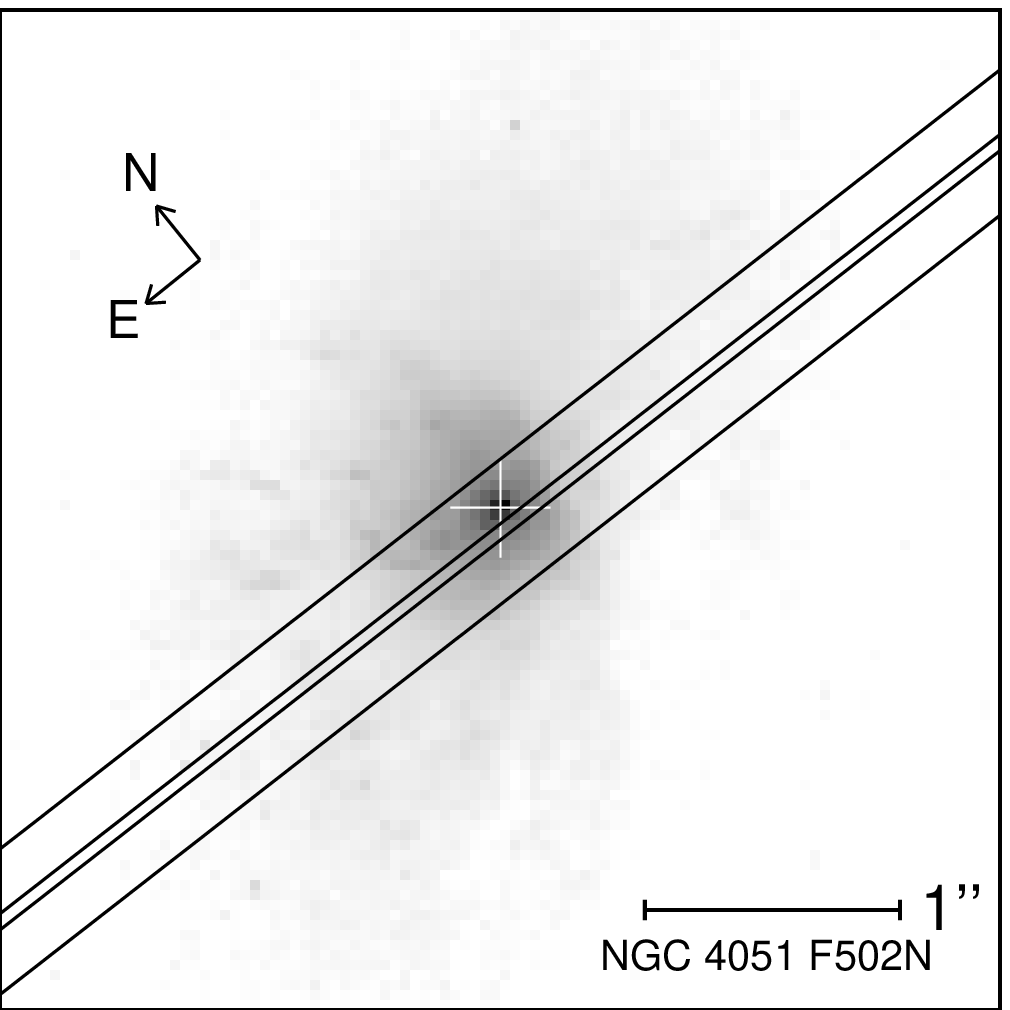} &
\includegraphics[width=.55\linewidth]{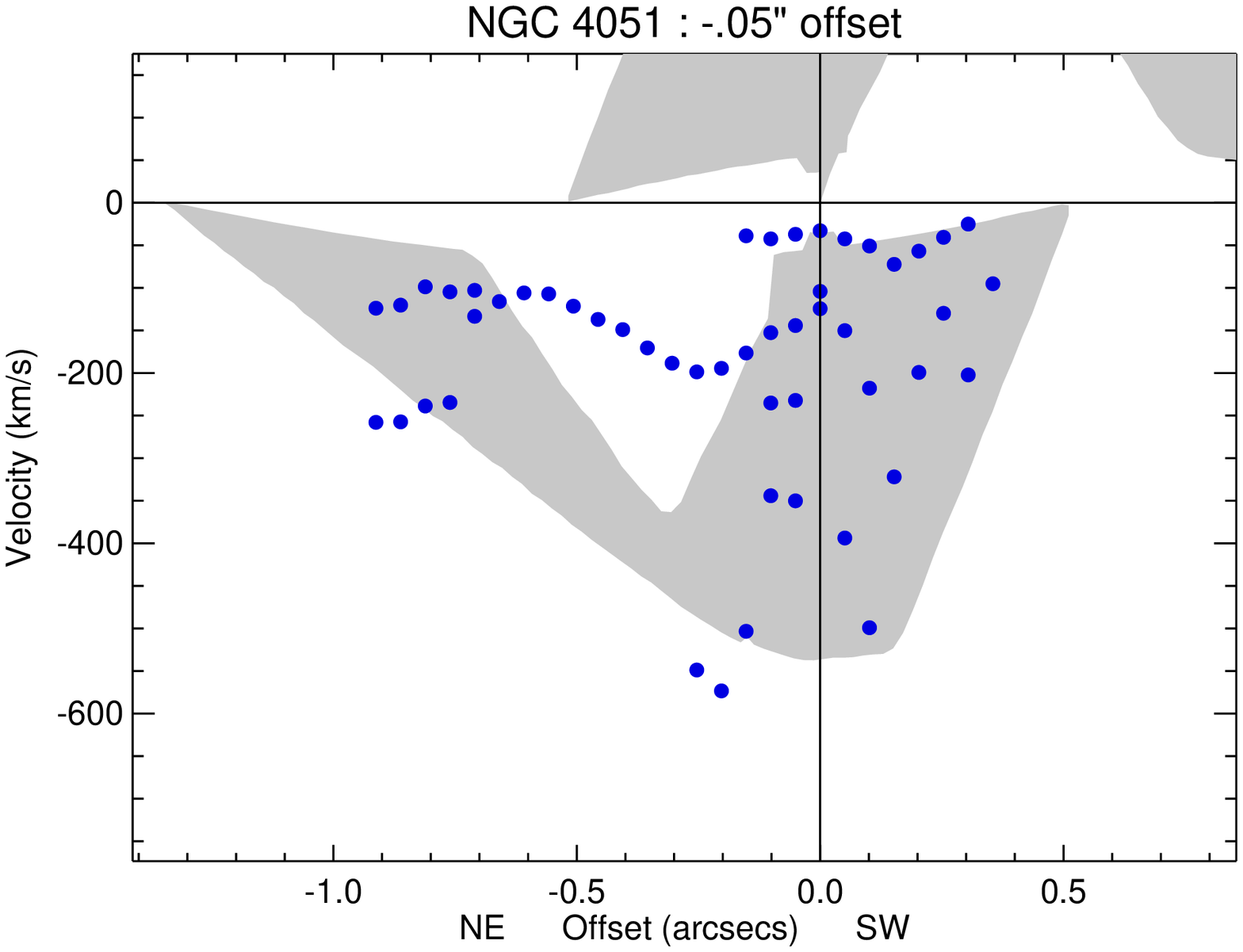} \\
\includegraphics[width=.4\linewidth]{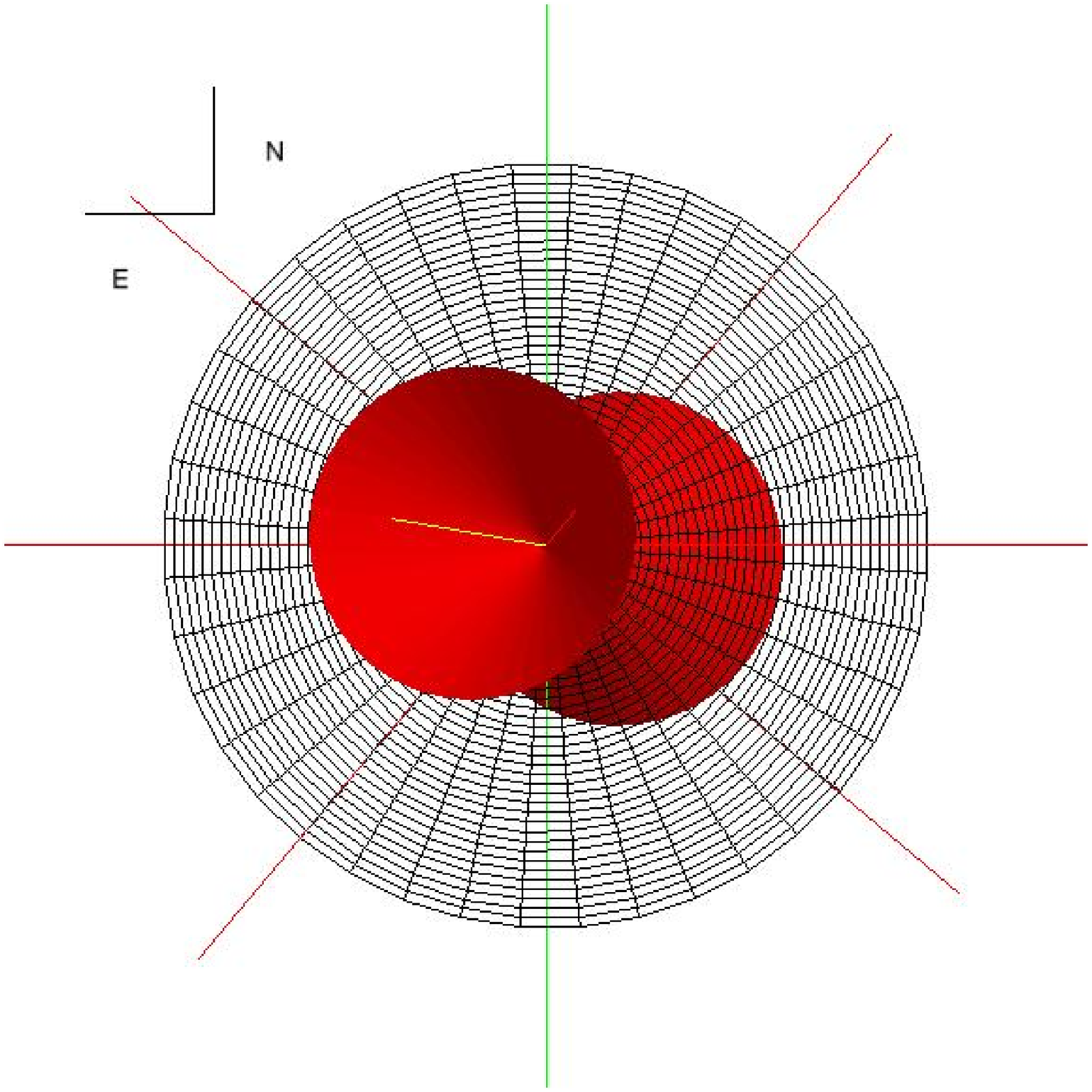} &
\includegraphics[width=.55\linewidth]{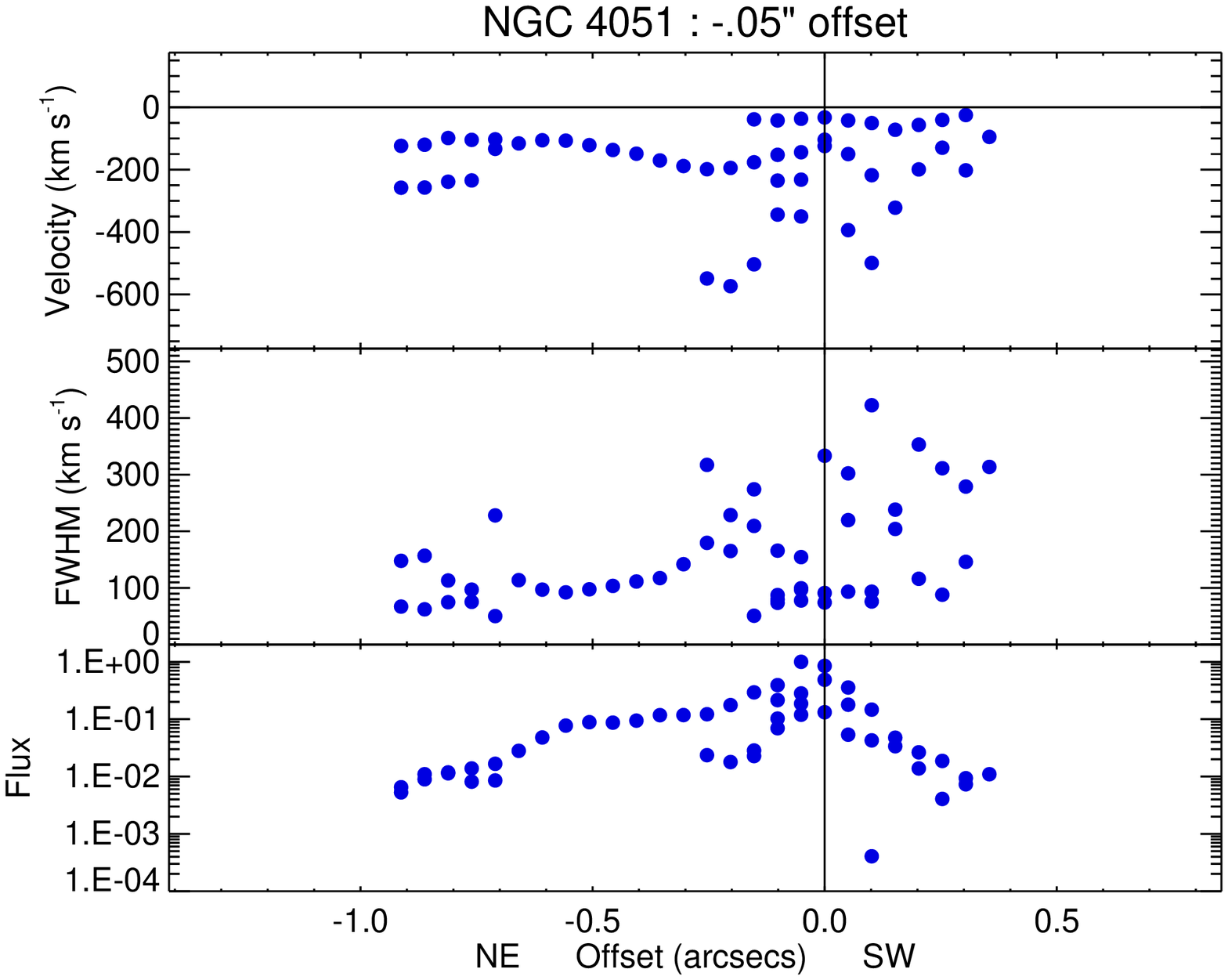}
\end{tabular}
\caption[NGC 4051 Kinematic and Geometric Models]{Same as Figure \ref{circinus}, but for NGC 4051 fitting two kinematic components.}
\label{ngc4051}
\end{figure}

The modeled kinematics provide a fair qualitative fit to the data (Figure \ref{ngc4051}), 
matching well to the inner accelerating blueshifted outflows. Using our final model 
parameters, we find that the axis of the bicone is inclined 12$^{\circ}$ away from 
pole-on, with our line of sight running near the edge of the NLR, between the inner 
and outer opening angles of the outflow. As we assume the highly blueshifted radial 
velocities are due to a biconical outflow, there is a noticeable lack of corresponding 
highly redshifted velocities near the nucleus. Combining the outer opening angle of 
the kinematic model with the inner disk geometry \citep{Bar09} suggests that the 
lack of redshifted outflow is likely due to disk obscuration. 

We find that the inclination of NGC 4051 is near pole-on at 12$^{\circ}$, 
and that the angle between the outer edge and our line of sight to be 
nearly the same. Analysis done for the three additional available unobscured NLS1s in our sample, 
Akn 564, Mrk 766 and Mrk 1040, show similar highly blueshifted velocities near their nuclei, 
suggest that a near pole-on orientation may be common. Unfortunately, the emission 
for these targets is too compact to fit with an accurate outflow model. Further study 
on the asymmetrical distribution of outflow velocities on either side of nuclei for 
both NLS1s and BLS1s will be required to test the pole-on hypothesis for NLS1s.

\clearpage

\subsection{{\it NGC 4507}}

\begin{figure}[h]
\centering
\begin{tabular}{cc}
\includegraphics[width=.4\linewidth]{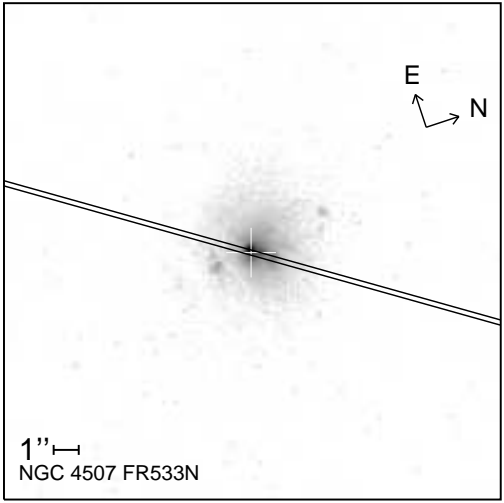} &
\includegraphics[width=.55\linewidth]{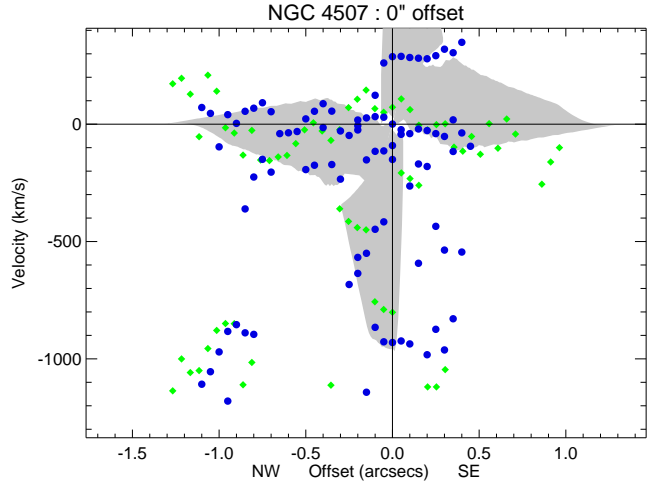} \\
\includegraphics[width=.4\linewidth]{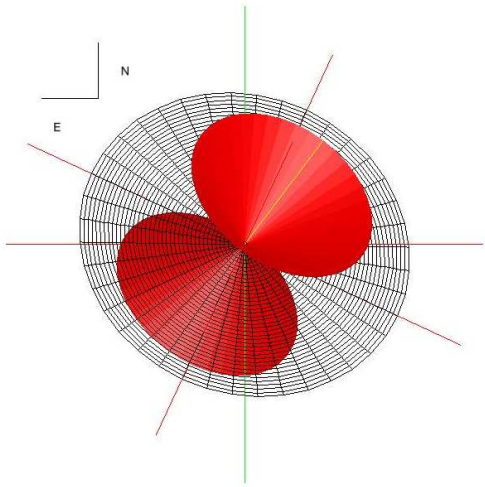} &
\includegraphics[width=.55\linewidth]{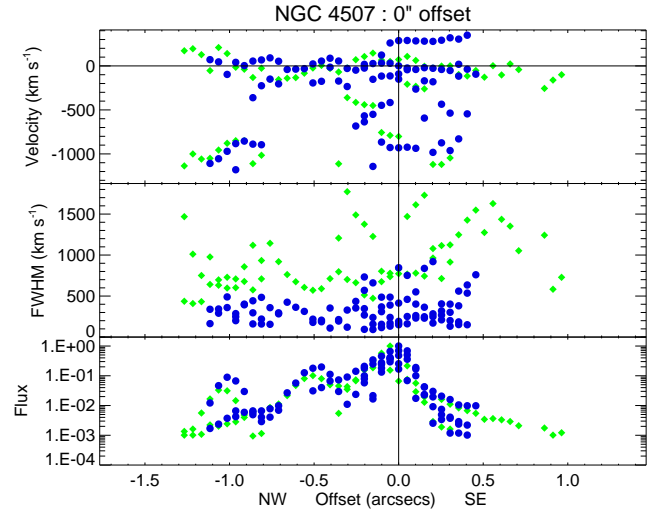}
\end{tabular}
\caption[NGC 4507 Kinematic and Geometric Models]{Same as Figure \ref{circinus}, but for NGC 4507 fitting three kinematic components.}
\label{ngc4507}
\end{figure}

This is a Seyfert 2 galaxy that features two extended knots of emission. The first, 
$0.5''$ northwest of the nucleus along the STIS slit, is traveling near systemic 
velocity and fits well in the final kinematic model. The second knot, at $1''$ northwest 
of the nucleus, is highly blueshifted at 1000 km s$^{-1}$. This second knot is abnormal 
in that it is a spatially resolved, high velocity knot at a large distance from the 
nucleus such that it cannot be fit with our model. Rogue high velocity clouds such as 
this have also been documented in the kinematics of NGC 4151 \citep{Das05}, though 
they are located at radii near the peak of the velocity curve versus the end of the 
curve in this case. Two other kinematic components exist in the blueshifted quadrant to 
the northwest of the nucleus, thus it is unlikely to be a part of the typical NLR outflow 
and is possibly a cloud near the axis of the bicone. Though many highly blueshifted 
velocities exist near the nucleus, the blueshifted component was fit such that the higher 
flux velocities (Figure \ref{ngc4507}) took priority. Additionally, opening the bicone 
to accommodate the high, southeast velocities would allow for the viewing of the central 
engine, which does not currently occur as our line of sight runs in between the inner 
and outer opening angles of the NLR. 

%Slit positions and corresponding spectral images of 
%NGC 4507 STIS obserations are available in Figures \ref{fig1g} and \ref{fig2l}.

\clearpage

\subsection{{\it NGC 5506}}\label{5506}

\begin{figure}[h]
\centering
\begin{tabular}{cc}
\includegraphics[width=.4\linewidth]{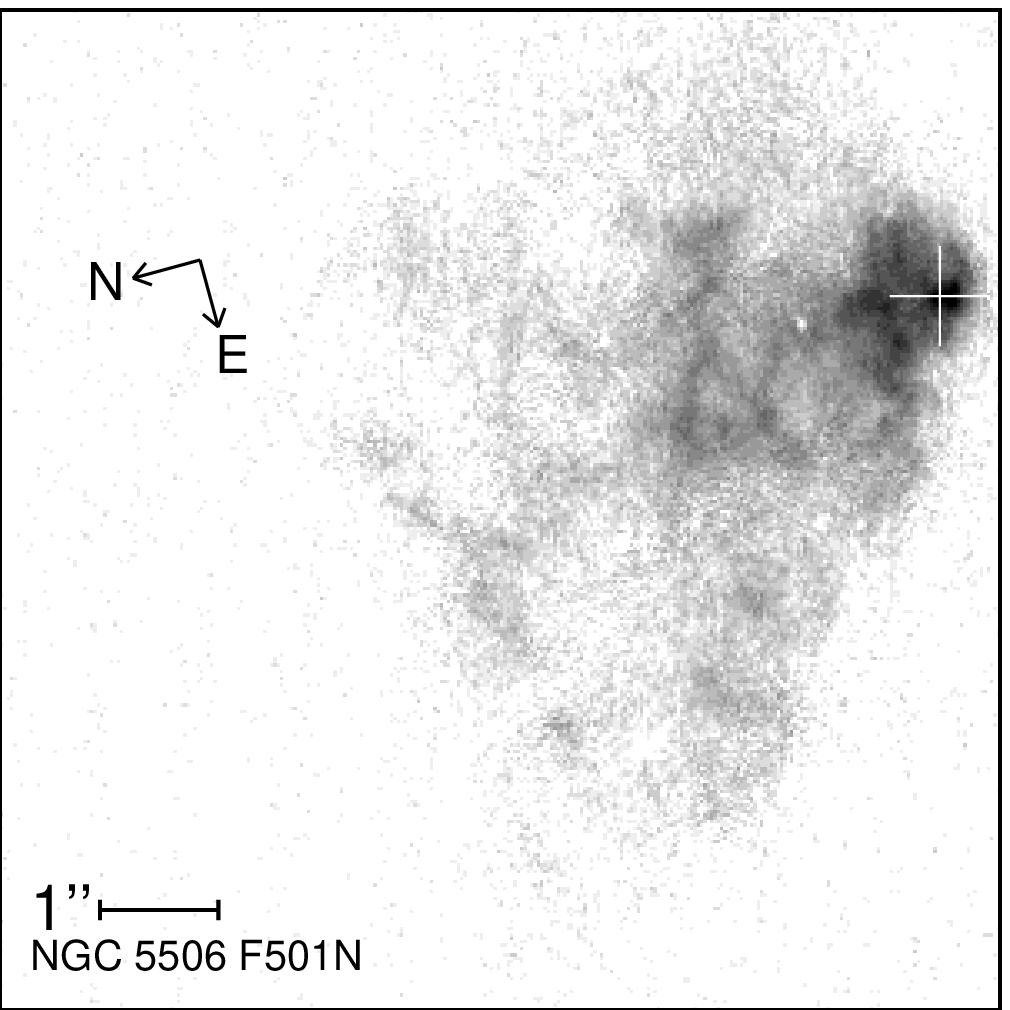} &
\includegraphics[width=.55\linewidth]{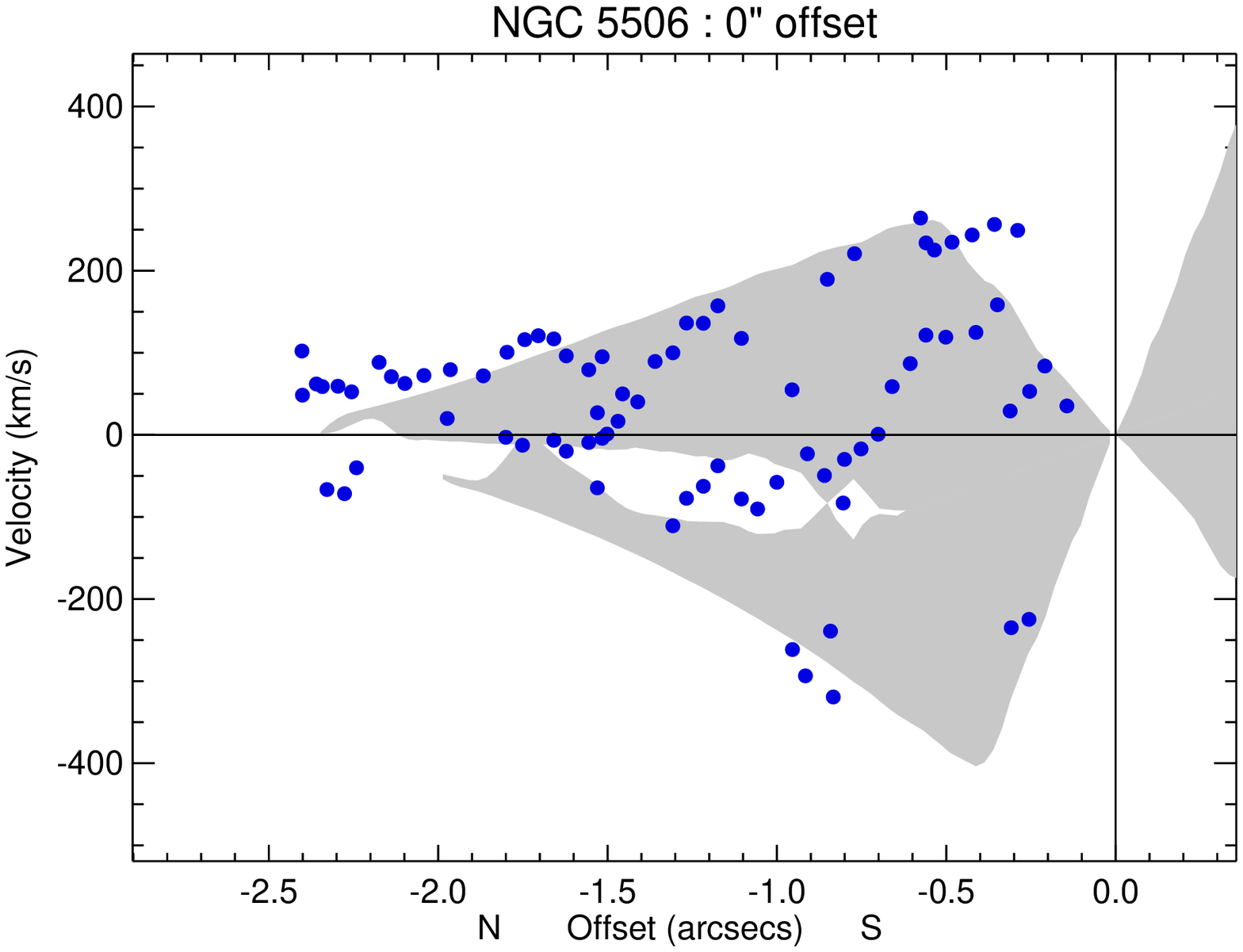} \\
\includegraphics[width=.4\linewidth]{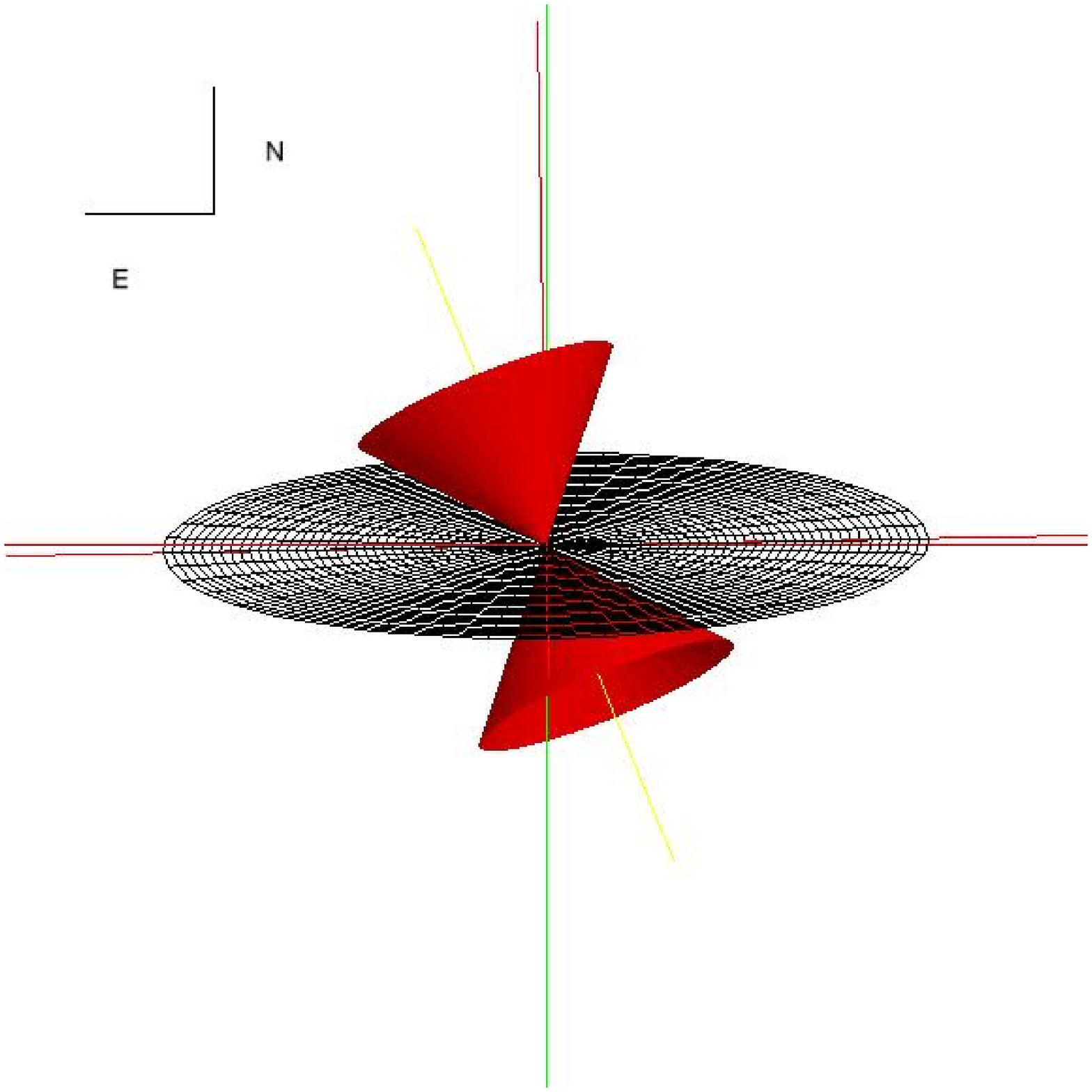} &
\end{tabular}
\caption[NGC 5506 Kinematic and Geometric Models]{Same as Figure \ref{circinus}, but for NGC 5506 fitting two kinematic components.
Kinematics are from \citet{Rui01} data set, no FWHM or fluxes available.}
\label{ngc5506}
\end{figure}

This is a debated NLS1 \citep{Gua10} / Seyfert 1.9 \citep{Mai95} / Seyfert 2 \citep{Tri10}, whose 
classification is likely muddled by the fact that it resides in a near edge-on ($i = 76^{\circ}$) 
host. From imaging (Figure \ref{ngc5506}, see also \citet{Mal98}), a single, well-defined 
NLR cone is visible with its axis appearing to be near the plane of the sky and perpendicular 
to the host disk. Spectra available for this target consist solely of the slitless \citet{Rui05} observations 
which can be fit well with a model that also agrees with the imaging. However, this edge-on AGN orientation 
does not conform to the unified model. With a majority of the extended emission being redshifted, it is possible 
that a stronger blueshifted component is obscured by the host disk \citep{Ima00}. As this is the only AGN with 
an edge-on host disk in our modeled sample, it is difficult to determine if the NLR of this AGN is also 
truly edge-on or if the host disk is extinguishing a large portion of the NLR emission, disguising a natural 
Type 1 AGN as an observed quasi-Type 2, and
we are left modeling what remains of the NLR peering through the host disk. Additional evidence that the 
NLR model of NGC 5506 is suspect lies in a correlation between inclination and neutral hydrogen column density (N$_H$)
(Fischer et al. in prep.) where AGN observed further from the axis of their NLR and closer to the toroidal structure 
surrounding the central engine have larger column densities. With a column density of 
$2.78 \times 10^{22}$ cm$^{-3}$ \citep{Win09}, NGC 5506 has one of the smallest densities in our sample, 
two to three orders of magnitude smaller than all other AGN with highly inclined NLRs. This implies a shallower 
inclination, more toward our LOS, than what is observed. As BLR emission has already 
successfully been detected in the near-IR \citep{Nag02}, further study of the NLR in the IR regime would avoid 
current extinction problems present at optical wavelengths and likely allow us to observe the full extent of the 
biconical outflow.

\clearpage

\subsection{{\it NGC 5643}}

\begin{figure}[h]
\centering
\begin{tabular}{cc}
\includegraphics[width=.4\linewidth]{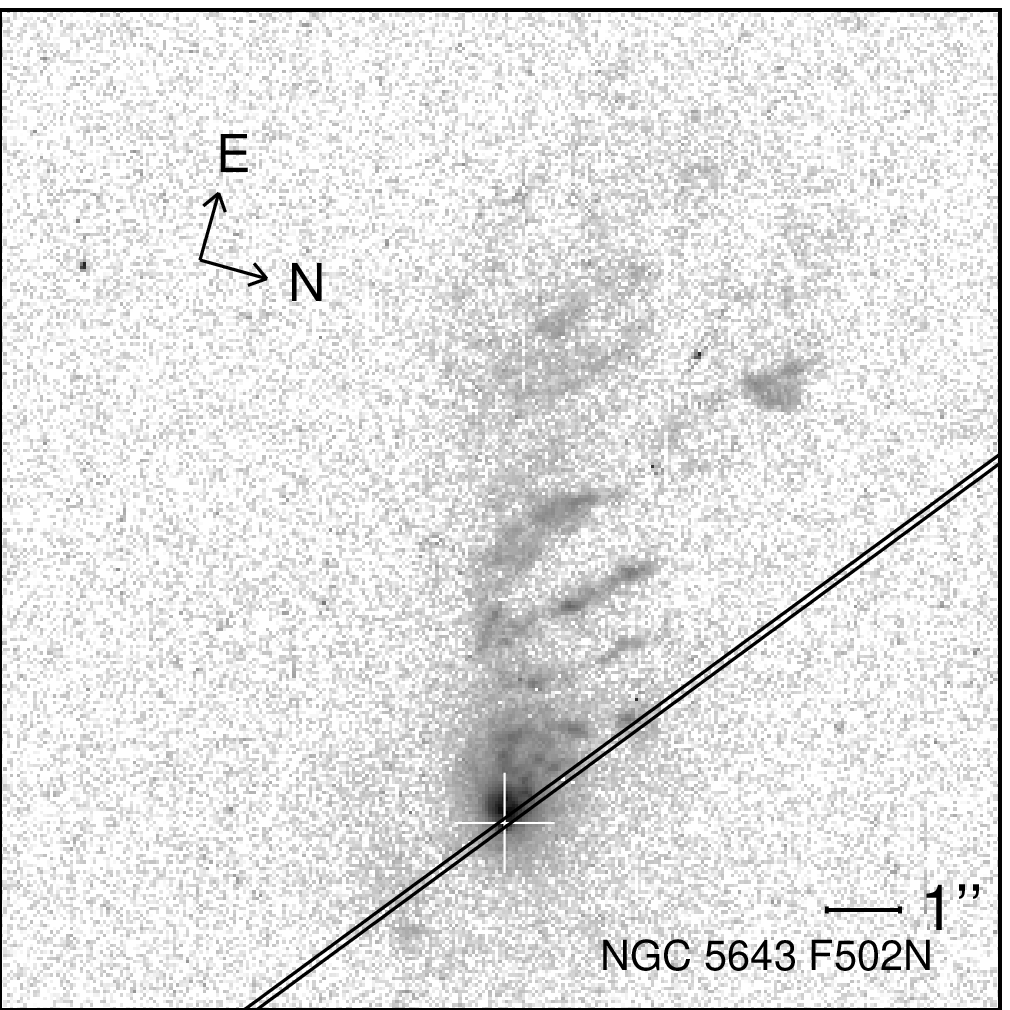} &
\includegraphics[width=.55\linewidth]{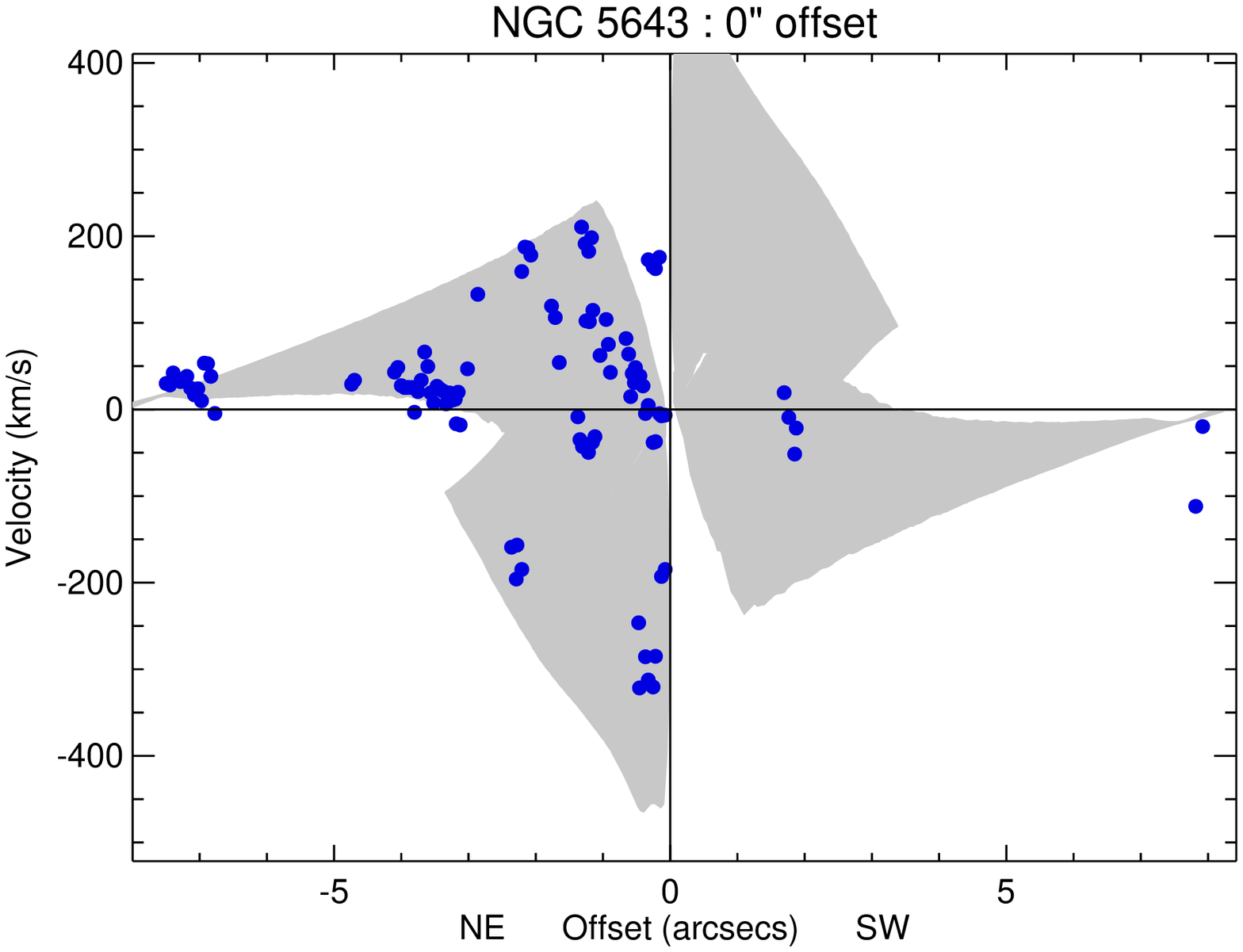} \\
\includegraphics[width=.4\linewidth]{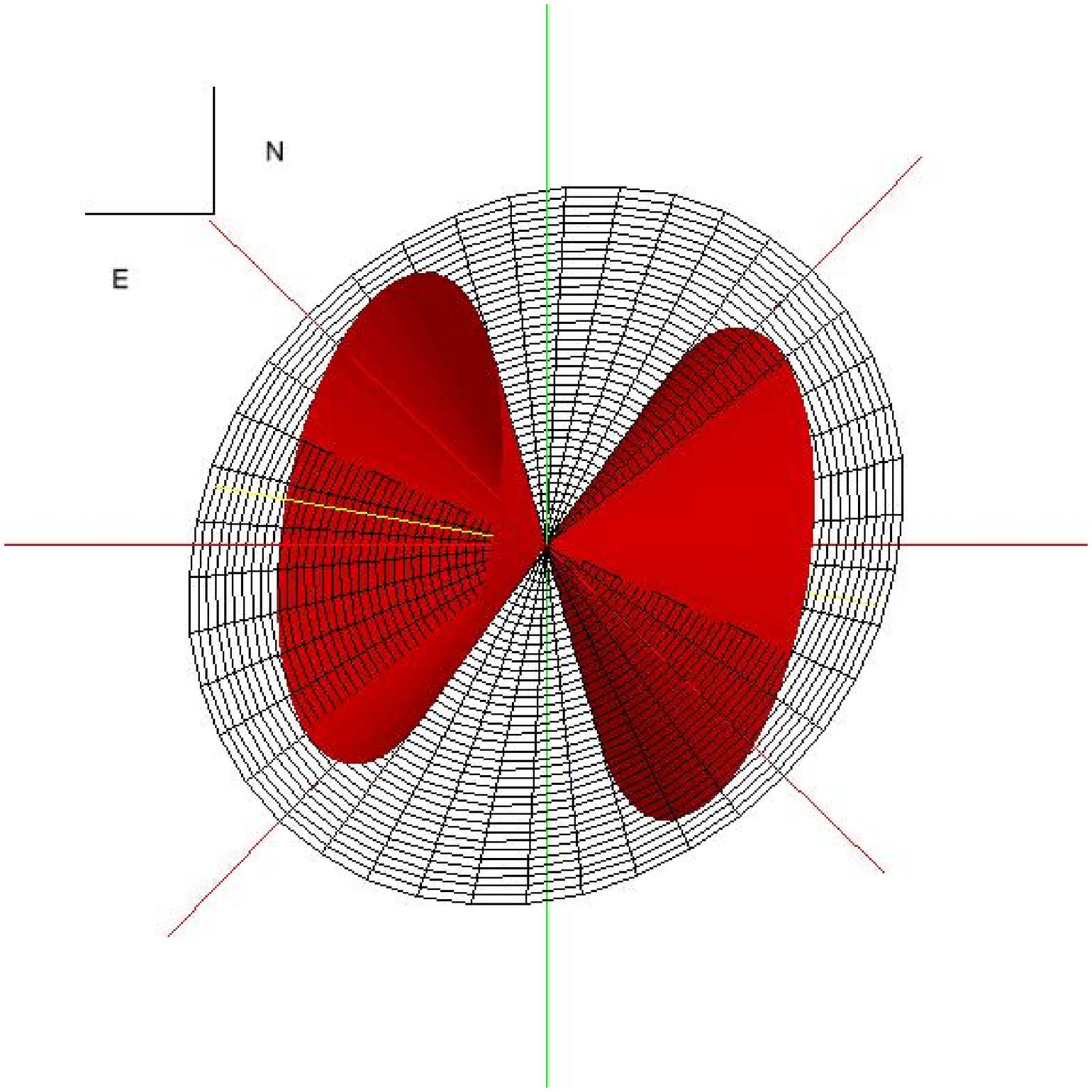} &
\includegraphics[width=.55\linewidth]{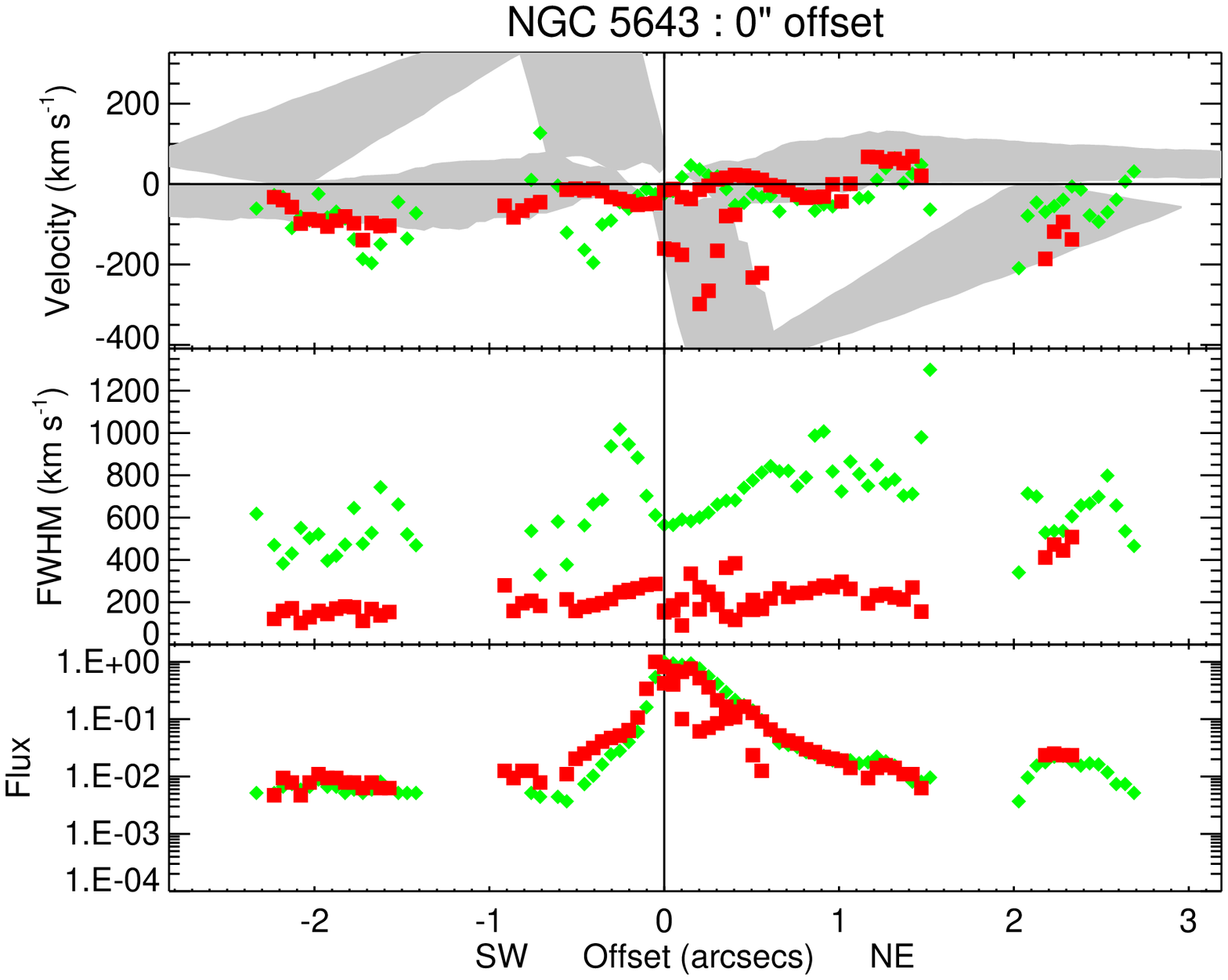}
\end{tabular}
\caption[NGC 5643 Kinematic and Geometric Models]{Top Left: {\it HST} imaging of NGC 5643 with STIS observation positions overplotted. 
Top Right: Kinematic model for the slitless kinematics of NGC 5643 fitting two kinematic 
components. Slitless kinematics are from \citet{Rui01} data set, no FWHM or fluxes available. Bottom Left: Corresponding geometric model of the NLR with 
disk geometry. The bicone axis is illustrated as a yellow line. Black axes illustrate the extended plane of the host galaxy. Red 
axes illustrate the plane of the sky. Bottom Right: Velocities with overlayed kinematic model, FWHMs, and fluxes normalized to the 
highest measured flux for the long-slit observation of NGC 5643. Green diamonds, blue circles, and red squares corresond to G430L, 
G430M, and G750M grating data respectively.}
\label{ngc5643}
\end{figure}

This is a Seyfert 2 which has a well-defined triangular emission region east of the 
nucleus, as seen in Figure \ref{ngc5643}. Filamentary structure within this ionized region 
appear to follow spiral arms in the host disk \citep{Mor85}. Kinematics from slitless observations 
show that the majority of the extended emission is redshifted, with blueshifted velocities 
only visible near the nucleus. We fit a model to these kinematics that corresponds to a 
single, wide opening-angle cone inclined 65$^{\circ}$ from our line of sight. 
Available G430L and G750M long-slit observations have a position angle of -128$^{\circ}$, placing them 
outside a majority of the NLR. Nonetheless, extracting components along the position angle of these 
observations shows that the model also fits these kinematics as well. Similar to 
Mrk 573, the extended [O~III] emission seen in the imaging appears to result from the 
intersection between the host disk and the ionization bicone, providing a much narrower 
apparent opening angle that that of the model bicon. Our geometric model 
suggests that extended blueshifted emission should be observed west of the nucleus. This 
emission may be quenched by a warped disk or the presence of a dust lane to the 
southwest of the nucleus where traces of emission 4.5$''$ and 8.2$''$ west of the 
nucleus could represent emission seen through less obscured portions of the disk 
\citep{Sim97}. Spectroscopic observations by \citet{Sch94} support this hypothesis as 
line emission to the west of the nucleus is more heavily reddened than that to the east. 

\clearpage

\subsection{{\it NGC 7674}}

NGC 7674 is a Seyfert 2 with a distribution of radial velocities similar to Circinus with 
asymmetric extended emission to the northeast with high blueshifts and near systemic 
redshifts. The modeled kinematics provide an excellent fit to the data, clearly fitting 
the two observed components and what remains of the other velocities. Our model suggests 
the extended emission is due to a single cone, with the other cone obscured by the host 
disk. Radio data \citep{Mom03} concurs with our model as jet axes are projected along the same 
position angle as the NLR bicone in our model, which also suggests a qualitatively good fit.

%Slit positions and corresponding spectral images of NGC 4507 STIS obserations are 
%available in Figures \ref{fig1k} and \ref{fig2p}.

\begin{figure}[h]
\centering
\begin{tabular}{cc}
\includegraphics[width=.4\linewidth]{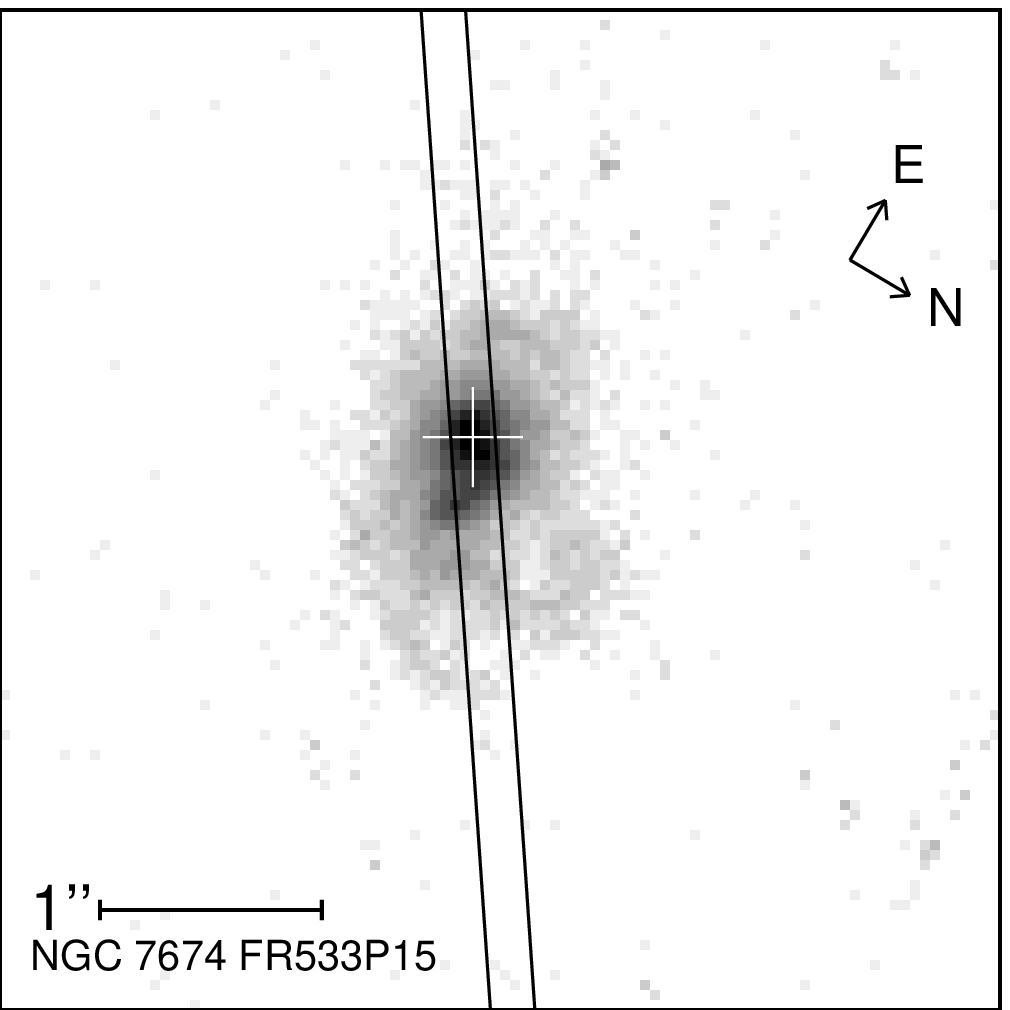} &
\includegraphics[width=.55\linewidth]{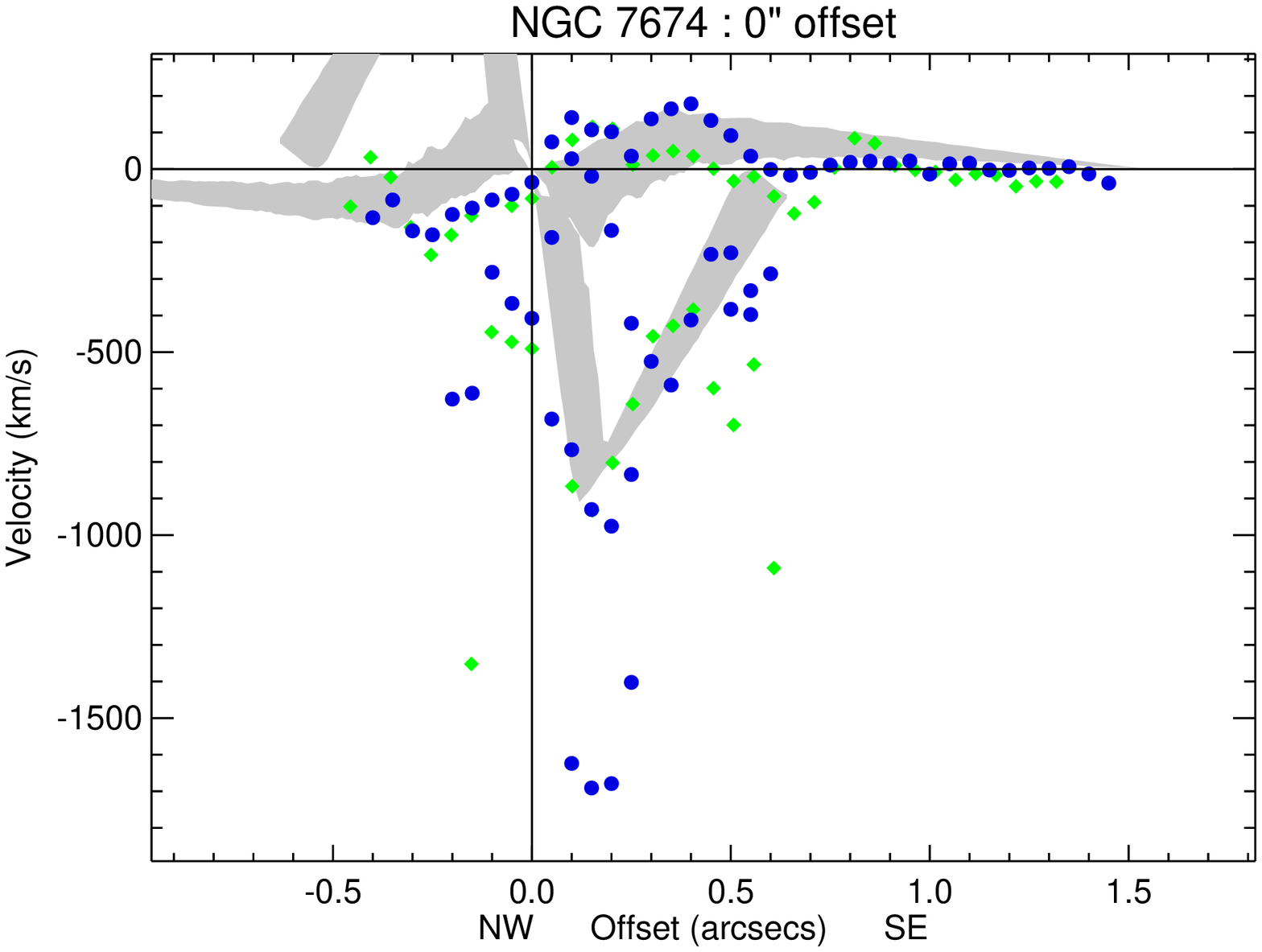} \\
\includegraphics[width=.4\linewidth]{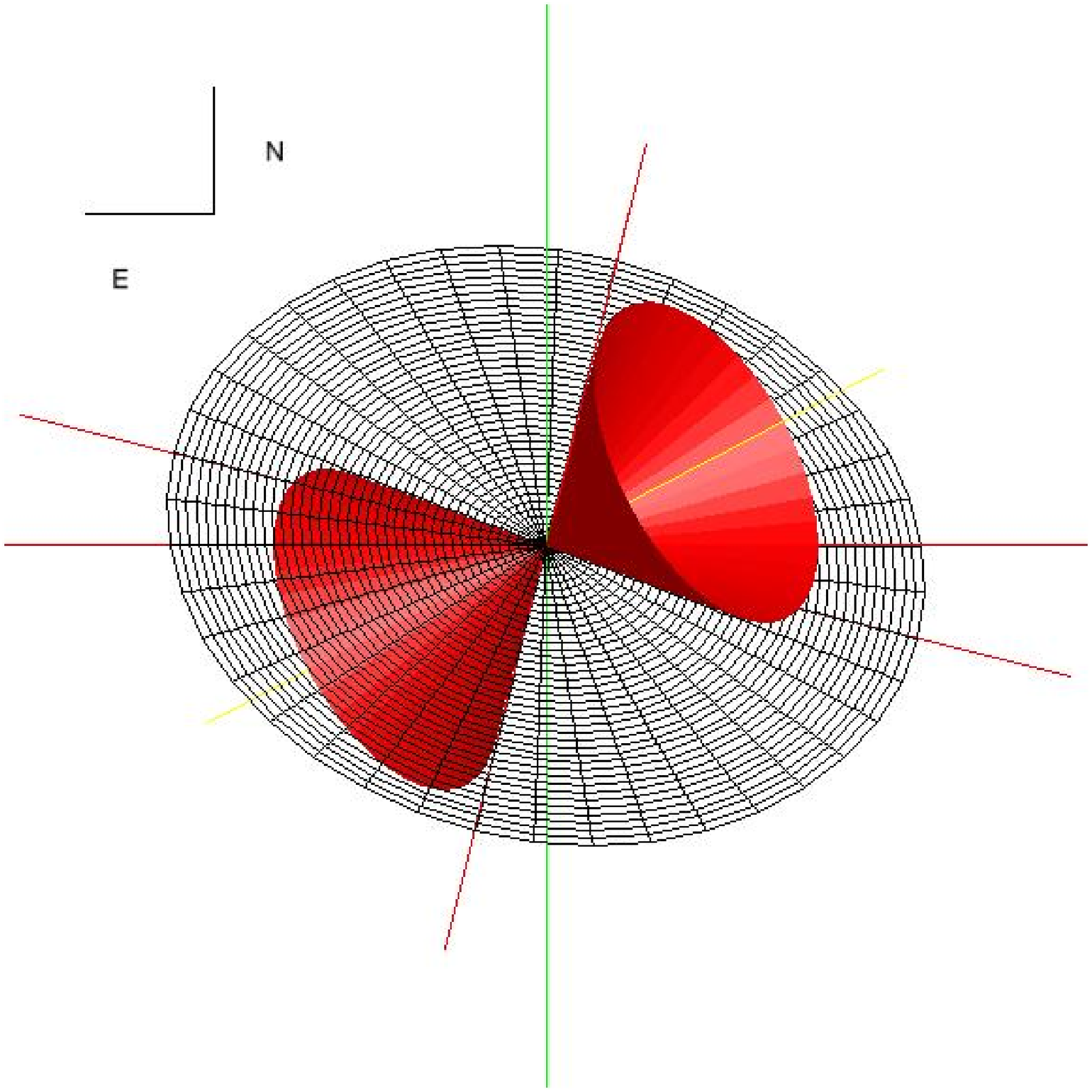} &
\includegraphics[width=.55\linewidth]{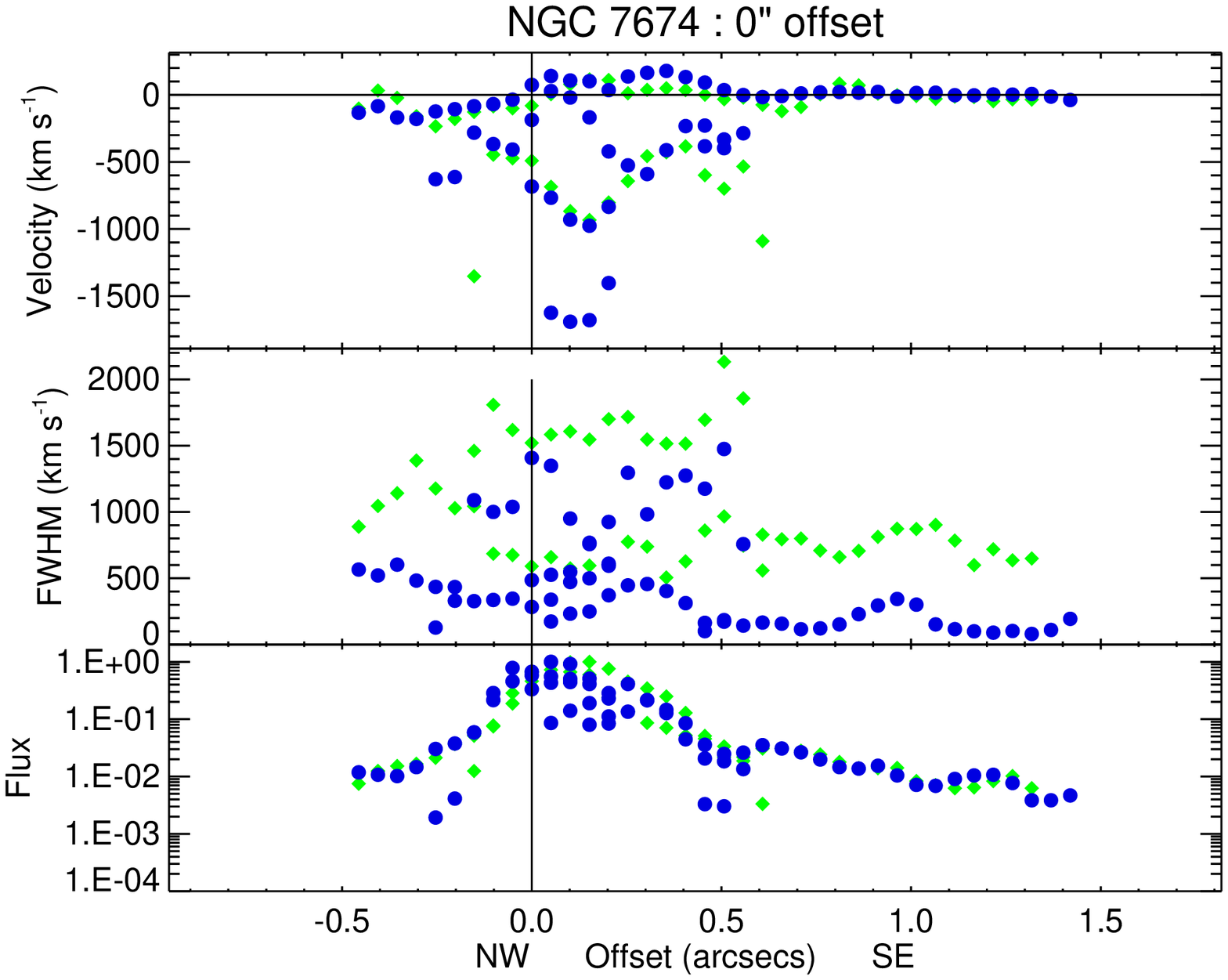}
\end{tabular}
\caption[NGC 7674 Kinematic and Geometric Models]{Same as Figure \ref{circinus}, but for NGC 7674 fitting two kinematic components.}
\label{ngc7674}
\end{figure}

\clearpage

\section{Distinct Unmodeled Targets}
\label{sec7}

\subsection{{\it IRAS 11058-1131}}

\begin{figure}[h]
\centering
\begin{tabular}{cc}
\includegraphics[width=.4\linewidth]{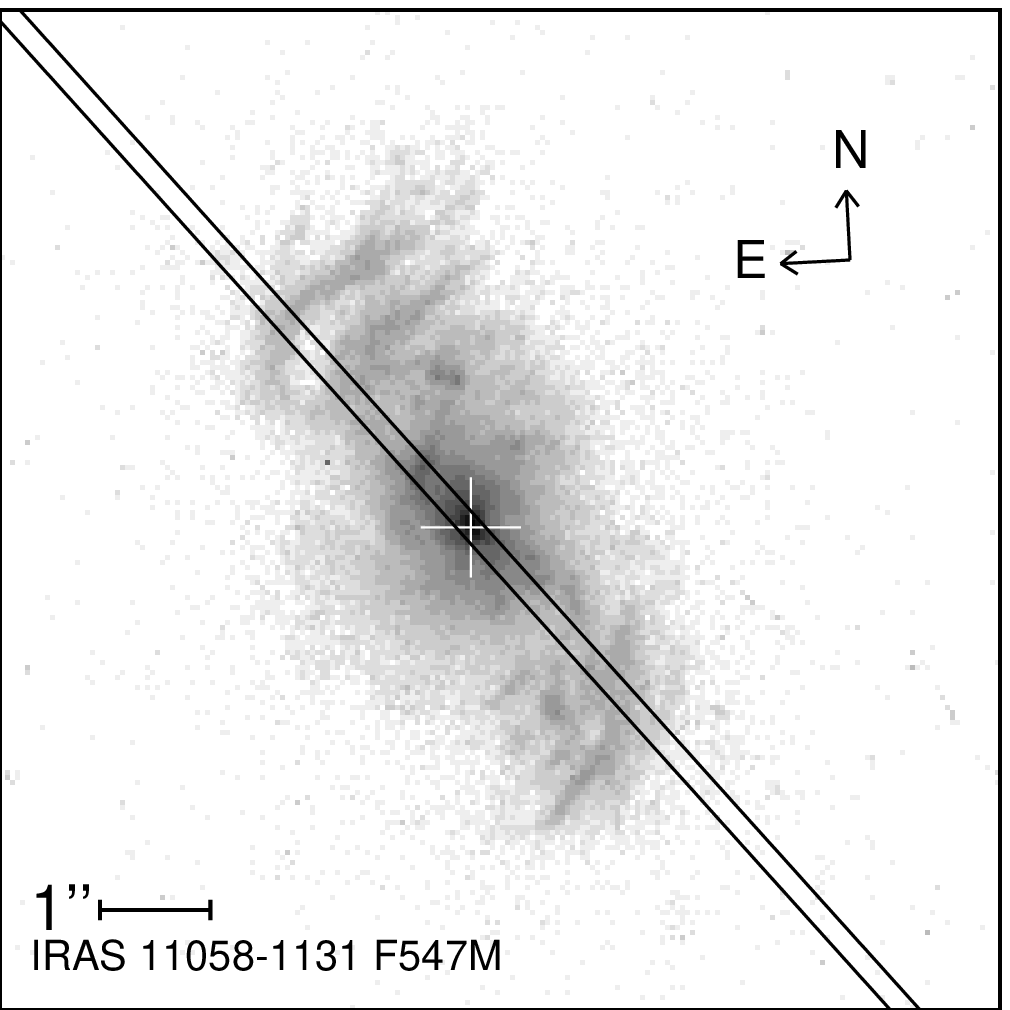} &
\includegraphics[width=.55\linewidth]{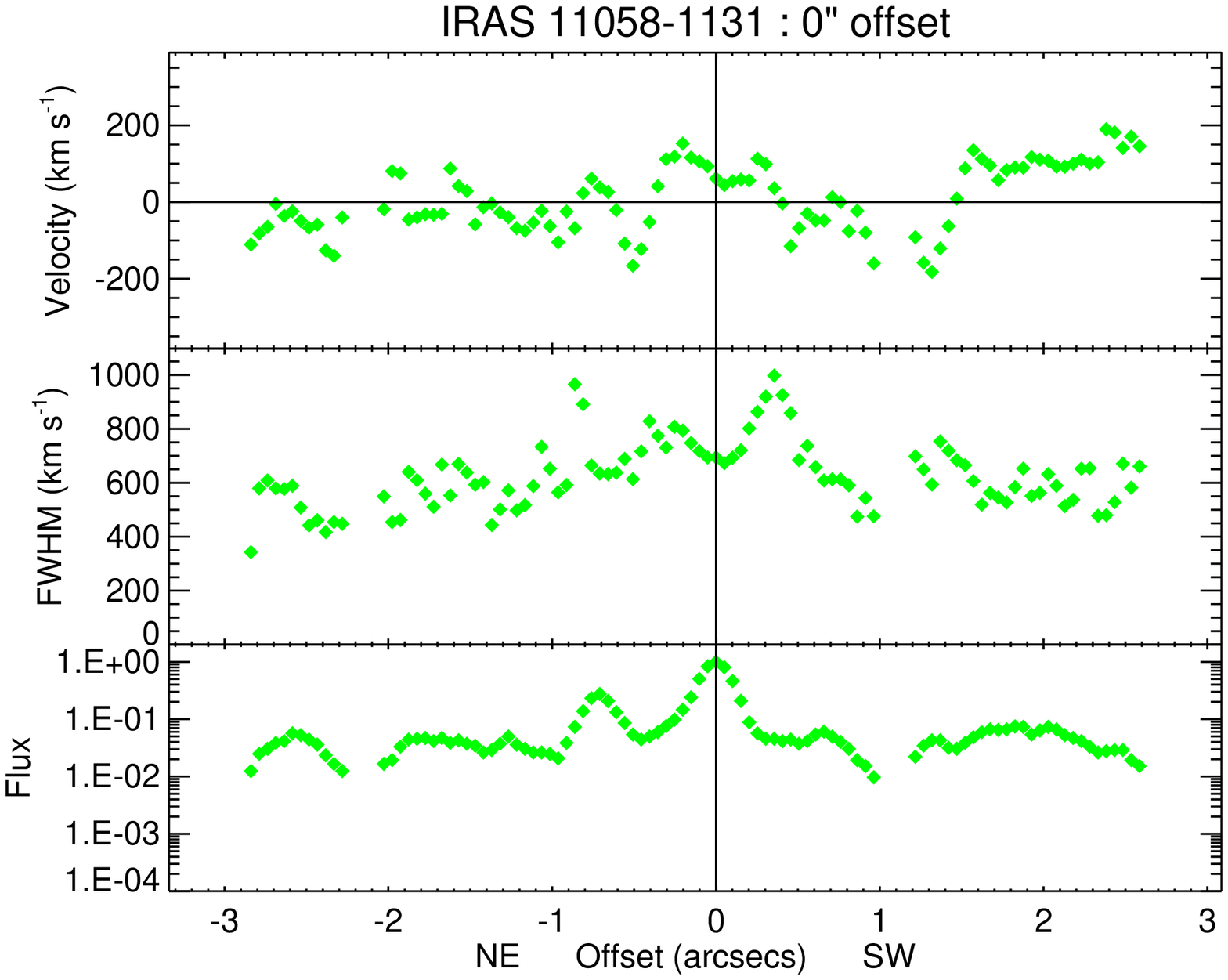}
\end{tabular}
\caption[IRAS 11058-1131 Kinematics]{Left: {\it HST} 
imaging of IRAS 11058-1131 with STIS observation position overplotted. 
Right: Velocities, FWHMs, and normalized fluxes across the slit position.}
\label{iras11058}
\end{figure}

This is a Seyfert 2 that shows a likely intersection between the NLR and host disk in the 
visible S-shaped NLR seen in Figure \ref{iras11058}. The kinematics of this target have been 
deemed 'complex' as no individual kinematic components can be seen in the radial velocity 
data. Velocities appear to be roughly rotational, as velocities 
further from the nucleus do not decelerate back toward systemic, but end with blueshifted 
values to the northeast and redshifted values to the southwest, reaching maximum velocities 
of $\sim$200 km s$^{-1}$. This implies a counter-clockwise rotation, which agrees with the 
apparent winding-up of the spiral arms that are illuminated. Velocity gradients responsible 
for the knotty, 'complex' kinematics observed may be due to in situ acceleration of ambient material 
similar to those seen in dust lanes of Mrk 573 \citep{Fis10}, though these gradients fail 
to align with specific flux peaks in the data.

\clearpage

\subsection{{\it Mrk 509}}

\begin{figure}[h]
\centering
\begin{tabular}{cc}
\includegraphics[width=.4\linewidth]{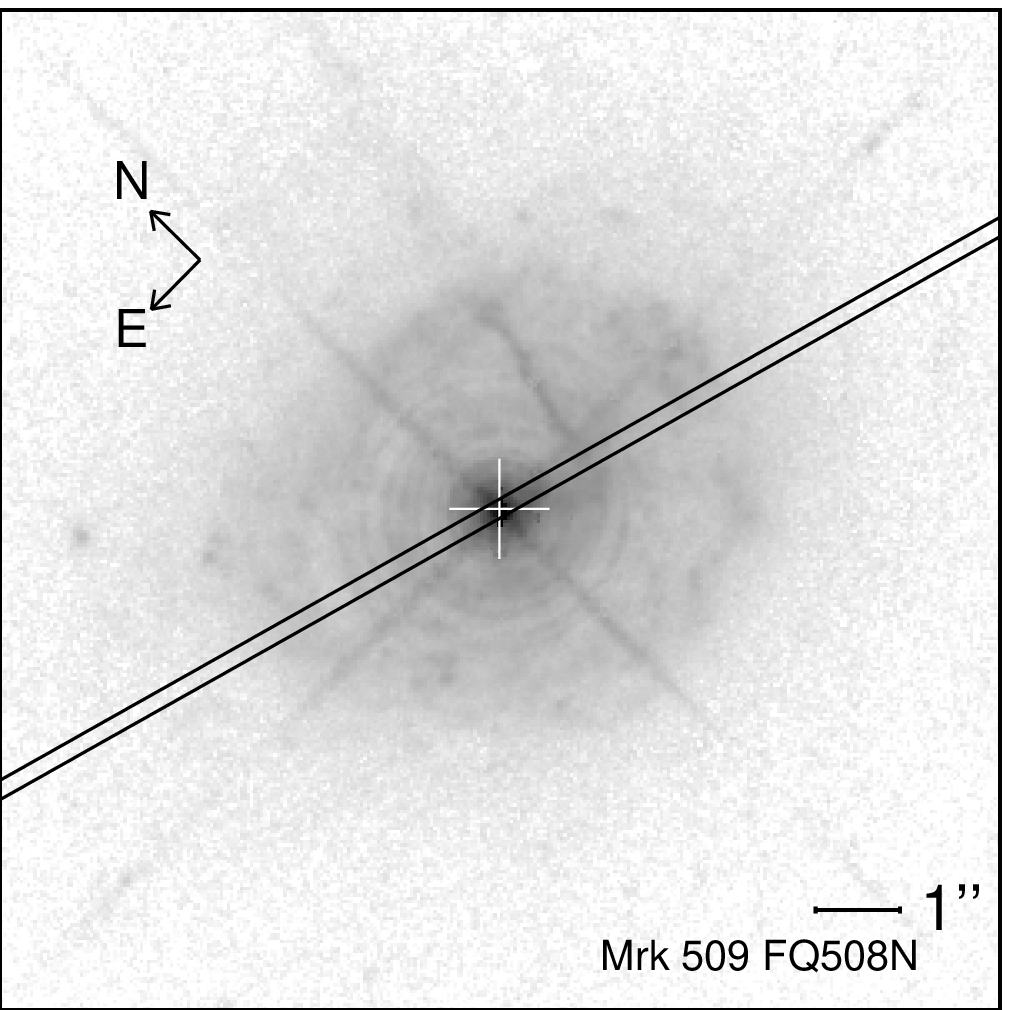} &
\includegraphics[width=.55\linewidth]{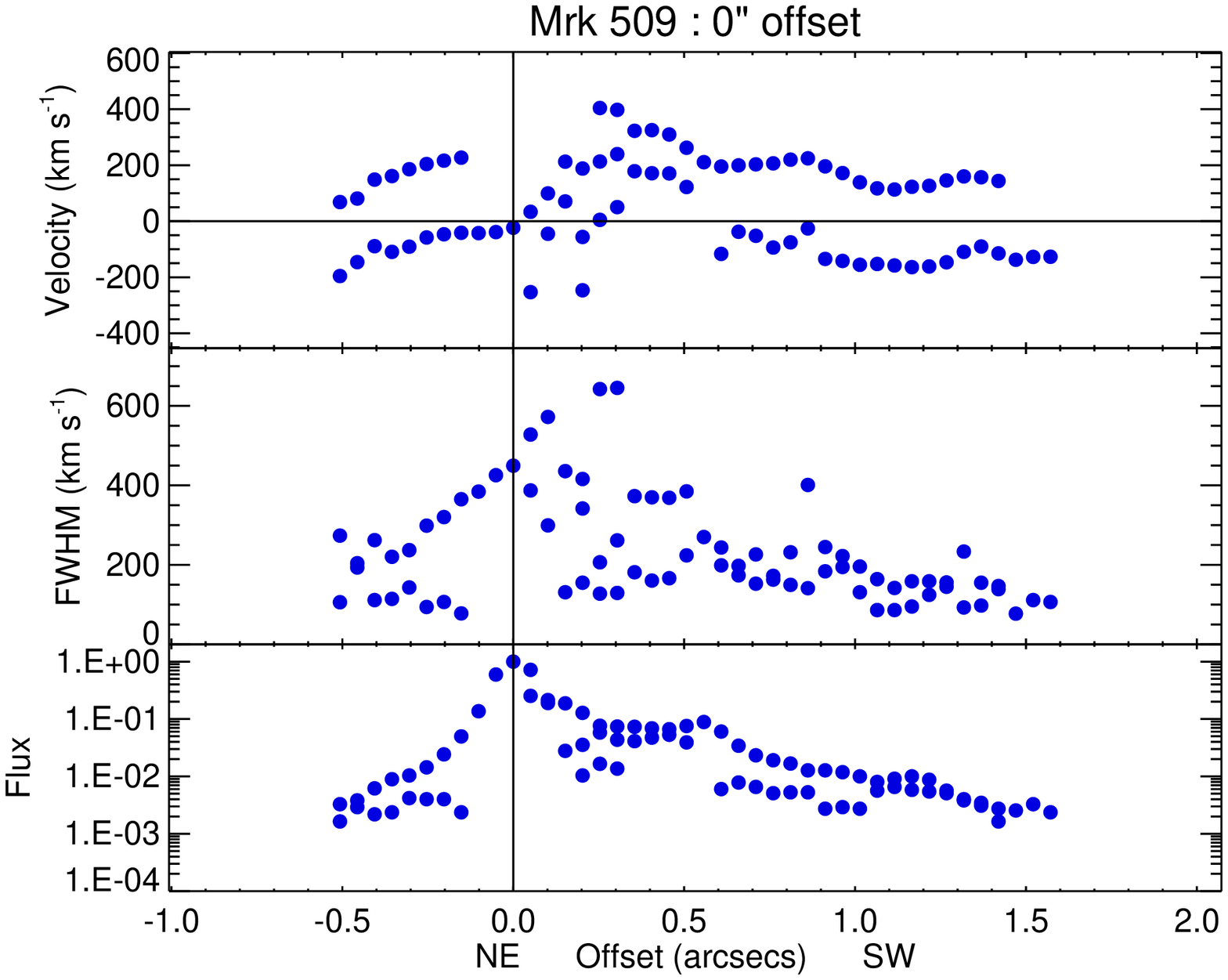}
\end{tabular}
\caption[Mrk 509 Kinematics]{Same as Figure \ref{iras11058}, but for Mrk 509.}
\label{mrk509}
\end{figure}

This is a Seyfert 1 that likely contains a tidal tail in its central regions. Seen in both 
optical continuum and [O~III] imaging (Figure \ref{mrk509}), this feature is composed of 
several knots of total length 2.2$''$ at a position angle of -13$^{\circ}$ before jutting towards 
the nucleus from the southwest at a position angle of -110$^{\circ}$. As the {\it HST} STIS 
slit was positioned favorably at 75$^{\circ}$, we are able to observe the inner portion of 
the tail and analyze its kinematics. Southwest of the nucleus are 
redshifted velocities peaking near 400 km s$^{-1}$, which correspond to the highest flux 
emission lines at those positions. Coupling this data with the apparent projection of the tidal tail 
above the host galaxy disk suggests that the bright, inner portion of the tidal tail is 
inflowing. Should this be the case, we may be viewing a minor merger with a dwarf galaxy, 
and this system would provide a great opportunity to study the fueling of an AGN by a 
minor merger in progress. Assuming that the redshifted kinematic component southwest of the 
nucleus is an inflowing tidal tail, the remaining velocities do not provide enough 
information to fit a model to the true outflowing kinematics. Further observations of the 
entire tail, particularly the extended linear feature, would clarify which direction the 
feature is moving and whether or not it is inflowing.

\clearpage

\subsection{{\it NGC 3393}}

\begin{figure}[h]
\centering
\begin{tabular}{cc}
\includegraphics[width=.4\linewidth]{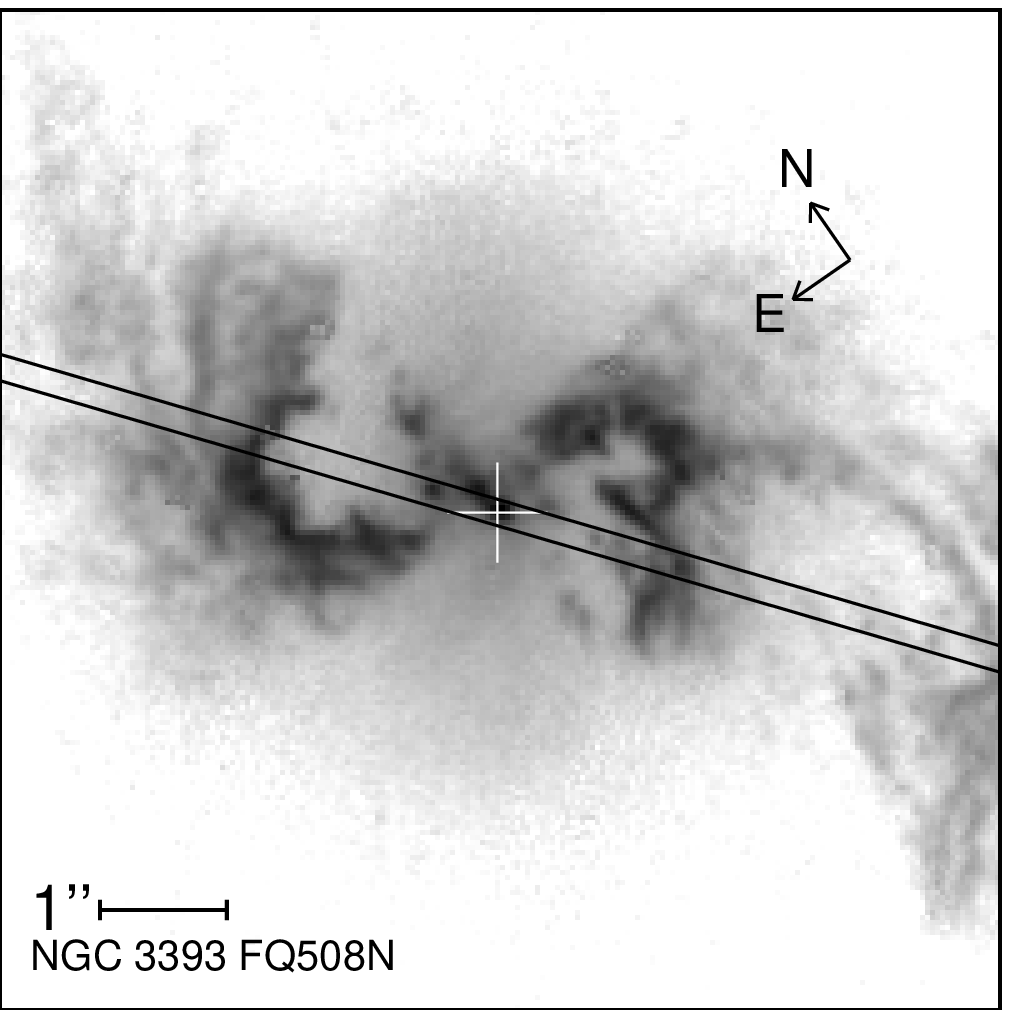} &
\includegraphics[width=.55\linewidth]{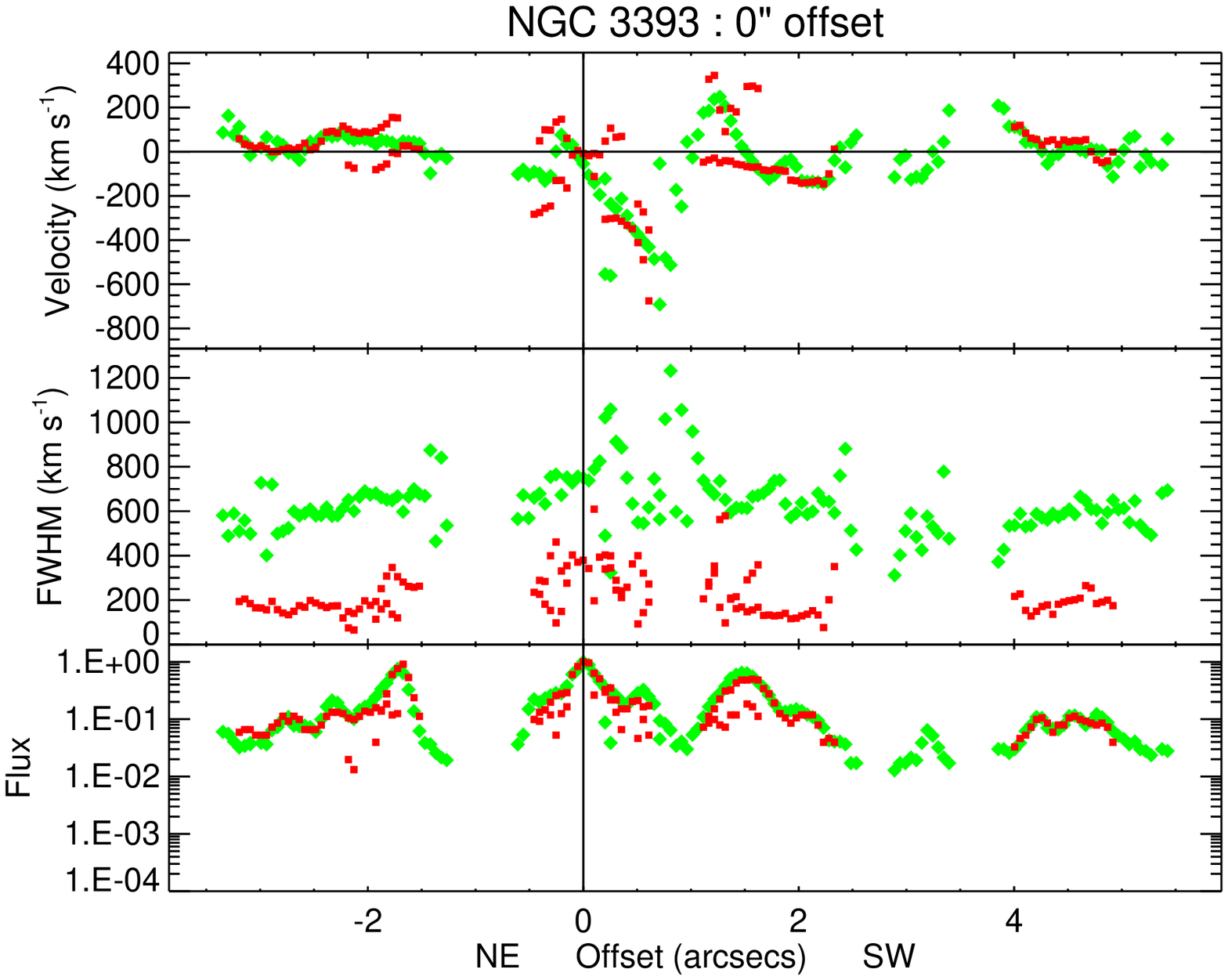}
\end{tabular}
\caption[NGC 3393 Kinematics]{Same as Figure \ref{iras11058}, but for NGC 3393.}
\label{ngc3393}
\end{figure}

NGC 3393 is another Seyfert 2 that contains an S-shaped NLR, shown in Figure \ref{ngc3393} 
that shows turbulent kinematics across several knots of emission, similar to Mrk 573 and 
IRAS 11058-1131. Though the NLR is well resolved (distinct knots visible over several 
arcseconds), it is unclear whether any distinct kinematic components are present. Two sets 
of high velocities exist to the southwest of the nucleus, one blueshifted component peaking 
at 0.75$''$ and one redshifted component peaking at 1.25$''$, which for a Seyfert 2 AGN would 
be two components for one cone to the southwest, inclined close to the plane of the 
sky. This is unlikely as imaging shows symmetric lobes of emission on either side 
of the nucleus which would not be replicated in the kinematic data. An alternative method 
to fitting the kinematics is to move the central position of the nucleus southwest 
$\sim$0.9$''$. This would imply that the location of the continuum peak in the spectral image 
would not suffice as the true position of the AGN, possibly due to obscuration and reflection 
of the central engine. Peaks in flux to either side of the new center support the possibility 
of obscuration over the AGN. At this new position, the two previous high velocity peaks are 
now on each side of the center, combined with the extended, near-systemic velocities further 
out to provide four kinematic components to model. However, there is not enough evidence to 
justify the altered placement of the kinematic center. 

\clearpage

\subsection{{\it NGC 3516}}

\begin{figure}[h]
\centering
\begin{tabular}{cc}
\includegraphics[width=.4\linewidth]{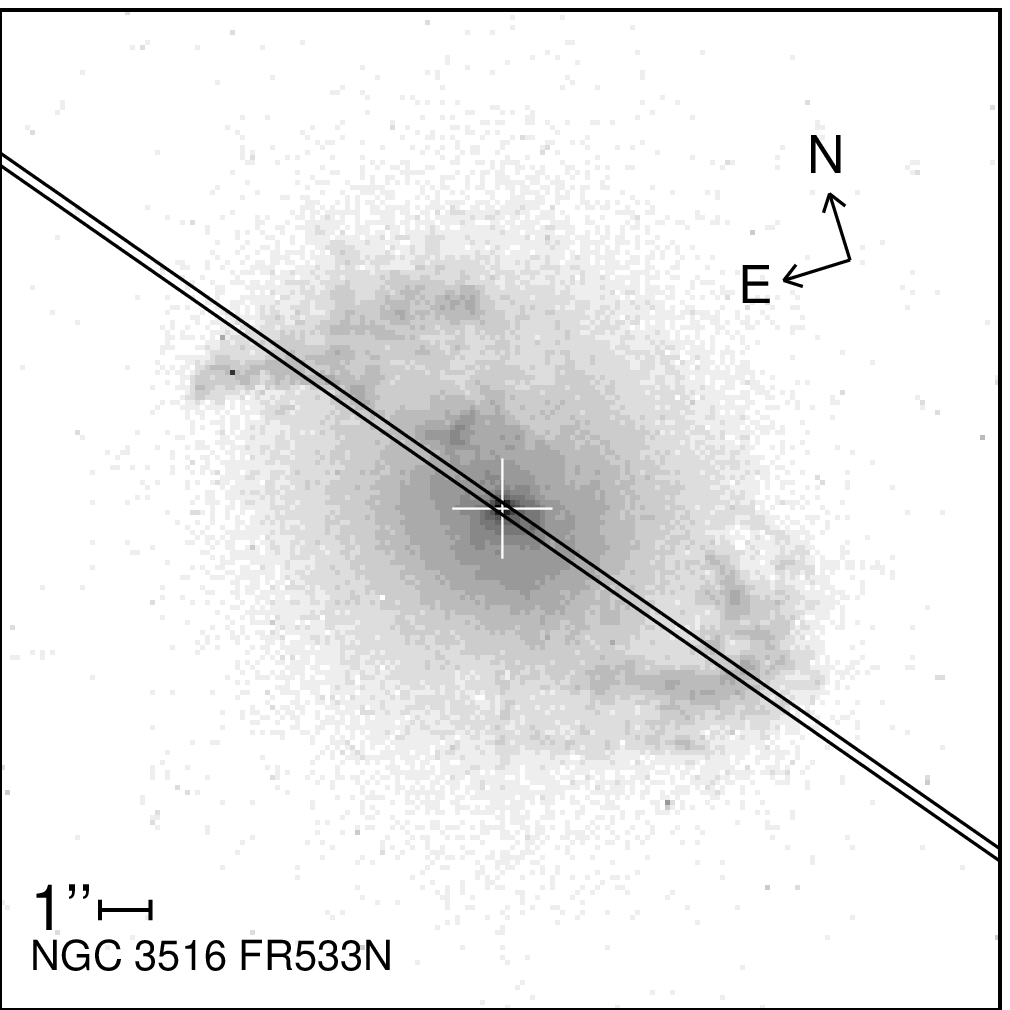} &
\includegraphics[width=.55\linewidth]{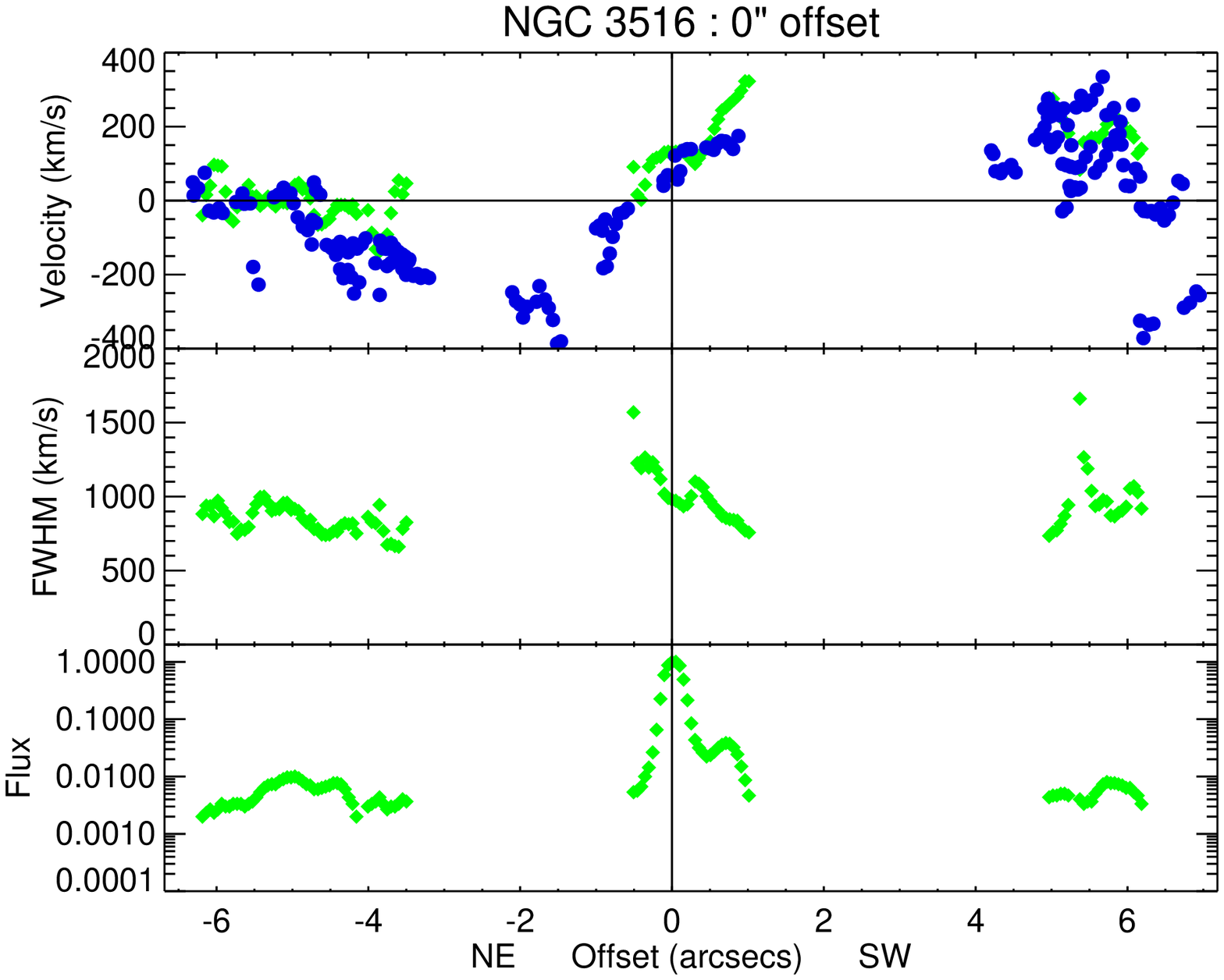}
\end{tabular}
\caption[NGC 3516 Kinematics]{Same as Figure \ref{iras11058}, but for NGC 3516. G430M kinematics are from \citet{Rui01} data set, 
no FWHM or fluxes available.}
\label{ngc3516}
\end{figure}

An atypical Seyfert 1 AGN as it contains an extended, S-shaped NLR (Figure \ref{ngc3516}) more often seen in Type 2 AGN, 
NGC 3516 is our prototypical $''$Ambiguous$''$ outflow system. Similar to Mrk 3 and Mrk 573, this S-shape is likely due to dust lanes 
of the host disk becoming ionized as they intercept the NLR bicone. A well-defined symmetric 
kinematic component on either side of the nucleus traveling in opposite directions corresponds to the extended emission 
seen in imaging. While the kinematics seen in these components match signatures of outflow employed by our model, we cannot 
fit the data as the components are not adjacent to each other (see Section \ref{sec5}). It is not clear why components 
that correspond to the opposite side of each cone are not present, though it may be possible that they reside near the nucleus and 
are outshone by its flux or that there is a lack of an ionizing medium to detect where the components exist. 

Though we cannot fit a kinematic model to NGC 3516, we attempted to recreate the NLR geometrically using the following 
assumptions: 

1.) The side of each cone nearer the plane of the sky must intercept the host disk (using a \citet{Kin00} disk geometry) 
such that the resultant intersection geometry matches that seen in the [O~III] imaging.

2.) The opening angle must be wide enough such that we are viewing the central engine (i.e. the geometric model must be 
a Type 1 AGN).

Using these parameters, we were able to construct a geometric model that agreed with the [O~III] imaging using a bicone 
position angle, inclination, and outer opening angle of 40$^{\circ}$, 50$^{\circ}$, and 50$^{\circ}$ respectively.
Unfortunately, the resultant kinematic model did not agree with the observed radial velocities and attempts to reconcile the two models 
have been unsuccessful.

\section{Online Figures}

\subsection{{\it HST} STIS Slit Positions}

The appendix contains {\it HST} images of the extended sample. 
Filters centered on [O~III] were used if available, otherwise 
F606W or F547M continuum filters are shown.  Plus signs give the 
location of the optical continuum peaks. Solid lines outline the 
position of each STIS slit. Details for each observation are listed 
in Table \ref{images}. Images without slits are for targets using 
slitless STIS observations.

%%%%%%%%%%%%%%%%%%%%%%%%%%%%%%%%%%%%%%%%%%%%%%%%%%%%%%%%%%%%%%%

% Figure 1
\begin{figure}
\centering
\begin{tabular}{cc}
\includegraphics[angle=0,scale=0.7]{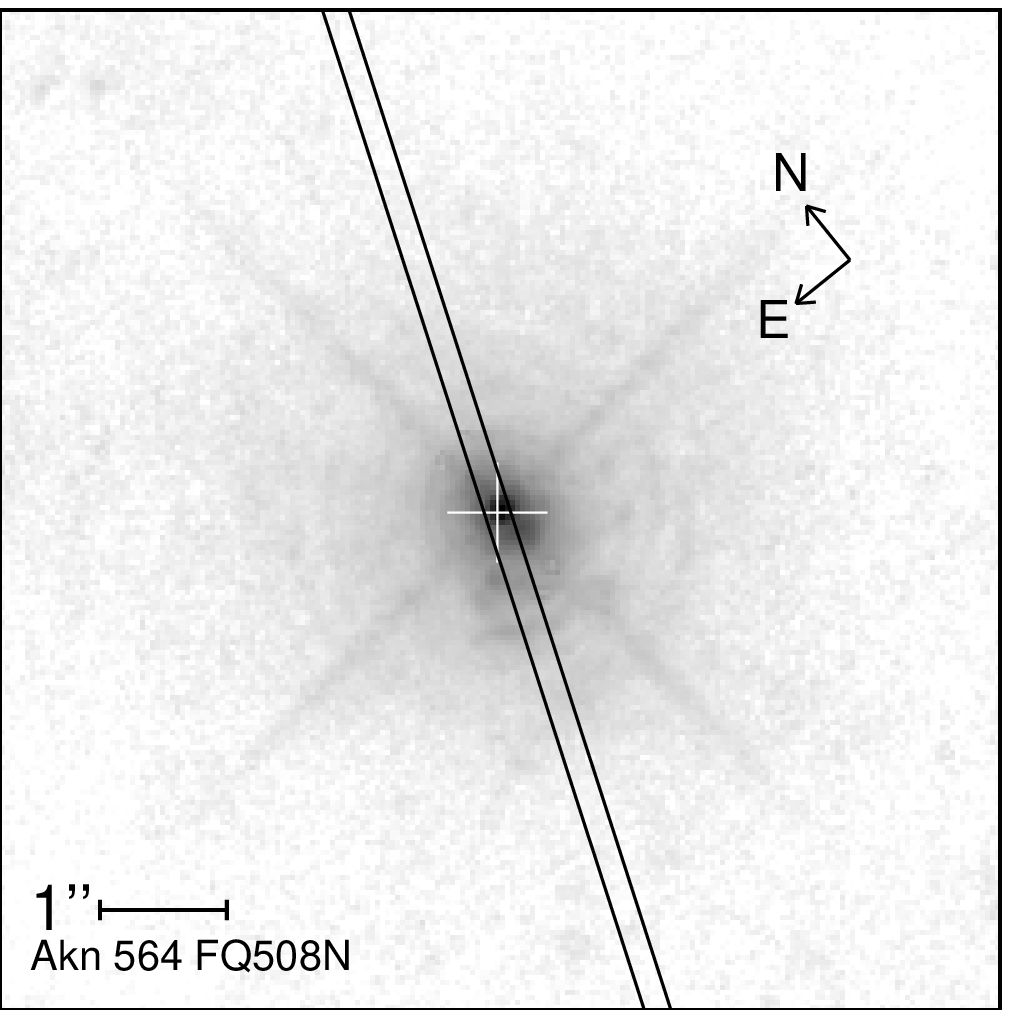} &
\includegraphics[angle=0,scale=0.7]{circinus.eps} \\
\includegraphics[angle=0,scale=0.7]{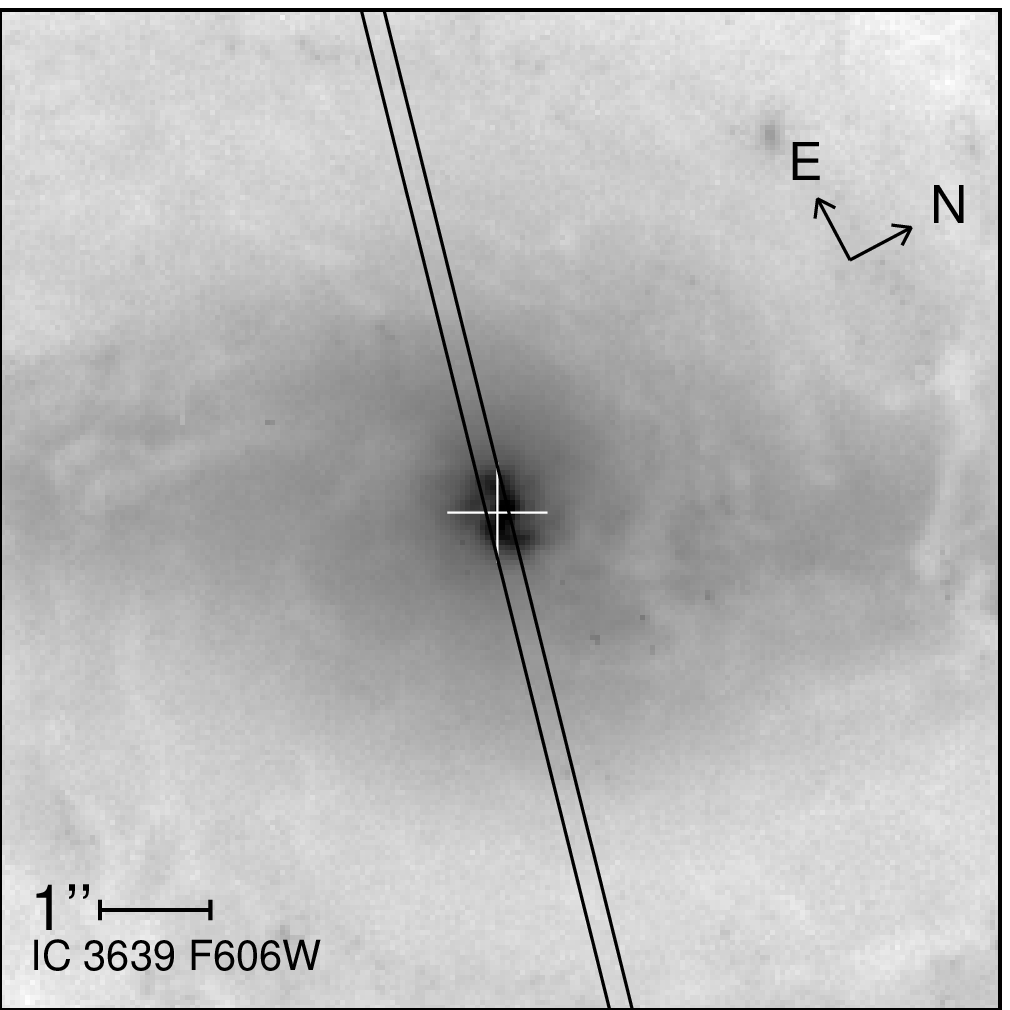} &
\includegraphics[angle=0,scale=0.7]{iras11058.eps}
\end{tabular}
\label{fig1a}
\end{figure}
%%%%%%%%%%%%%%%%%%%%%%%%%%%%%%%%%%%%%%%%%%%%%%%%%%%%%%%%%%%%%%%

\clearpage

% Figure 2
\begin{figure}
\centering
\begin{tabular}{cc}
\includegraphics[angle=0,scale=0.7]{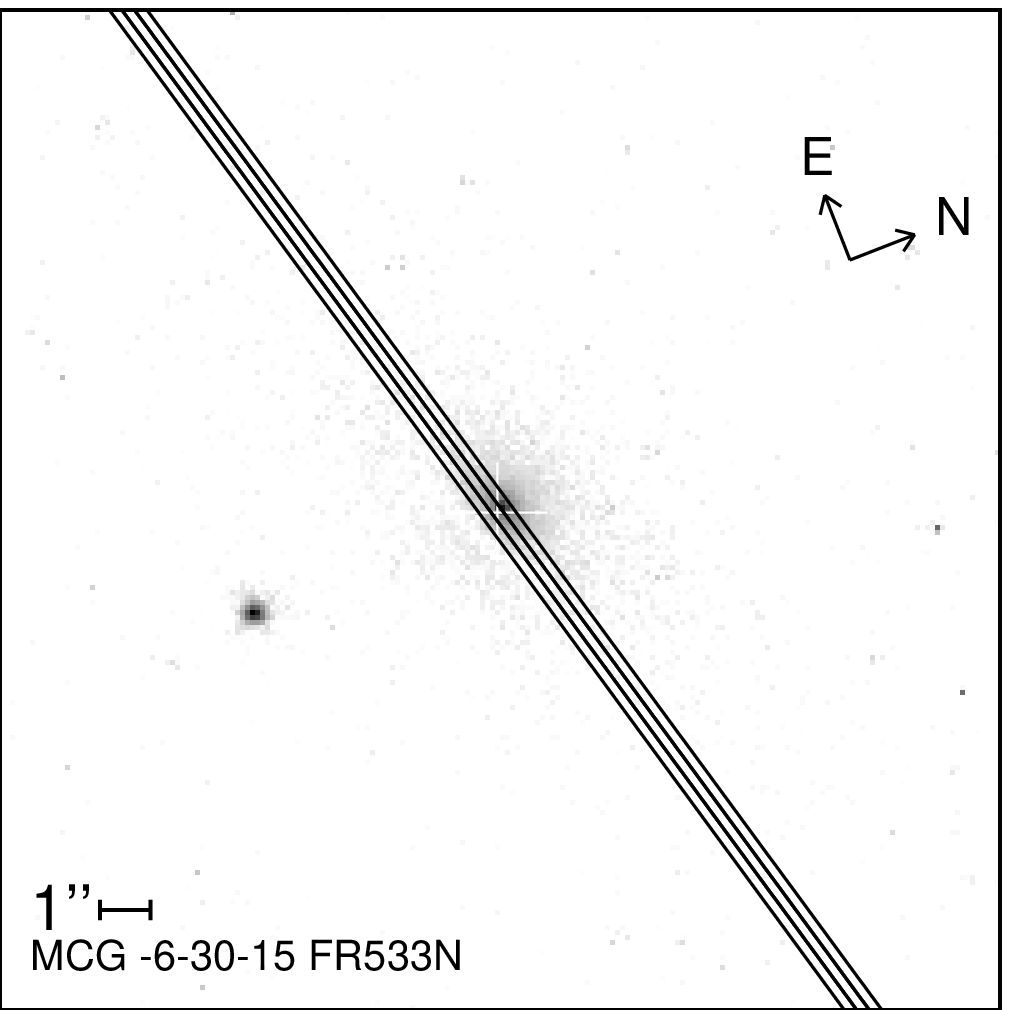} &
\includegraphics[angle=0,scale=0.7]{mrk34.eps} \\
\includegraphics[angle=0,scale=0.7]{mrk279.eps} &
\includegraphics[angle=0,scale=0.7]{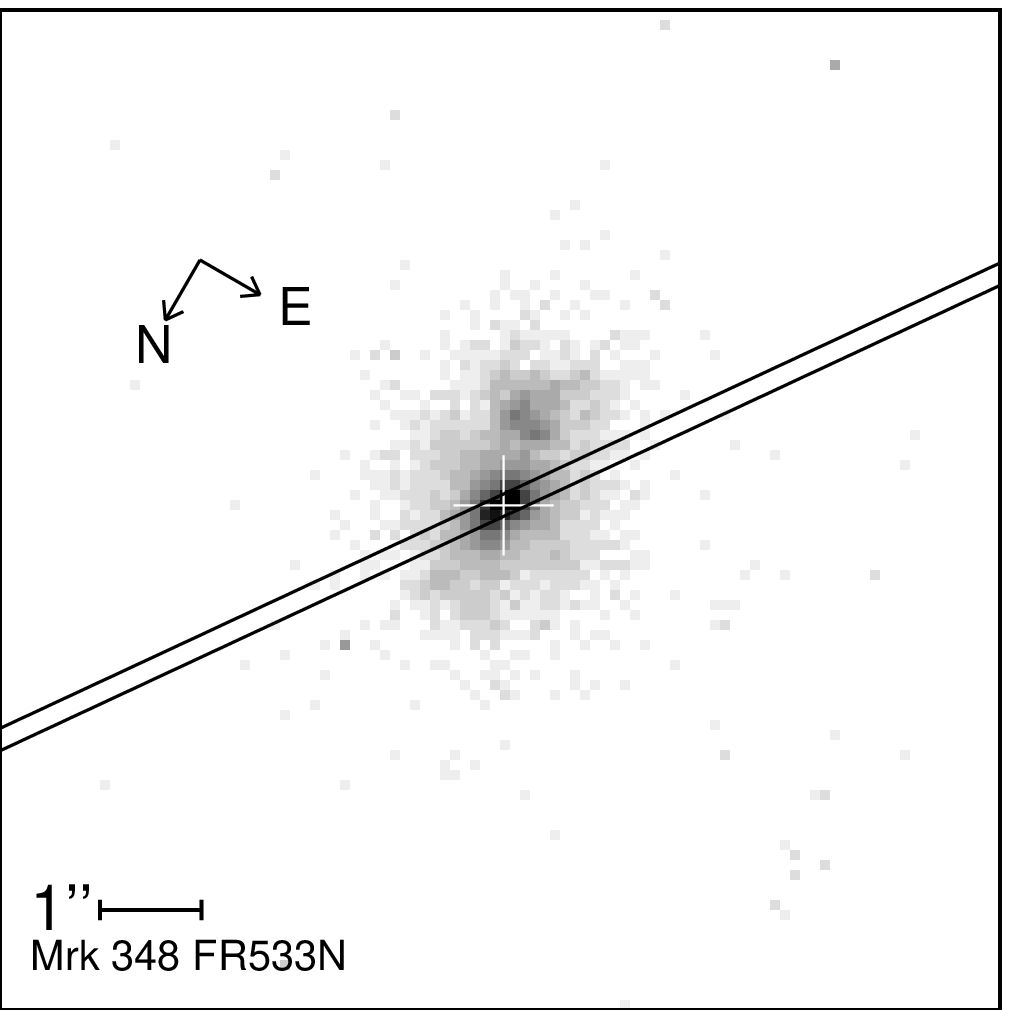}
\end{tabular}
\label{fig1b}
\end{figure}
%%%%%%%%%%%%%%%%%%%%%%%%%%%%%%%%%%%%%%%%%%%%%%%%%%%%%%%%%%%%%%%

\clearpage

% Figure 3
\begin{figure}
\centering
\begin{tabular}{cc}
\includegraphics[angle=0,scale=0.7]{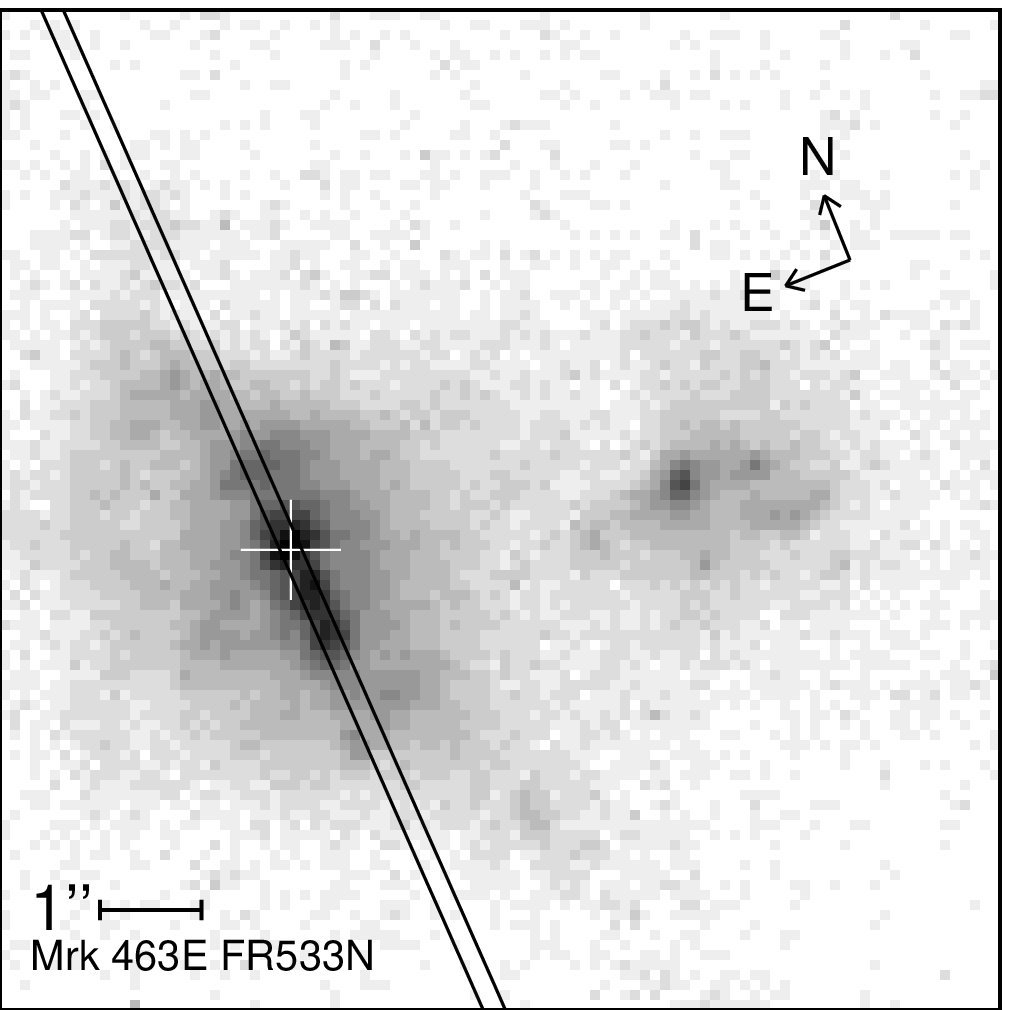}&
\includegraphics[angle=0,scale=0.7]{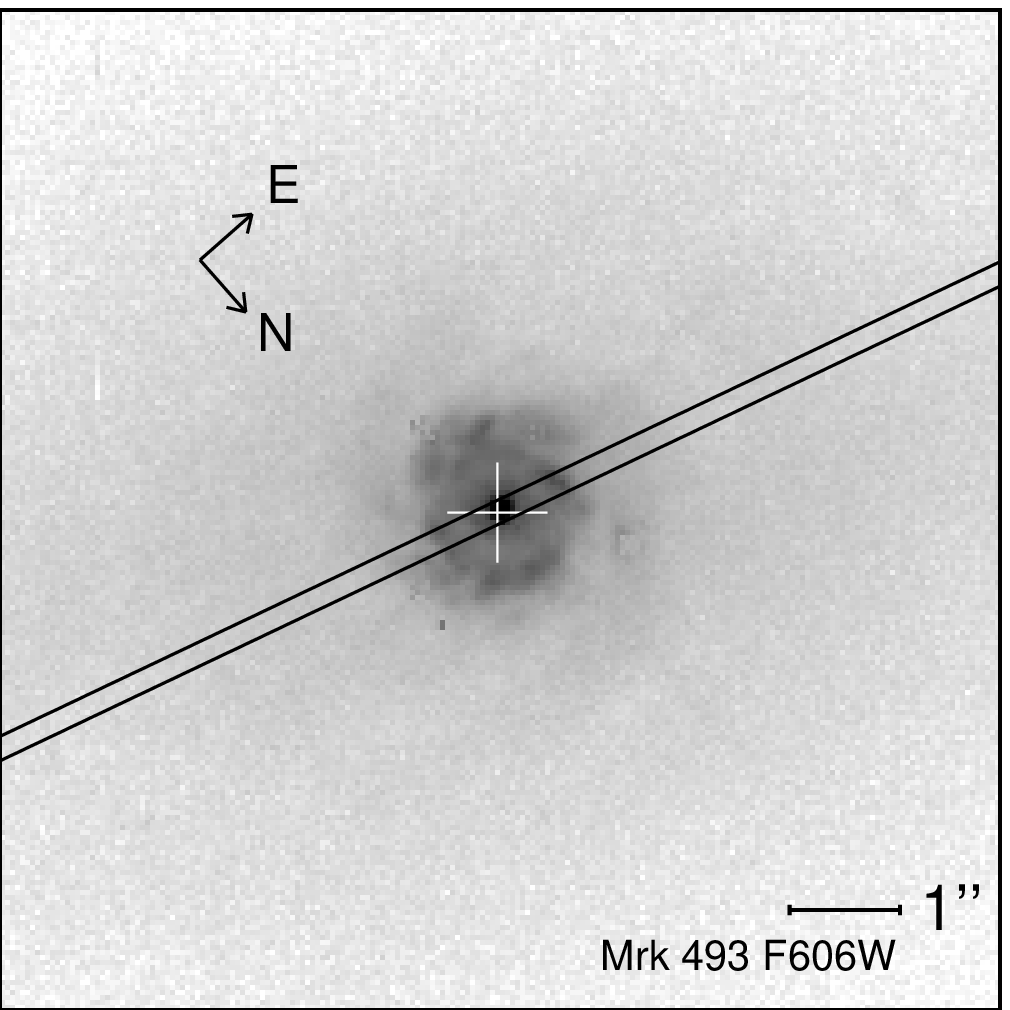} \\
\includegraphics[angle=0,scale=0.7]{mrk509.eps} & 
\includegraphics[angle=0,scale=0.7]{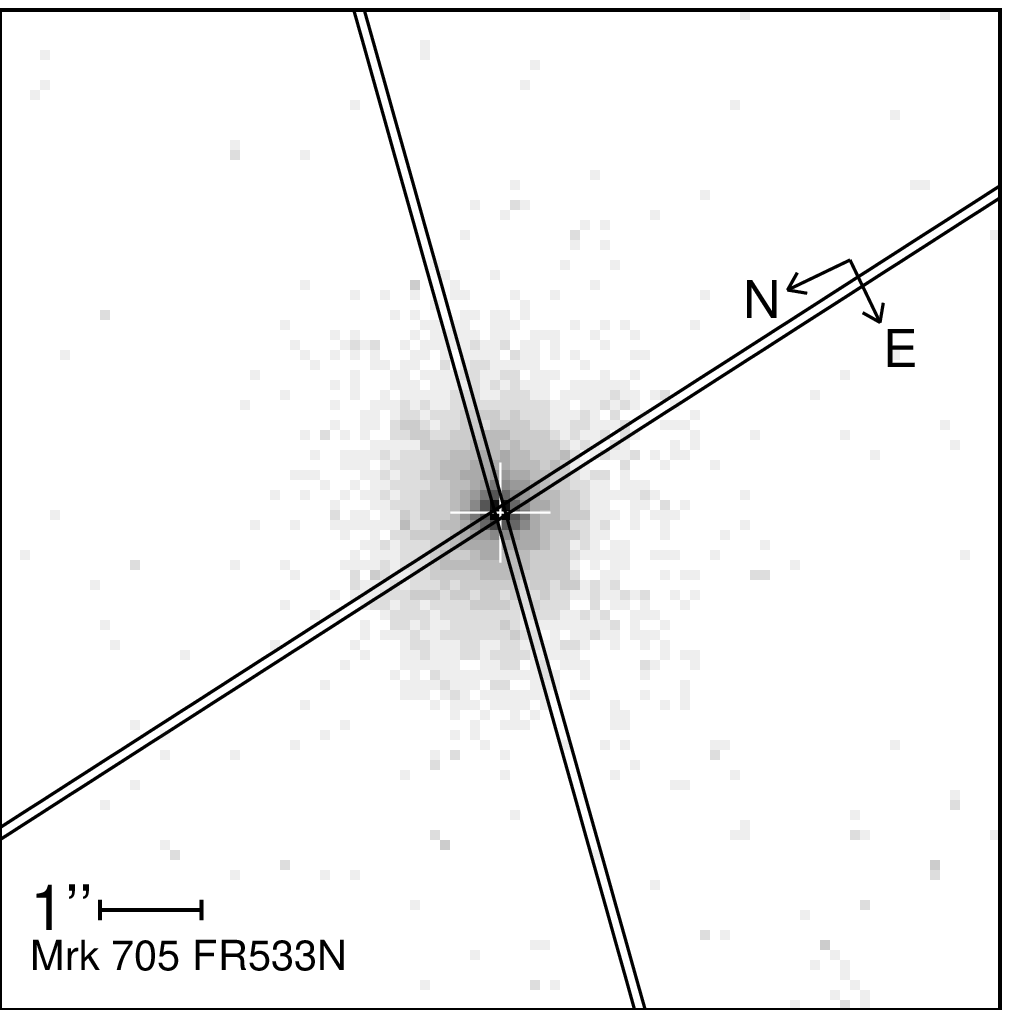}
\end{tabular}
\label{fig1c}
\end{figure}
%%%%%%%%%%%%%%%%%%%%%%%%%%%%%%%%%%%%%%%%%%%%%%%%%%%%%%%%%%%%%%%

\clearpage

% Figure 4
\begin{figure}
\centering
\begin{tabular}{cc}
\includegraphics[angle=0,scale=0.7]{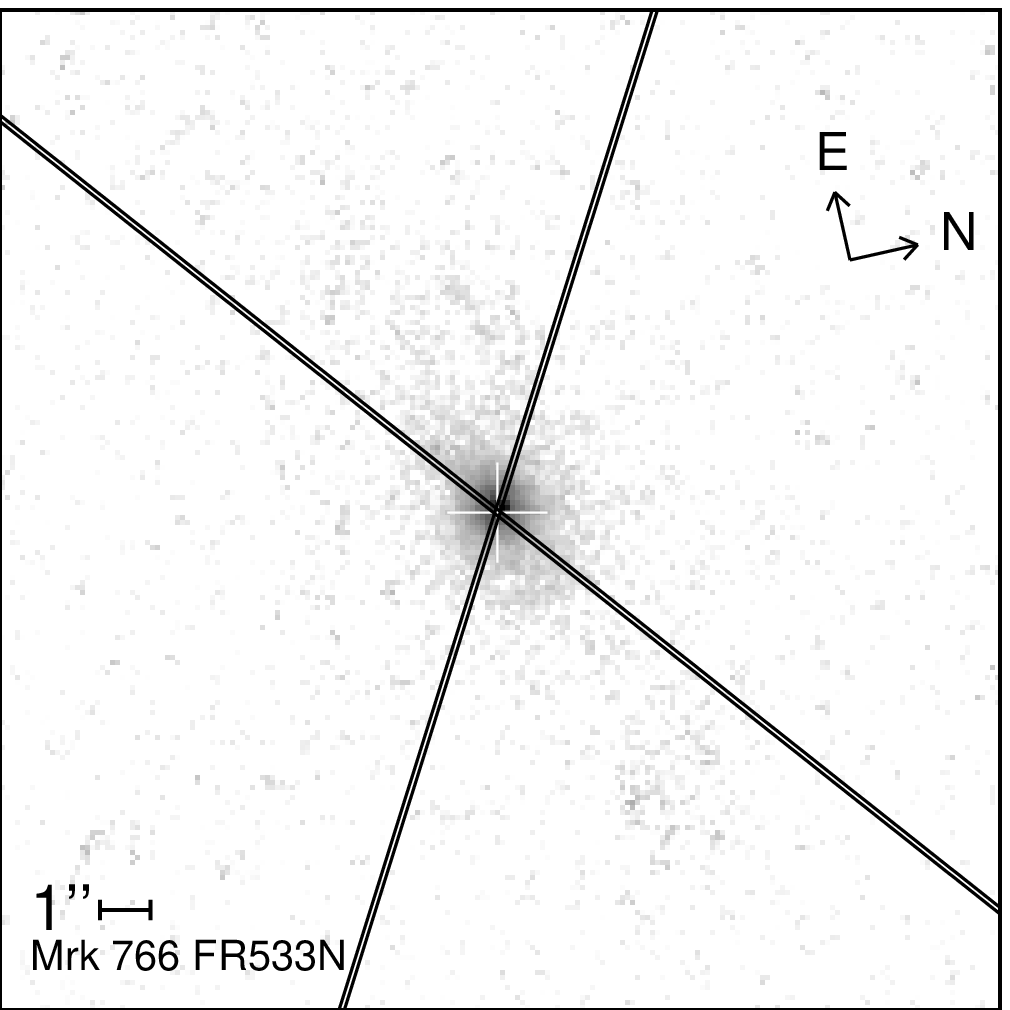} &
\includegraphics[angle=0,scale=0.7]{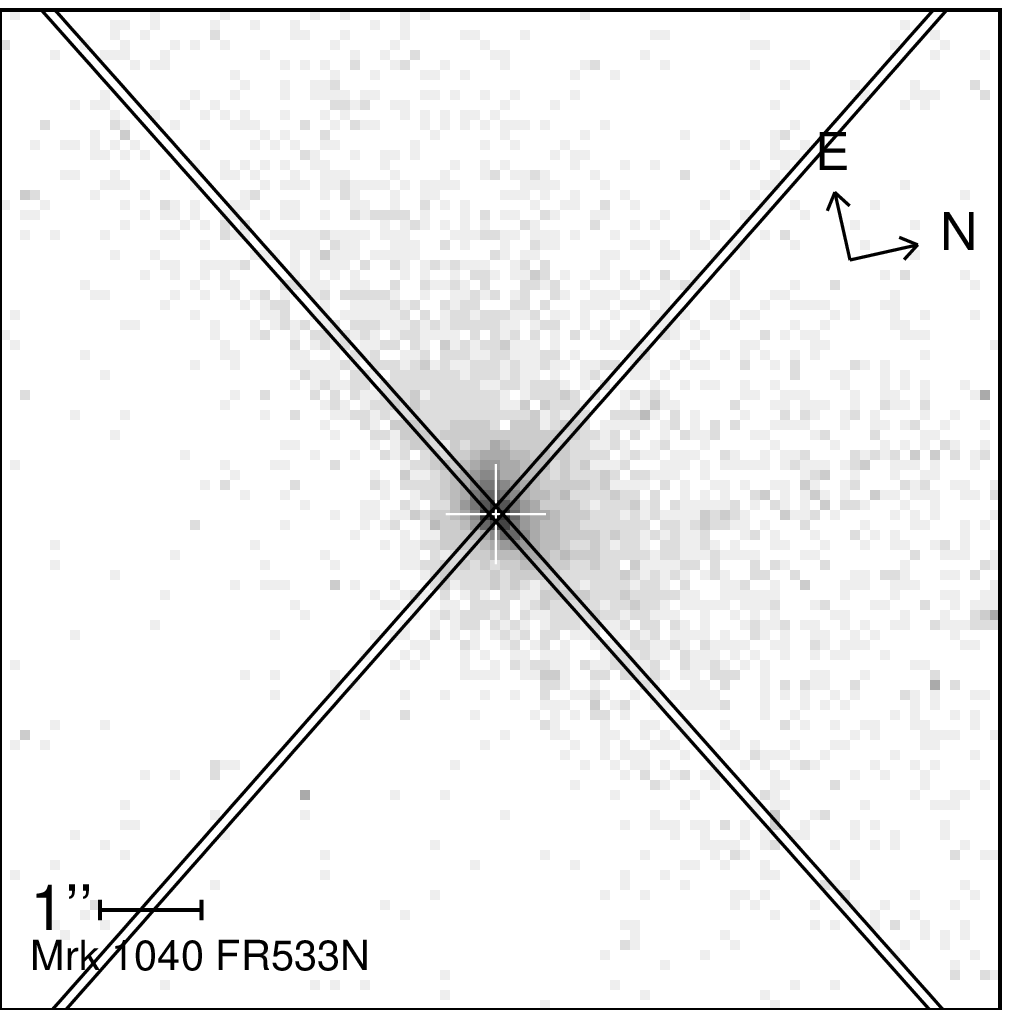} \\
\includegraphics[angle=0,scale=0.7]{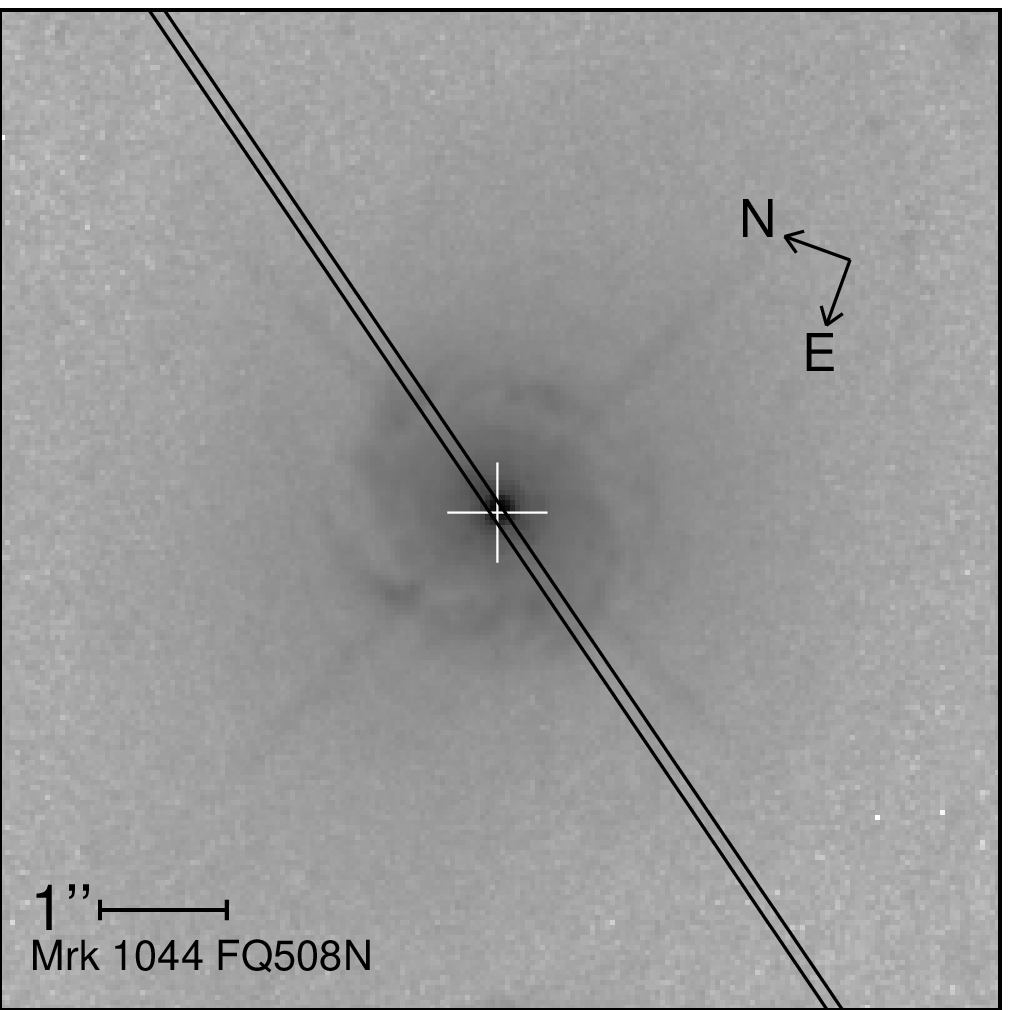}&
\includegraphics[angle=0,scale=0.7]{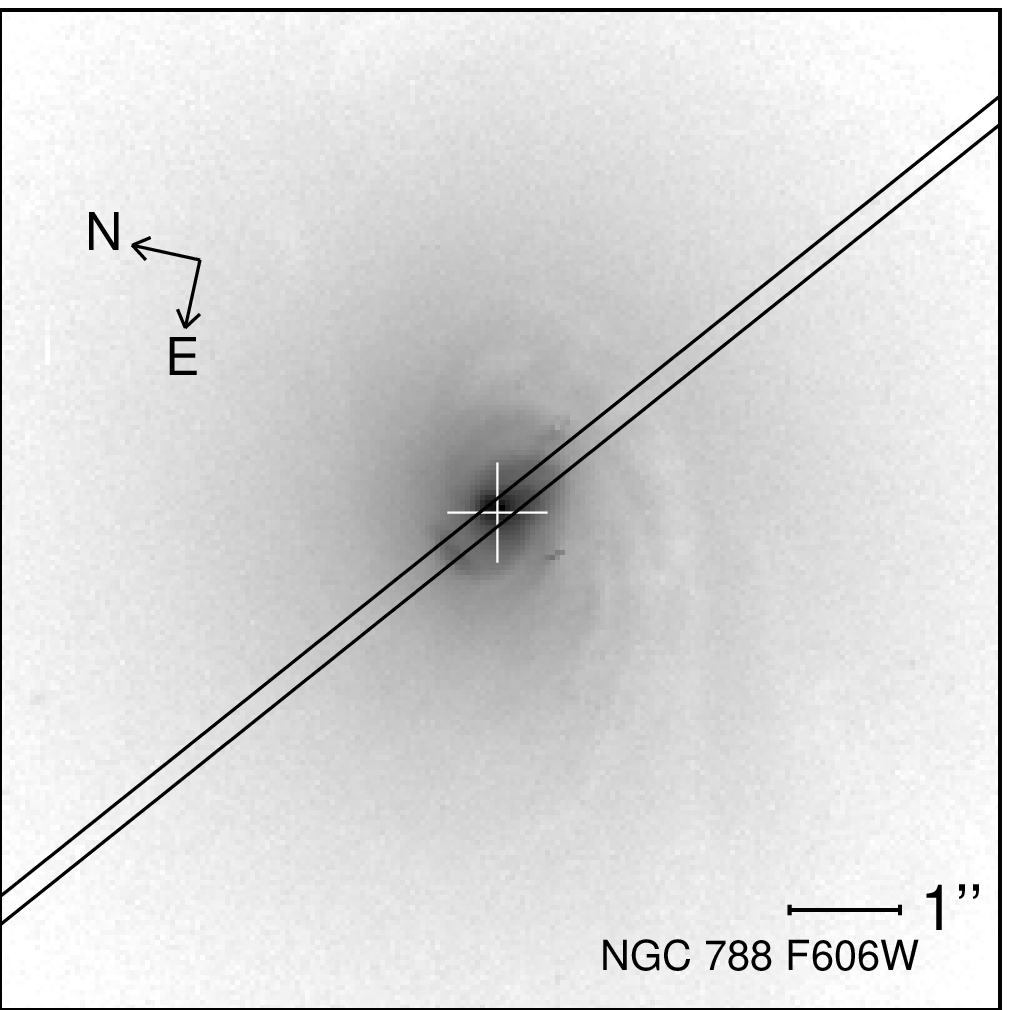} 
\end{tabular}
\label{fig1d}
\end{figure}
%%%%%%%%%%%%%%%%%%%%%%%%%%%%%%%%%%%%%%%%%%%%%%%%%%%%%%%%%%%%%%%

\clearpage

% Figure 5
\begin{figure}
\centering
\begin{tabular}{cc}
\includegraphics[angle=0,scale=0.7]{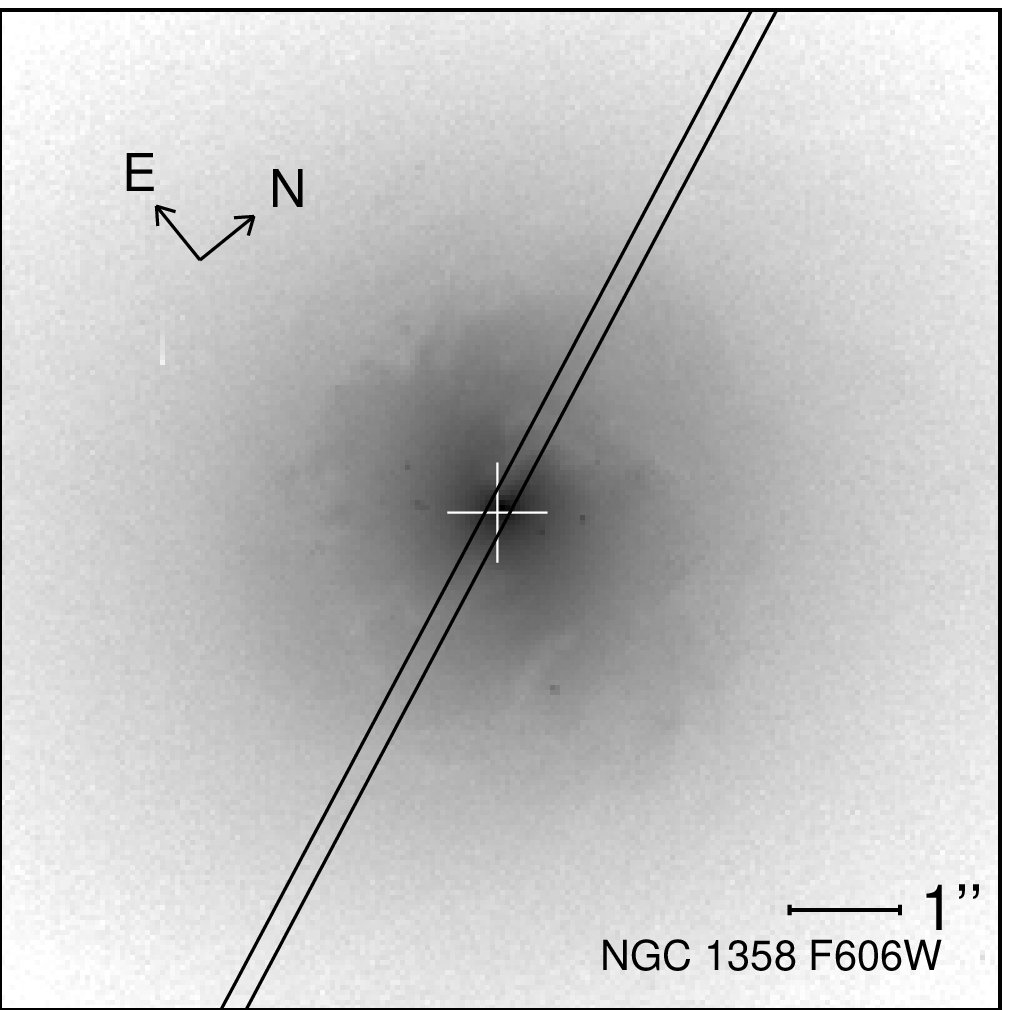}& 
\includegraphics[angle=0,scale=0.7]{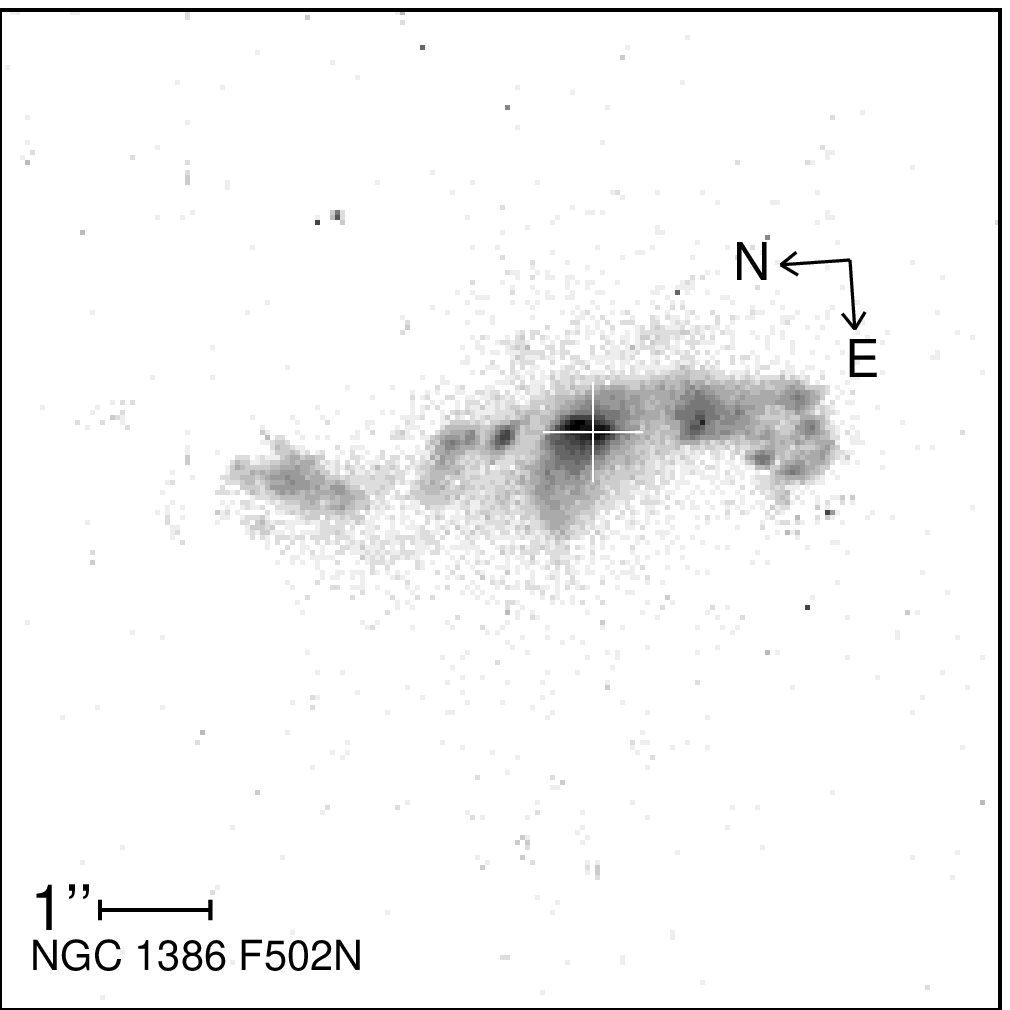}\\ 
\includegraphics[angle=0,scale=0.7]{ngc1667.eps}&  
\includegraphics[angle=0,scale=0.7]{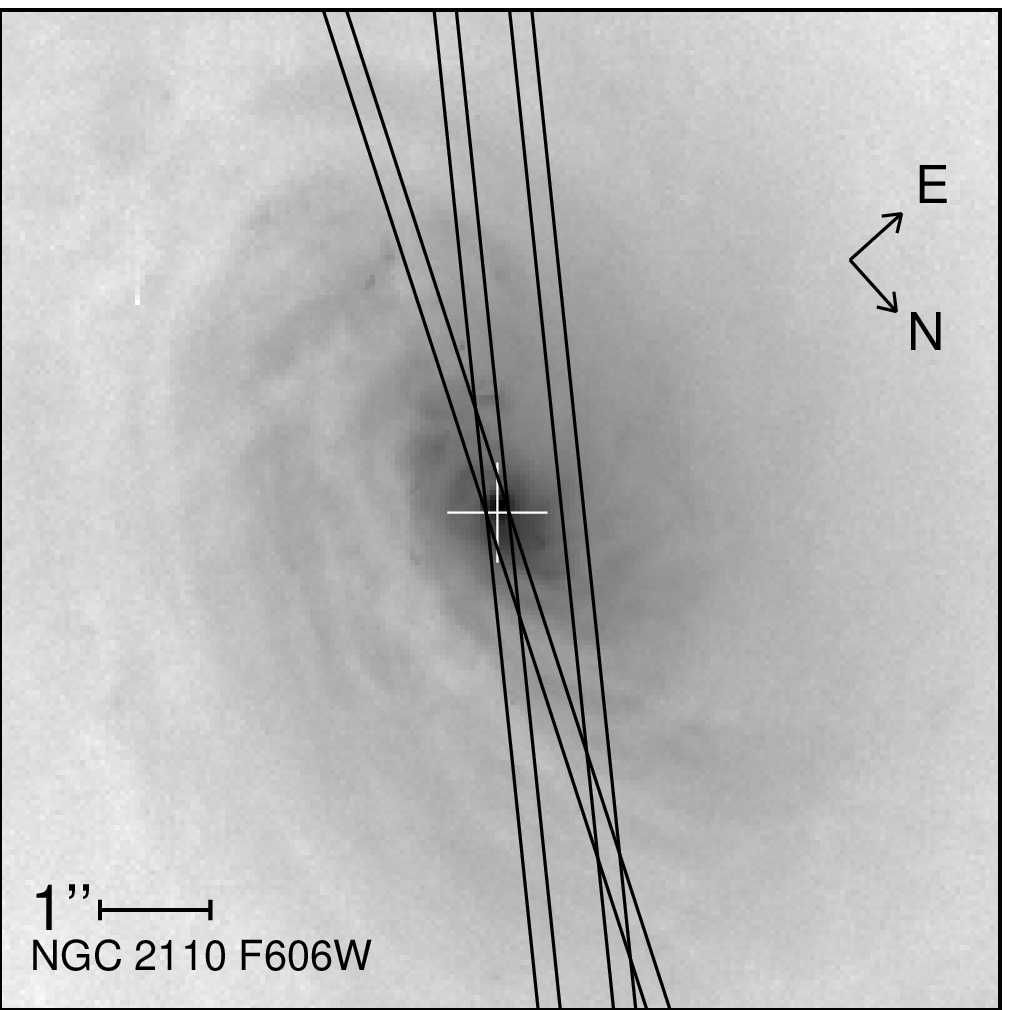} 

\end{tabular}
\label{fig1e}
\end{figure}
%%%%%%%%%%%%%%%%%%%%%%%%%%%%%%%%%%%%%%%%%%%%%%%%%%%%%%%%%%%%%%%

\clearpage

% Figure 6
\begin{figure}
\centering
\begin{tabular}{cc}
\includegraphics[angle=0,scale=0.7]{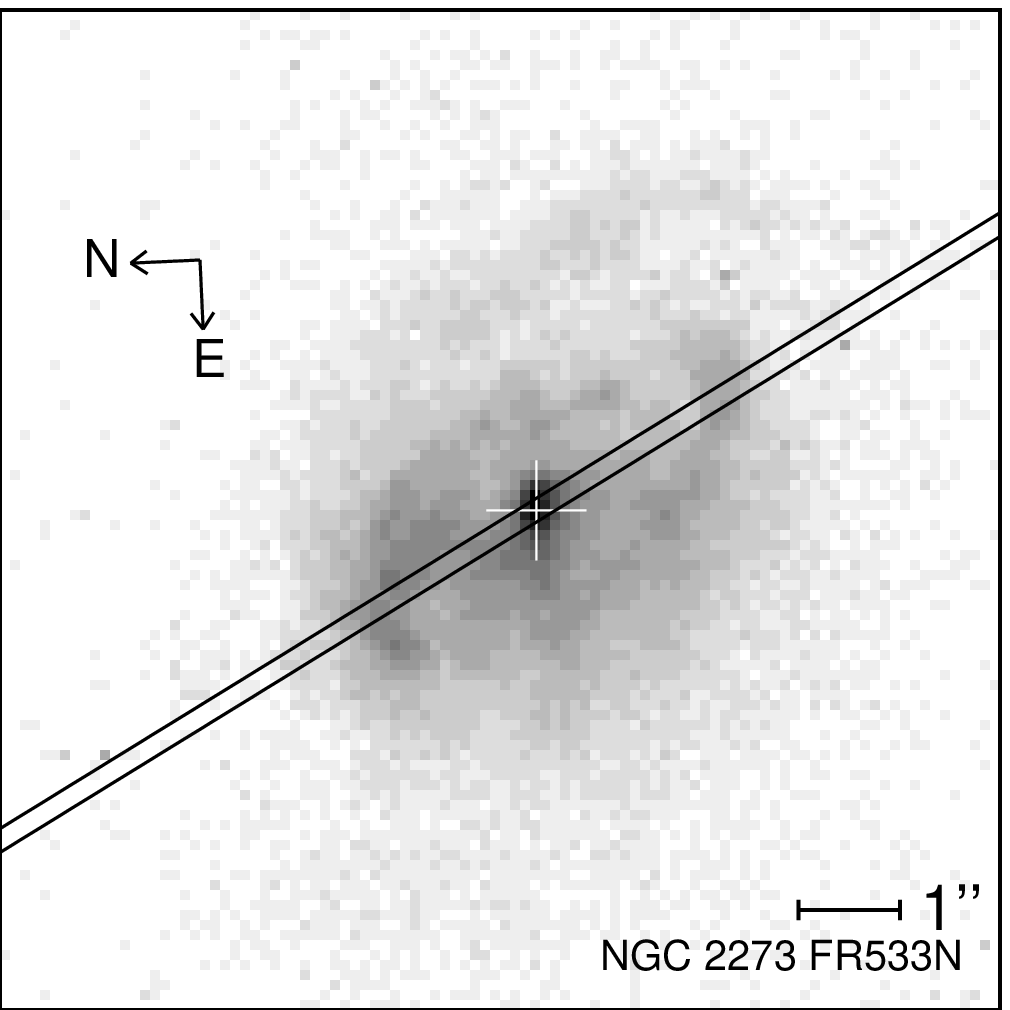}& 
\includegraphics[angle=0,scale=0.7]{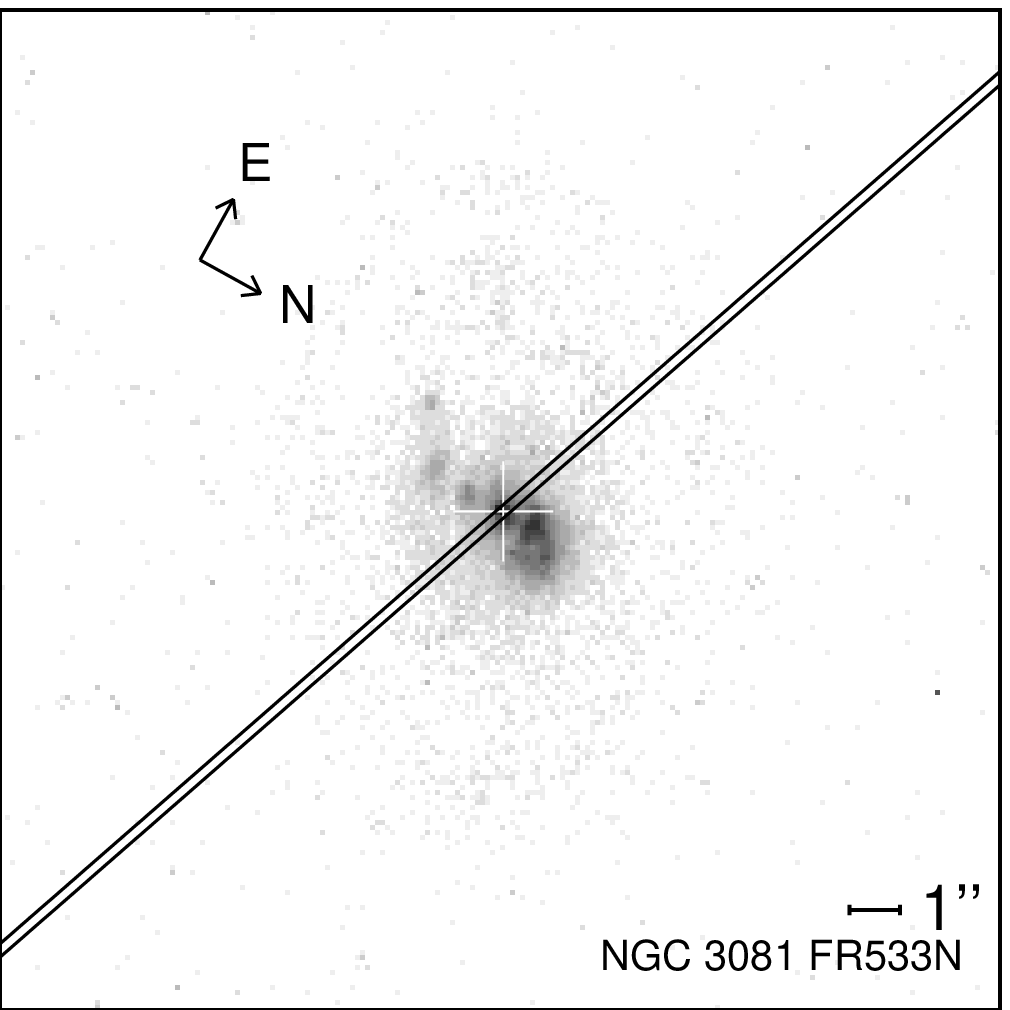}\\ 
\includegraphics[angle=0,scale=0.7]{ngc3227.eps}&  
\includegraphics[angle=0,scale=0.7]{ngc3393.eps} 

\end{tabular}
\label{fig1f}
\end{figure}
%%%%%%%%%%%%%%%%%%%%%%%%%%%%%%%%%%%%%%%%%%%%%%%%%%%%%%%%%%%%%%%

\clearpage

% Figure 7
\begin{figure}
\centering
\begin{tabular}{cc}
\includegraphics[angle=0,scale=0.7]{ngc3516.eps}& 
\includegraphics[angle=0,scale=0.7]{ngc3783.eps}\\ 
\includegraphics[angle=0,scale=0.7]{ngc4051.eps}&  
\includegraphics[angle=0,scale=0.7]{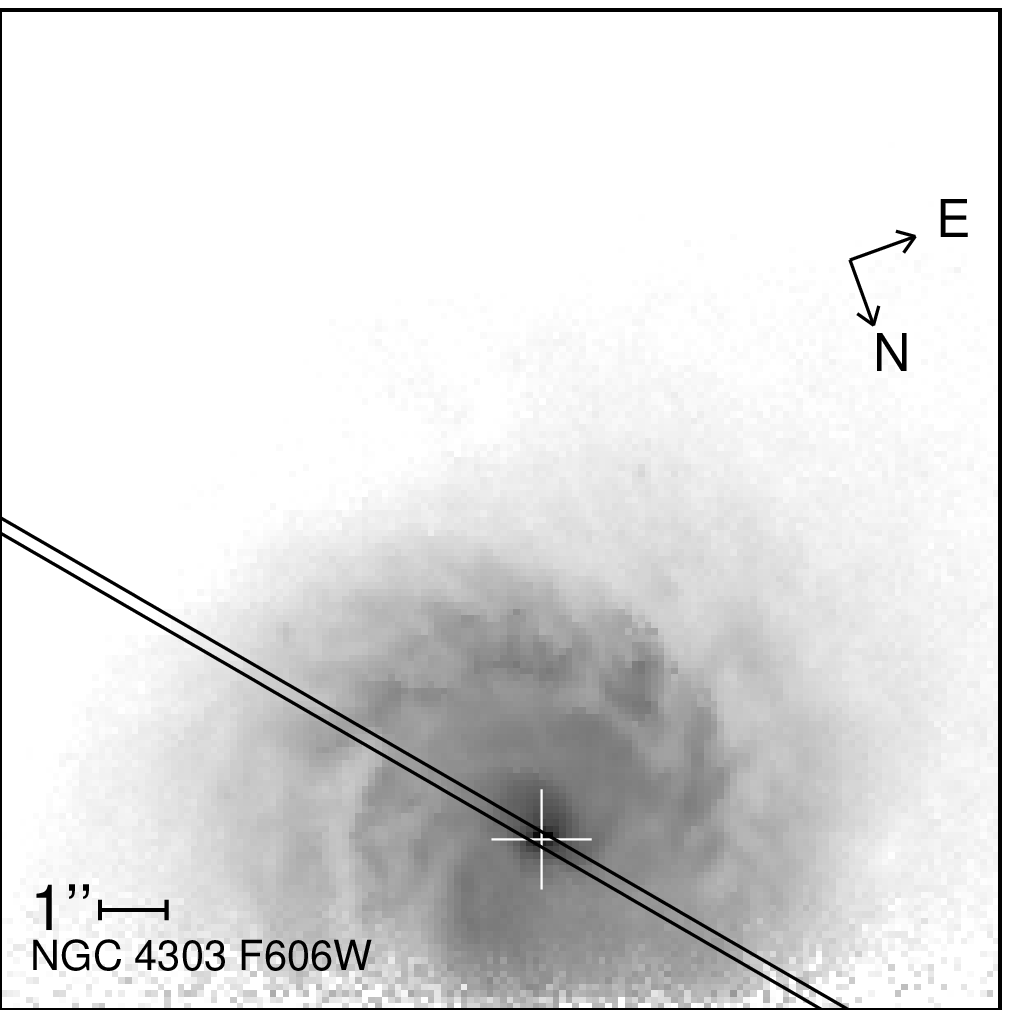} 

\end{tabular}
\label{fig1g}
\end{figure}
%%%%%%%%%%%%%%%%%%%%%%%%%%%%%%%%%%%%%%%%%%%%%%%%%%%%%%%%%%%%%%%

\clearpage

% Figure 8
\begin{figure}
\centering
\begin{tabular}{cc}
\includegraphics[angle=0,scale=0.7]{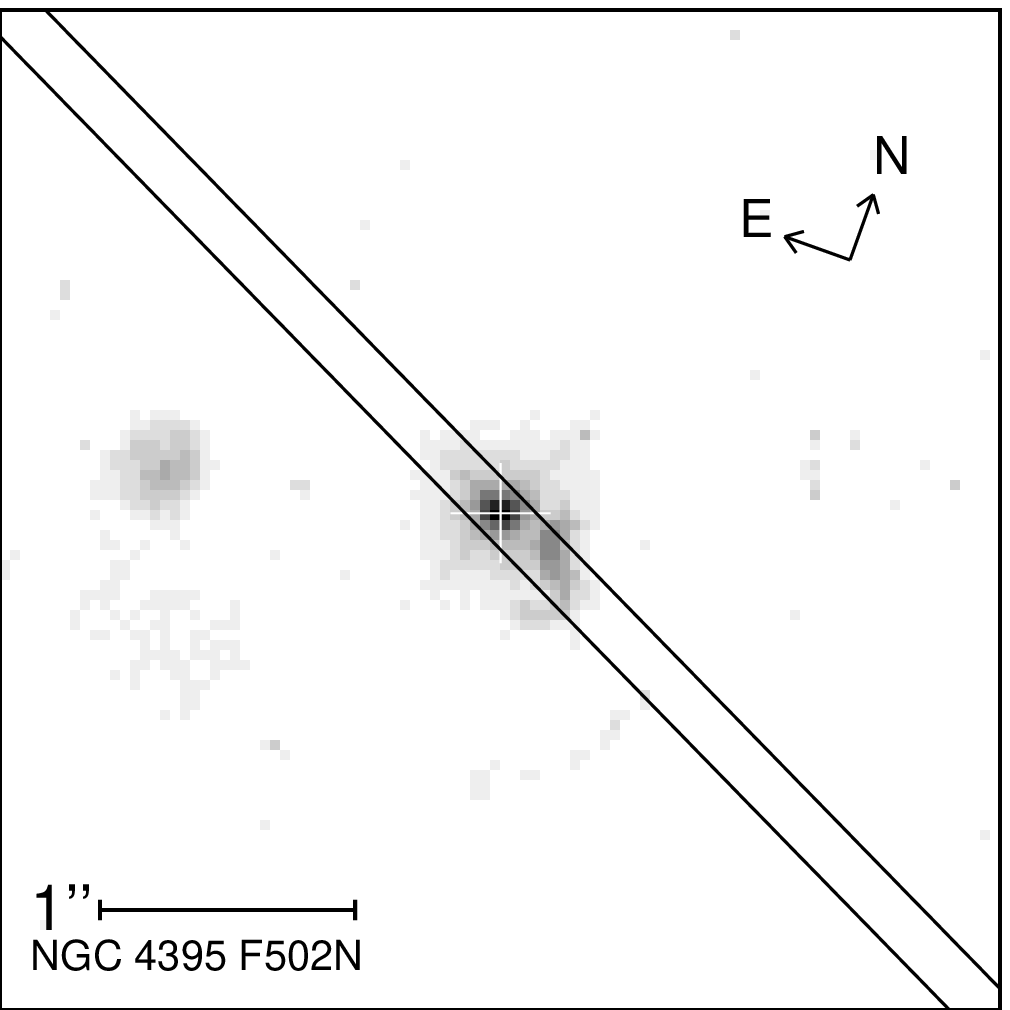}& 
\includegraphics[angle=0,scale=0.7]{ngc4507.eps}\\  
\includegraphics[angle=0,scale=0.7]{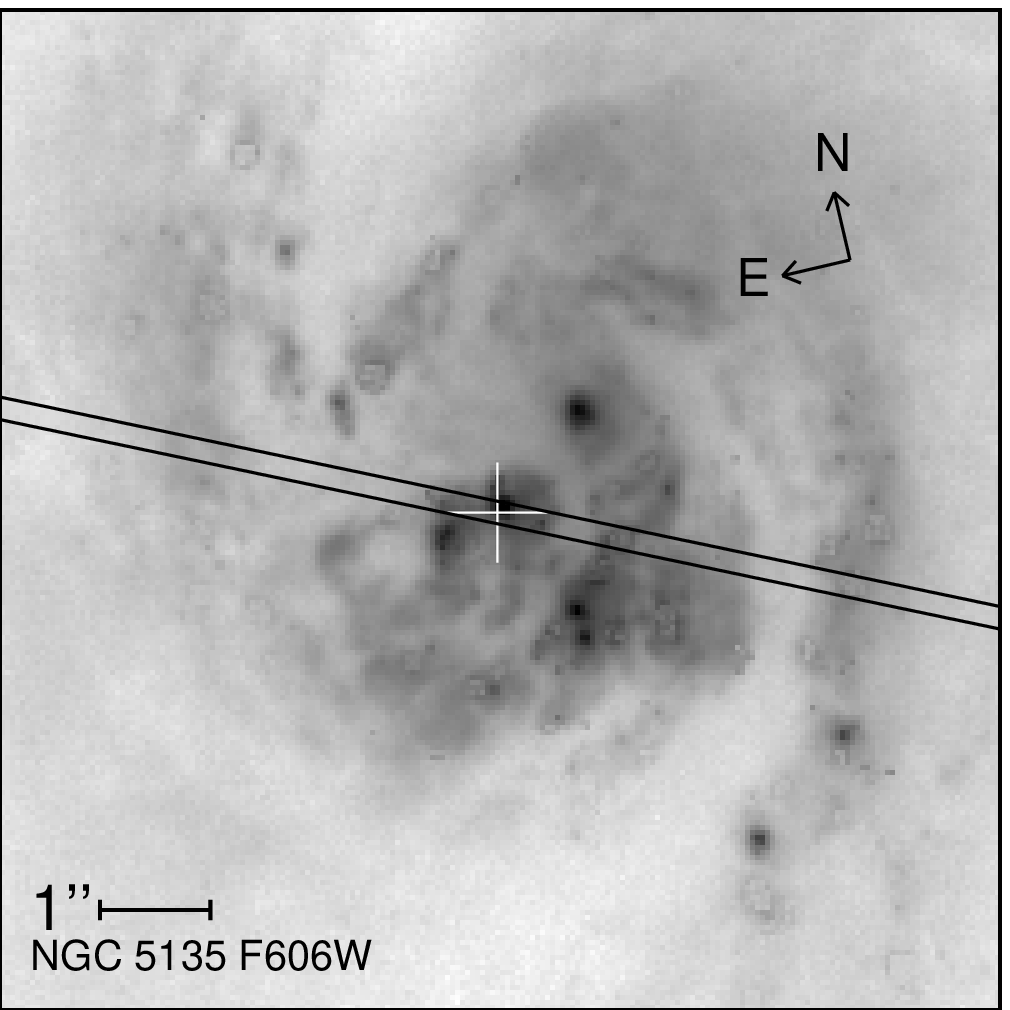}&   
\includegraphics[angle=0,scale=0.7]{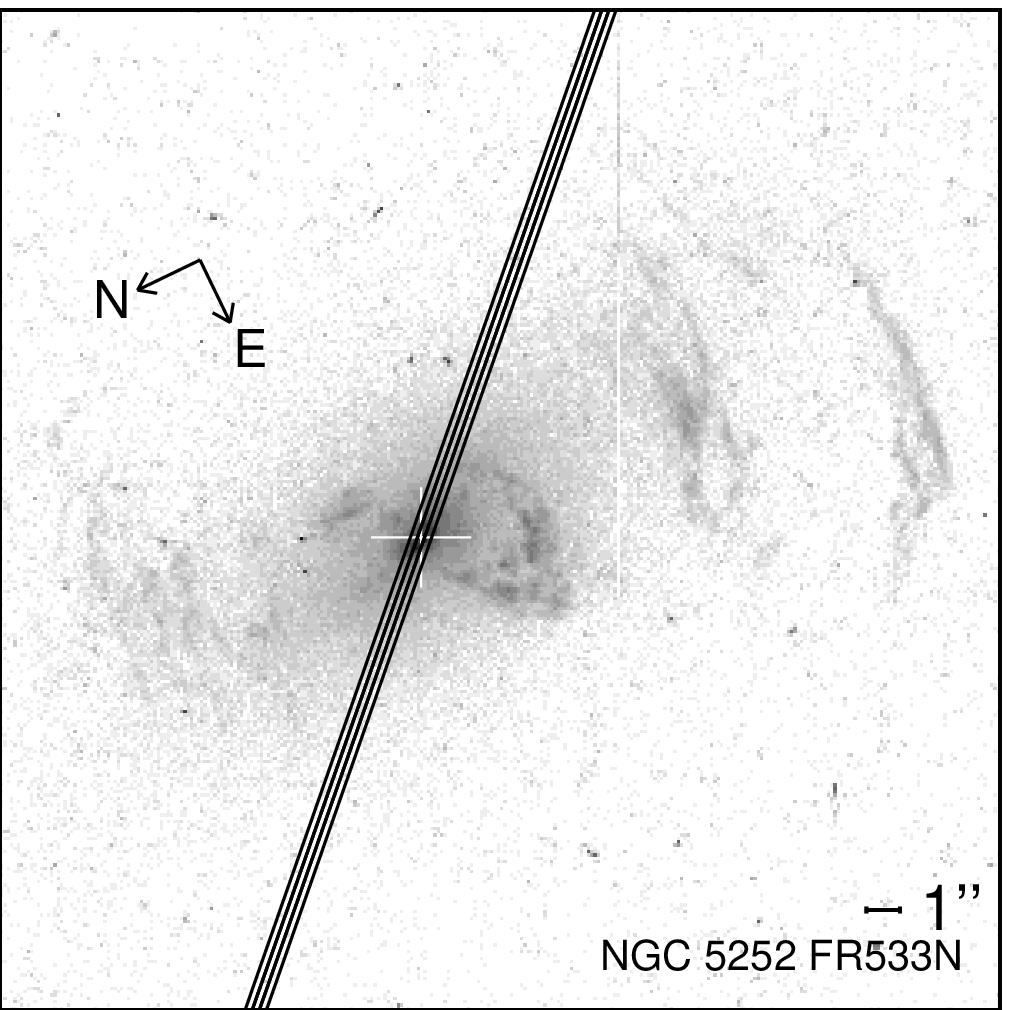}

\end{tabular}						 						   
\label{fig1h}						 						   
\end{figure}						 
%%%%%%%%%%%%%%%%%%%%%%%%%%%%%%%%%%%%%%%%%%%%%%%%%%%%%%%%%%%%
							 						   
\clearpage						 							   
							 						   
% Figure 9						 						   
\begin{figure}						 						   
\centering						 					   
\begin{tabular}{cc}			
\includegraphics[angle=0,scale=0.7]{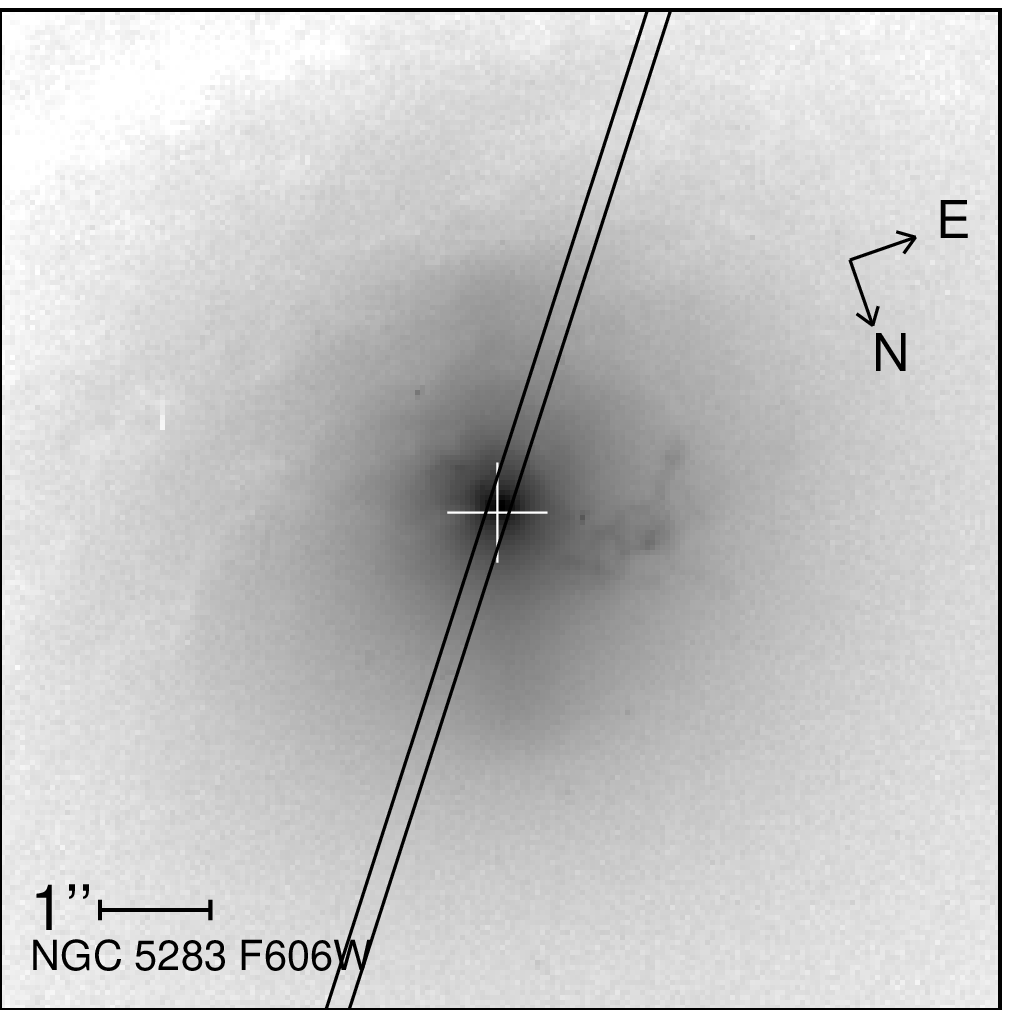}& 
\includegraphics[angle=0,scale=0.7]{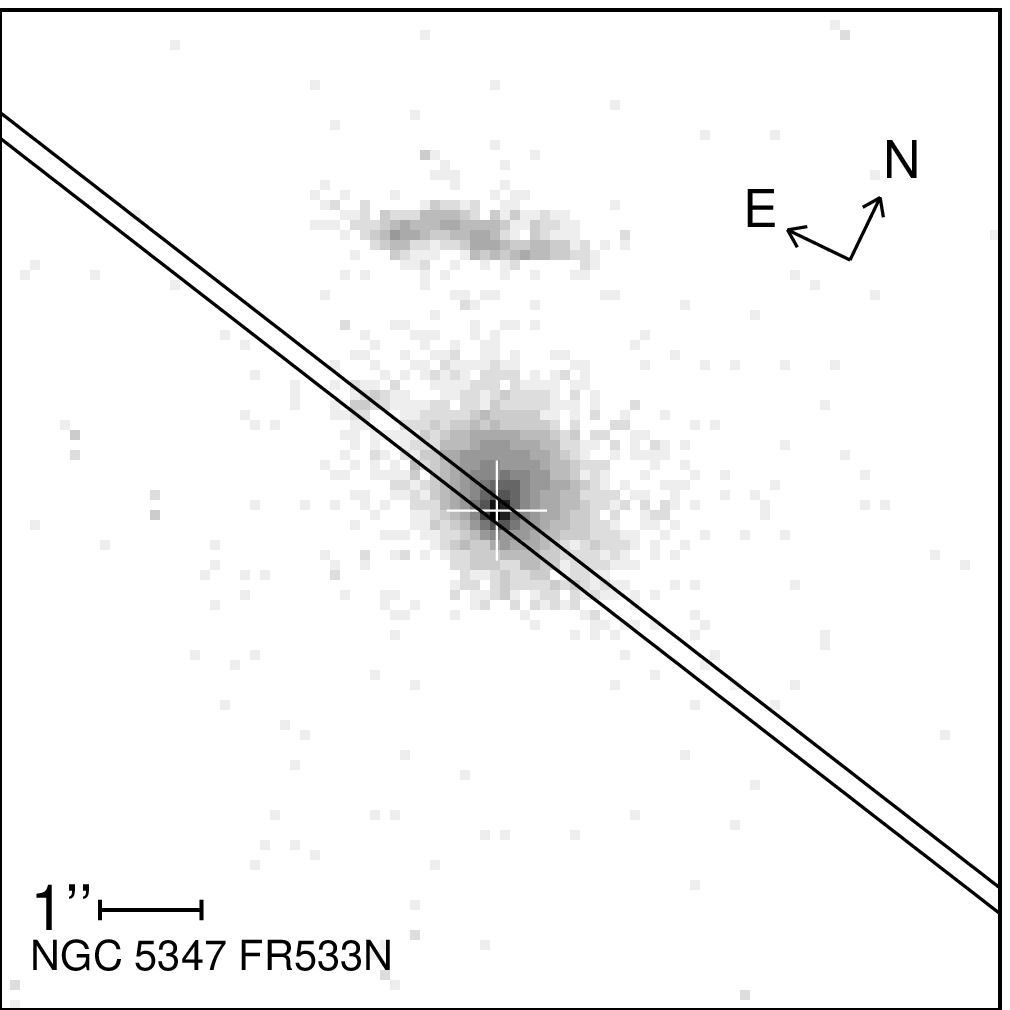}\\  
\includegraphics[angle=0,scale=0.7]{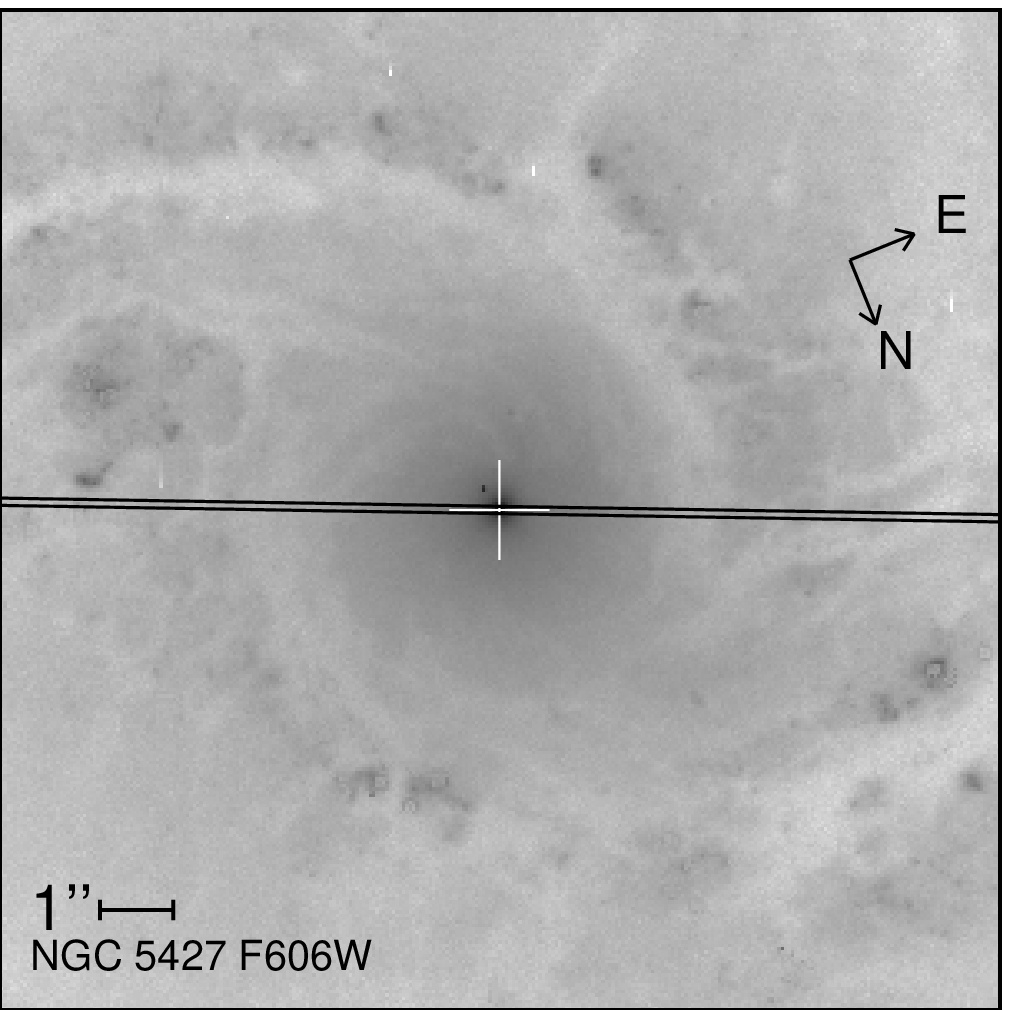}&   
\includegraphics[angle=0,scale=0.7]{ngc5506.eps}  
 						   
\end{tabular}						 						   
\label{fig1i}						 						   
\end{figure}						 
%%%%%%%%%%%%%%%%%%%%%%%%%%%%%%%%%%%%%%%%%%%%%%%%%%%%%%%%%%%%							 
							 						   
\clearpage						 							   
							 						   
% Figure 10						 						   
\begin{figure}						 						   
\centering						 					   
\begin{tabular}{cc}
\includegraphics[angle=0,scale=0.7]{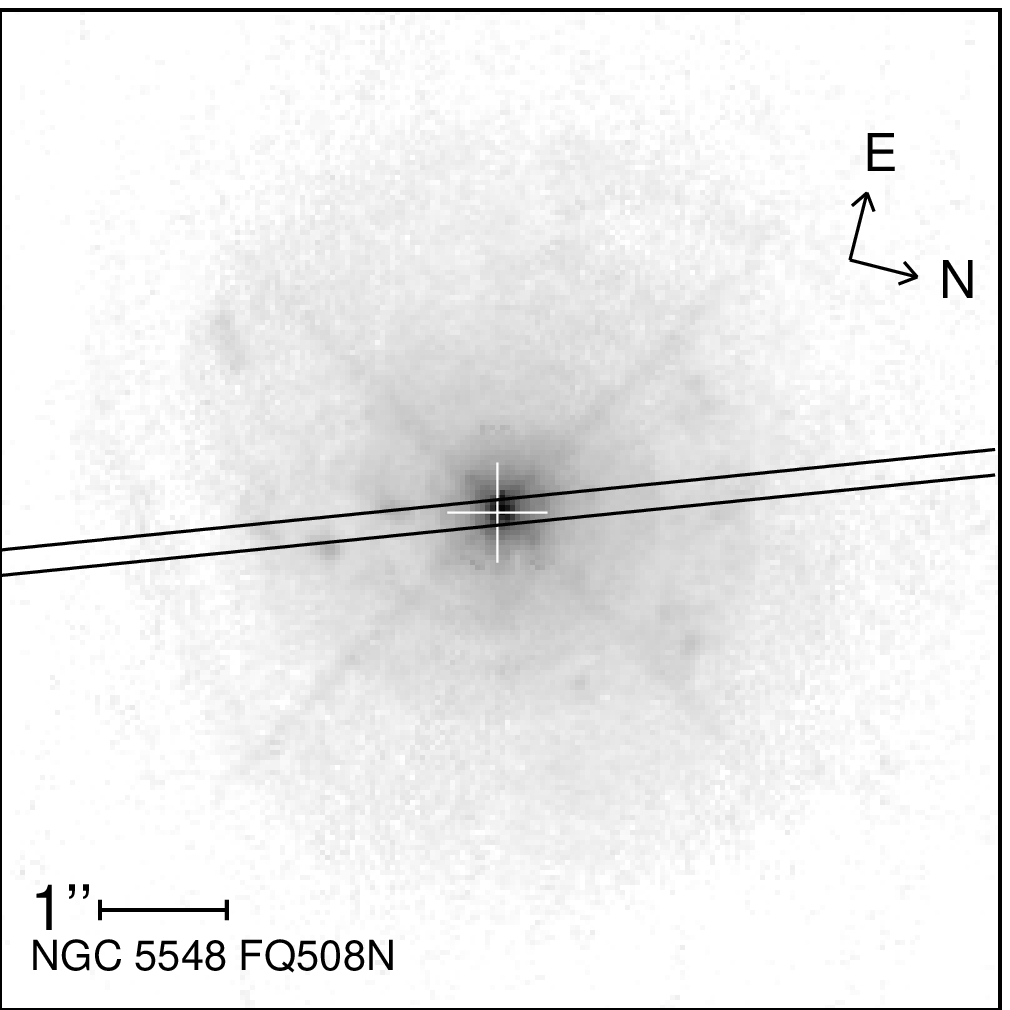}& 				    
\includegraphics[angle=0,scale=0.7]{ngc5643.eps}\\  
\includegraphics[angle=0,scale=0.7]{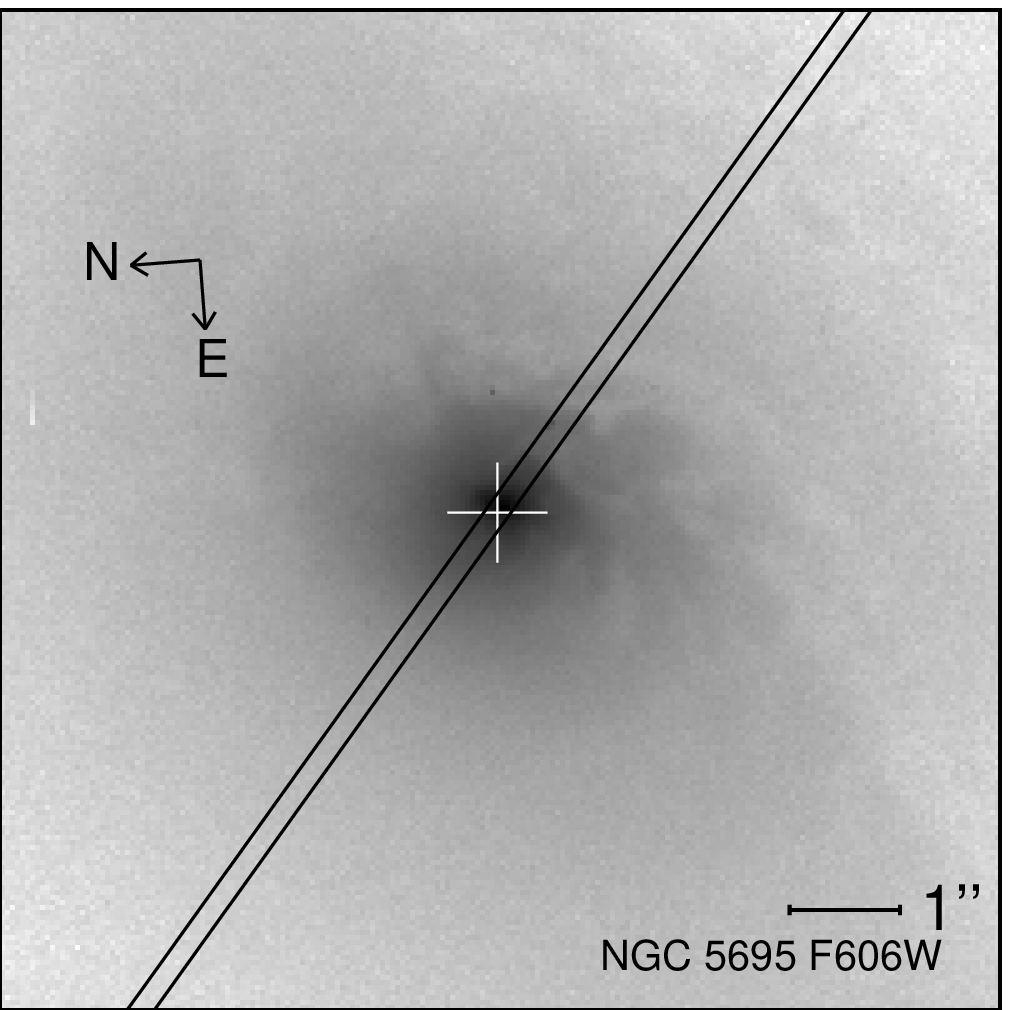}&   
\includegraphics[angle=0,scale=0.7]{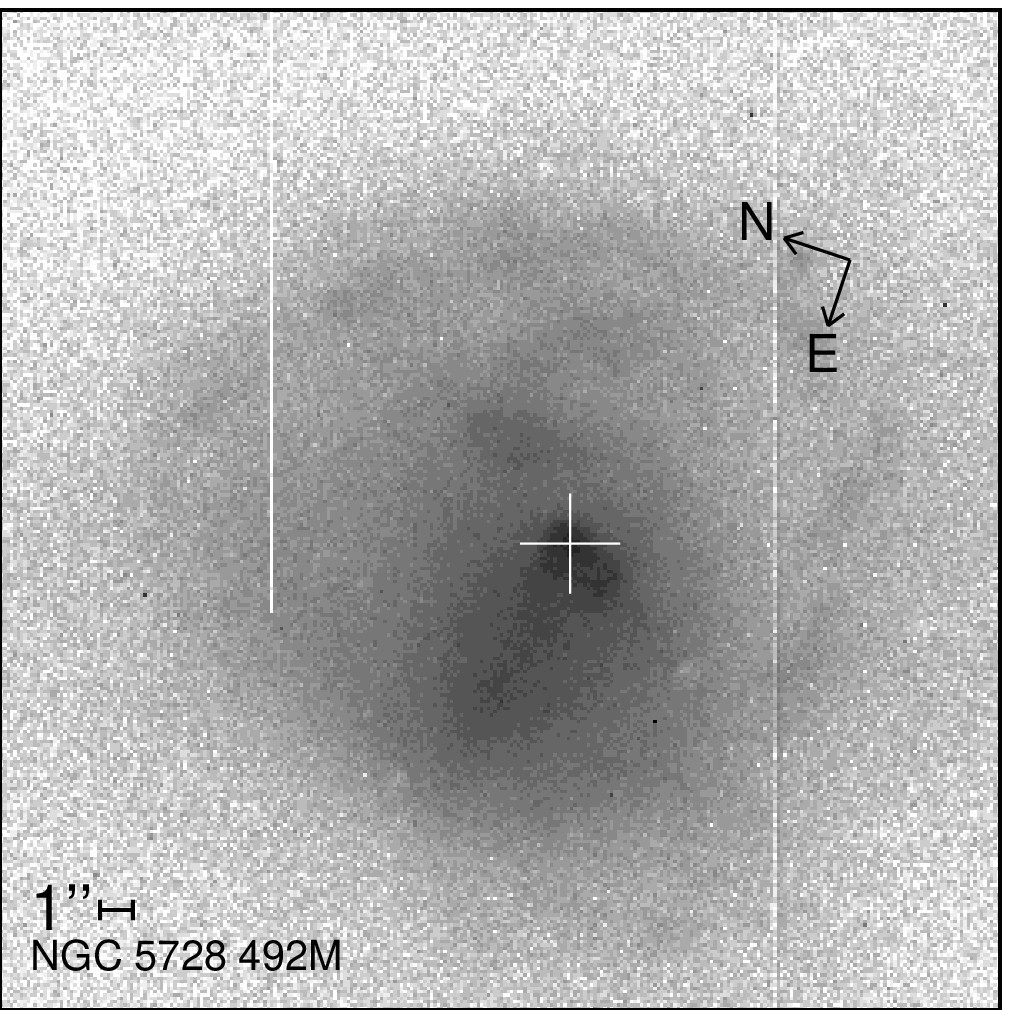}  
 						   
\end{tabular}						 						   
\label{fig1j}						 						   
\end{figure}						 
%%%%%%%%%%%%%%%%%%%%%%%%%%%%%%%%%%%%%%%%%%%%%%%%%%%%%%%%%%%%
							 						   
\clearpage						 							   
							 						   
% Figure 11						 						   
\begin{figure}						 						   
\centering						 					   
\begin{tabular}{cc}
\includegraphics[angle=0,scale=0.7]{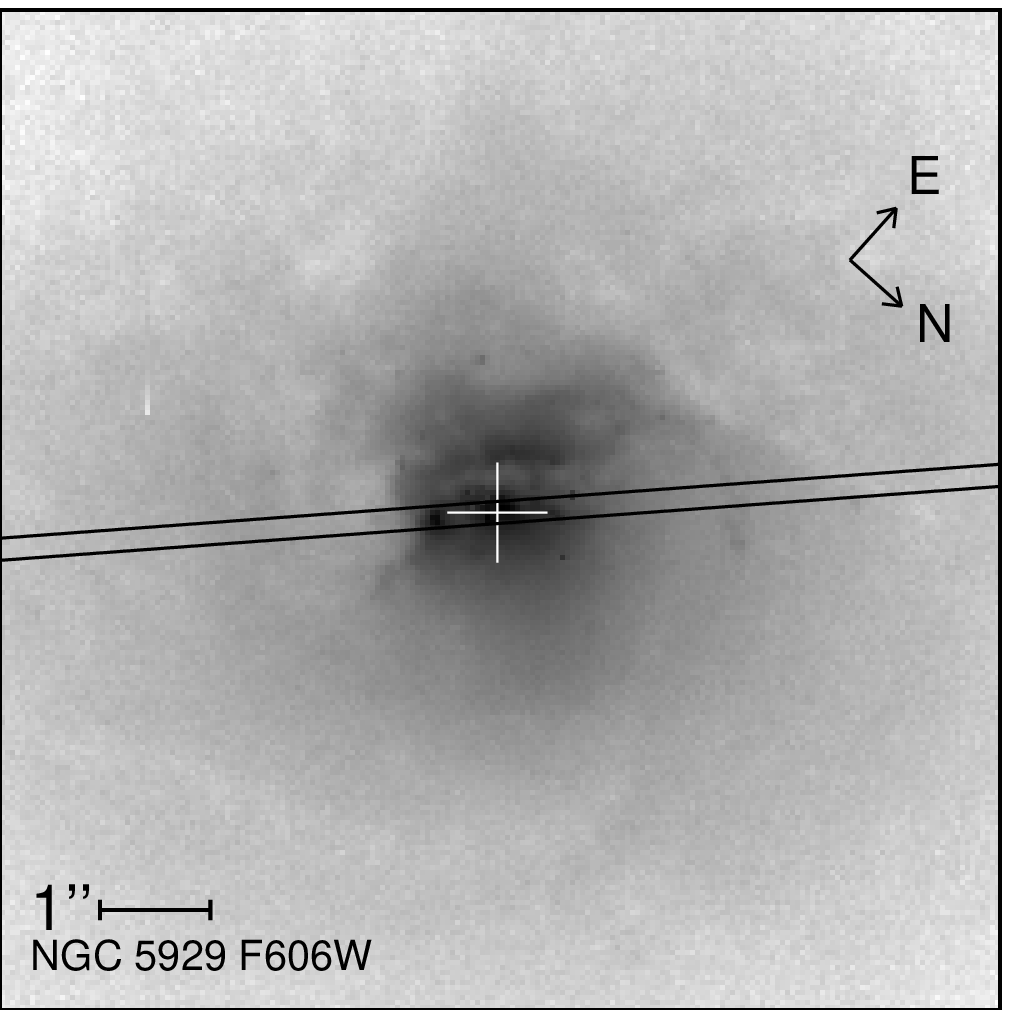}& 			    
\includegraphics[angle=0,scale=0.7]{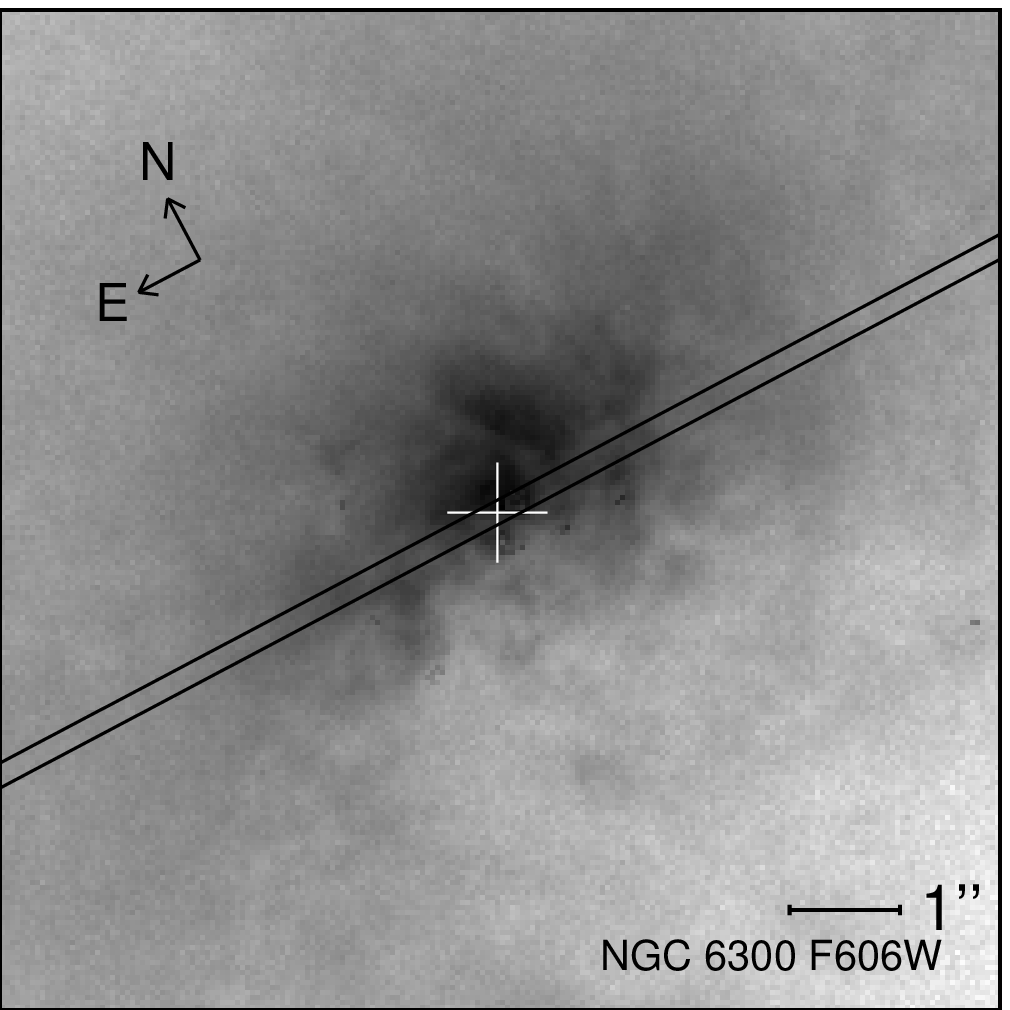}\\  
\includegraphics[angle=0,scale=0.7]{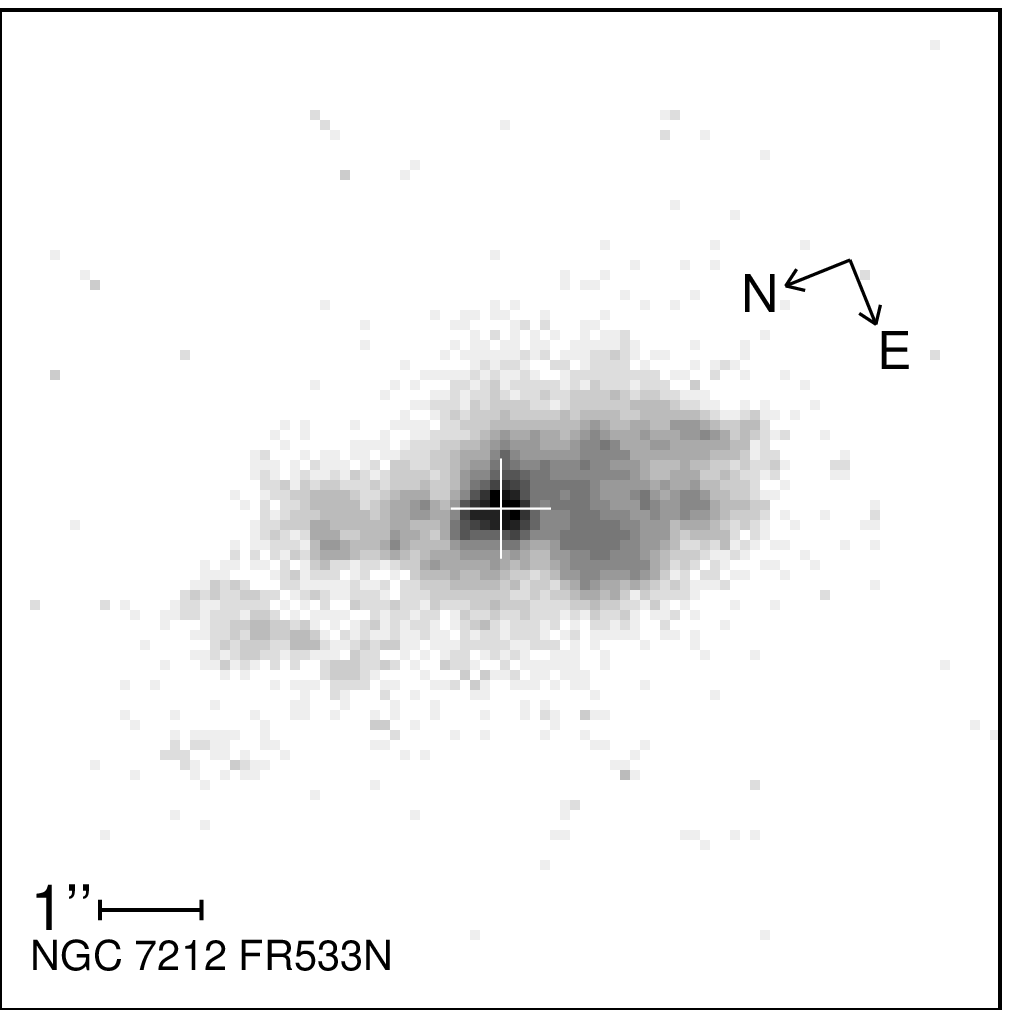}&  
\includegraphics[angle=0,scale=0.7]{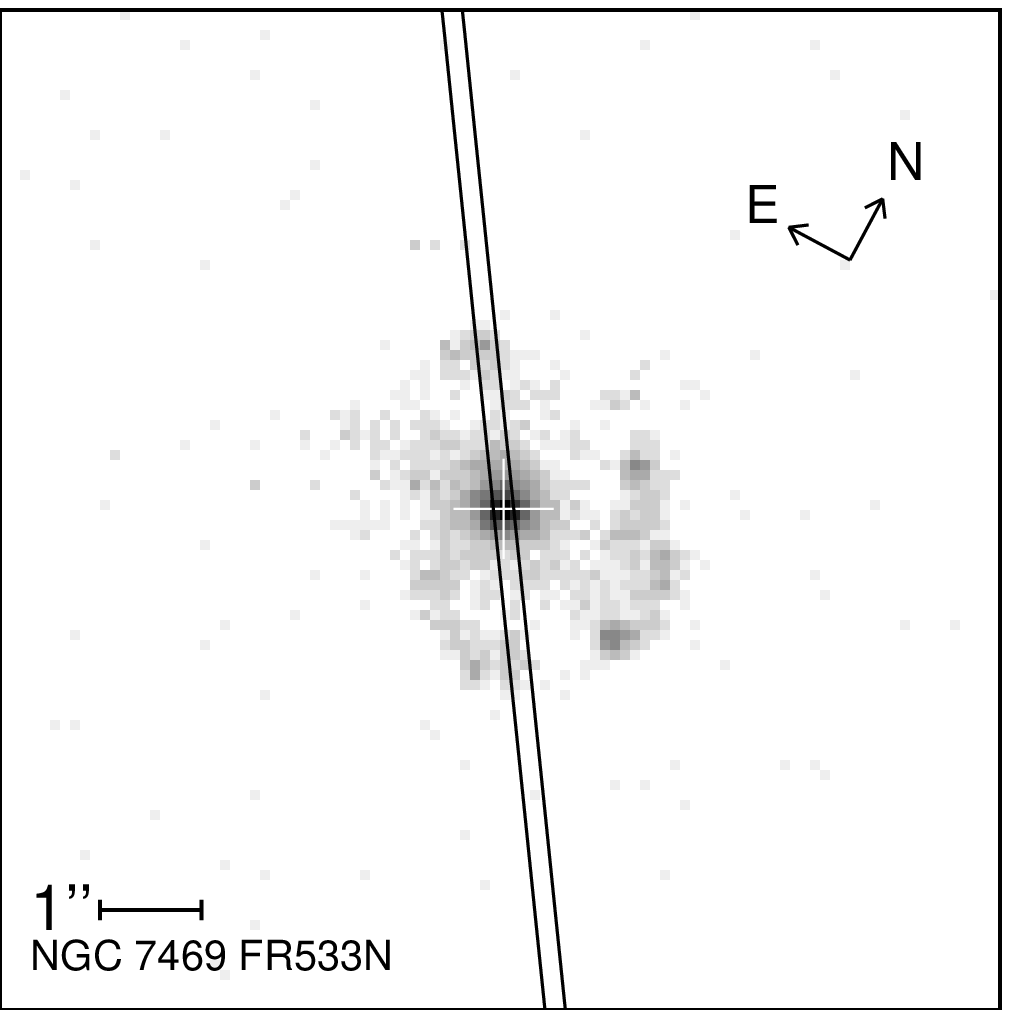}   

\end{tabular}
\label{fig1k}
\end{figure}
%%%%%%%%%%%%%%%%%%%%%%%%%%%%%%%%%%%%%%%%%%%%%%%%%%%%%%%%%%%%%%%

\clearpage

% Figure 12
\begin{figure}
\centering
\begin{tabular}{cc}
\includegraphics[angle=0,scale=0.7]{ngc7674.eps}& 
\includegraphics[angle=0,scale=0.7]{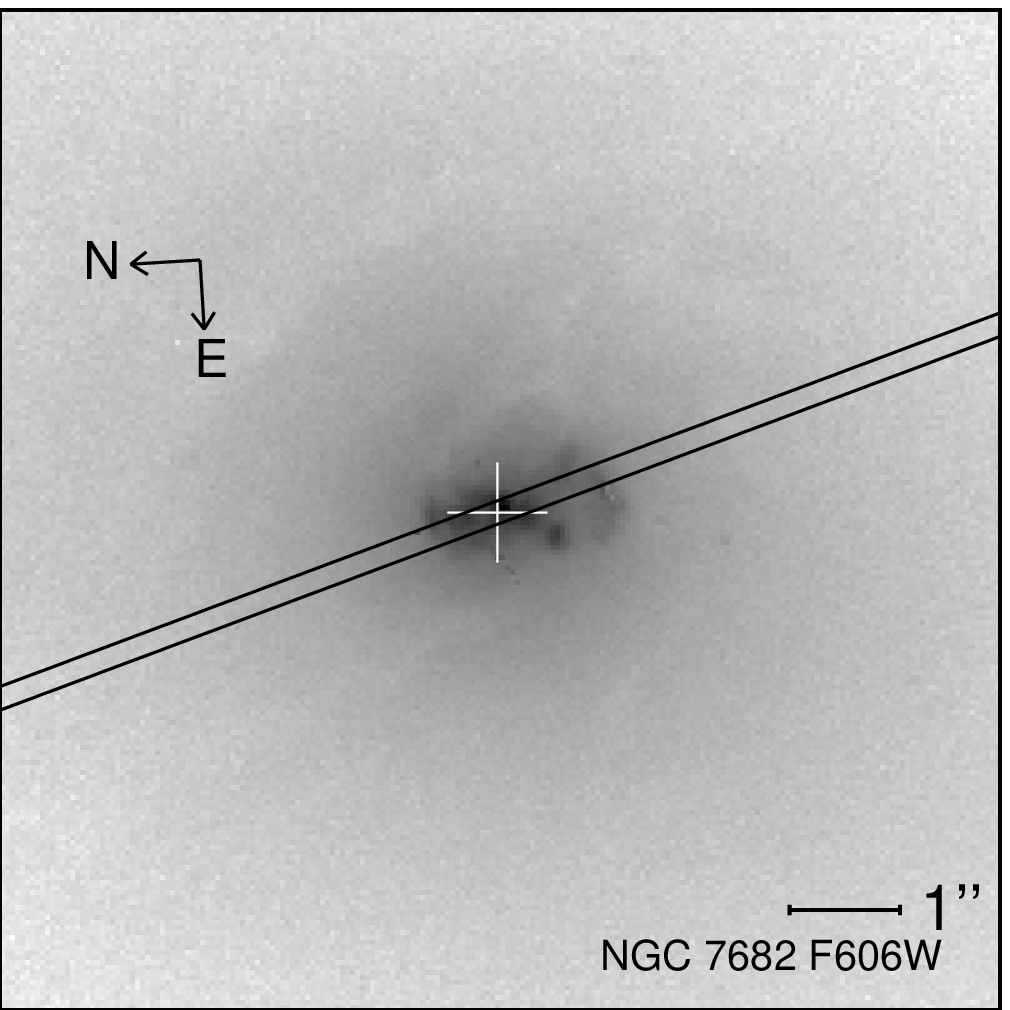}\\  
\includegraphics[angle=0,scale=0.7]{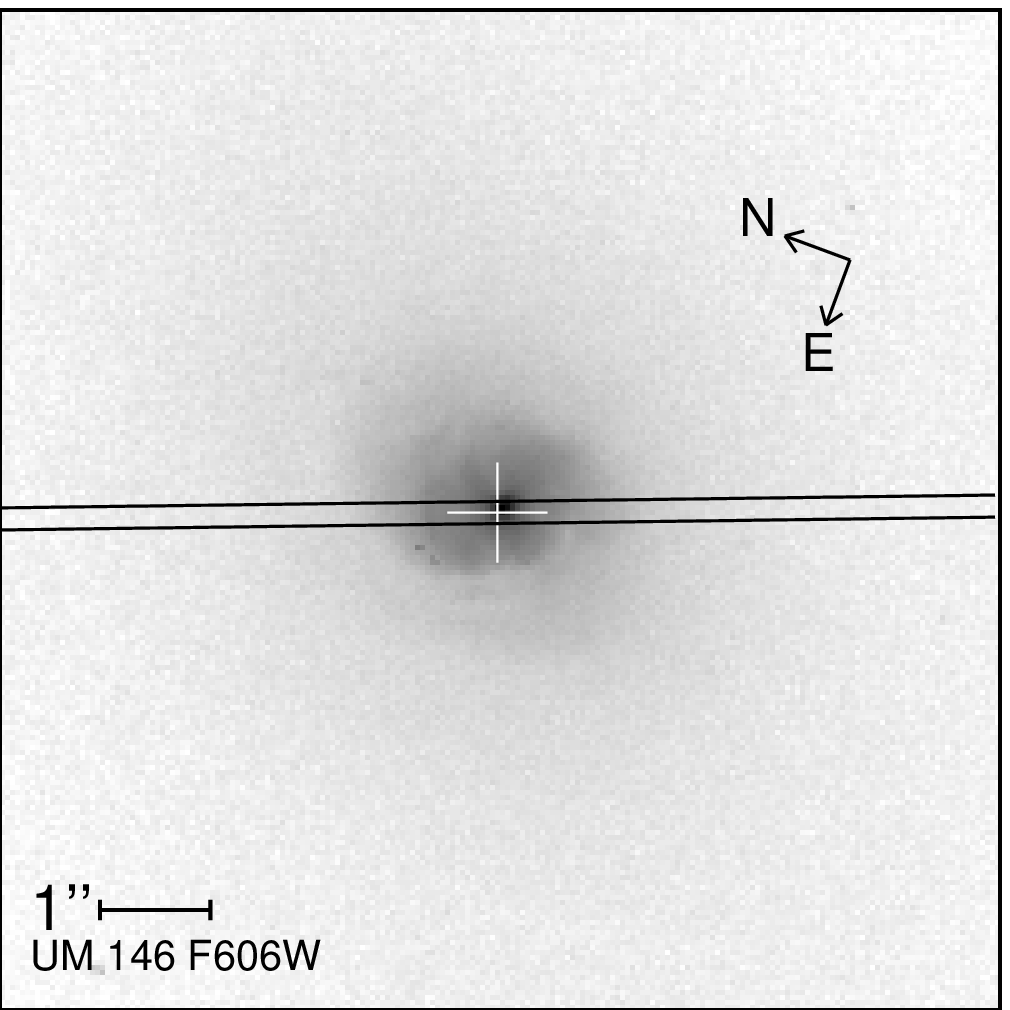} 	 
\end{tabular}
\label{fig1l}
\end{figure}
%%%%%%%%%%%%%%%%%%%%%%%%%%%%%%%%%%%%%%%%%%%%%%%%%%%%%%%%%%%%%%%

\clearpage

\subsection{AGN Kinematic Data}

%Models
This appendix contains velocities, FWHMs, and normalized fluxes of each spectrum collected
to created our expanded sample. Figures containing multiple data sets signify observations using
separate gratings at an identical position. Kinematic data for slitless observations are available 
in \citet{Rui05}.

Kinematics data correspond to {\it HST} gratings as follows:

Green diamonds: [O III] 5007 emission line using G430L grating

Blue circles: [O III] 5007 emission line using G430M grating

Red squares: H$\alpha$ 6564 emission line using G750M grating

%%%%%%%%%%%%%%%%%%%%%%%%%%%%%%%%%%%%%%%%%%%%%%%%%%%%%%%%%%%%%%%

\begin{figure}
\centering
\includegraphics[angle=0,scale=0.7]{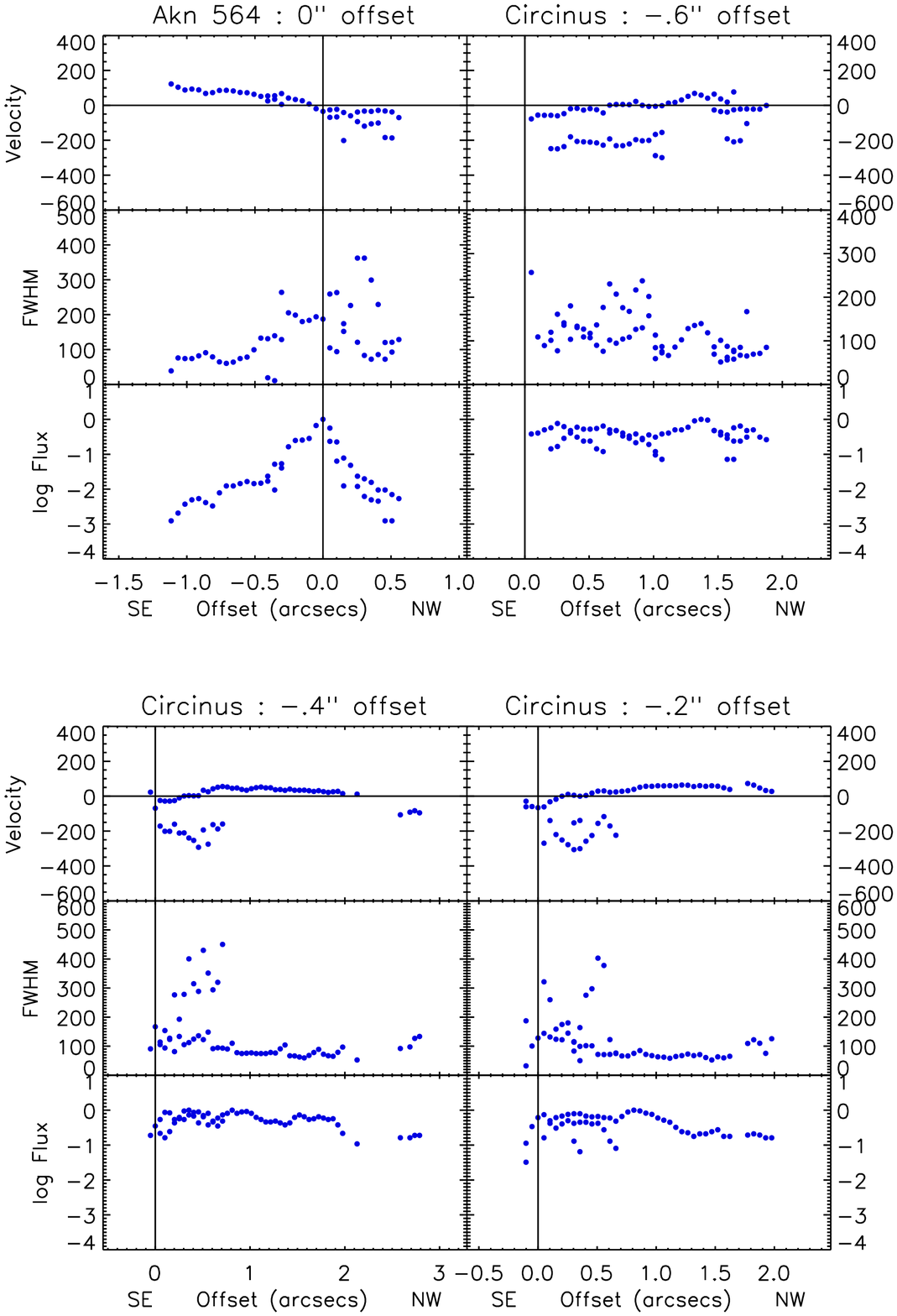}
\end{figure}
\clearpage
%%%%%%%%%%%%%%%%%%%%%%%%%%%%%%%%%%%%%%%%%%%%%%%%%%%%%%%%%%%%%%%

\begin{figure}
\centering
\includegraphics[angle=0,scale=0.7]{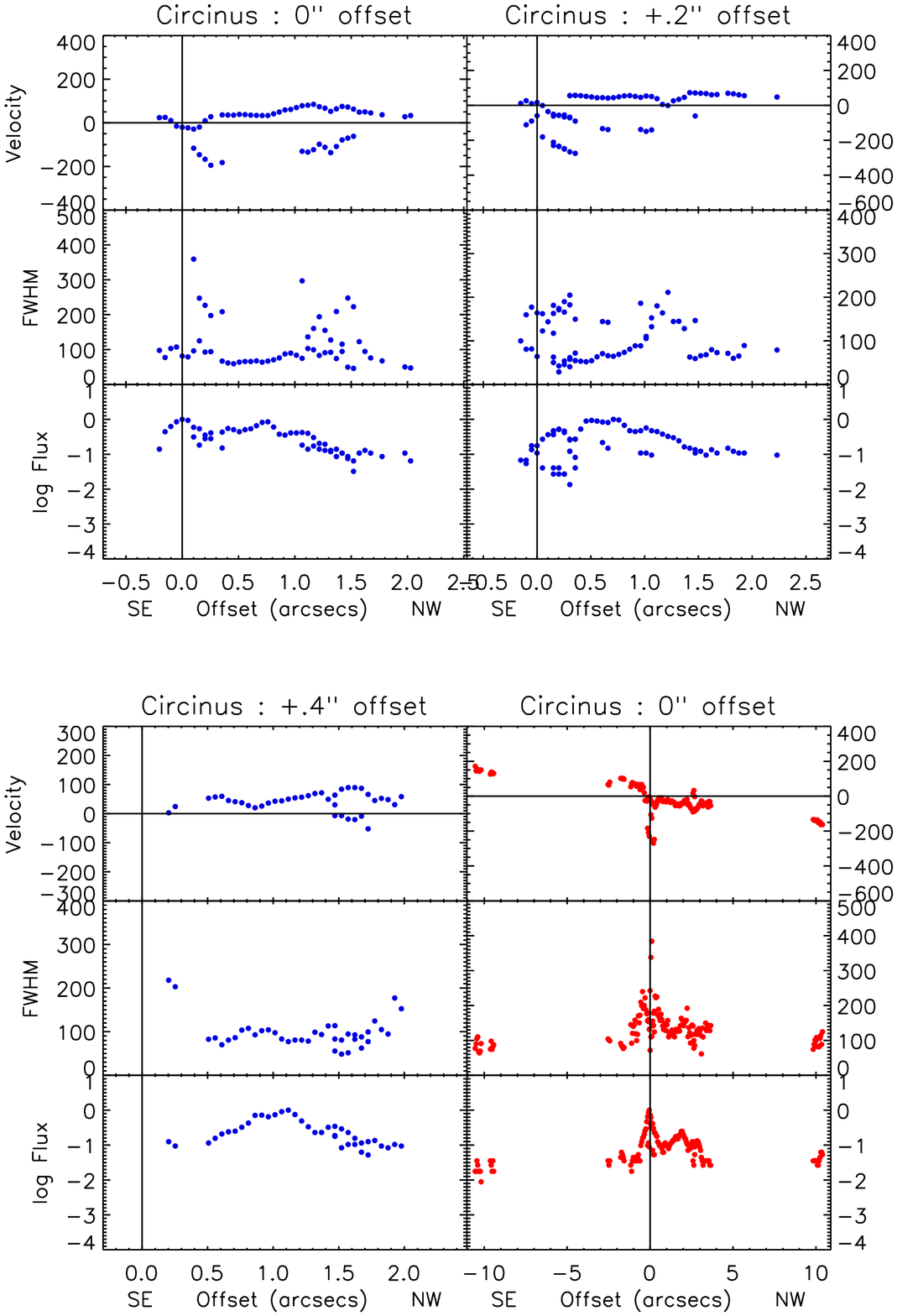}
\end{figure}
\clearpage
%%%%%%%%%%%%%%%%%%%%%%%%%%%%%%%%%%%%%%%%%%%%%%%%%%%%%%%%%%%%%%%

\begin{figure}
\centering
\includegraphics[angle=0,scale=0.7]{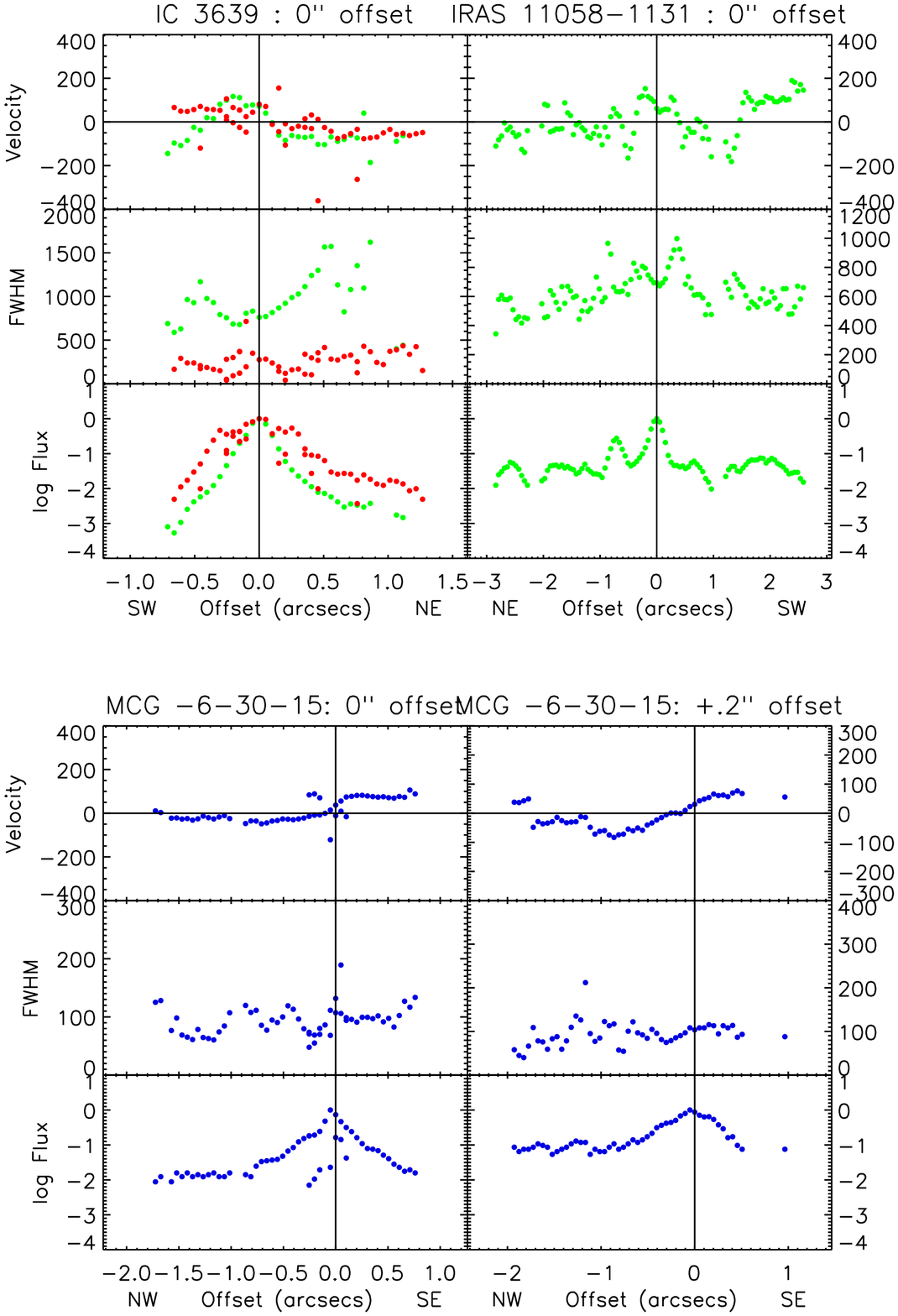}
\end{figure}
\clearpage
%%%%%%%%%%%%%%%%%%%%%%%%%%%%%%%%%%%%%%%%%%%%%%%%%%%%%%%%%%%%%%%

\begin{figure}
\centering
\includegraphics[angle=0,scale=0.7]{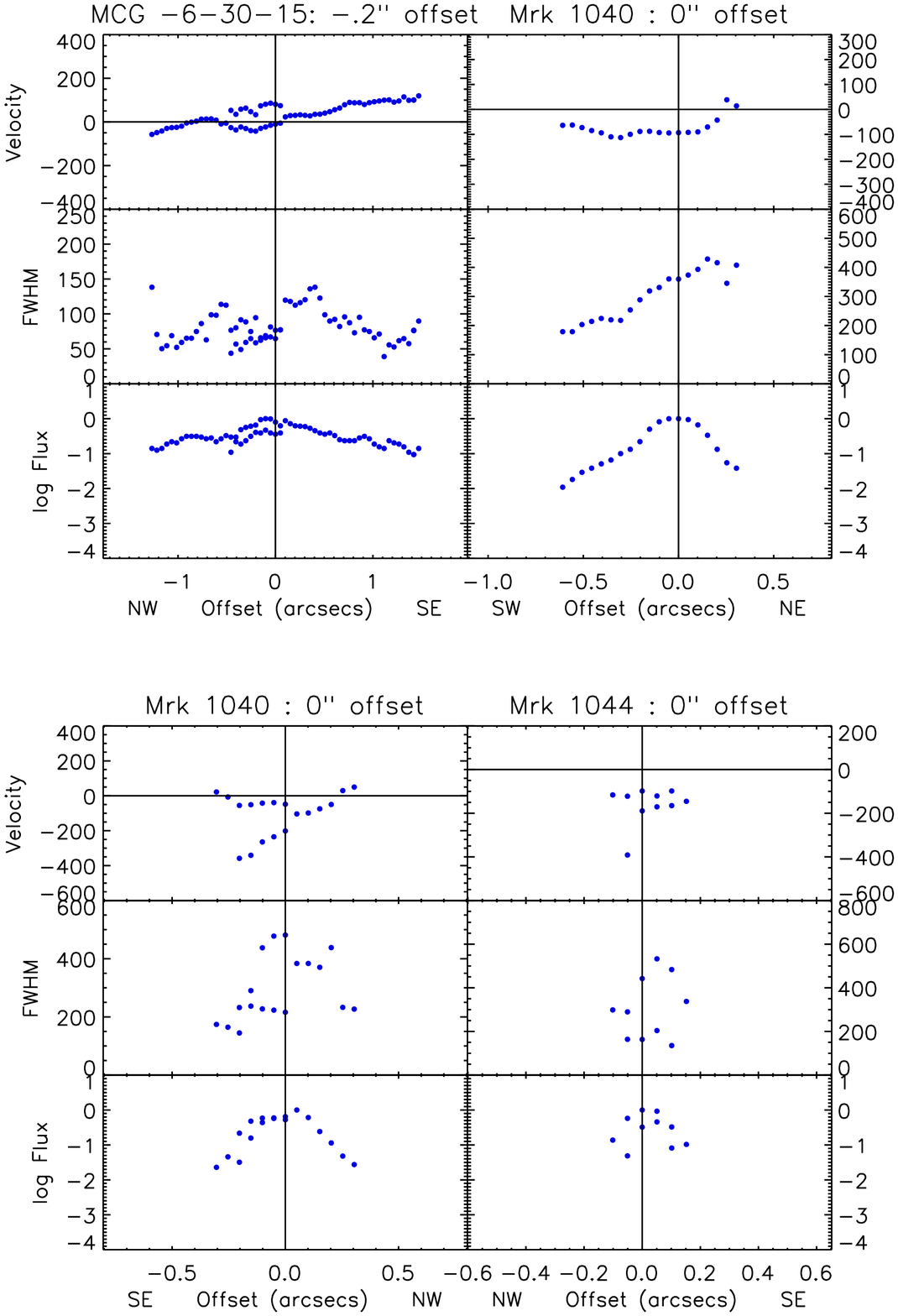}
\end{figure}
\clearpage
%%%%%%%%%%%%%%%%%%%%%%%%%%%%%%%%%%%%%%%%%%%%%%%%%%%%%%%%%%%%%%%

\begin{figure}
\centering
\includegraphics[angle=0,scale=0.7]{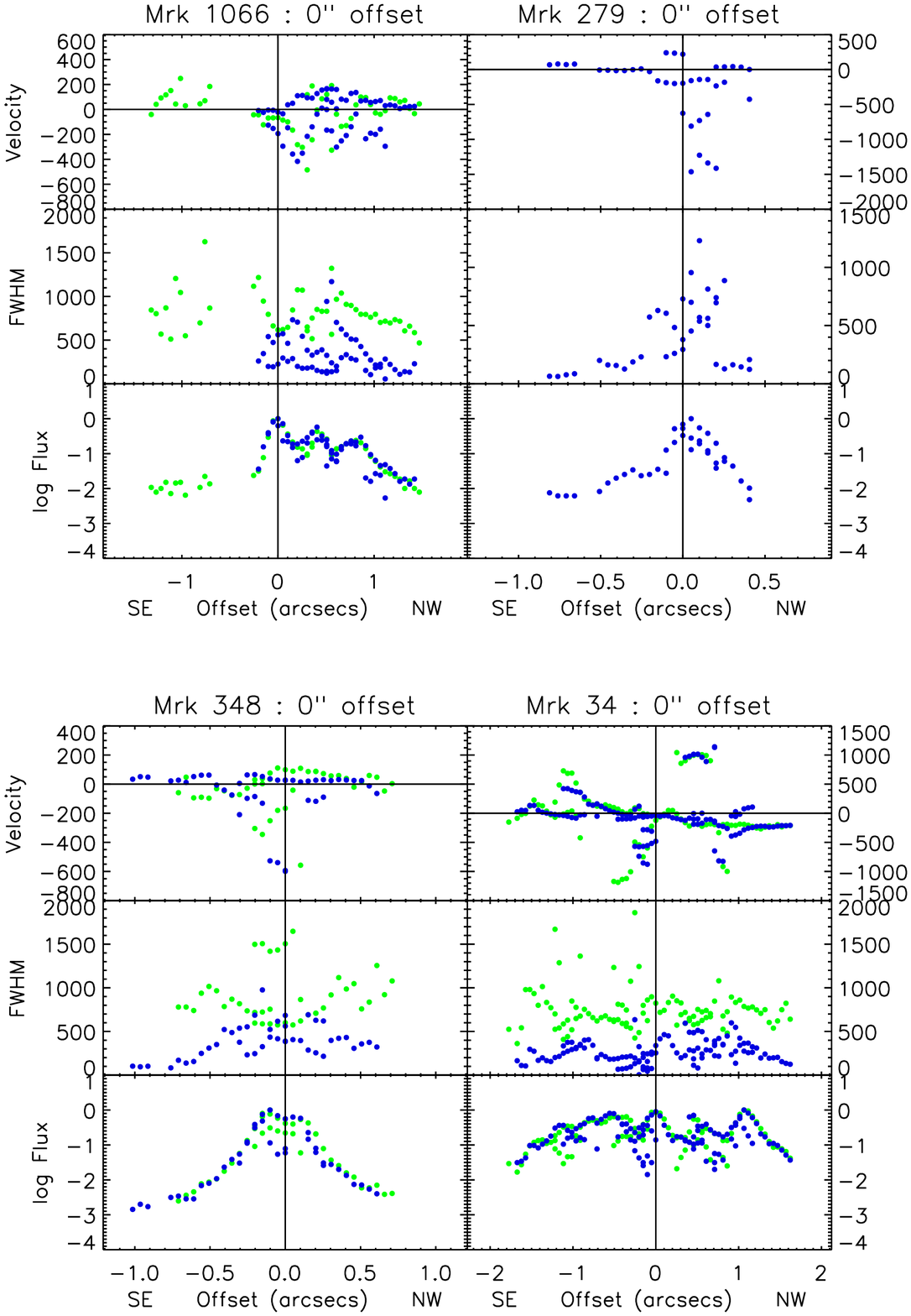}
\end{figure}
\clearpage
%%%%%%%%%%%%%%%%%%%%%%%%%%%%%%%%%%%%%%%%%%%%%%%%%%%%%%%%%%%%%%%

\begin{figure}
\centering
\includegraphics[angle=0,scale=0.7]{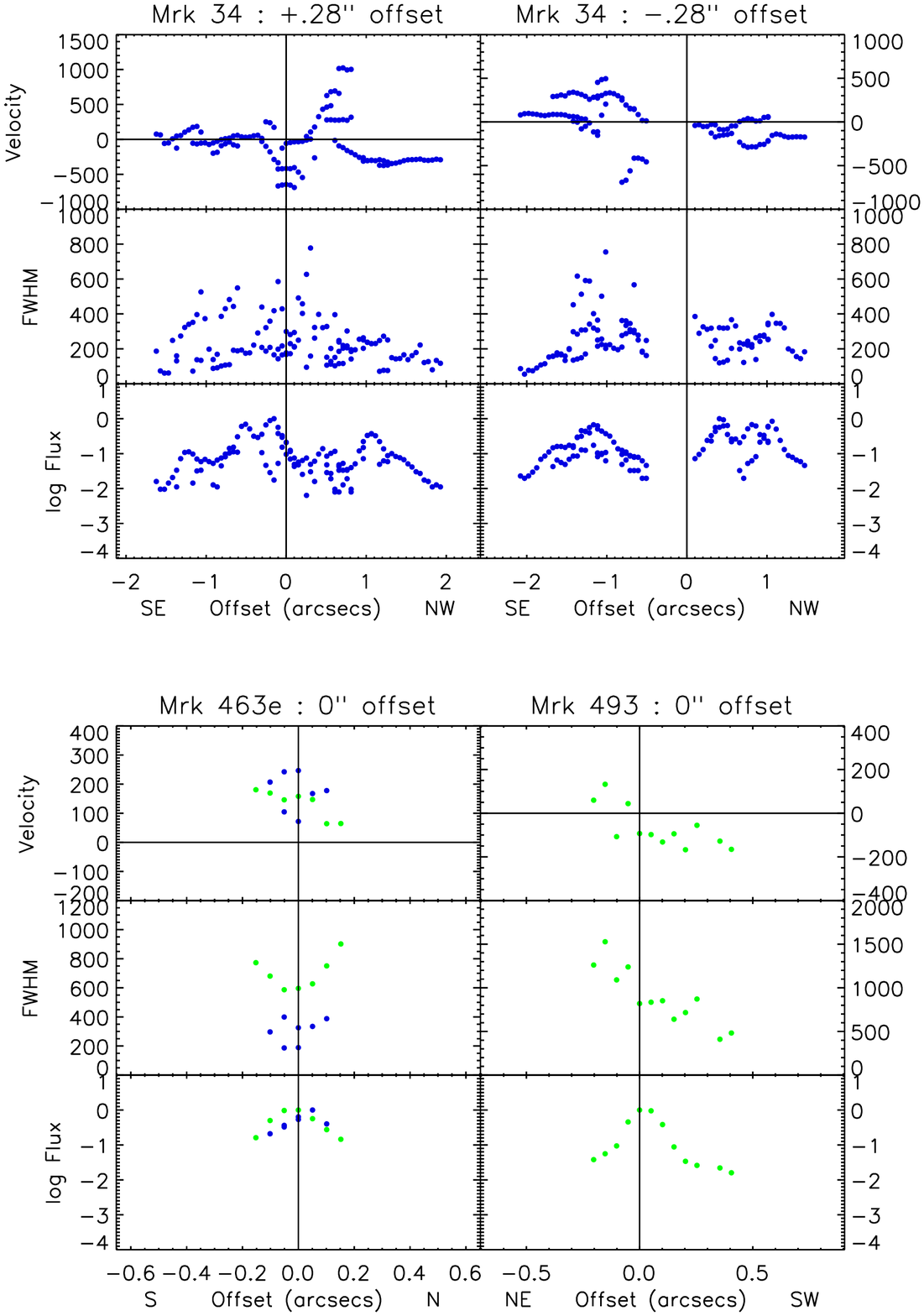}
\end{figure}
\clearpage
%%%%%%%%%%%%%%%%%%%%%%%%%%%%%%%%%%%%%%%%%%%%%%%%%%%%%%%%%%%%%%%

\begin{figure}
\centering
\includegraphics[angle=0,scale=0.7]{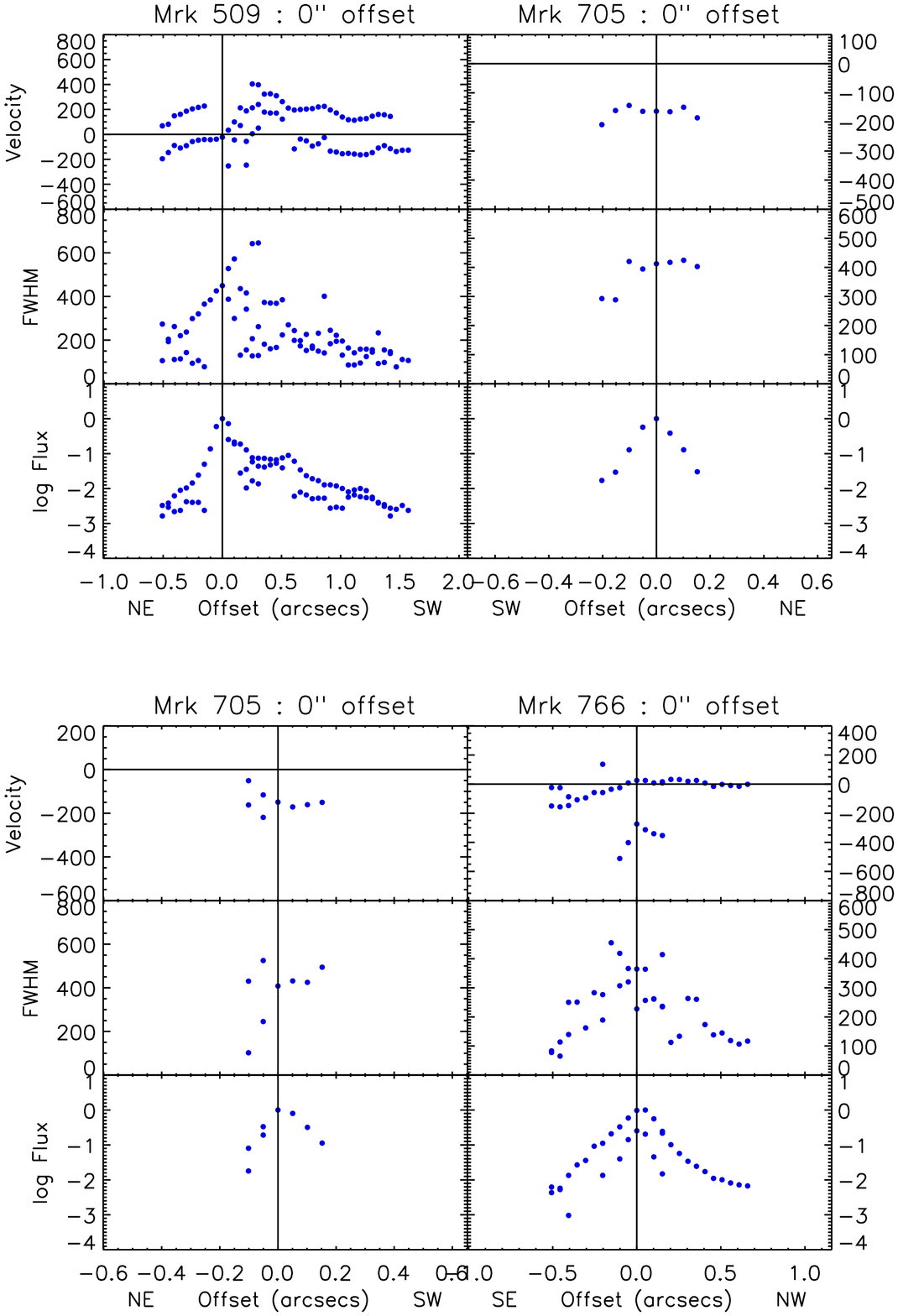}
\end{figure}
\clearpage
%%%%%%%%%%%%%%%%%%%%%%%%%%%%%%%%%%%%%%%%%%%%%%%%%%%%%%%%%%%%%%%

\begin{figure}
\centering
\includegraphics[angle=0,scale=0.7]{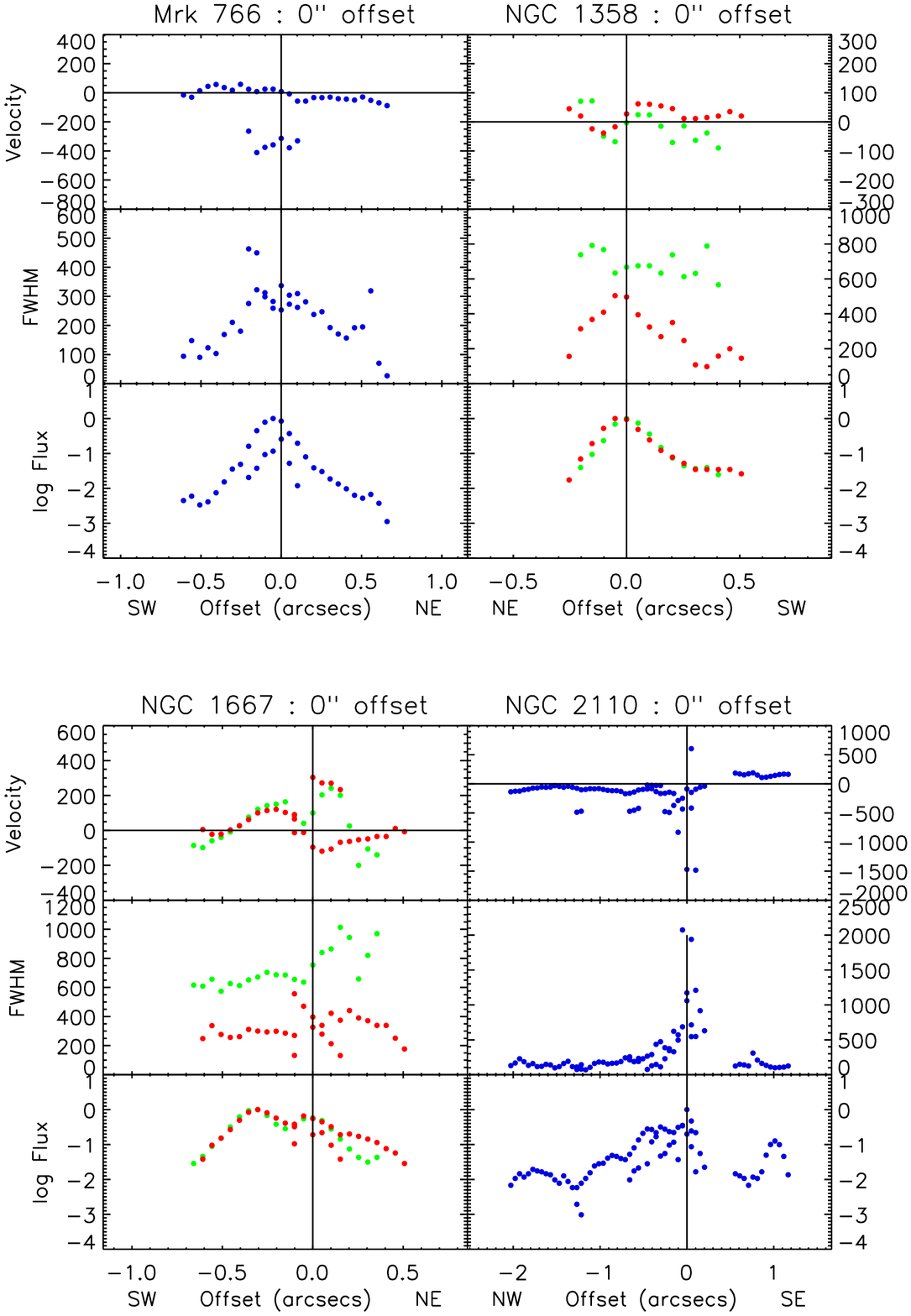}
\end{figure}
\clearpage
%%%%%%%%%%%%%%%%%%%%%%%%%%%%%%%%%%%%%%%%%%%%%%%%%%%%%%%%%%%%%%%

\begin{figure}
\centering
\includegraphics[angle=0,scale=0.7]{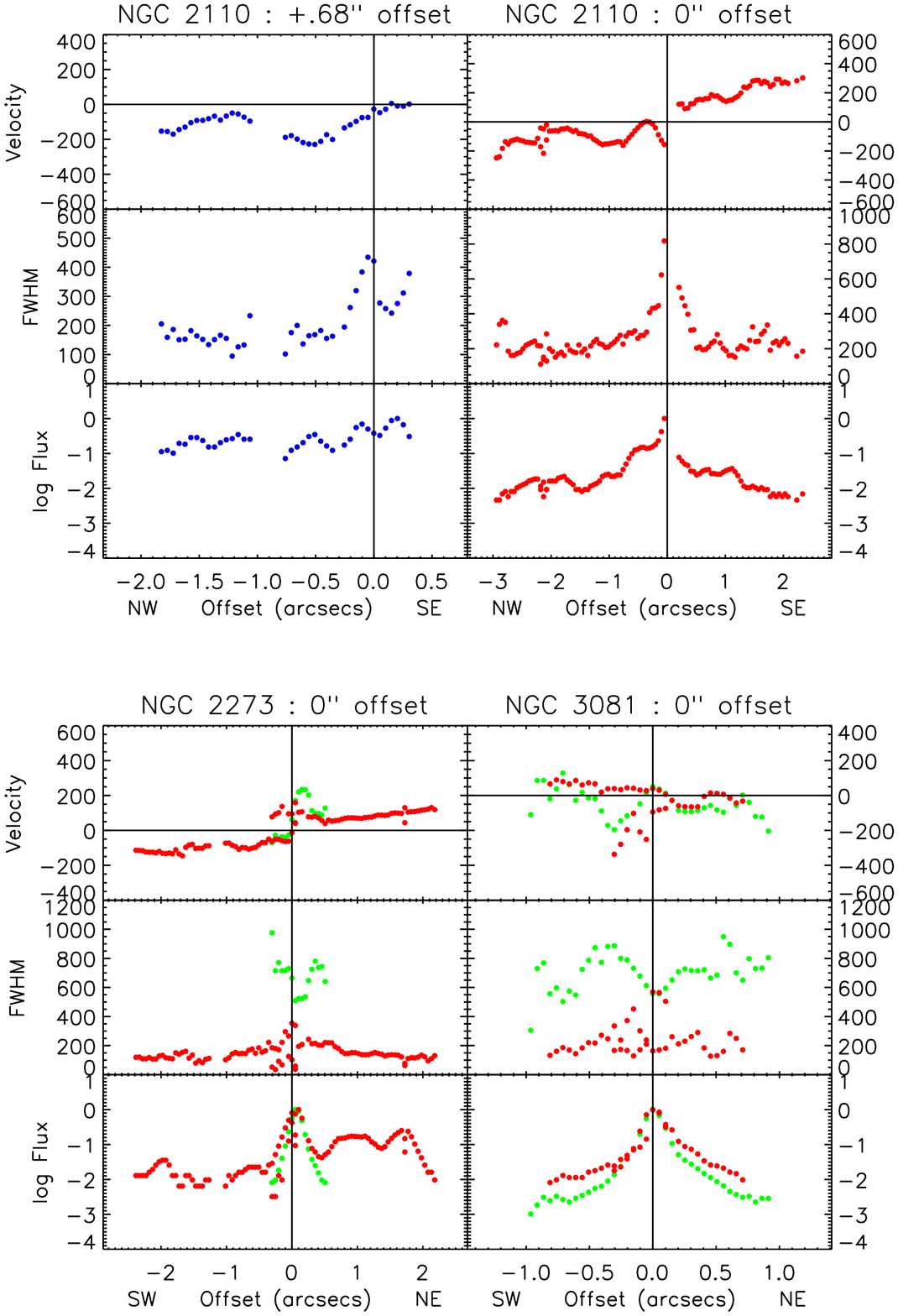}
\end{figure}
\clearpage
%%%%%%%%%%%%%%%%%%%%%%%%%%%%%%%%%%%%%%%%%%%%%%%%%%%%%%%%%%%%%%%

\begin{figure}
\centering
\includegraphics[angle=0,scale=0.7]{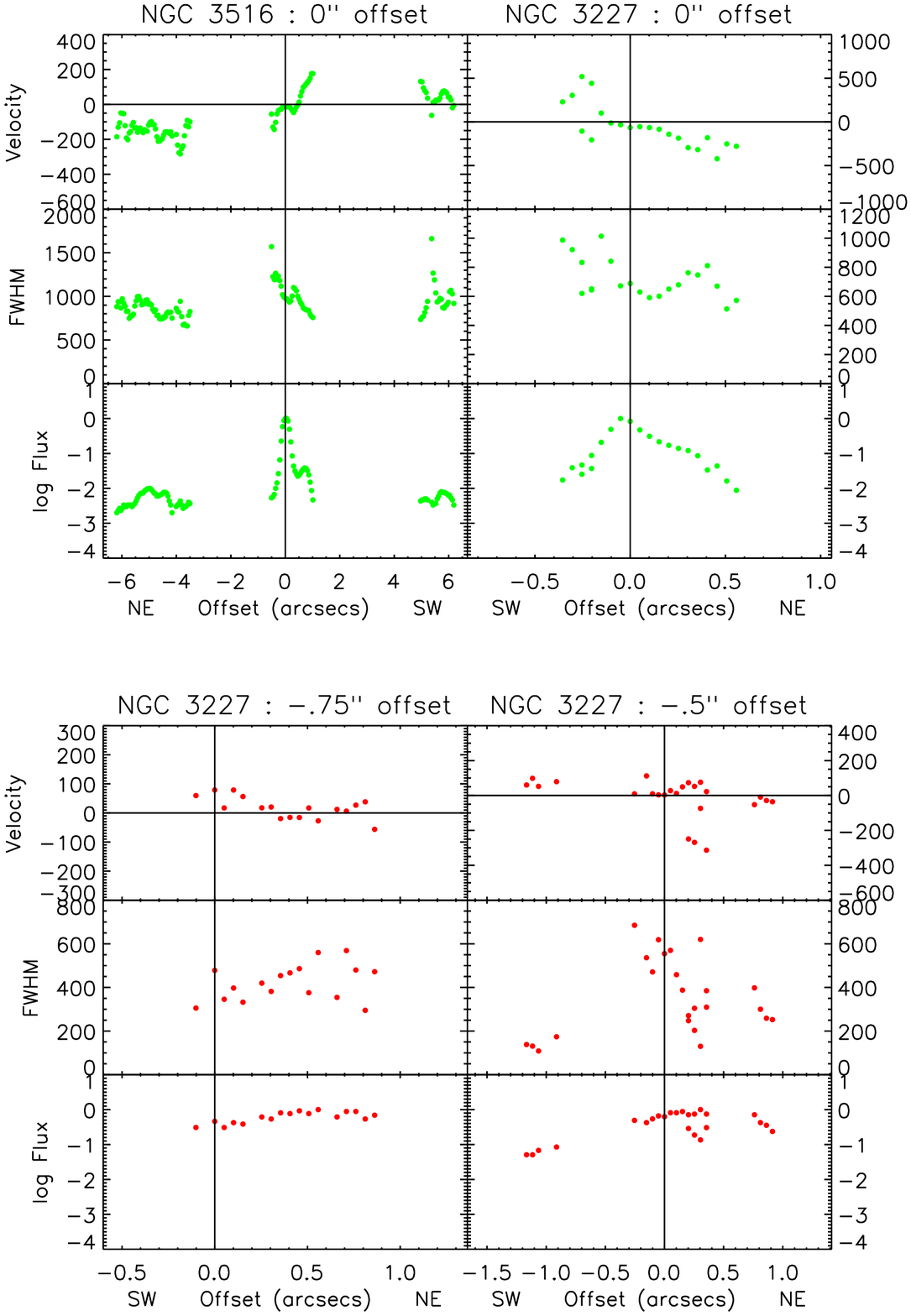}
\end{figure}
\clearpage
%%%%%%%%%%%%%%%%%%%%%%%%%%%%%%%%%%%%%%%%%%%%%%%%%%%%%%%%%%%%%%%

\begin{figure}
\centering
\includegraphics[angle=0,scale=0.7]{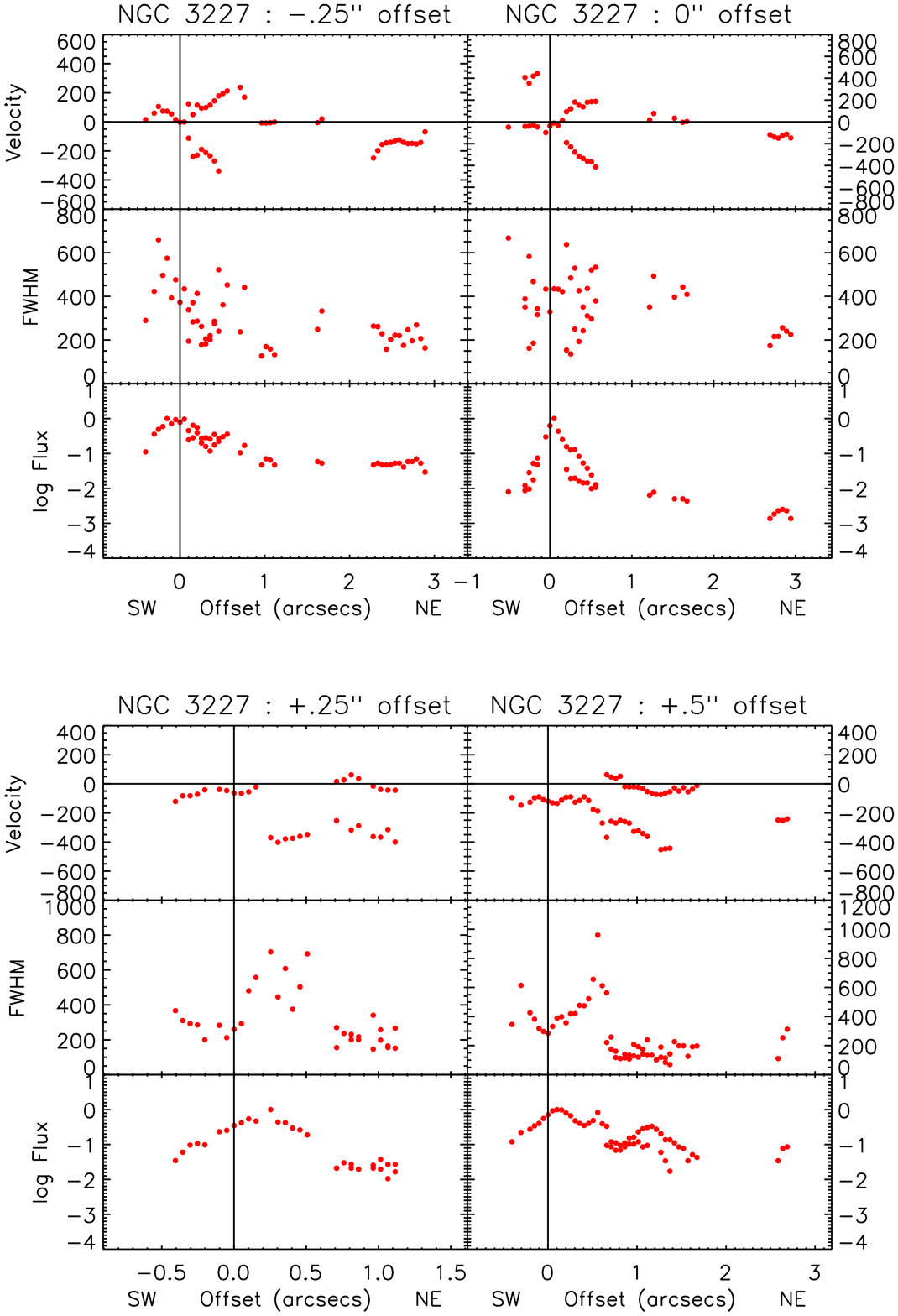}
\end{figure}
\clearpage
%%%%%%%%%%%%%%%%%%%%%%%%%%%%%%%%%%%%%%%%%%%%%%%%%%%%%%%%%%%%%%%

\begin{figure}
\centering
\includegraphics[angle=0,scale=0.7]{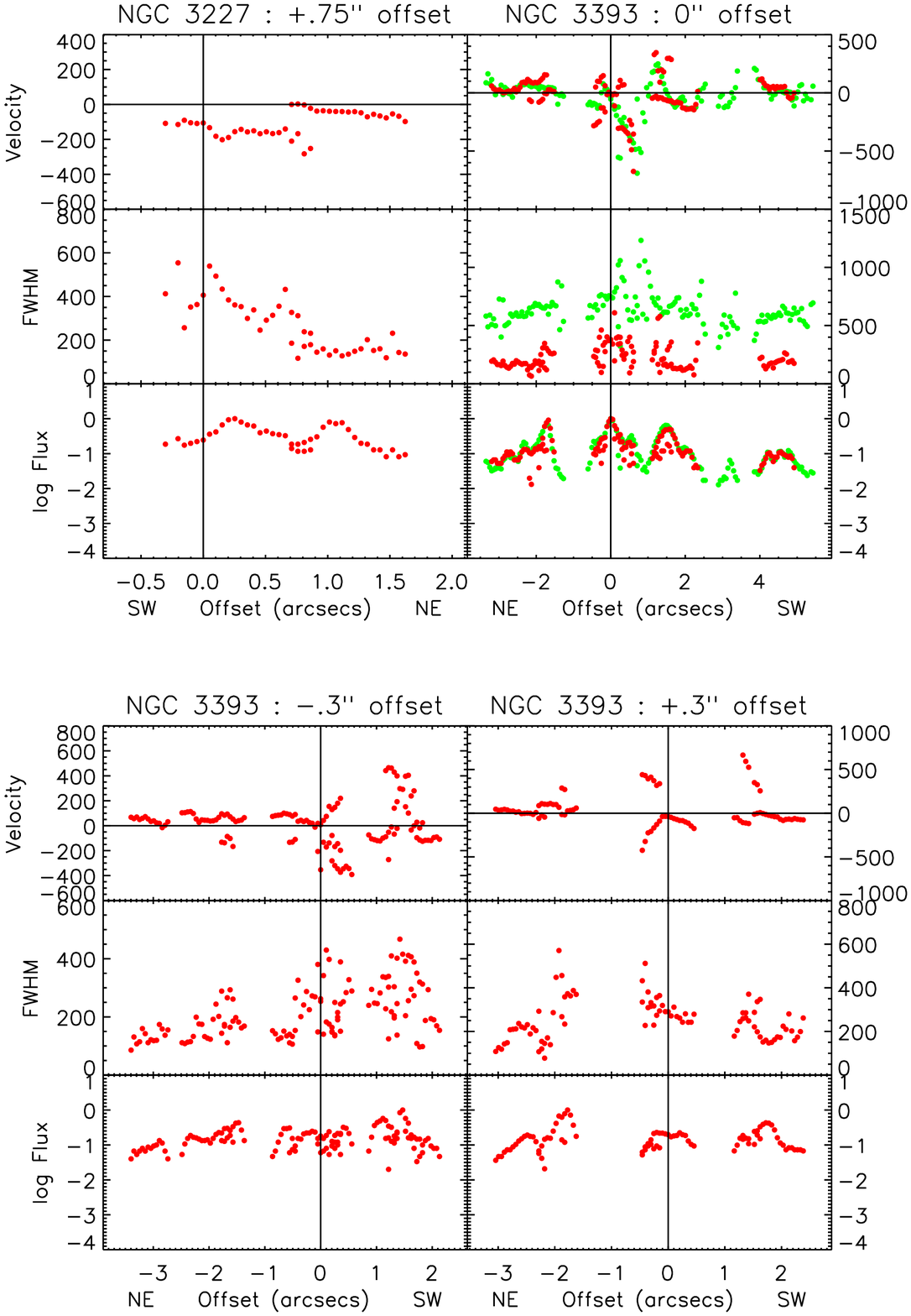}
\end{figure}
\clearpage
%%%%%%%%%%%%%%%%%%%%%%%%%%%%%%%%%%%%%%%%%%%%%%%%%%%%%%%%%%%%%%%

\begin{figure}
\centering
\includegraphics[angle=0,scale=0.7]{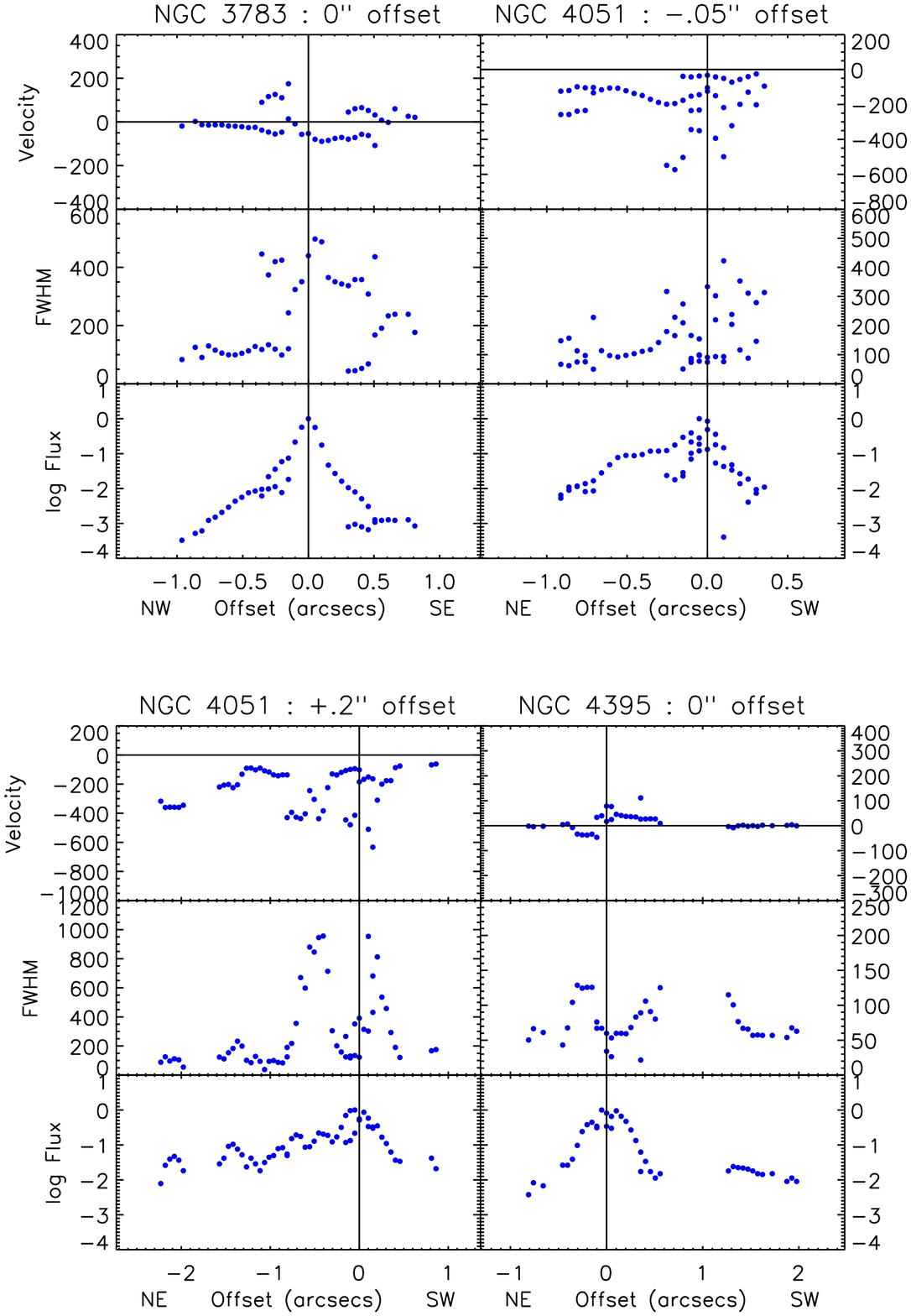}
\end{figure}
\clearpage
%%%%%%%%%%%%%%%%%%%%%%%%%%%%%%%%%%%%%%%%%%%%%%%%%%%%%%%%%%%%%%%

\begin{figure}
\centering
\includegraphics[angle=0,scale=0.7]{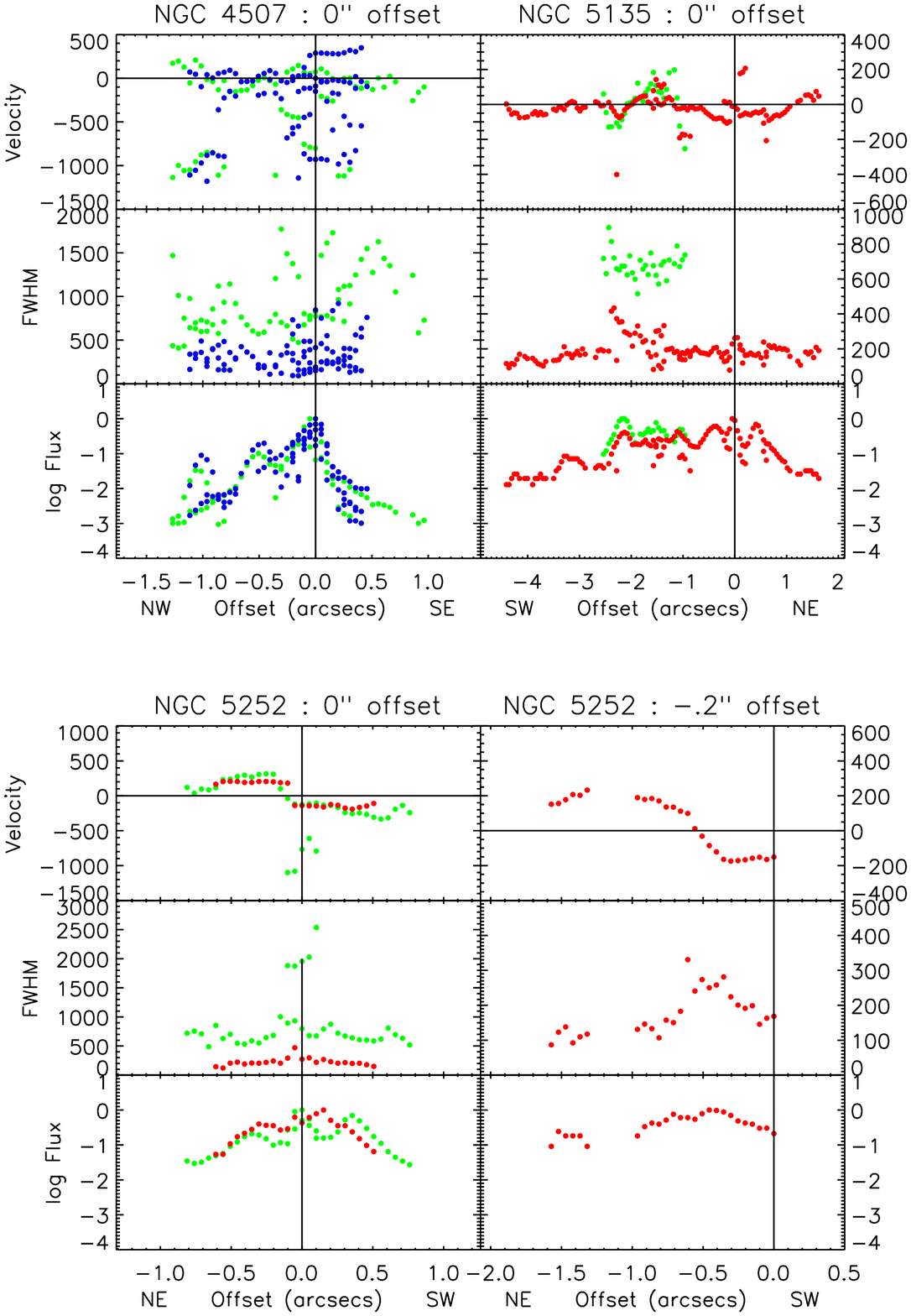}
\end{figure}
\clearpage
%%%%%%%%%%%%%%%%%%%%%%%%%%%%%%%%%%%%%%%%%%%%%%%%%%%%%%%%%%%%%%%

\begin{figure}
\centering
\includegraphics[angle=0,scale=0.7]{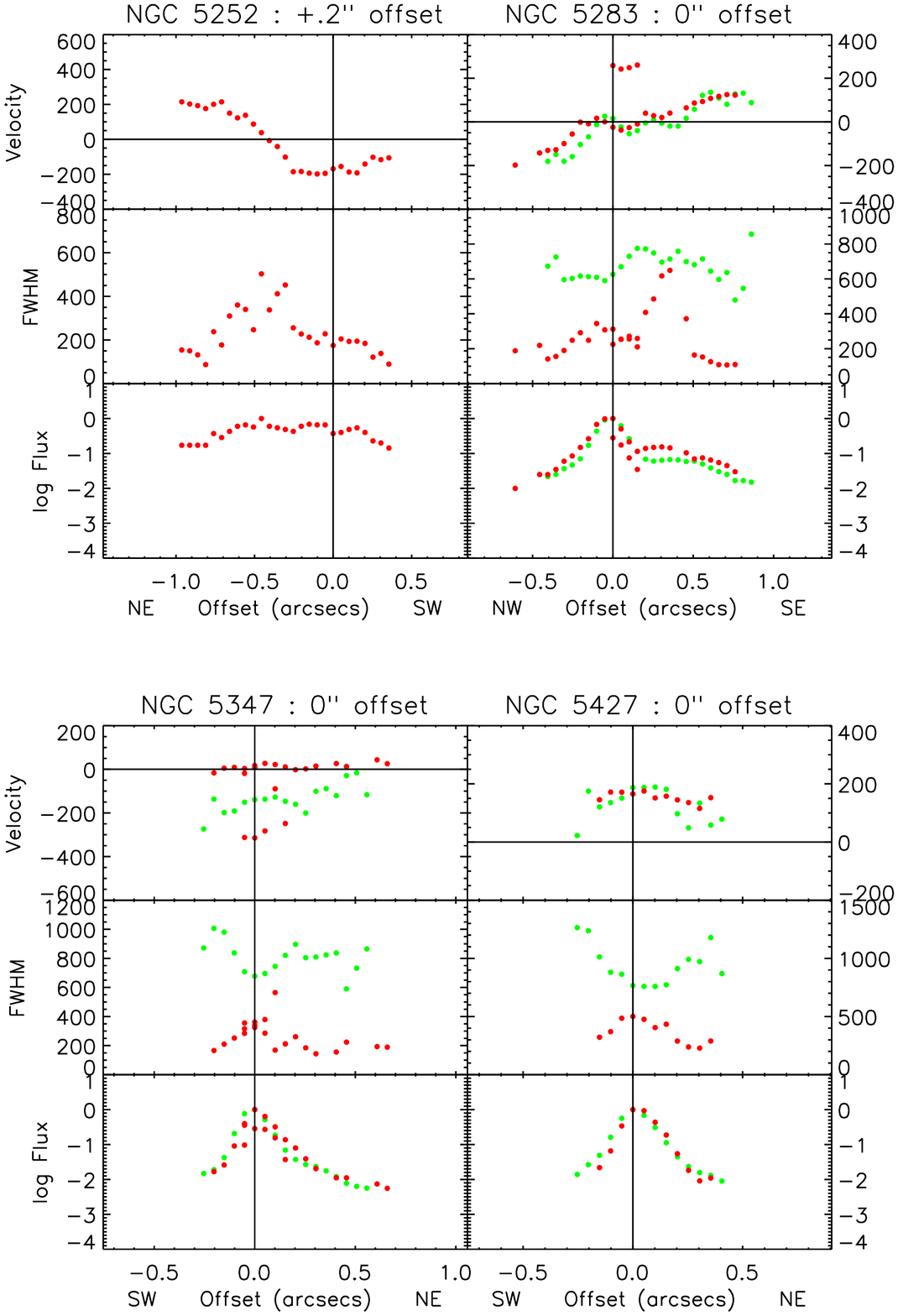}
\end{figure}
\clearpage
%%%%%%%%%%%%%%%%%%%%%%%%%%%%%%%%%%%%%%%%%%%%%%%%%%%%%%%%%%%%%%%

\begin{figure}
\centering
\includegraphics[angle=0,scale=0.7]{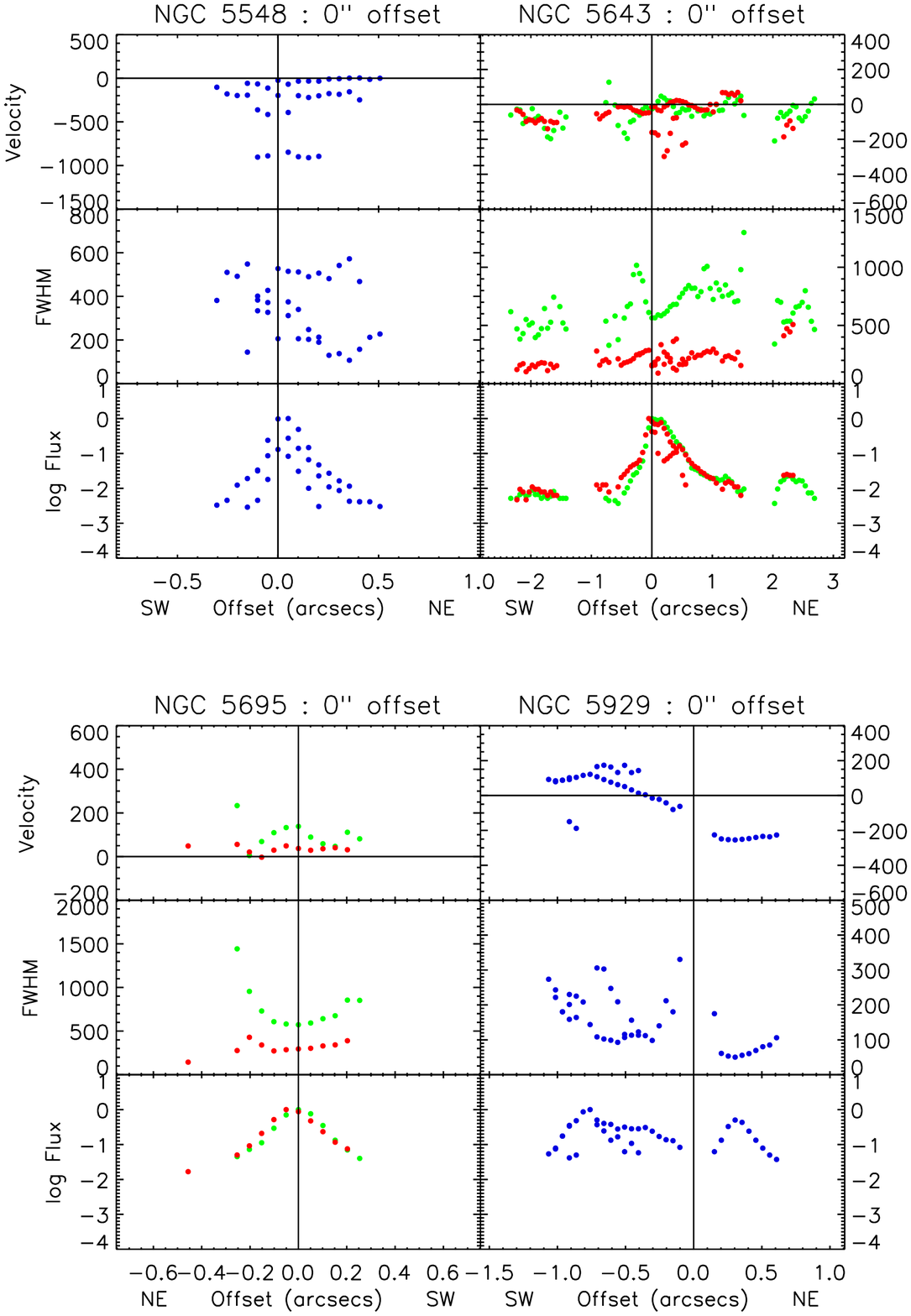}
\end{figure}
\clearpage
%%%%%%%%%%%%%%%%%%%%%%%%%%%%%%%%%%%%%%%%%%%%%%%%%%%%%%%%%%%%%%%

\begin{figure}
\centering
\includegraphics[angle=0,scale=0.7]{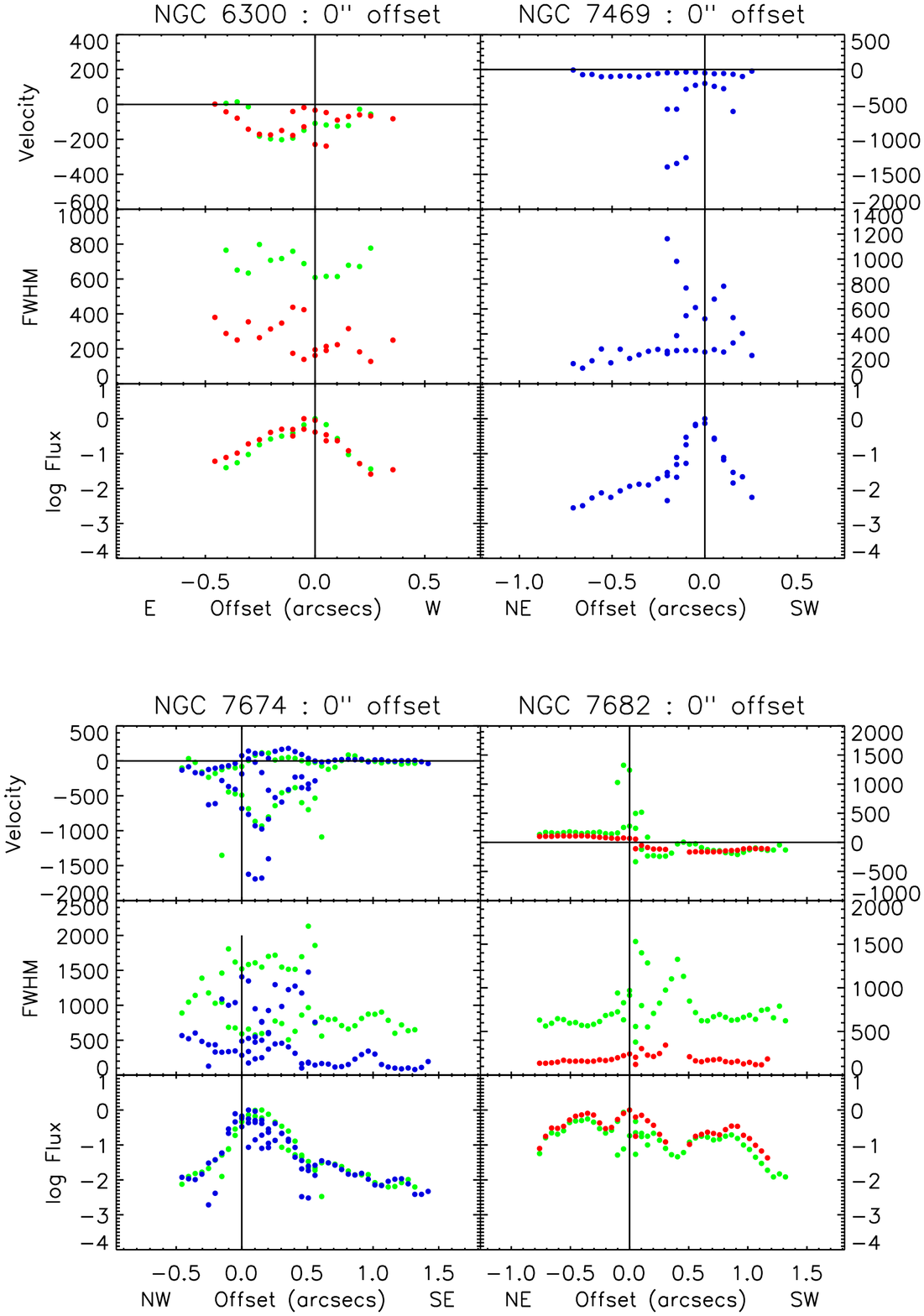}
\end{figure}
\clearpage
%%%%%%%%%%%%%%%%%%%%%%%%%%%%%%%%%%%%%%%%%%%%%%%%%%%%%%%%%%%%%%%

\begin{figure}
\centering
\includegraphics[angle=0,scale=0.7]{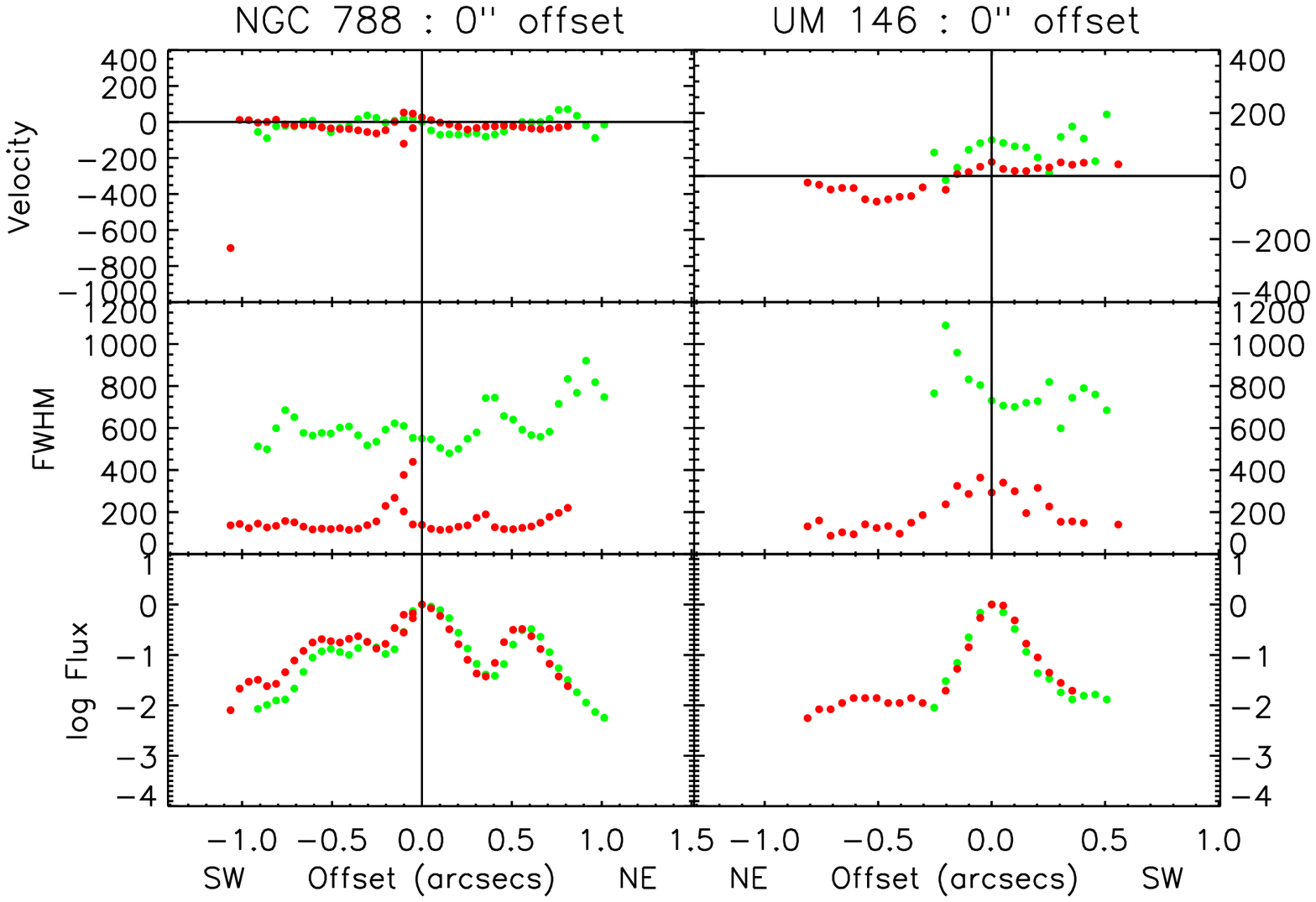}
\end{figure}
\clearpage
%%%%%%%%%%%%%%%%%%%%%%%%%%%%%%%%%%%%%%%%%%%%%%%%%%%%%%%%%%%%%%%

%\begin{figure}
%\centering
%\includegraphics[angle=0,scale=0.7]{ngc788_g430l1_mul_quad.eps}
%\end{figure}

%%%%%%%%%%%%%%%%%%%%%%%%%%%%%%%%%%%%%%%%%%%%%%%%%%%%%%%%%%%%%%%

\clearpage

\subsection{AGN Kinematic Models}

%Models
Kinematics model chosen as the best fit for our radial velocity data 
set for modeled AGN. Parameters used to create this model are given 
in Table \ref{modelvals}.

Kinematics data correspond to {\it HST} gratings as follows:

Green diamonds: [O III] 5007 emission line using G430L grating

Blue circles: [O III] 5007 emission line using G430M grating

Red squares: H$\alpha$ 6564 emission line using G750M grating

%%%%%%%%%%%%%%%%%%%%%%%%%%%%%%%%%%%%%%%%%%%%%%%%%%%%%%%%%%%%%%%

\begin{figure}
\centering
 \includegraphics[angle=0,scale=0.6]{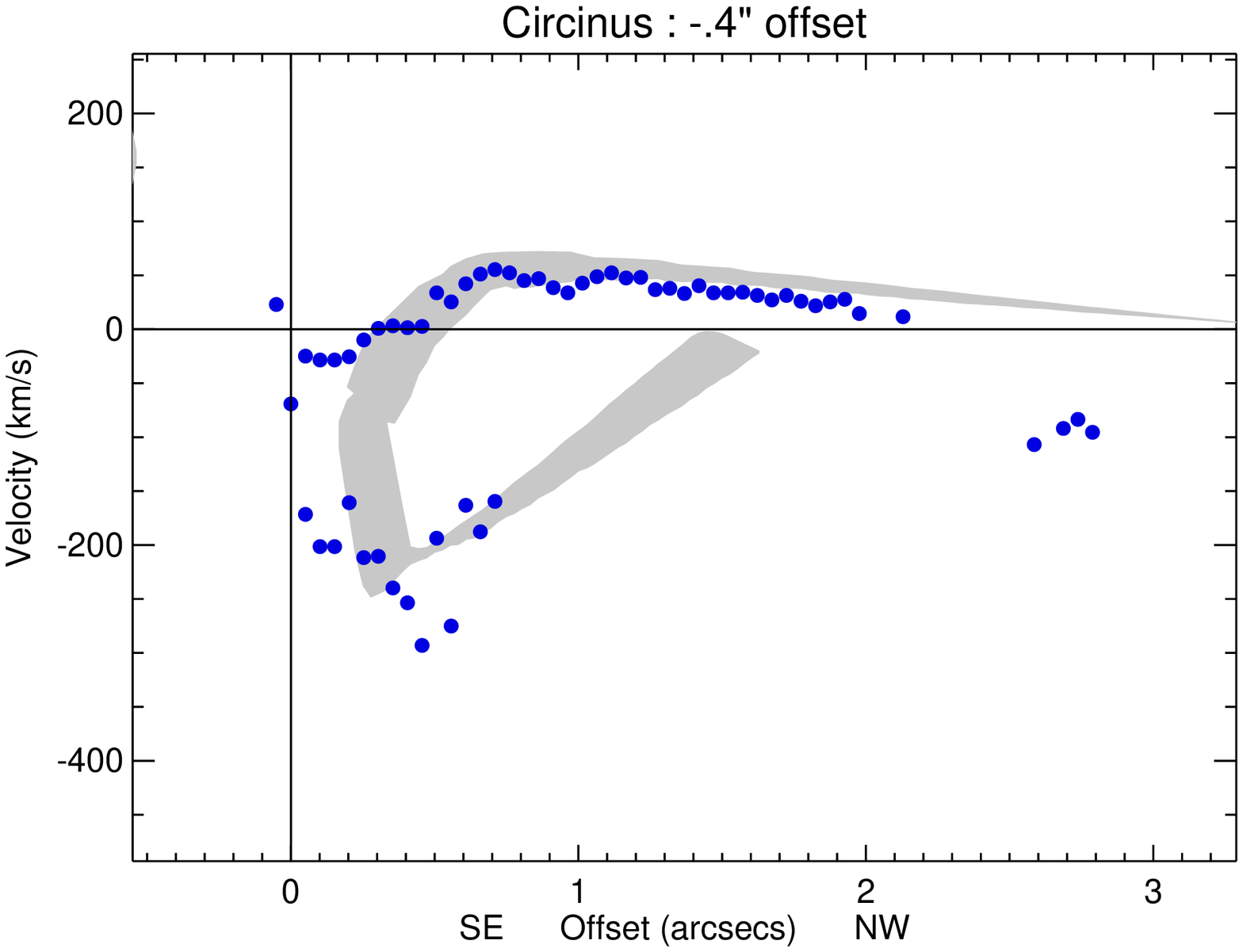}\\
 \includegraphics[angle=0,scale=0.6]{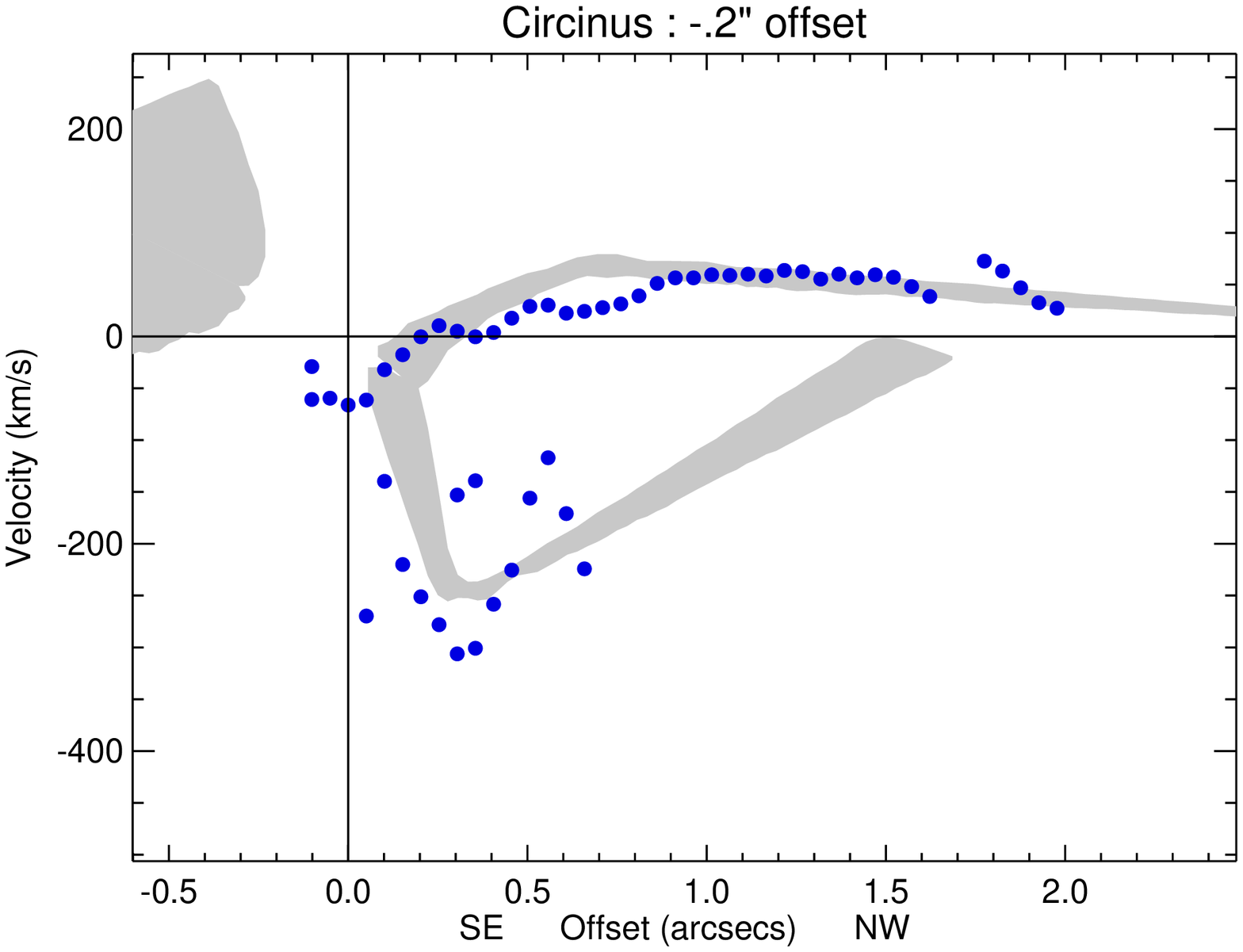}
\label{model1}
\end{figure}
\clearpage
%%%%%%%%%%%%%%%%%%%%%%%%%%%%%%%%%%%%%%%%%%%%%%%%%%%%%%%%%%%%%%%

\begin{figure}
\centering
 \includegraphics[angle=0,scale=0.6]{circinus_g430m4_sha.eps}\\
 \includegraphics[angle=0,scale=0.6]{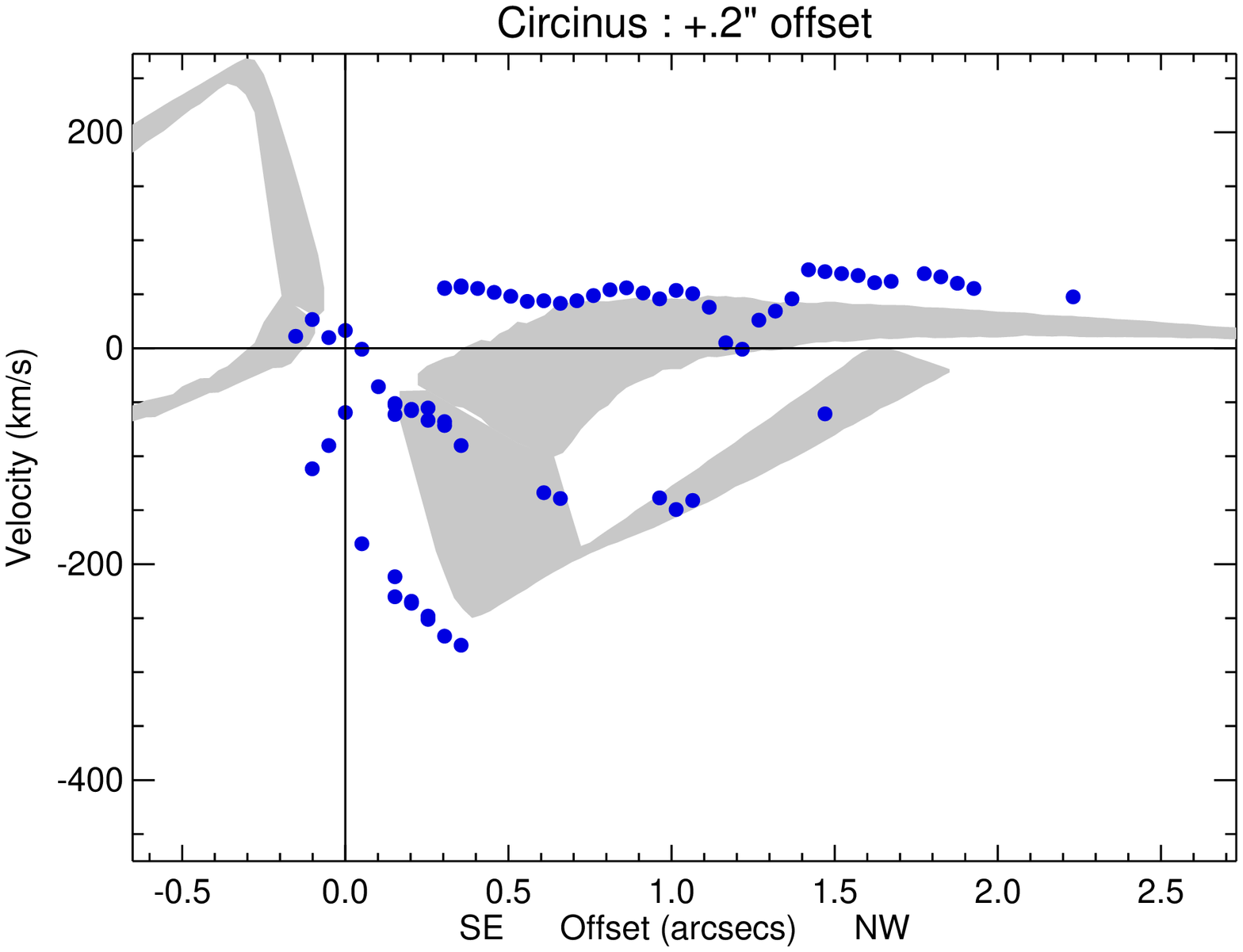}
\label{model2}
\end{figure}
\clearpage
%%%%%%%%%%%%%%%%%%%%%%%%%%%%%%%%%%%%%%%%%%%%%%%%%%%%%%%%%%%%%%%

\begin{figure}
\centering
 \includegraphics[angle=0,scale=0.6]{mrk34_g430l1_sha.eps}\\
 \includegraphics[angle=0,scale=0.6]{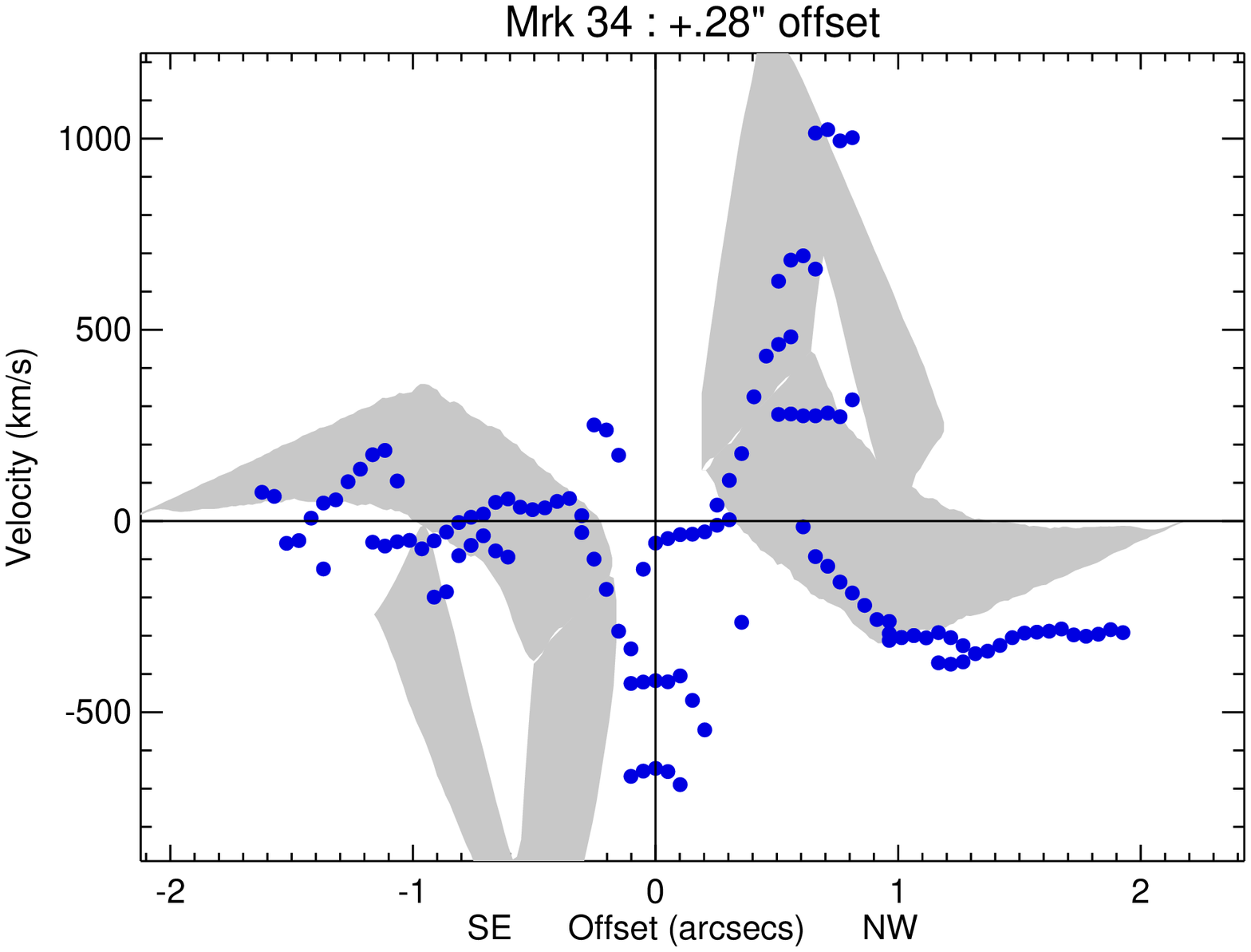}
\label{model3}
\end{figure}
\clearpage
%%%%%%%%%%%%%%%%%%%%%%%%%%%%%%%%%%%%%%%%%%%%%%%%%%%%%%%%%%%%%%%

\begin{figure}
\centering
 \includegraphics[angle=0,scale=0.6]{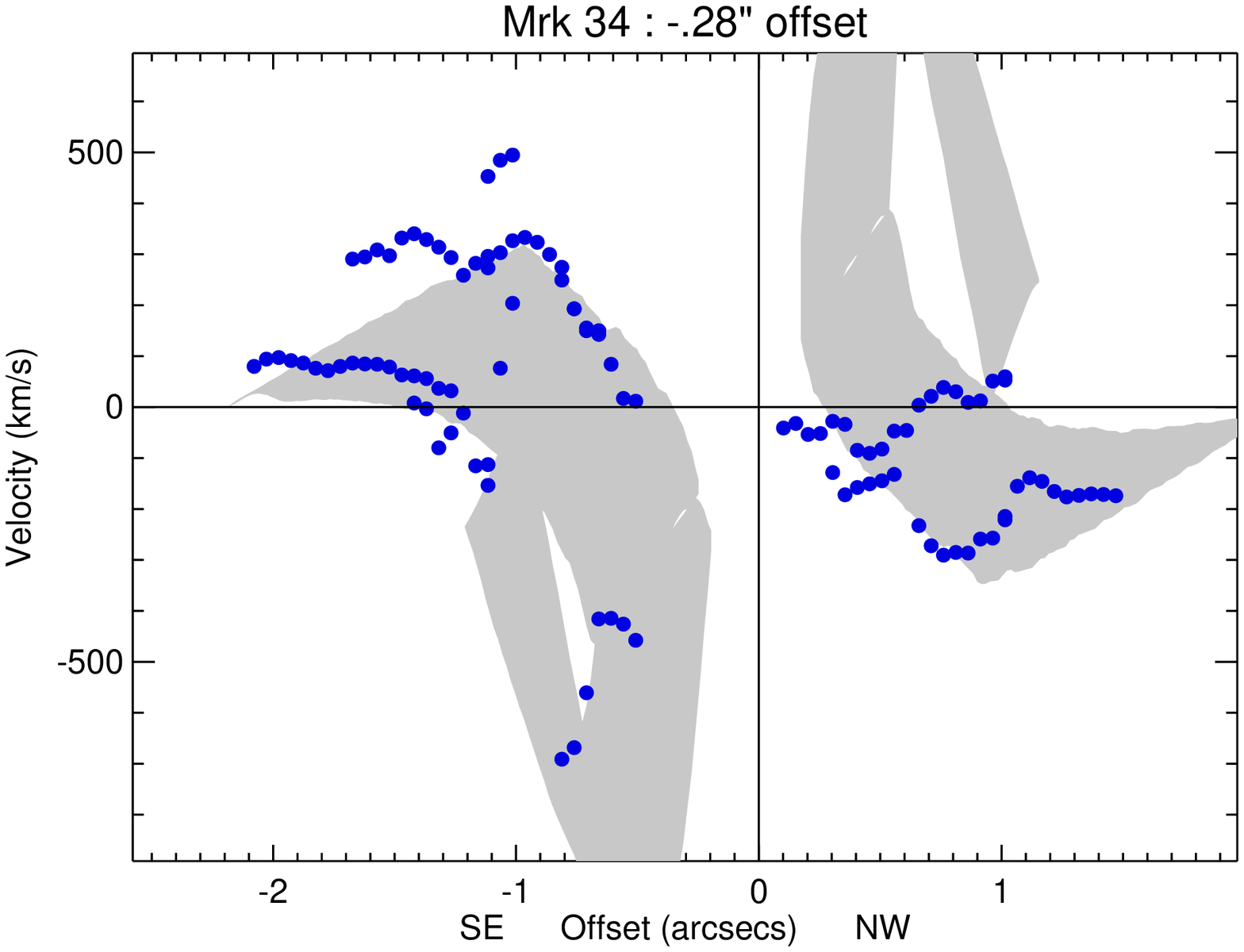}\\
 \includegraphics[angle=0,scale=0.6]{mrk279_g430m1_sha.eps}
\label{model4}
\end{figure}
\clearpage
%%%%%%%%%%%%%%%%%%%%%%%%%%%%%%%%%%%%%%%%%%%%%%%%%%%%%%%%%%%%%%%

\begin{figure}
\centering
 \includegraphics[angle=0,scale=0.6]{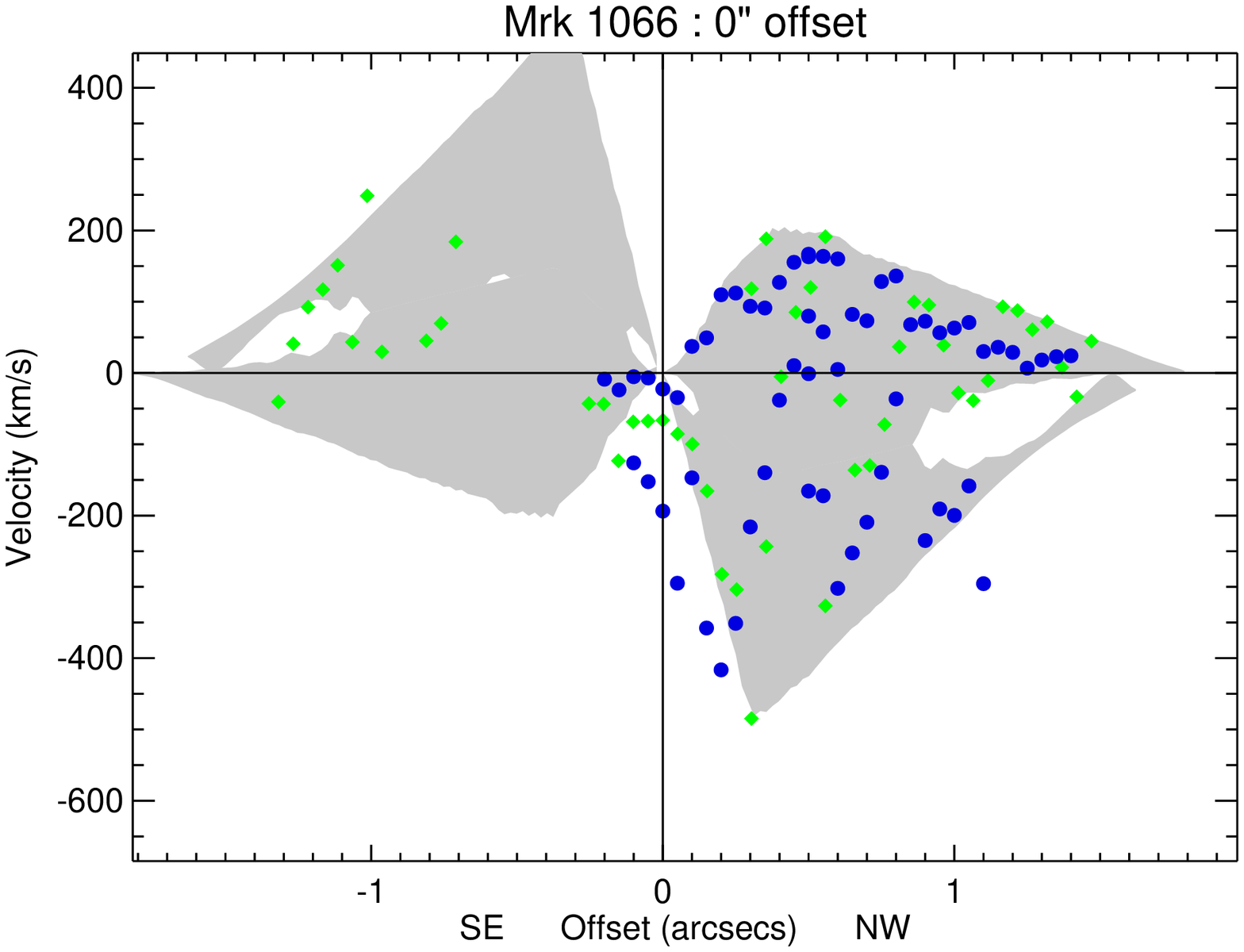}\\
 \includegraphics[angle=0,scale=0.6]{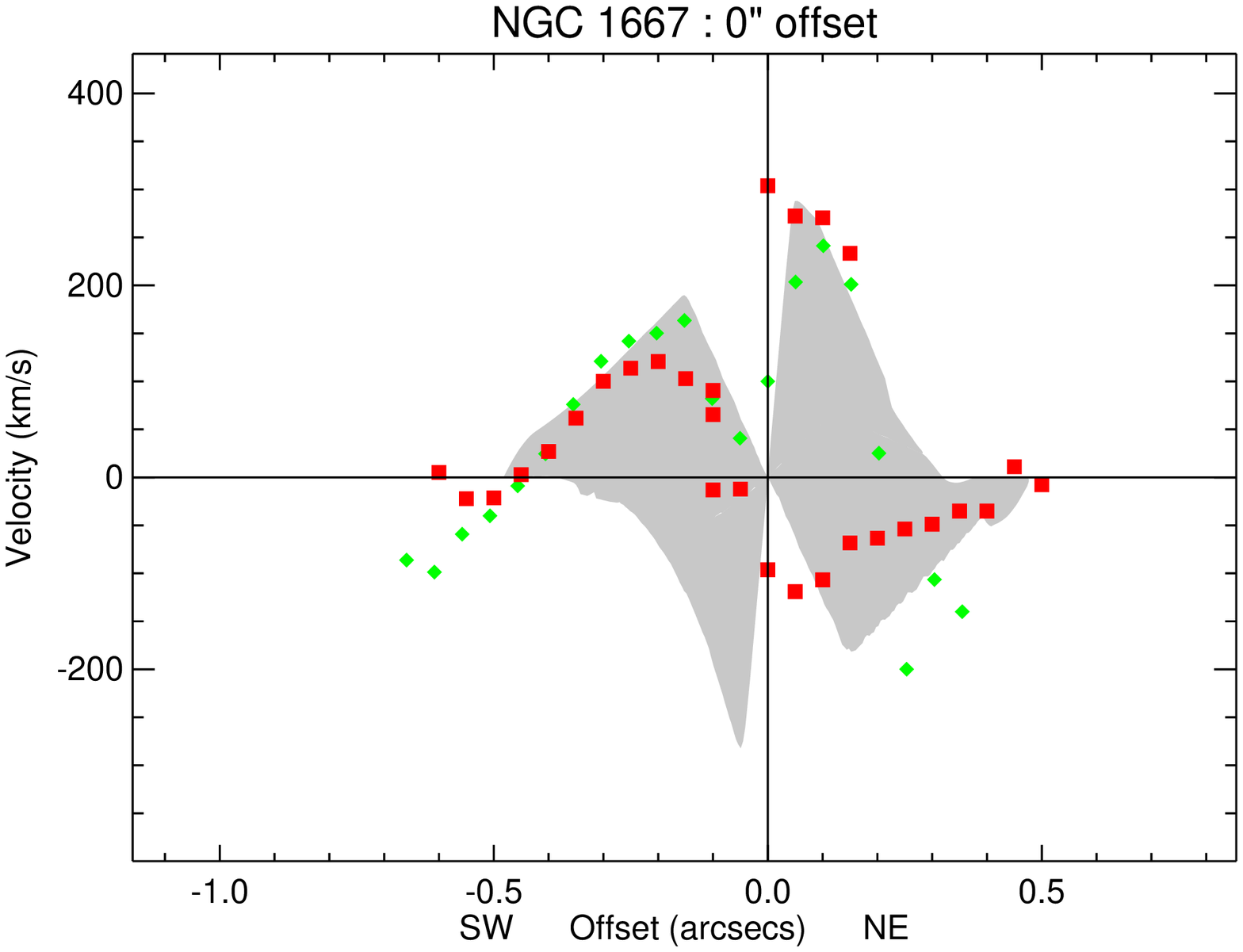}
\label{model5}
\end{figure}
\clearpage
%%%%%%%%%%%%%%%%%%%%%%%%%%%%%%%%%%%%%%%%%%%%%%%%%%%%%%%%%%%%%%%

\begin{figure}
\centering
 \includegraphics[angle=0,scale=0.6]{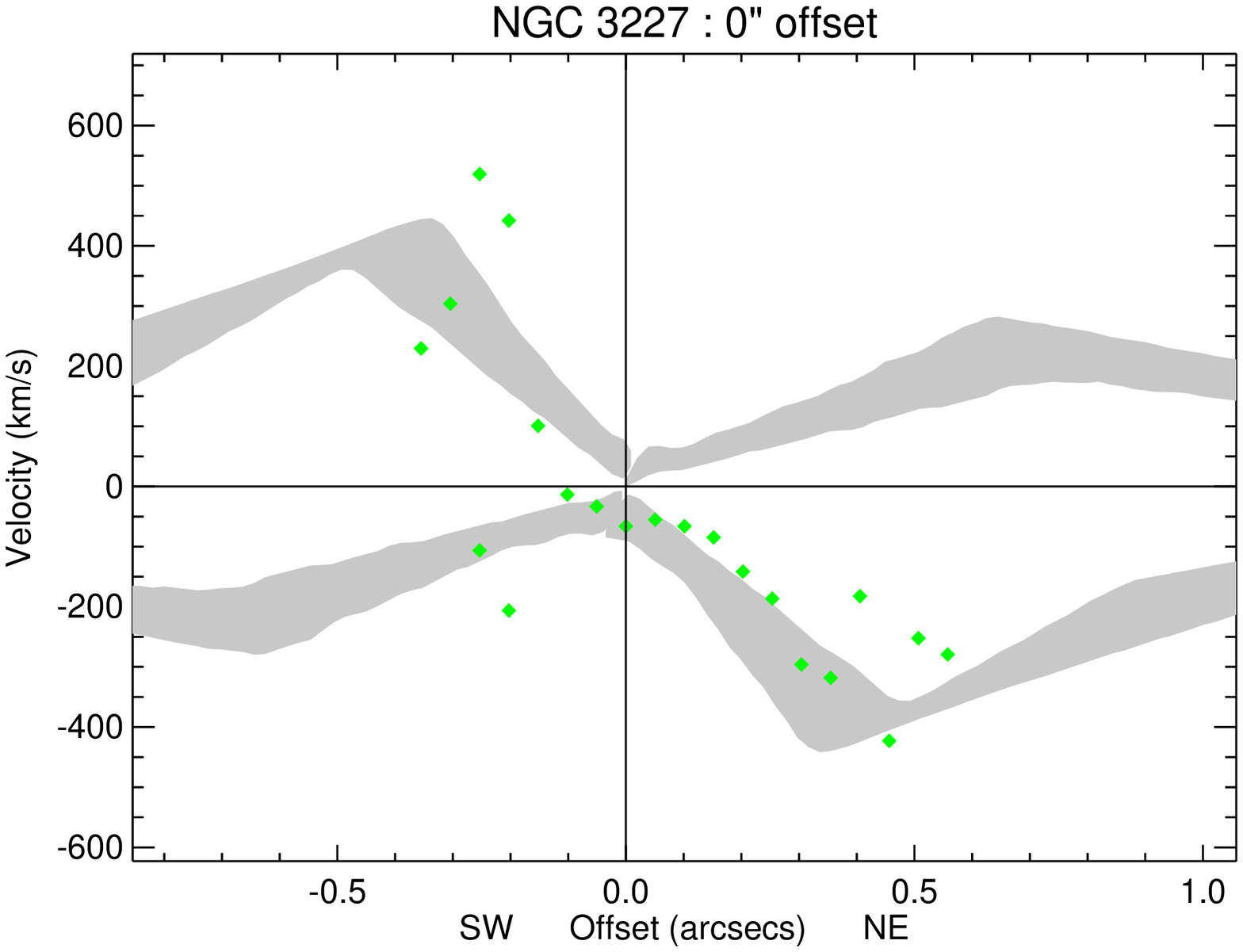}\\
 \includegraphics[angle=0,scale=0.6]{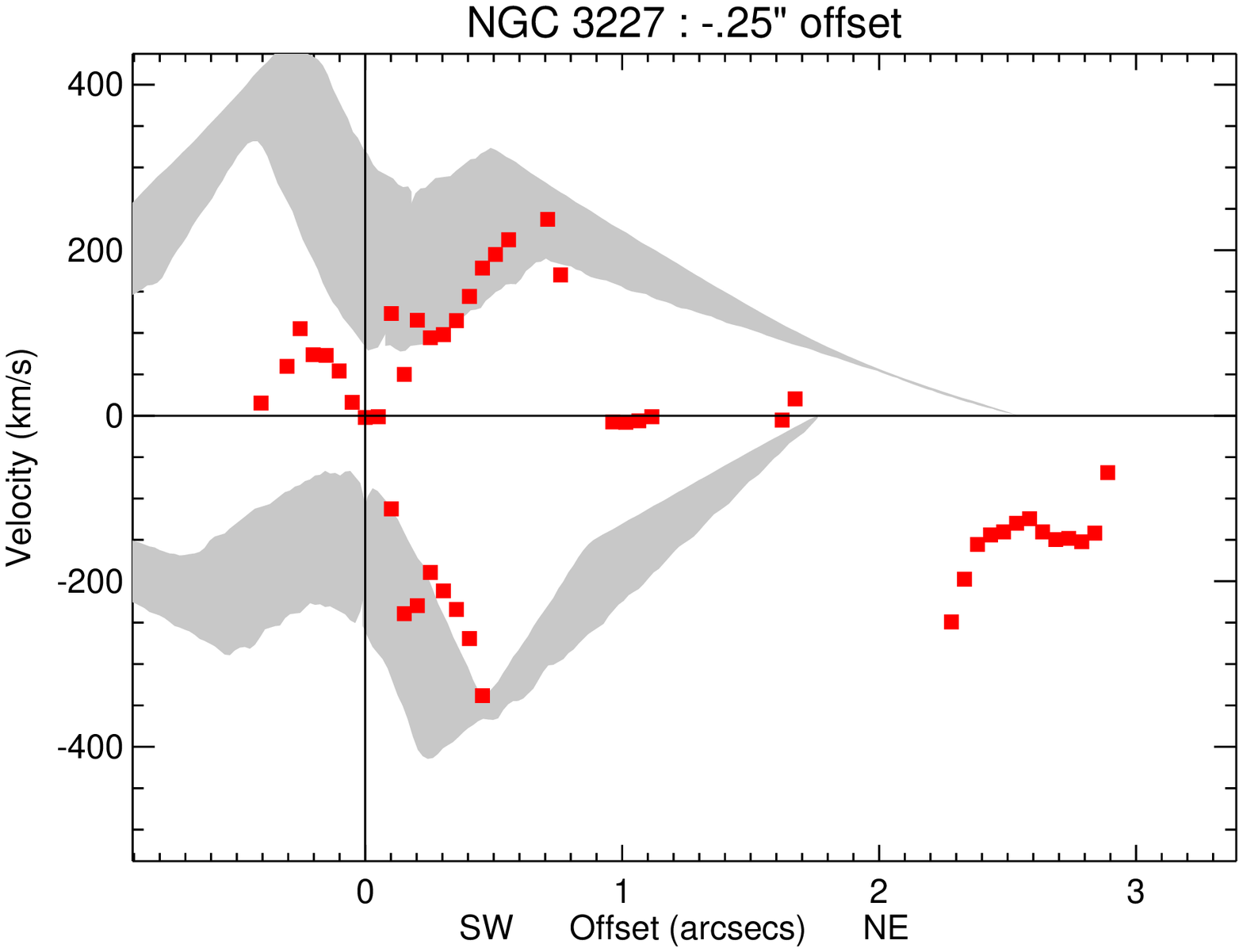}
\label{model6}
\end{figure}
\clearpage
%%%%%%%%%%%%%%%%%%%%%%%%%%%%%%%%%%%%%%%%%%%%%%%%%%%%%%%%%%%%%%%

\begin{figure}
\centering
 \includegraphics[angle=0,scale=0.6]{ngc3227_g750m4_sha.eps}\\
 \includegraphics[angle=0,scale=0.6]{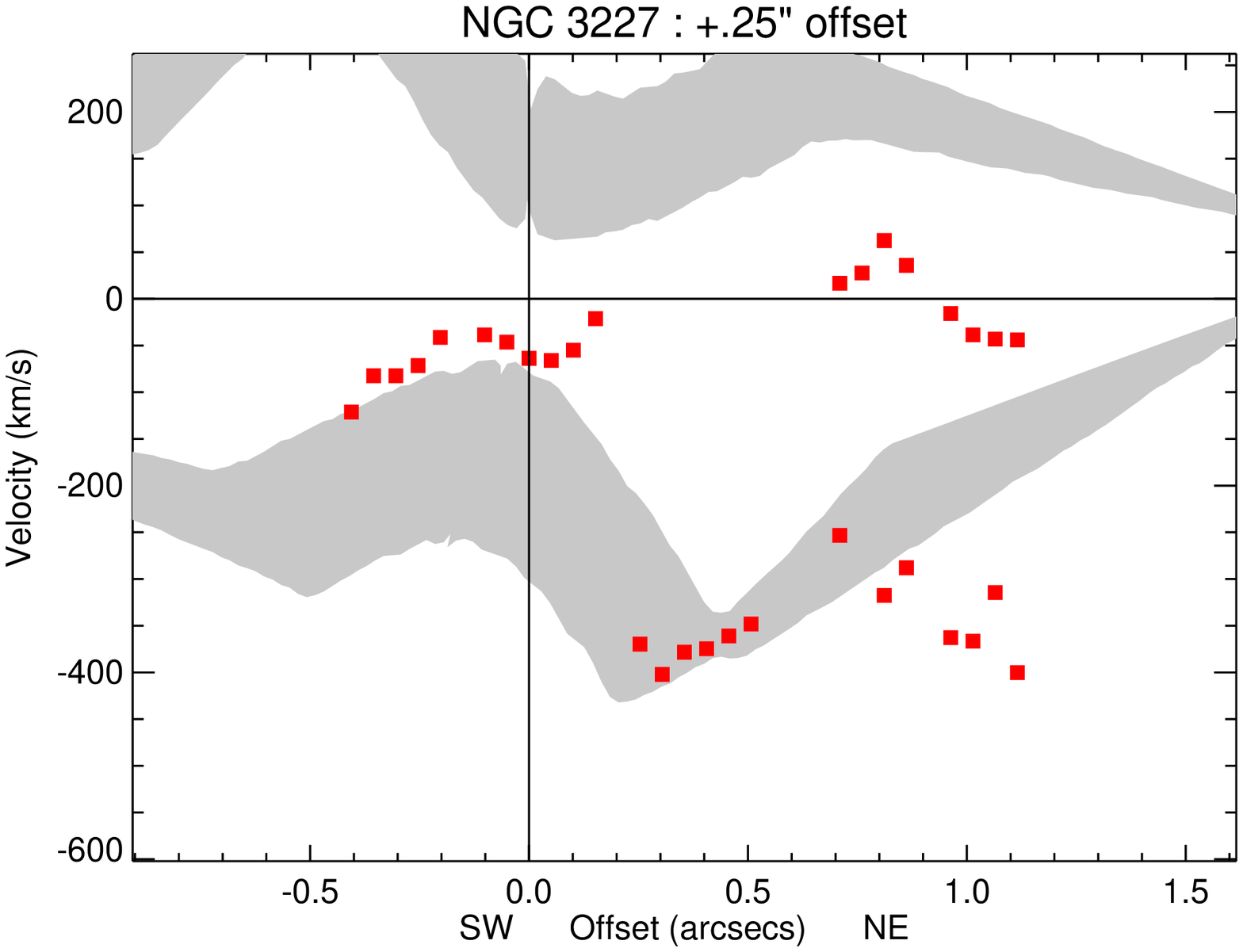}
\label{model7}
\end{figure}
\clearpage
%%%%%%%%%%%%%%%%%%%%%%%%%%%%%%%%%%%%%%%%%%%%%%%%%%%%%%%%%%%%%%%

\begin{figure}
\centering
 \includegraphics[angle=0,scale=0.6]{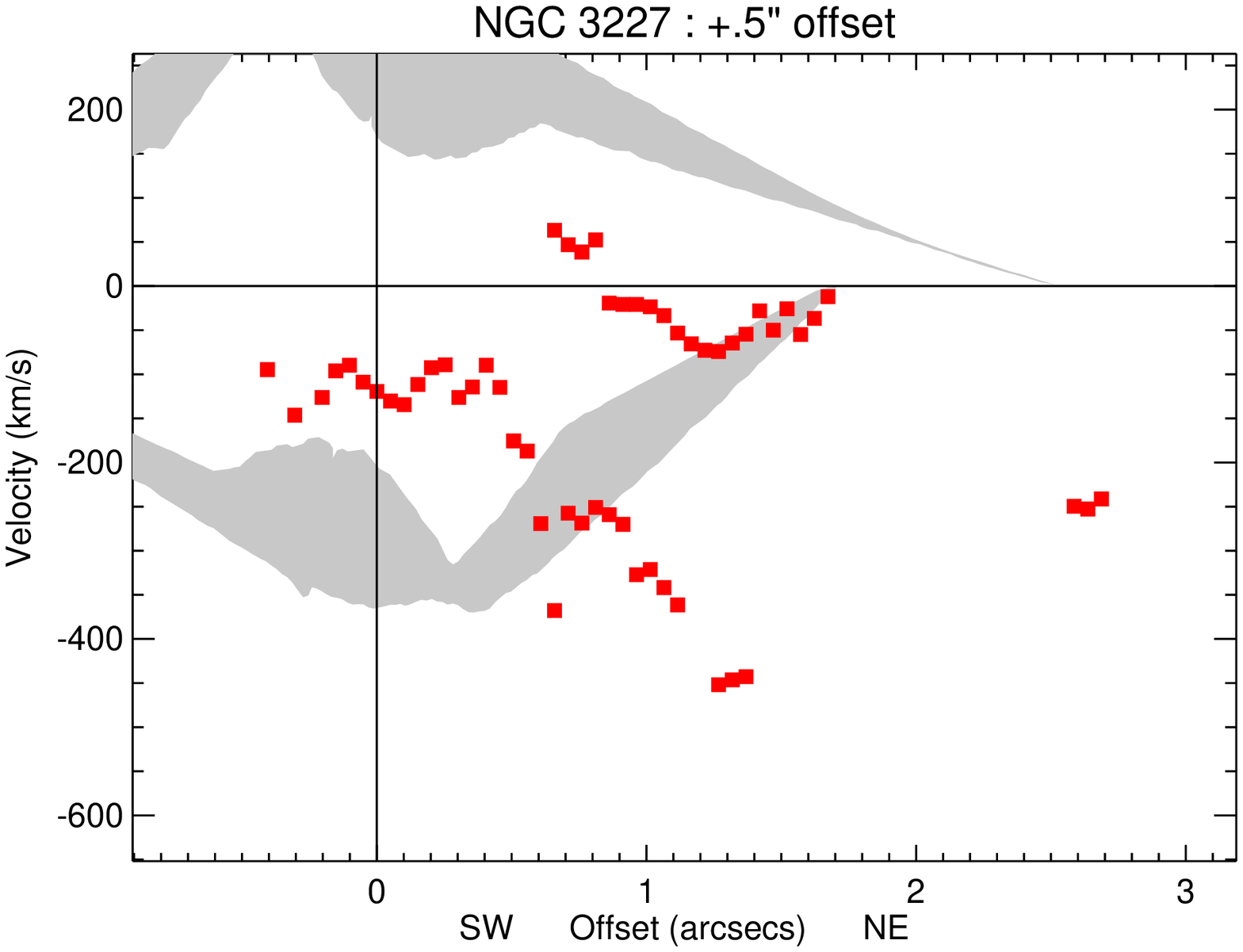}\\
 \includegraphics[angle=0,scale=0.6]{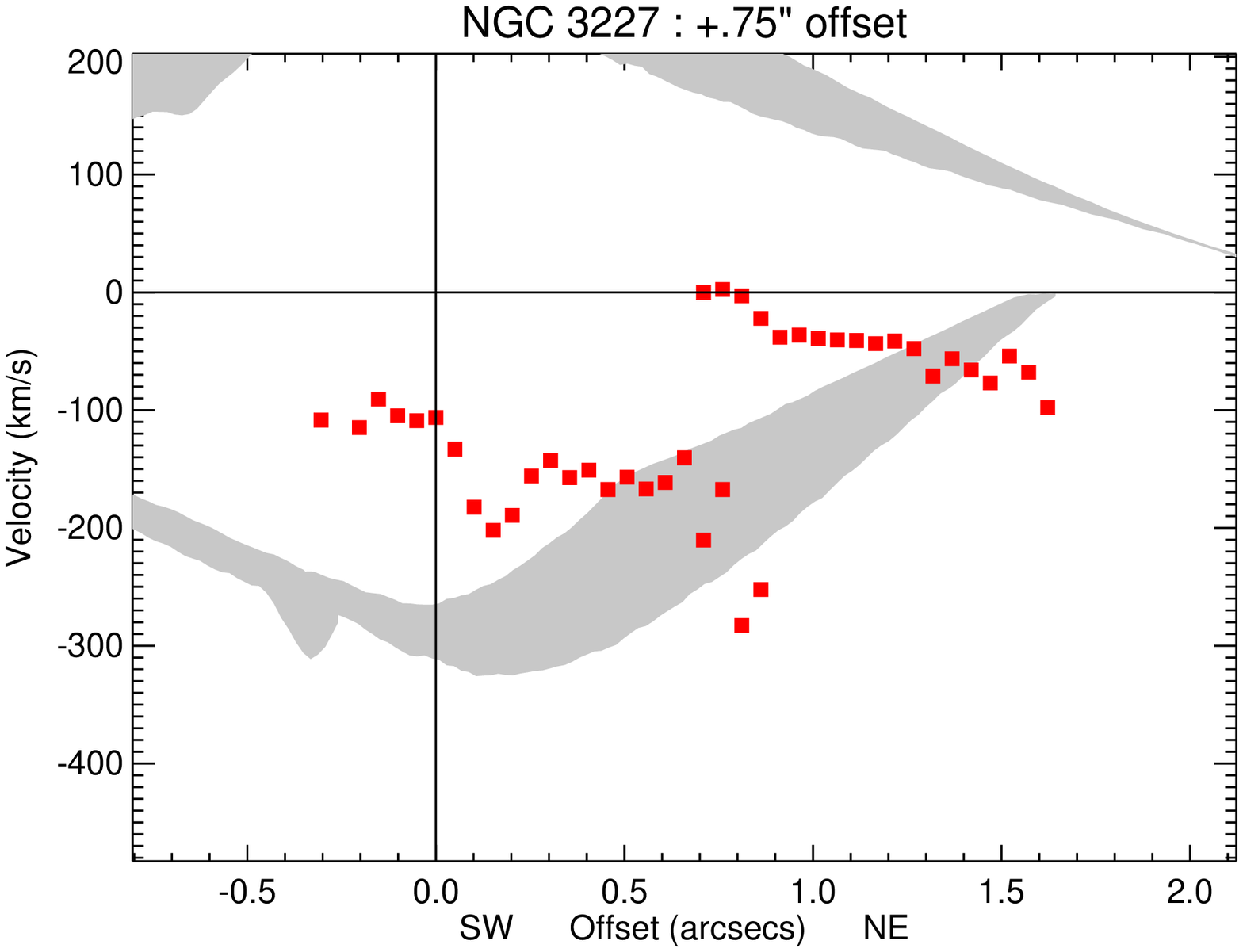}
\label{model8}
\end{figure}
\clearpage
%%%%%%%%%%%%%%%%%%%%%%%%%%%%%%%%%%%%%%%%%%%%%%%%%%%%%%%%%%%%%%%

\begin{figure}
\centering
 \includegraphics[angle=0,scale=0.6]{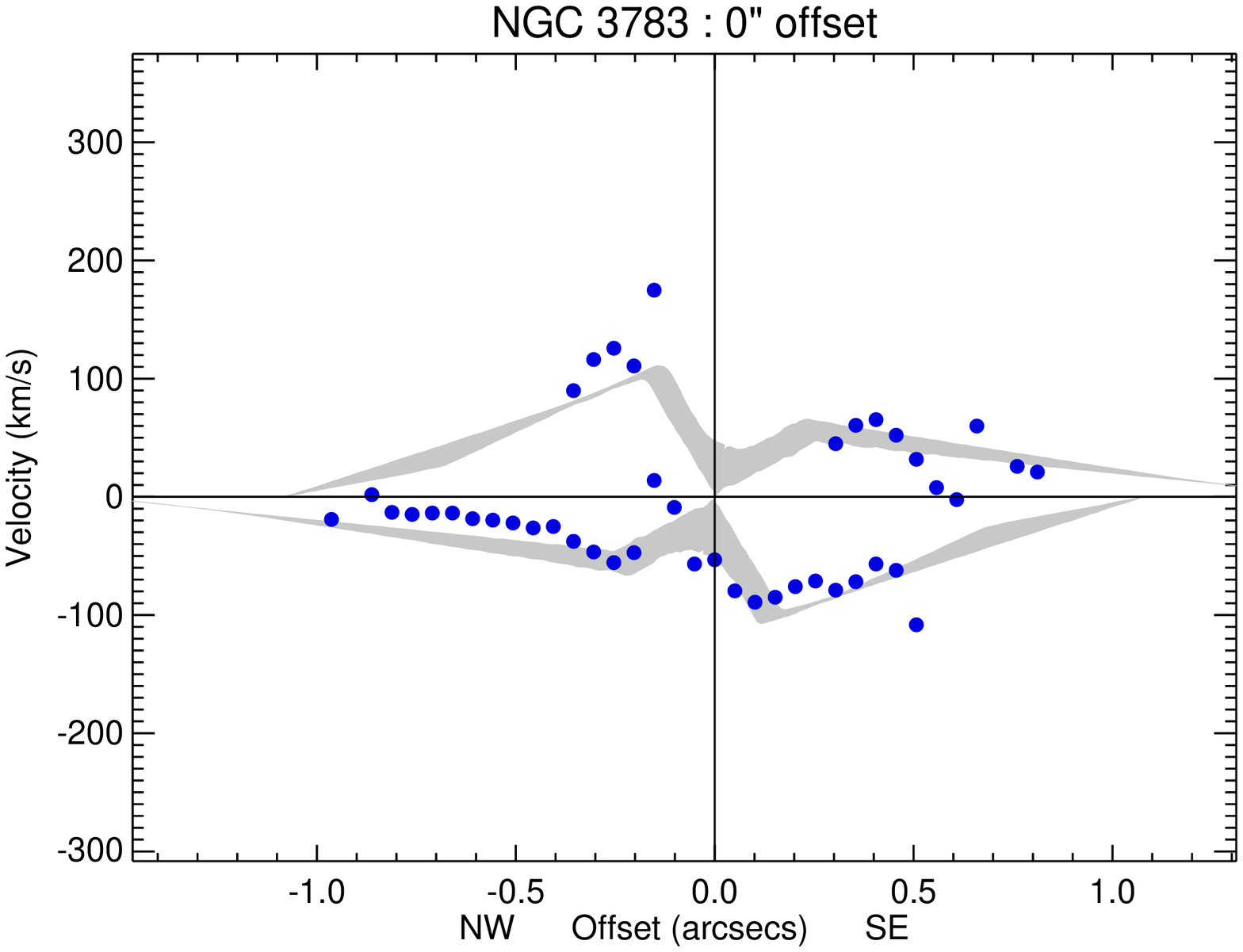}\\
 \includegraphics[angle=0,scale=0.6]{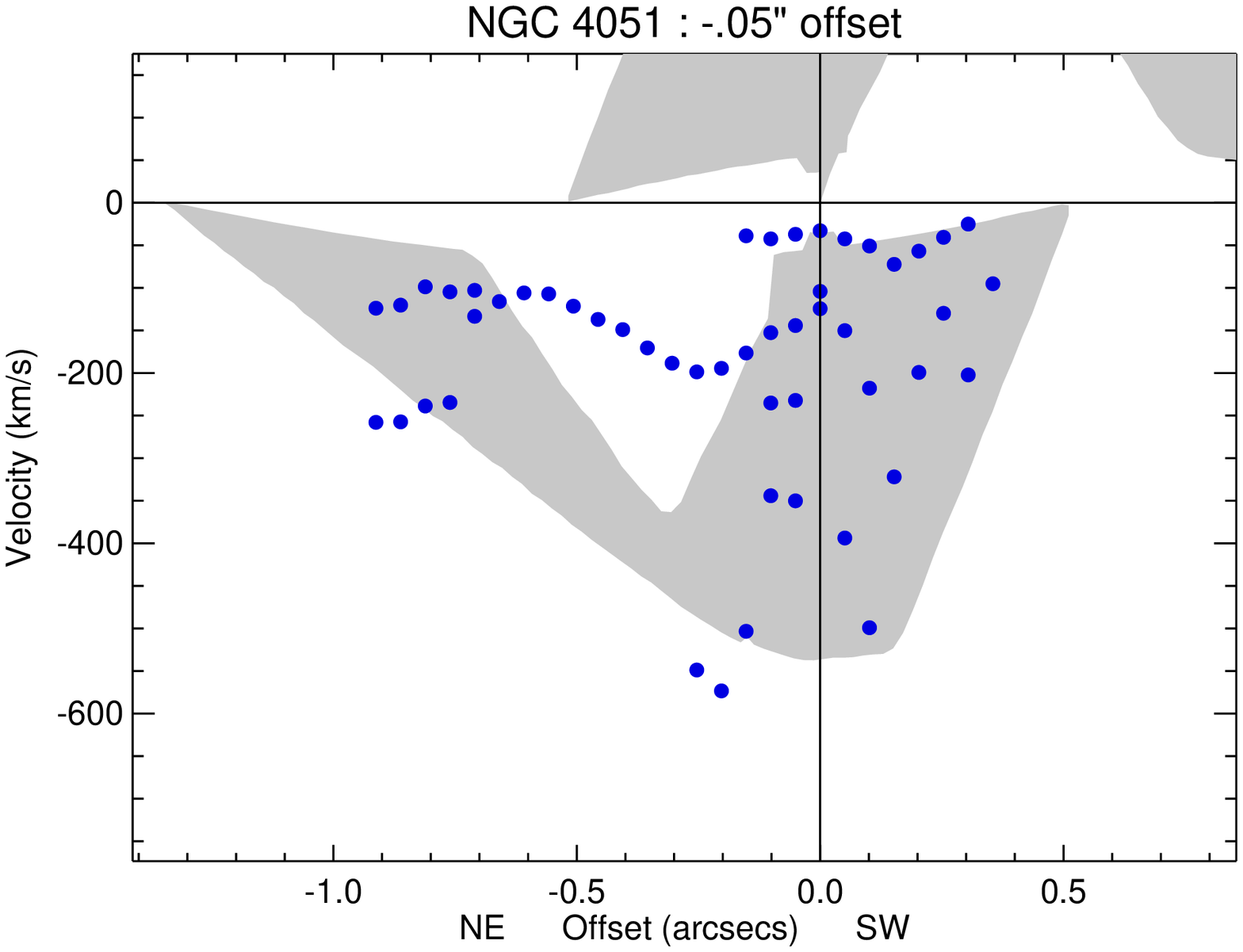}
\label{model9}
\end{figure}
\clearpage
%%%%%%%%%%%%%%%%%%%%%%%%%%%%%%%%%%%%%%%%%%%%%%%%%%%%%%%%%%%%%%%

\begin{figure}
\centering
 \includegraphics[angle=0,scale=0.6]{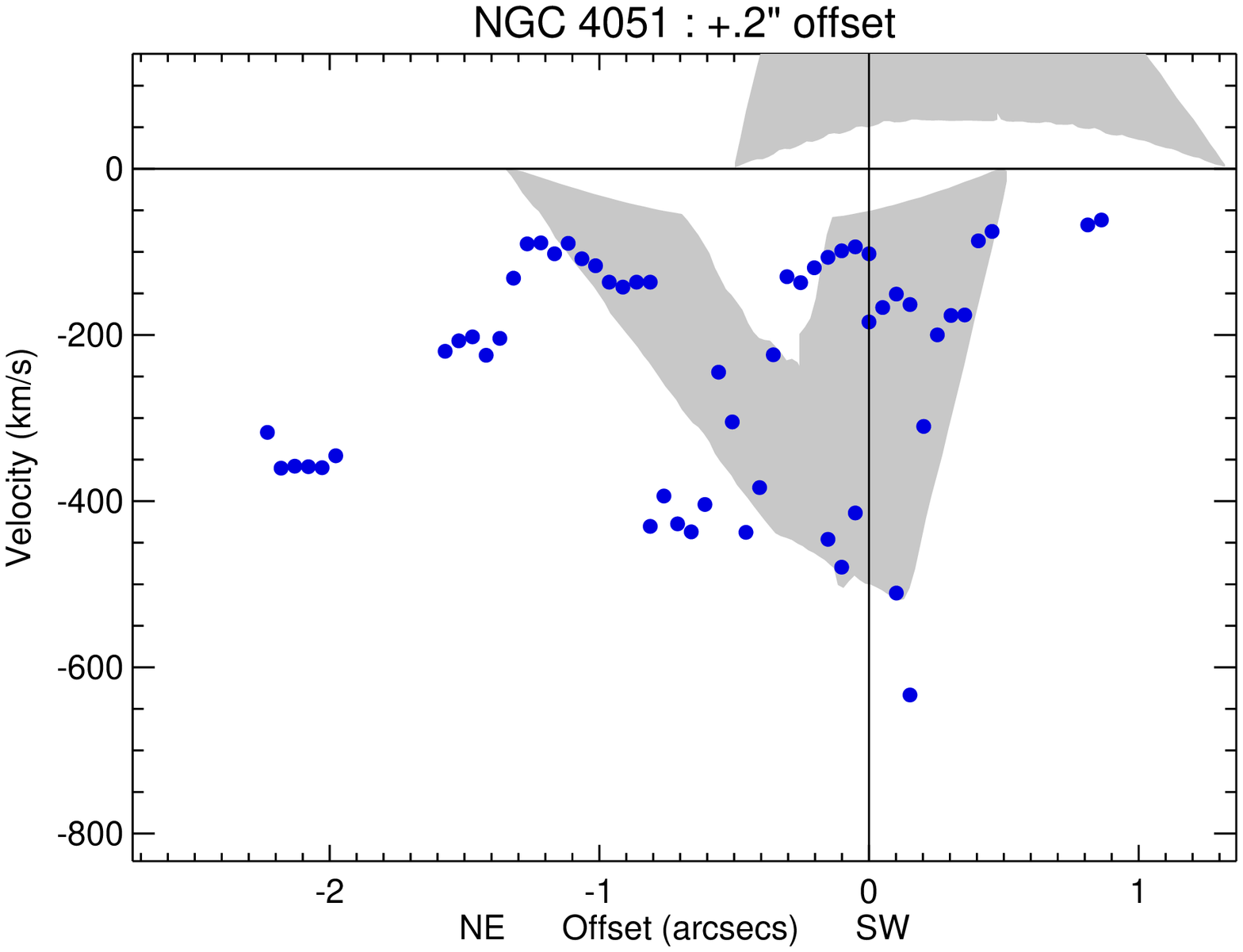}\\
 \includegraphics[angle=0,scale=0.6]{ngc4507_g430l1_sha.eps}
\label{model10}
\end{figure}
\clearpage
%%%%%%%%%%%%%%%%%%%%%%%%%%%%%%%%%%%%%%%%%%%%%%%%%%%%%%%%%%%%%%%

\begin{figure}
\centering
 \includegraphics[angle=0,scale=0.6]{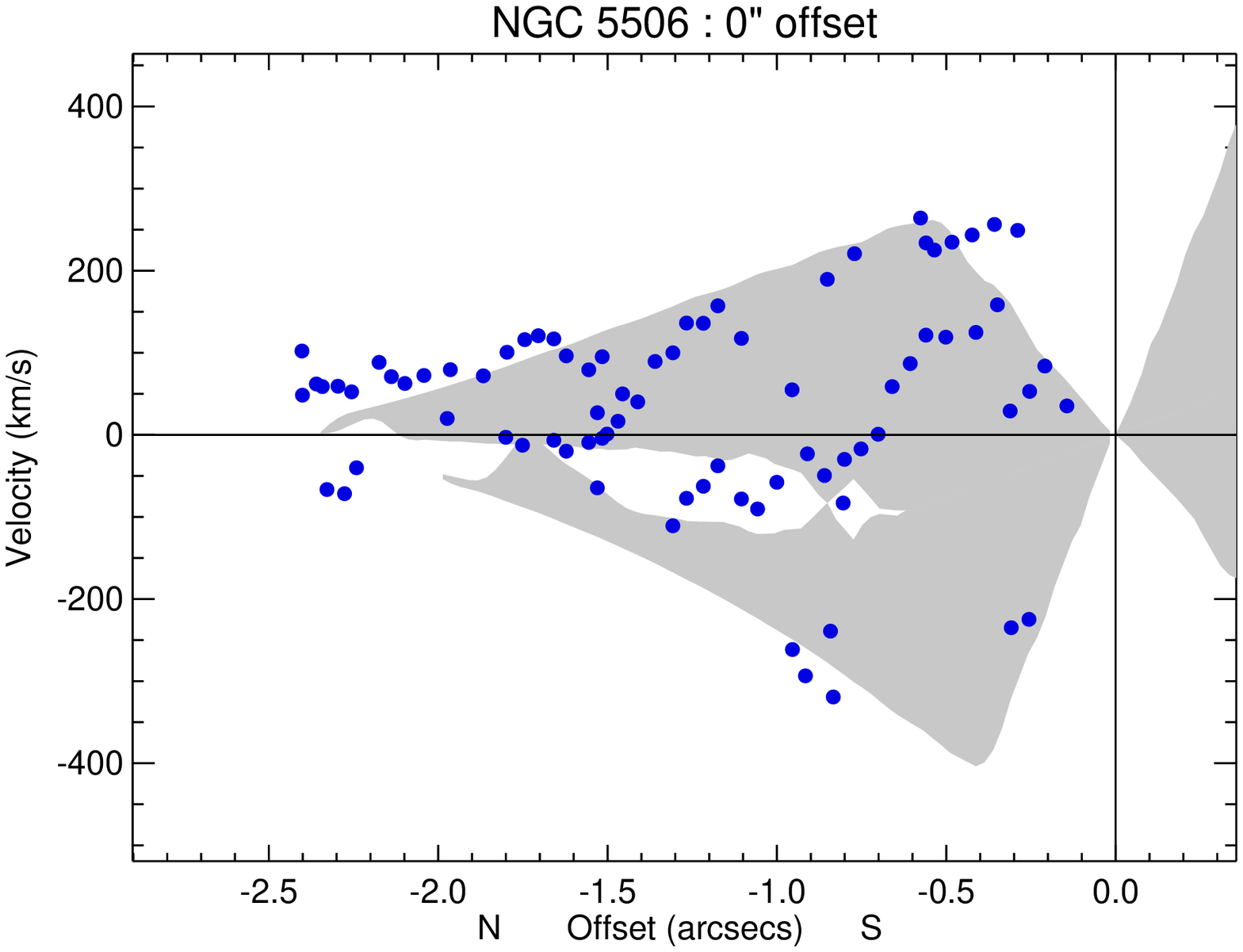}\\
 \includegraphics[angle=0,scale=0.6]{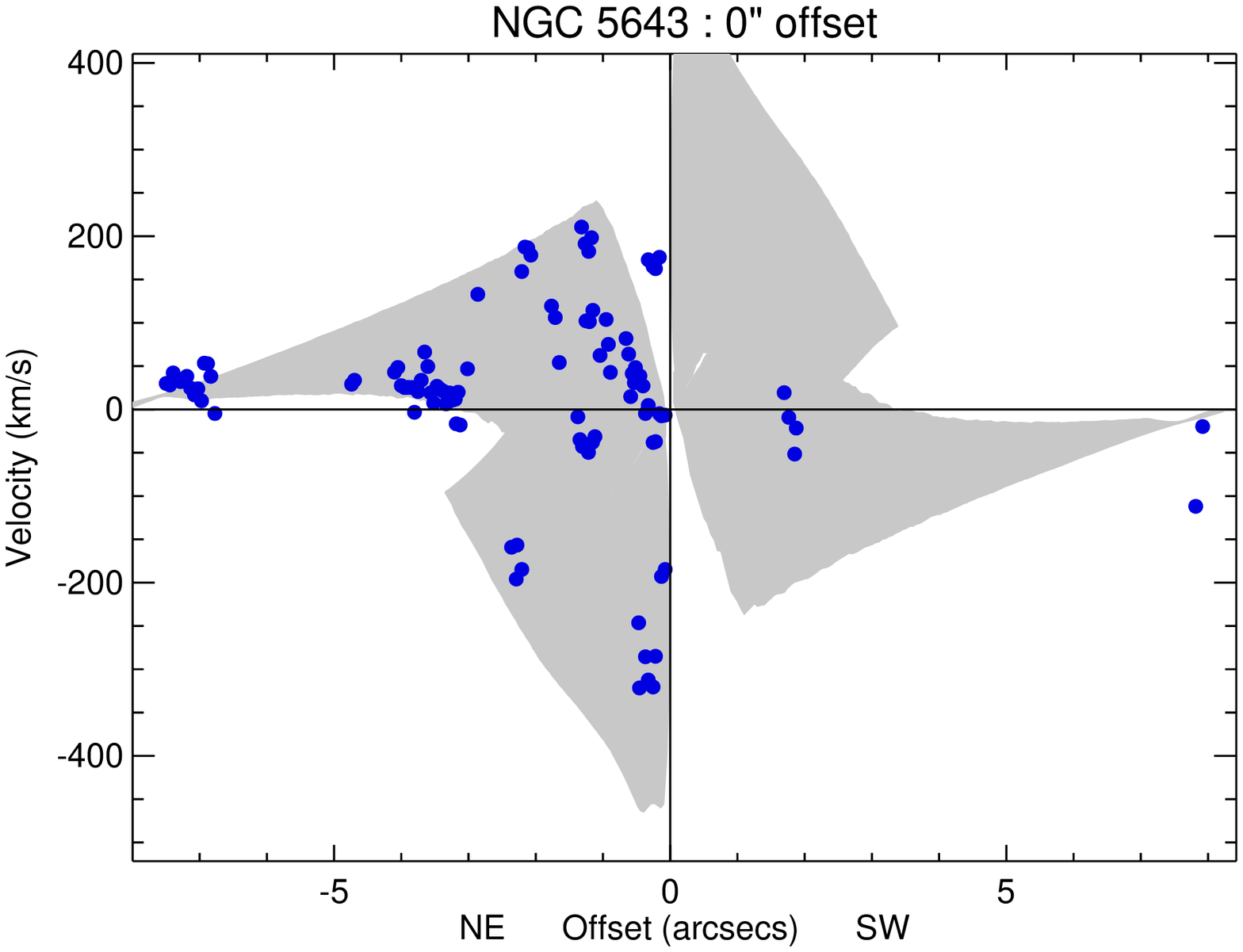}
\label{model11}
\end{figure}
\clearpage
%%%%%%%%%%%%%%%%%%%%%%%%%%%%%%%%%%%%%%%%%%%%%%%%%%%%%%%%%%%%%%%

\begin{figure}
\centering
 \includegraphics[angle=0,scale=0.6]{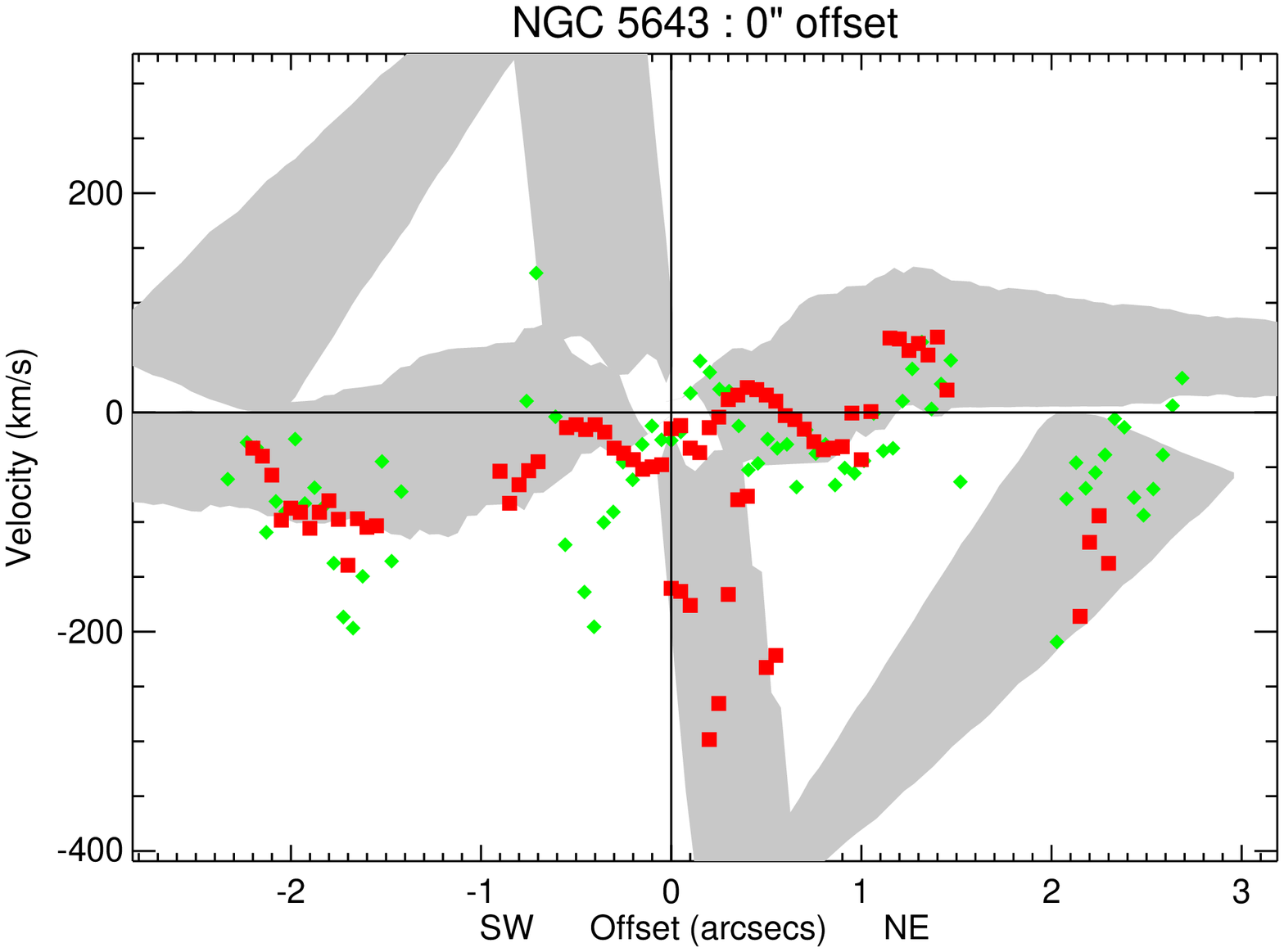}\\
 \includegraphics[angle=0,scale=0.6]{ngc7674_g430l1_sha.eps}
\label{model12}
\end{figure}

%%%%%%%%%%%%%%%%%%%%%%%%%%%%%%%%%%%%%%%%%%%%%%%%%%%%%%%%%%%%%%%

%\appendices
%\include{appendix-slitpos}
%\include{appendix-specimg}
%\include{appendix-kin}
%\endappendices

\end{document}